\documentclass[useAMS,usenatbib]{mn2e}

  \usepackage{graphicx}
   \usepackage{lscape}
 \usepackage{rotating}
  \usepackage{array}
  \usepackage{flafter}

\title{The Bulge-Disk Decomposed Evolution of Massive Galaxies at $1<z<3$ in CANDELS }
\author[V. A. Bruce]
{V.A. Bruce$^{1}$\thanks{E-mail: vab@roe.ac.uk}, J.S. Dunlop$^{1}$, R.J. McLure$^{1}$, M. Cirasuolo$^{1,2}$, 
F. Buitrago$^{1}$,
\newauthor
R.A.A. Bowler$^{1}$, T.A. Targett$^{3}$, E.F. Bell$^{4}$, D.H. McIntosh$^{5}$, A. Dekel$^{6}$, 
\newauthor
S.M. Faber$^{7}$, H.C. Ferguson$^{8}$, N.A. Grogin$^{8}$, W. Hartley$^{9}$,  D.D. Kocevski$^{10}$,  
\newauthor
 A.M. Koekemoer$^{8}$, D.C. Koo$^{7}$, E.J. McGrath$^{11}$ \\
$^1$SUPA\thanks{Scottish Universities Physics Alliance} Institute for Astronomy, University of Edinburgh, Royal Observatory, Edinburgh EH9 3HJ\\
$^2$UK Astronomy Technology Centre, Science and Technology Facilities Council, Royal Observatory, Edinburgh EH9 3HJ\\  
$^3$Department of Physics and Astronomy, Sonoma State University,1801 East Cotati Avenue, Rohnert Park, CA 94928-3609, USA\\
$^4$Department of Astronomy, University of Michigan, Ann Arbor, MI 48109, USA\\
$^5$Department of Physics \& Astronomy, University of Missouri-Kansas City, 5110 Rockhill Road, Kansas City, MO 64110, USA\\
$^6$Racah Institute of Physics, The Hebrew University, Jerusalem 91904, Israel\\
$^7$UCO/Lick Observatory, Department of Astronomy and Astrophysics, University of California, Santa Cruz, CA 95064, USA\\
$^8$Space Telescope Science Institute, 3700 San Martin Drive, Baltimore, MD 21218, USA\\
$^9$ETH Z{\"u}rich, Institut f{\"u}r Astronomie, HIT J 11.3, Wolfgang-Pauli-Str. 27, 8093 Z{\"u}rich\\
$^{10}$Department of Physics and Astronomy, University of Kentucky, Lexington, KY 40506, USA\\
$^{11}$Department of Physics and Astronomy, Colby College, Waterville,ME 0490, USA }

\begin{document}

\date{}

\pagerange{\pageref{firstpage}--\pageref{lastpage}} \pubyear{2013}

\pagestyle{myheadings}
\markboth{V. A. Bruce et al.} {The Morphological Evolution of Massive Galaxies at $1 < z < 3$}

\maketitle

\label{firstpage}

\vspace{-5cm}

\begin{abstract}

We present the results of a new and improved study of the morphological and spectral evolution of massive galaxies over the redshift range $1 < z < 3$. Our analysis is based on a bulge-disk decomposition of 396 galaxies with $M_* > 10^{11}\,{\rm M_{\odot}}$
uncovered from the CANDELS {\it WFC3/IR} imaging within the COSMOS and UKIDSS UDS survey fields. We find that, by modelling the $H_{160}$ image of each galaxy with a combination of a de Vaucouleurs bulge (Sersic index $n=4$)
and an exponential disk ($n = 1$), we can then lock all derived morphological parameters for the bulge and disk components, and successfully reproduce the shorter-wavelength $J_{125}$, $i_{814}$, $v_{606}$ {\it HST} images simply
by floating the magnitudes of the two components. This then yields sub-divided 4-band {\it HST} photometry for the bulge and disk components which, with no additional priors, is well described by spectrophotometric models of galaxy evolution. Armed with this information we are able to properly determine the masses and star-formation rates for the bulge and disk components, and find that: i) from $z = 3$ to $z = 1$ the galaxies move from disk-dominated to increasingly bulge-dominated, but very few galaxies are pure bulges/ellipticals by $z = 1$; ii) while most passive galaxies are bulge-dominated, and most star-forming
galaxies disk-dominated, $18 \pm 5$\% of passive galaxies are disk-dominated, and $11 \pm 3$\% of star-forming galaxies are bulge-dominated, a result which needs to be explained by any model
purporting to connect star-formation quenching with morphological transformations; iii) there exists a small but significant population
of pure passive disks, which are generally flatter than their star-forming counterparts (whose axial ratio distribution peaks at $b/a \simeq 0.7$); iv) flatter/larger disks re-emerge at the highest star-formation
rates, consistent with recent studies of sub-mm galaxies, and with the concept of a maximum surface-density for star-formation activity.
\end{abstract}

\begin{keywords} galaxies: evolution - galaxies: structure - galaxies: spiral -  galaxies: elliptical and lenticular - cD,  galaxies: high-redshift
\end{keywords}

\section{Introduction}
Although detailed morphological studies of galaxies have been conducted since the 1920s, the underlying physical processes responsible for the formation and subsequent evolution of galaxy morphologies remain to be fully understood. In particular, it is still unclear how, or even if, transformations in morphology are linked to star-formation history. 

Locally, galaxy morphologies are complex, with high-resolution studies finding that disk-dominated galaxies often possess additional structure such as bars, and central cores with exponential, disk-like, profiles ( classified as pseudo-bulges \citealt{Kormendy2004}) which are proposed to originate through secular processes. Despite this, the recent local population morphological studies conducted using SDSS data, have revealed that the overall galaxy populations can still be broadly represented by two classes: bulge-dominated (e.g. elliptical) and disk-dominated (e.g. spiral) systems, with a strong colour bimodality comprising a prominent red-sequence of bulge-dominated galaxies and a blue-cloud of  disk-dominated galaxies \citep{Baldry2004,Driver2006, Drory2007}. However, in more recent years, these studies have also led to the the discovery of a more puzzling population of blue bulges and red disks \citep{Bamford2009,Masters2010}. These blue bulges and red disks appear to be at odds with the merger-driven hierarchical evolution paradigm where it is proposed that gas-poor major mergers at intermediate redshifts transform galaxies from disk to bulge-dominated systems (e.g. \citealt{Robertson2006a}) while simultaneously quenching star formation. Thus, the prevalence of quenching via major mergers in comparison to alternative quenching mechanisms (e.g. \citealt{Schawinski2014}) and the role of S0 galaxies during these processes (e.g. \citealt{Johnston2014}) remains unclear.

Increasingly detailed morphological studies are now able to push further back in cosmic time, into the key cosmic epoch at $1<z<3$ where the star-formation rate density of the Universe peaks. These studies have found evidence that, while the Hubble sequence is observed to already be in place at these high redshifts, a larger fraction of galaxies are disk-dominated or composite systems \citep{Buitrago2011,vanderWel2011, Chevance2012, McLure2013} and there is an increase in the fraction of visually disturbed morphologies \citep{Mortlock2013}. In addition, disk structures are also observed to be ``clumpier'' \citep{Mozena2013, Wuyts2012} and ``puffier'' \citep{ForsterSchreiber2009}. Intriguingly, the presence of passive disks has also been observed at higher redshifts ($z>1$) (e.g. \citealt{McGrath2008,Stockton2008,Cameron2011,vanderWel2011, McLure2013}), providing further evidence that the physical processes which quench star formation may be distinct from those that drive morphological transformations.

Several of the currently-proposed quenching mechanisms can account for the cessation of star-formation and the retention of a massive disk, or the presence of star-forming bulges, including: models of violent disk instabilities (e.g. \citealt{Dekel2009b, Dekel2014}), the ``hot halo'' quenching scenario (e.g. \citealt{Dekel2009a}) new observational evidence for which has been reported recently by \citet{Hartley2013}, and potentially the phenomenological mass and environment models of \citet{Peng2010}, where new evidence suggests that both of these processes may act on galaxy morphologies in the same way \citep{Carollo2014}. However, as yet, it is unclear which of these or which combination of these processes is most viable. In fact, recent focus has been placed on explicitly exploring the role that $B/T$ light and mass fractions play in quenching fractions (e.g. \citealt{Bell2012}), with the recent local study of \citet{Omand2014} suggesting that the quenching fraction of central galaxies is not simply dependent on stellar mass but is also strongly correlated with $B/T$ fractions such that in the stellar mass range $9<log_{10}(M_{*}/\rm{M_{\odot}})<11.5$ almost all galaxies with $B/T>0.3$ are quenched. This connection between quiescence and $B/T$ fraction has also been observed by \citet{Lang2014} at $z\sim2.5$ using both light and mass fractions from decompositions conducted in a similar manner to those presented in \citet{Bruce2012}, and is consistent with both the VDI model of \citet{Dekel2009b} and \citet{Ceverino2010} and,  given the natural connection between central bulge and black hole growth, AGN feedback. Within the context of these large statistical studies the role of the sub-dominant populations of passive disks and star-forming bulges is particularly interesting, as it is within the $1<z<3$ redshift range that we see a considerable build-up of the red-sequence and observe galaxies undergoing significant morphological transformations. Therefore, further study of this informative regime is crucial for improving our understanding of the main drivers of galaxy morphology evolution and the processes responsible for star-formation quenching.

In order to best conduct stxudies of the morphological evolution of galaxies at these redshifts it is instructive to decompose the galaxies into their bulge and disk components. Bulge-disk decompositions have been conducted extensively in the local Universe (e.g. \citealt{deJong1996,Allen2006,Cameron2009,Simard2011,Lackner2012}), however \citet{Bruce2012} was the first study to attempt bulge-disk decomposition in the redshift range $1<z<3$ for a large sample (by utilising the latest high-resolution imaging from HST WFC3 provided by the CANDELS survey in the UDS field).
In this paper we further develop this work by effectively doubling the sample size (by including an analysis of the CANDELS COSMOS field), and extending the multiple-component analysis to the additional $J_{125}$, $i_{814}$ and $v_{606}$-bands provided by the CANDELS survey. Crucially, this has allowed us to construct separate bulge and disk model photometry across four bands, which has then enabled us to successfully isolate and model the SEDs of the bulge and disk components. The new multiple-component SED fitting technique presented in this paper yields separate stellar-mass estimates and star-formation rates for the bulge and disk components of the massive galaxies within our sample. This has then enabled new exploration of the extent to which morphological transformations are linked to star-formation quenching. Particular emphasis has been placed on addressing the dominance of the bulge and disk masses, the star-formation activity displayed by the individual components, and the axial ratio distributions of the bulge and disk components.

This paper is structured as follows. In Section 2 we summarize our data-sets and sample properties obtained from SED fitting. This is followed in Section 3 by a description of our multiple-component, multi-wavelength, morphology fitting technique, alongside which we present the results from our mock galaxy simulations, and discuss the extension to decomposed SED fitting in Section 4.
In Section 5 we utilise our decomposed morphology and stellar-mass results to explore the overall morphological evolution of the galaxies in our $1<z<3$ sample and in Section 6 we additionally incorporate the decomposed star-formation rates of these objects to study the links between morphology and star-formation history, which includes a revised, stricter, identification of a population of passive disks and star-forming bulges. Section 7 focusses on a study of the axial ratios of our sub-divided galaxy sample and in Section 8 we extend this work to a comparison with (sub-)mm selected galaxies.
Finally, we conclude with a discussion of the results from our decompositions  within the context of current models of galaxy evolution and quenching and summarise our conclusions in Sections 9 and 10.

Throughout we quote magnitudes in the AB system, and calculate all physical
quantities assuming a $\Lambda$CDM universe with $\Omega_{m}=0.3$, $\Omega_{\Lambda}=0.7$ and $H_{0}=70\rm{kms^{-1}Mpc^{-1}}$.

\section{Data and Sample Properties}
We have used the {\it HST} WFC3/IR data from the CANDELS multi-cycle treasury programme \citep{Grogin2011, Koekemoer2011} centred on the UKIDSS Ultra Deep Survey (UDS; \citealt{Lawrence2007}) and COSMOS \citep{Scoville2007} fields. Both the CANDELS UDS and COSMOS near-infrared data comprise $4 \times 11$ WFC3/IR tiles covering a total area of 187\,arcmin$^2$ in each field, in both the F125W and F160W filters. The point-source depths are 27.1 and 27.0 (AB mag, 5-$\sigma$) for the UDS and COSMOS fields respectively. In addition to the near-infrared data, for this extended analysis we have also made use of the accompanying CANDELS {\it HST} ACS parallel imaging in the F814W and F606W filters (hereafter $i_{814}$ and $v_{606}$). The corresponding 5-$\sigma$ point-source depths are 28.4 for both the $i_{814}$ and $v_{606}$ bands in UDS and  28.5 in COSMOS, due to the inclusion of existing ACS legacy data in COSMOS \citep{Scoville2007,Koekemoer2007}. The WFC3 and ACS cameras are offset by 6 arcmin in the {\it HST} focal plane and during the CANDELS observations were oriented to provide the maximum area of contiguous WFC3+ACS coverage (which is $\sim80\%$ of the area of both fields).

\subsection{Supporting multi-wavelength data}
In order to constrain the SED fitting and determine the physical properties for the galaxies in our sample we have also utilised the multi-wavelength data-sets available in each field. In the UDS these comprise: $u'$-band imaging obtained with MegaCam on CFHT; deep optical imaging in the $B$, $V$, $R$, $i'$ and $z'$-band  
filters from the Subaru XMM-Newton Deep Survey (SXDS; \citealt{Sekiguchi2005}; \citealt{Furusawa2008}); $J$, $H$ and $K$-band UKIRT WFCAM imaging from Data Release 8 (DR8) of the UKIDSS UDS; and {\it Spitzer} 3.6\,$\mu$m, 4.5\,$\mu$m IRAC and 24\,$\mu$m MIPS imaging from the SpUDS legacy programme (PI Dunlop). For COSMOS we use: optical imaging in  $u'$, $g'$, $r'$, $i'$ and $z'$-bands from MegaCam CFHTLS-D2; $z'$-band from Subaru; $Y$,$J$,$H$ \& $Ks$ from Ultra-VISTA (PI Dunlop); and {\it Spitzer} 3.6\,$\mu$m, 4.5\,$\mu$m IRAC and 24\,$\mu$m MIPS imaging from the S-COSMOS survey (PI Sanders).
  
\subsection{Mass estimation and sample selection}
We follow the same sample selection procedure detailed in \citet{Bruce2012} and have determined photometric redshifts for the UDS and COSMOS master samples using a $\chi^{2}$ 
fitting procedure based on the photometric redshift code HYPERZ from \citet{Bolzonella2000}, described in \citet{Cirasuolo2007}. 
These photometric redshifts were then used to determine stellar-mass estimates, however in comparison to \citet{Bruce2012} we now implement a slightly updated stellar-mass SED fitting procedure which we have applied to the COSMOS field and have used to re-define the \citet{Bruce2012} UDS sample. 
These stellar-mass estimates were based on the \citet{Bruzual2003} models with single-component exponentially decaying star-formation histories with e-folding times in the range $0.3\leq \tau {\rm (Gyr)}\leq 5$ and with a minimum model age limit of 50 Myr to ensure physically meaningful mass estimates. This results in a sample of 205 objects in the UDS and 191 objects in COSMOS with $M_*>1\times10^{11}\,{\rm M_{\odot}}$ in the $1<z<3$ redshift range.

\begin{figure*}
\begin{center}
\includegraphics[scale=0.75]{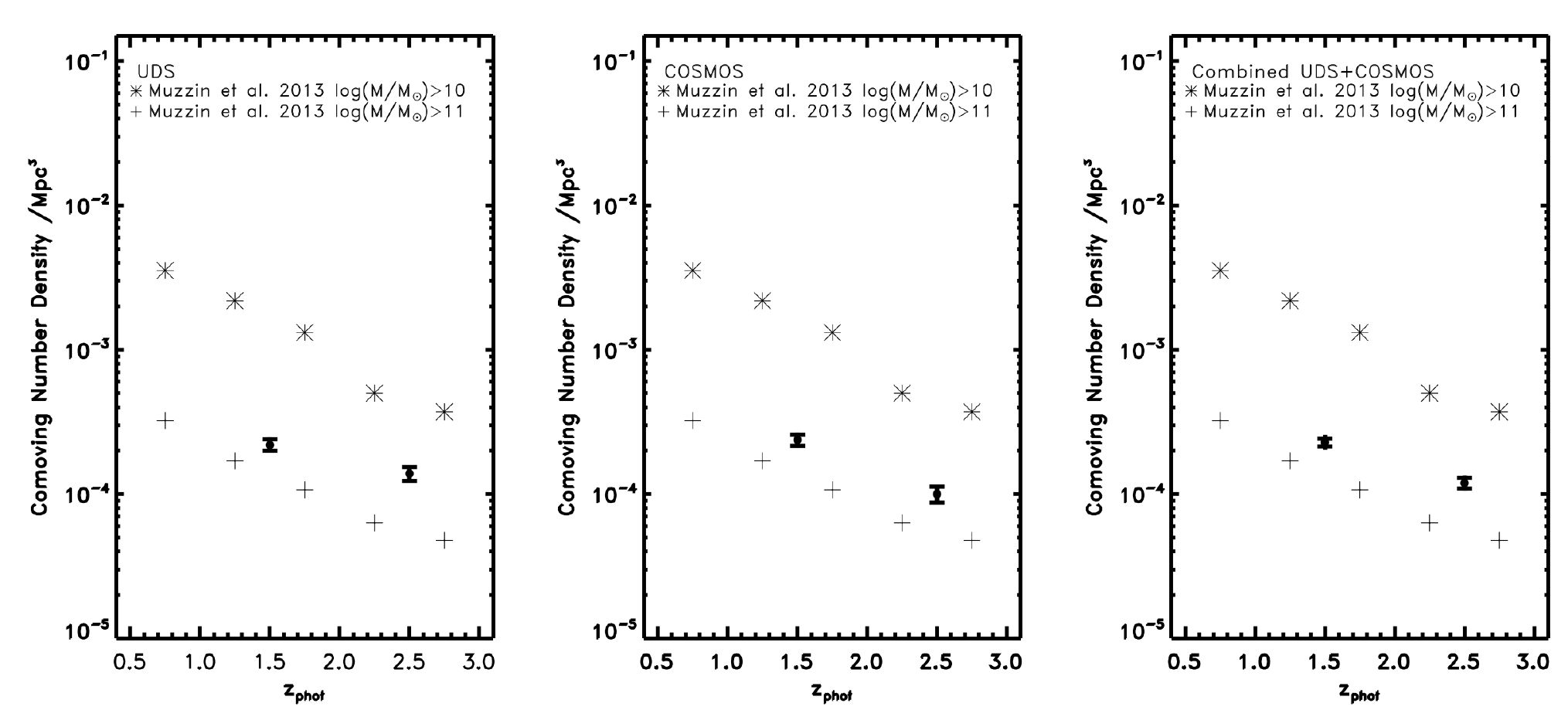} 
 \caption[Comparison of co-moving number densities between fields.]{Co-moving number densities of massive galaxies ($M_*>10^{11}\,{\rm M_{\odot}}$) in the CANDELS-UDS (left), CANDELS-COSMOS (middle) and combined fields (right). These number densities have been over-plotted with the values from \citet{Muzzin2013} for their $M_*>1\times10^{10}\,{\rm M_{\odot}}$ and $M_*>1\times10^{11}\,{\rm M_{\odot}}$ mass bins, given by asterisks and crosses respectively. As discussed, the slightly lower values reported by \citet{Muzzin2013} for galaxies within a similar mass range to this study is not unexpected given the uncertainties associated with stellar mass estimates from SED fitting, especially given the steepness of the stellar mass function in this high-mass regime.}
\label{fig:no_dense}
 \end{center}
\end{figure*}

The massive galaxy co-moving number densities in both fields, and the combined sample, are shown in Fig.\,\ref{fig:no_dense}. These have been over-plotted with the co-moving number densities determined from the latest stellar mass function study at $0.2<z<4$ by \citet{Muzzin2013}, which was conducted over the full $1.62\,{\rm deg^{2}}$ of COSMOS covered by UltraVISTA. These plots show the co-moving number densities for galaxies with derived stellar masses of $M_*>1\times10^{10}\,{\rm M_{\odot}}$ (in asterisks) and $M_*>1\times10^{11}\,{\rm M_{\odot}}$ (in crosses), illustrating how the steepness of the mass function at this high-mass end affects the number densities of objects. As can be seen, the number densities for  $M_*>1\times10^{11}\,{\rm M_{\odot}}$ galaxies from \citet{Muzzin2013} are a factor of $\approx1.6$ lower than the values for our combined sample. However, the factor of 1.6 can be accounted for by a $\sim 35\%$ decrease in the mass estimates derived by \citet{Muzzin2013} at $M_*\simeq1\times10^{11}\,{\rm M_{\odot}}$ which is consistent with the expected level of systematic uncertainty associated with stellar masses derived from SED fitting and different choices of star-formation histories and stellar population age ranges.

As the sample sizes and areas in the UDS and COSMOS fields are comparable and there is good agreement between the co-moving number densities of the two fields, in the following sections the science results are based on the combined UDS and COSMOS sample, except for a few cases where the results are presented individually, specifically to facilitate comparison between fields.

Finally, due to the difference in area coverage between the WFC3 pointings and the accompanying ACS parallels taken as part of CANDELS, the final sample which is subject to the multiple-component multi-wavelength morphological decomposition and SED fitting comprises 178 objects which have the required photometry across all four of the CANDELS {\it HST} bands and have a best-fit model with a bulge+disk.

\subsection{Star-formation rates}
The star-formation rates for the UDS and COSMOS samples were estimated from the best-fit SED models and $24\mu$m fluxes by adopting the convention of \citet{Wuyts2011}. Following this scheme if any of the objects in the sample has a $24\mu$m counterpart within a 2 arcsecond radius in the SpUDS and S-COSMOS catalogues, its star-formation rate is given by:
\begin{equation}
 SFR_{UV+IR}[{\rm M_{\odot}yr^{-1}}]=1.09\times10^{-10}(L_{IR}+3.3L_{2800})/{\rm L_{\odot}}
\end {equation}
where $L_{2800}=\nu L_{\nu}(2800\rm\AA)$ and $L_{IR}$ is the contribution from radiation from hot young stars re-radiated by dust over the wavelength range $8-1000\mu$m. Objects which do not have $24\mu$m counterparts, have star-formation rates calculated following \citet{Kennicutt1998}:
\begin{equation}
SFR_{UV, dust\,corrected}[{\rm M_{\odot}yr^{-1}}]=1.4\times10^{-28}{\rm L_{\nu}(ergs\,s^{-1}\,Hz^{-1}})
\end{equation}

\section{Multiple-Component Morphology Fitting}
The morphologies of the 396 objects in the combined UDS and COSMOS sample have been fitted with both single and multiple-S\'{e}rsic light profiles using GALFIT \citep{Peng2010galfit}. The explicit procedures adopted for this analysis have been presented in \citet{Bruce2012}. The fitting is conducted on $6\times6$\,arcsec image stamps of the objects and uses an empirical PSF generated from a median stack of the brightest (unsaturated) stars in the individual fields. A comparison with fits determined from adopting the latest CANDELS PSFs of \citep{vanderWel2012} (which are generated from Tiny Tim models at small radii and empirical stacks further out) reveals that all fitted parameters are consistent to within a few percent. The procedure also adopts an object-by-object background determination by calculating the median value within an annular aperture centred on each source with an inner radius of 3\,arcsec and an outer radius of 5\,arcsec. The multiple-component S\'{e}rsic modelling was done by fitting a set of nested models to each object comprising six models which are various combinations of $n=1$ exponential disks, $n=4$ de Vaucouleurs bulges and a centrally concentrated light profile component (here a PSF) to account for any nuclear starbursts or AGN. 

This analysis provided acceptable multiple component model fits for 163 objects in the COSMOS sample, and is combined with the further 184 objects with acceptable model fits in the UDS field.

\subsection{Mock Galaxy Simulations}
In order to better quantify the systematic and random uncertainties in the above decomposition procedure, we have generated a grid of 9216 mock galaxies of known morphological parameters and have attempted to  recover them with our fitting procedure. Similar simulations have been used previously to determine the uncertainties on fitted model parameters with GALFIT for both single-component \citep{Haussler2007,vanderWel2012, Newman2012} and multiple-component \citep{Davari2014} models. However, these have previously been limited to cases where the {\it fitted} morphologies use only a single-S\'{e}rsic model, with the exception of \citet{Lang2014} who conduct similar bulge-disk decompositions on mass maps, but in contrast to our fitting method they allow the centroid positions of their bulge and disk components to vary slightly and they explicitly limit their fits so that the disk component effective radius is always equal to or larger than that of the bulge component.
Thus, for this work, we conduct  our own mock galaxy simulations for $n=4+n=1$ models. 

All mock galaxies were generated to have the same total brightness, corresponding to the median magnitude of our combined CANDELS UDS and COSMOS sample at 21.8 mag [AB], and were constructed to have $B/T$ ratios 0.99, 0.95, 0.9, 0.75, 0.5, 0.25, 0.1, 0.05 and 0.01 to fully explore how our fitting method treats these different contributions. The models were also created with: bulge and disk effective radii of 1, 5, 10 and 20 pixels; axial ratios of 0.1, 0.3, 0.6 and 1.0; and relative position angles between the components of 0, 30, 60 and 90 degrees. 

\begin{table*}
\begin{tabular}{m{12cm}m{6.5cm}}
\hline
&
Model Component $R_{e}$\\
\end{tabular}
\begin{tabular}{m{3cm}m{2cm}m{3.5cm}m{2cm}m{2cm}m{2cm}m{2cm}}
Fractional
\newline Contribution&
Component&
Error&
 1 pixel &
 5 pixels &
10 pixels&
 20 pixels \\
\hline
$B/T>0.9$&
bulge&
systematic offset &
$1.6\%$&
$0.3\%$&
$1.6\%$&
$4.8\%$\\
&
&
random uncertainty&
$4.6\%$ &
$3.5\%$ &
$5.2\%$ &
$6.6\%$ \\
$B/T<0.1$&
disk&
systematic offset &
$0.9\%$&
$0.9\%$&
$2.0\%$&
$3.8\%$\\
&
&
random uncertainty&
$1.6\%$ &
$1.6\%$ &
$2.1\%$ &
$3.9\%$ \\
$0.1<B/T<0.9$&
bulge&
systematic offset &
$1.4\%$&
$2.7\%$&
$5.8\%$&
$8.3\%$\\
&
&
random uncertainty&
$12.6\%$ &
$17.6\%$ &
$19.1\%$ &
$17.2\%$ \\
&
disk&
systematic offset &
$1.8\%$&
$0.5\%$&
$2.4\%$&
$3.9\%$\\
&
&
random uncertainty&
$4.2\%$ &
$4.1\%$ &
$6.3\%$ &
$8.9\%$ \\
\end{tabular}
\caption{Tabulated results for the systematic and random fractional uncertainty on the recovered sizes of the bulge and disk components from our multiple-S\'{e}rsic mock galaxy simulations. This table is split into model components of different sizes to fully illustrate how reliable the results from our fitting procedure are for different cases.}
\end{table*}

The analysis of these simulated galaxies demonstrates that in $\sim80\%$ of cases we are able to recover $B/T$ ratios to within $10\%$ and that there is no significant systematic offset in the recovered $B/T$ ratios. Our tests also show that the sizes of the individual components can be measured to within the accuracies summarized in Table 1, for both pure bulge and disk, and mixed component systems.

 For the pure bulge and disk systems the fractional systematic offset in the measured bulge and disk sizes, respectively, ranges from $1-5\%$, increasing with the increasing model component size, with the offset resulting in an under-estimate of all sizes (except for the smallest 1 pixel model, where the sizes are consistently over-estimated by $\sim 1-2\%$). Systems which are a mixed bulge+disk systems with $0.1<B/T<0.9$ have similar systematic offsets in the measured disk sizes, between $1-5\%$, but the recovered bulge sizes of these systems are offset by $1-10\%$, again increasing with the increasing modelled sizes. The fractional random uncertainties on the bulge and disk size measurements, however, are roughly constant in the range $\sim 5-10\%$ for the pure bulges and disks, and the disk components of mixed systems. There is potential evidence that the disk sizes of pure disks can be recovered more accurately, however it is unclear whether this is significant. Again, the derived bulge sizes of the mixed systems carry a larger uncertainty of $10-20\%$, which further demonstrates that the bulge components of mixed systems appear to be the hardest to accurately constrain, both in terms of their contribution to the overall light and their effective radii. Given that the uncertainties on the bulge component measurements become larger for the more extended bulge models, this effect can be understood by considering that, in order to constrain the model parameters of an $n=4$ bulge component, any model fitting is more heavily influenced by flux at larger radii, where S/N is poorer, compared to an $n=1$ disk-model fit which can be constrained more accurately from the central flux from the object alone.

Throughout the rest of this paper, it should be kept in mind that all $B/T$ light fractions can be considered accurate to within $\sim10\%$ and quoted component sizes are robust to the level of $\sim10-20\%$, including any effects from systematic offsets. However, it should be noted that the accuracy with which we have been able to determine these fitted parameters results from a sample which has been selected to have $S/N\geq50$. 

In addition to the uncertainties on our fitted sizes and light fractions, the mock galaxy simulations have also allowed us to explore any degeneracies involved in fitting multiple-component models which contain a PSF component. The adoption of the additional PSF component was motivated by the high ($n>10$) single-S\'{e}rsic fits which were obtained in \citet{Bruce2012} and the effects of including a PSF component in the single-S\'{e}rsic fits are discussed in Bruce et al. (2014, in preparation). Here we limit our discussion to the more relevant cases of the multiple-S\'{e}rsic + PSF fits.

We find that in only $\sim3\%$ of cases the best-fit multiple-S\'{e}rsic model will adopt a PSF component despite the fact that no component was included in the model galaxy, although this fraction drops to $\sim0.3\%$ when we exclude those cases where one of the components was modelled by a single pixel effective radius. However, by explicitly examining these cases where one of the model components has an effective radius equal to one pixel, we find that in fact $90\%$ of these $r_{eff}$=1 pixel bulge models correctly avoid a best-fit PSF component and $95\%$ of 1 pixel disks do the same. 

Further exploration seems to suggest that the adoption of the PSF component for these small component cases does not appear to correlate with the flux or relative position angle of the small component, but in the case of small bulges may, understandably, be more prevalent for models with lower axial ratios, i.e. flatter, small bulge models.
Moreover, for these 1 pixel component models which adopt a PSF best-fit the flux of the small component is not attributed to the PSF (even within $\sim10\%$) in $>70\%$ of these cases. Instead, the majority of the flux of the small component is attributed to one or more of the bulge/disk components.

From these tests it is clear that the ``spurious'' adoption of a PSF component, despite being rare, is most common in cases where there is a single pixel component, but it is not purely confined to these cases.
Thus, it appears that there are some degeneracies in our fitting procedure when GALFIT is allowed to adopt an additional PSF component which are not easily quantified, although these occurrences are limited.

\subsection{Extension to additional bands}
In addition to the morphological analysis conducted on the $H_{160}$-band imaging, we have also utilised the accompanying WFC3 $J_{125}$ pointings and ACS  $i_{814}$ and $v_{606}$ parallels taken in both the CANDELS UDS and COSMOS fields in order to extend our decomposition across a broad wavelength range and explore how the component contributions vary in the different bands. 
As the aim of this study is to explore the morphology of the majority of the assembled stellar mass within our $1<z<3$ galaxy sample, we have adopted the $H_{160}$-band models as they sample the galaxy light long-ward of the $4000\rm\AA$ break and are therefore less affected by contributions from young stars. We then fixed all model parameters at the values determined for the  $H_{160}$-band fits and in the fitting procedure allowed only the magnitude of each component to vary as a free parameter. This required keeping the centroids of the objects fixed, along with the effective radius, axis ratio and position angle. During the fitting across the additional $J_{125}$, $i_{814}$ and $v_{606}$ bands the background for each object was calculated using the same procedure developed for the $H_{160}$ contribution, by taking the median value within an annulus between $3-5$\,arcsec in radius. The $\sigma$ maps for the $\chi^{2}$-fitting and PSFs used were also taken from the corresponding mosaics of the individual bands, while a single bad-pixel map was applied. This map was constructed from the segmentation map of the $H_{160}$ image, rather than from each of the individual mosaics, as this was deemed to be the best way to ensure consistent masking of companions in relation to the master model fits across all bands.
By fixing the rest of the model parameters we were able to scale the contributions of the separate bulge and disk (and PSF) components across all 4 of the bands and provide magnitudes for these components. During this fitting, $\sim 10-15\%$ of the sample were best-fit with a very faint second component or with a single component fit only. In order to fully explore the best-fit models for those objects we constructed a grid of different  component magnitudes ranging from $20$ to $35$ (well below the $5-\sigma$ depths of the 4 bands) AB magnitudes (in steps of 0.1 mags). Where the fitted magnitude of a component in one of the three accompanying bands fell below the $1-\sigma$ detection limit of the corresponding band we disregarded the model fitted to that component and set it as a non-detection in the subsequent SED fitting.

As a check of the validity of this approach of fixing the model parameters to the $H_{160}$-band fit and allowing only the magnitudes of the individual component to vary across the multi-wavelength fitting, we have conducted several tests of the effects of also allowing both the effective radius and the effective radius plus the axis ratio to float as free parameters. The full details for these tests are presented in Appendix A. In summary, despite the naive expectation that in bluer bands the disk component of a galaxy will become more prominent and affect the goodness of a fit locked at the $H_{160}$ parameters, the additional degrees of freedom introduced by allowing the $r_{e}$ and $b/a$ parameters to be freely fitted significantly improve the fits for only a small subset of the galaxies in our sample. Moreover, for objects which were best-fit with an $H_{160}$ bulge-only model we also experimented with allowing the addition of a disk component at the bluer wave-bands but again found that this did not deliver significant improvements in the model fits, although the number of these cases was limited.

\begin{figure*}
\begin{center}
\begin{tabular}{m{3cm}m{3cm}m{3cm}m{3cm}}
\includegraphics[scale=0.15 ]{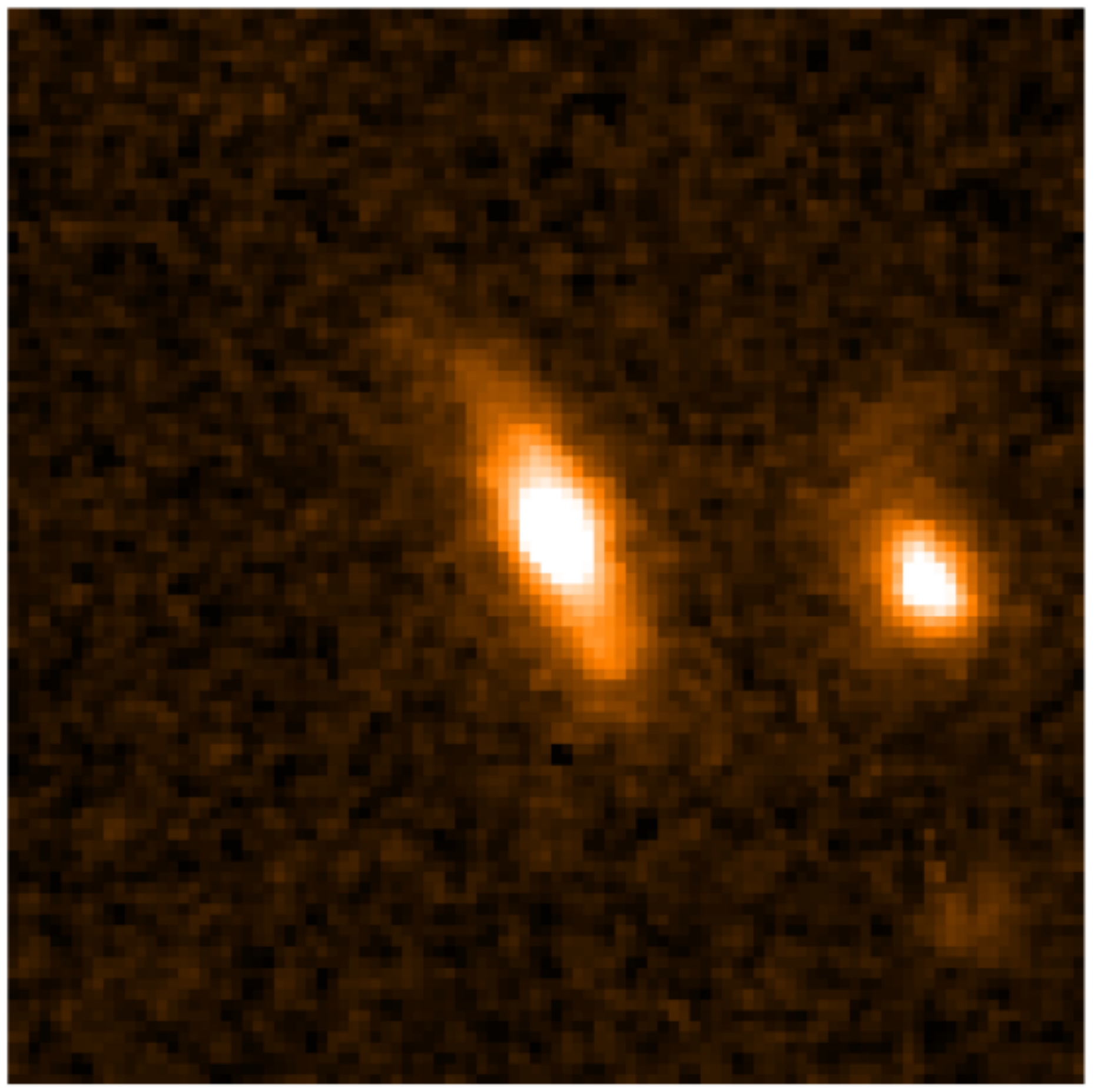}  &
\includegraphics[scale=0.15]{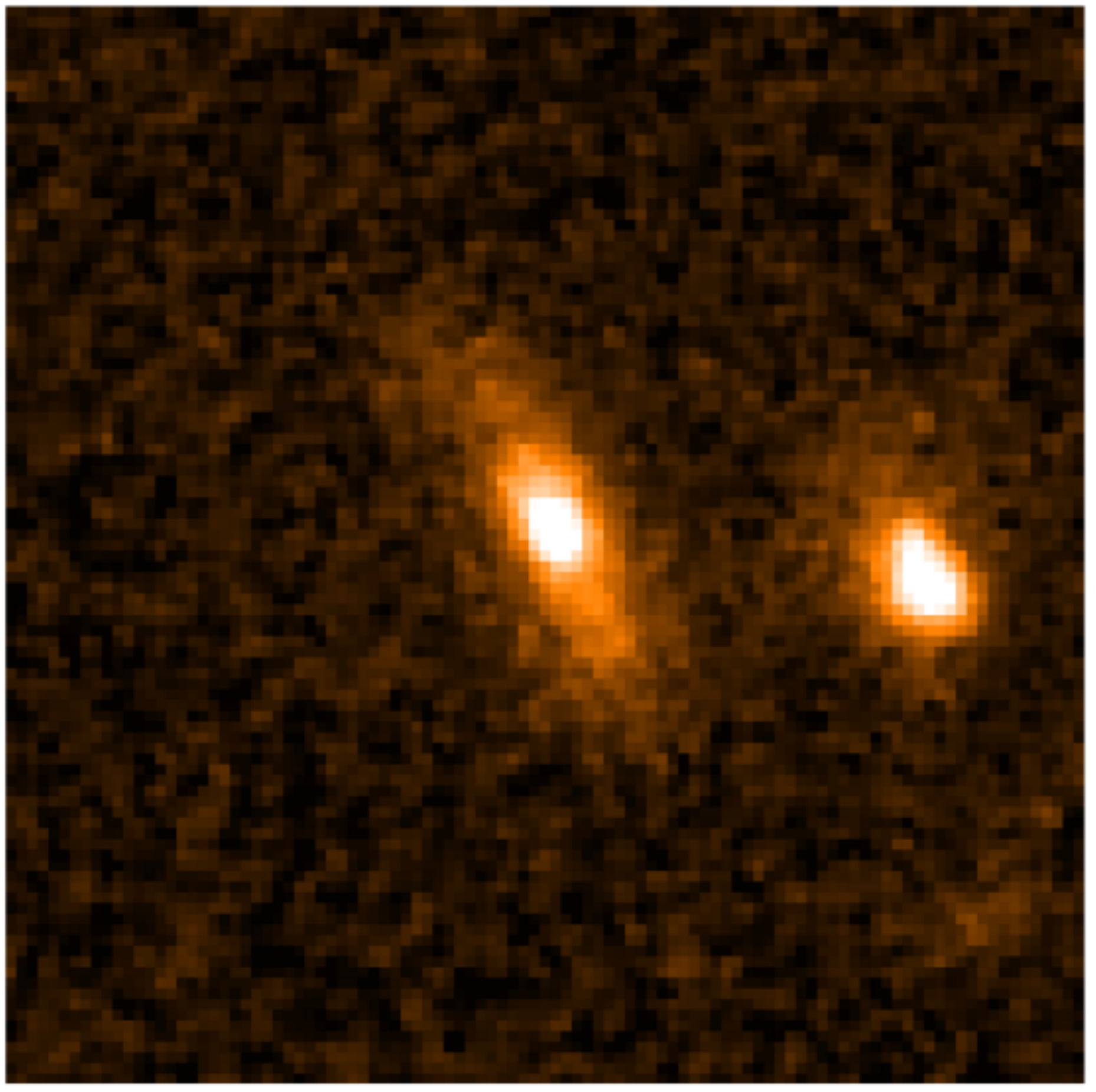}  &
\includegraphics[scale=0.15]{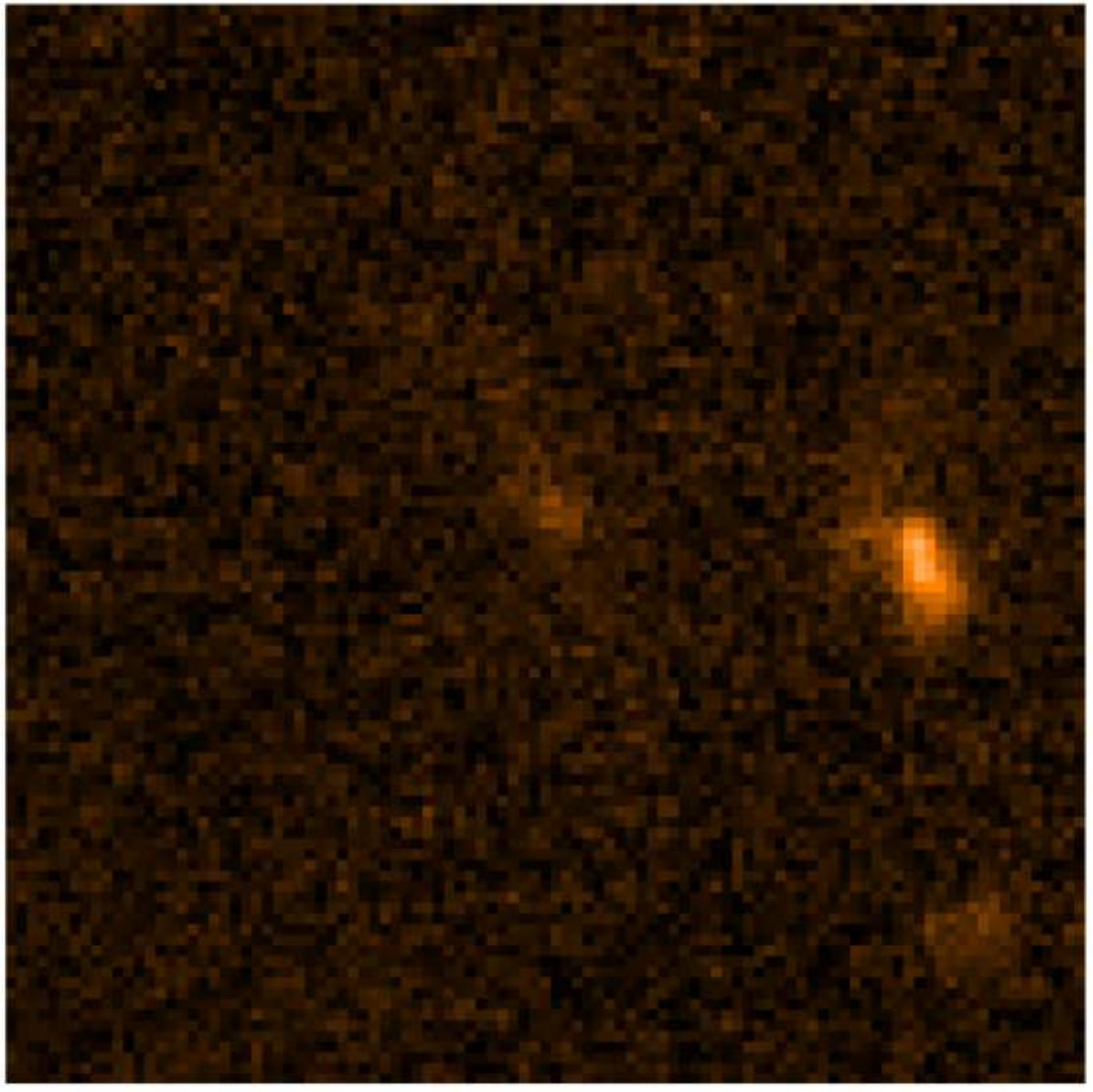}  &
\includegraphics[scale=0.15]{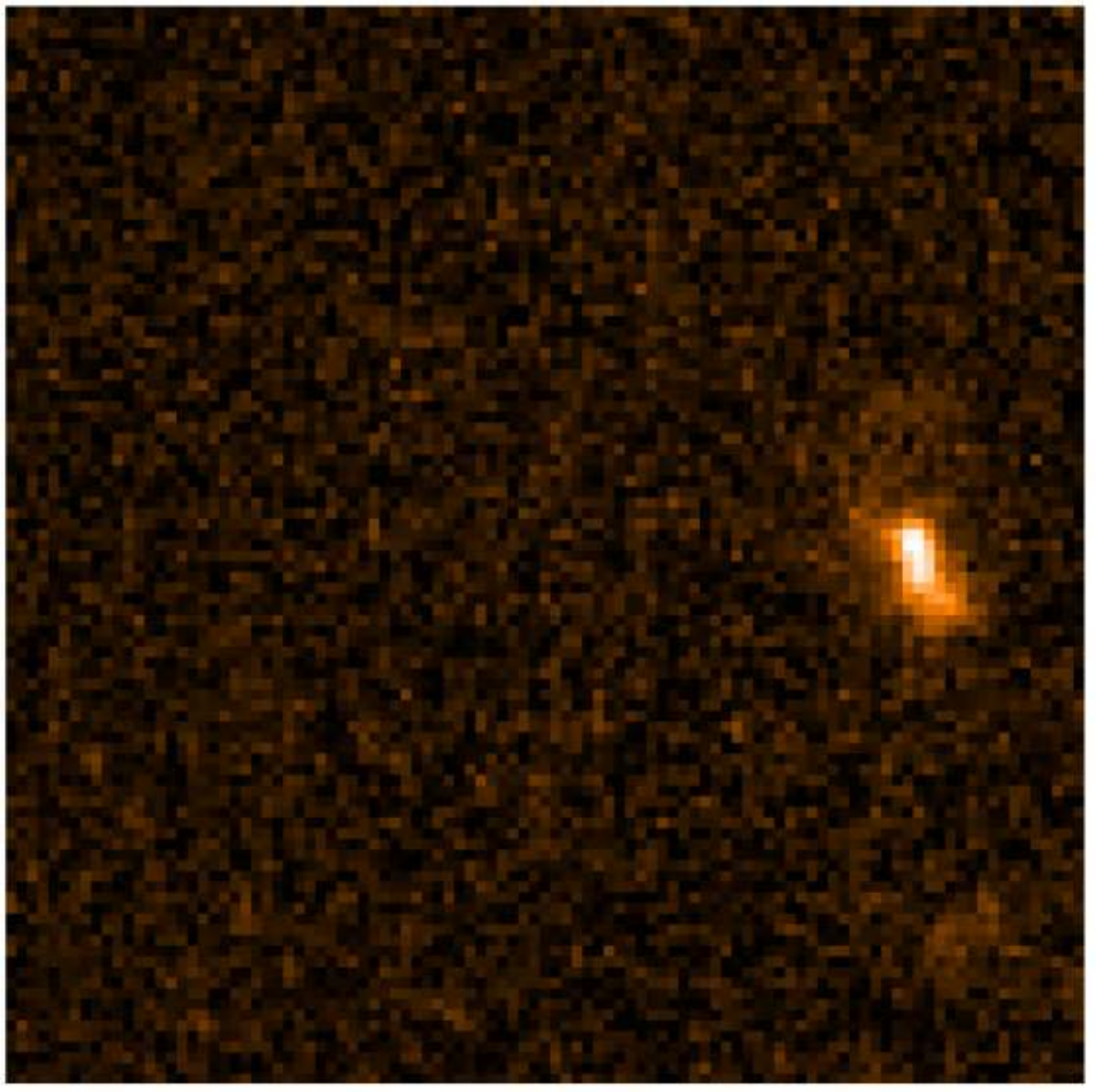}  \\
\includegraphics[scale=0.15]{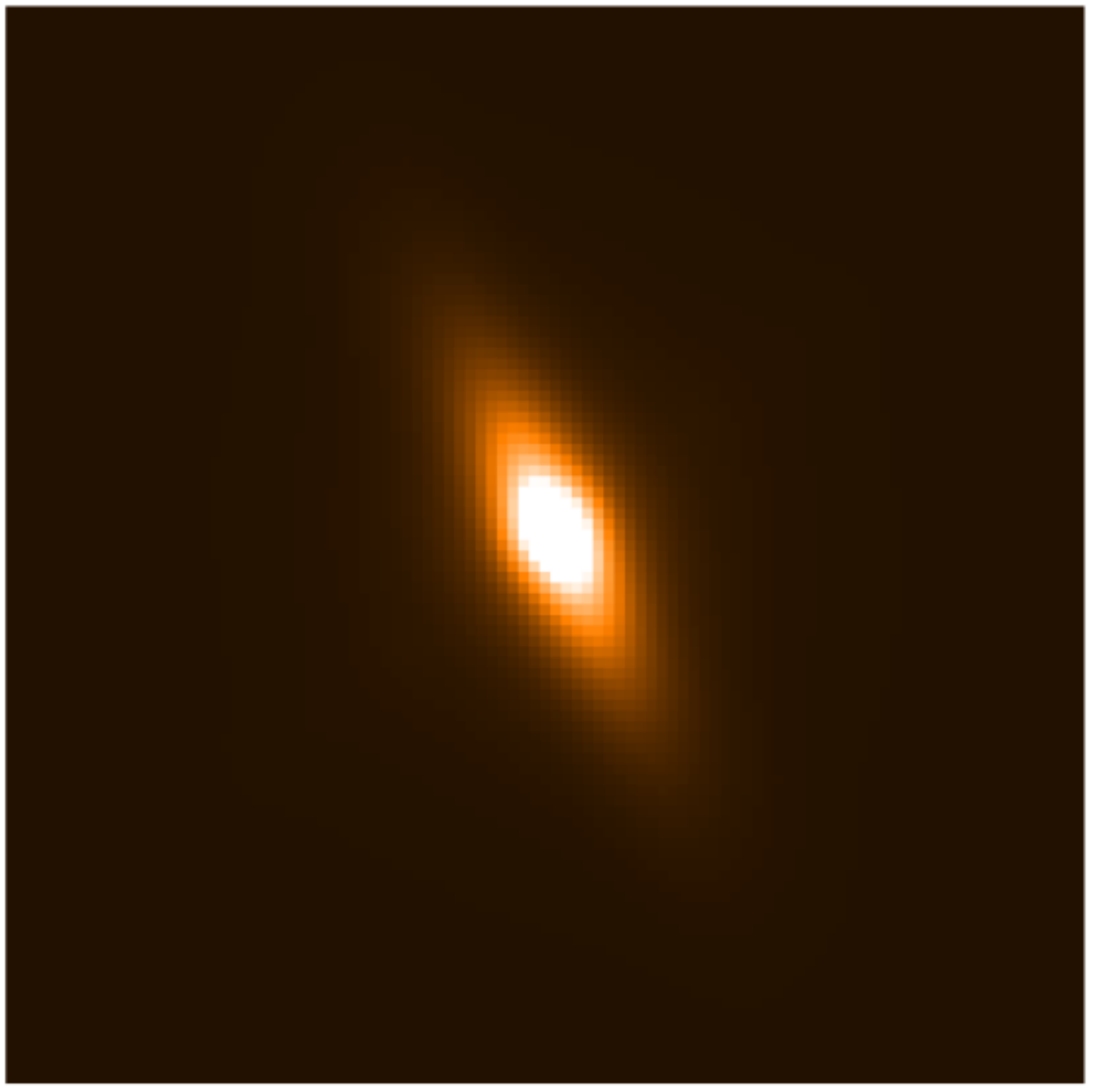}  &
\includegraphics[scale=0.15]{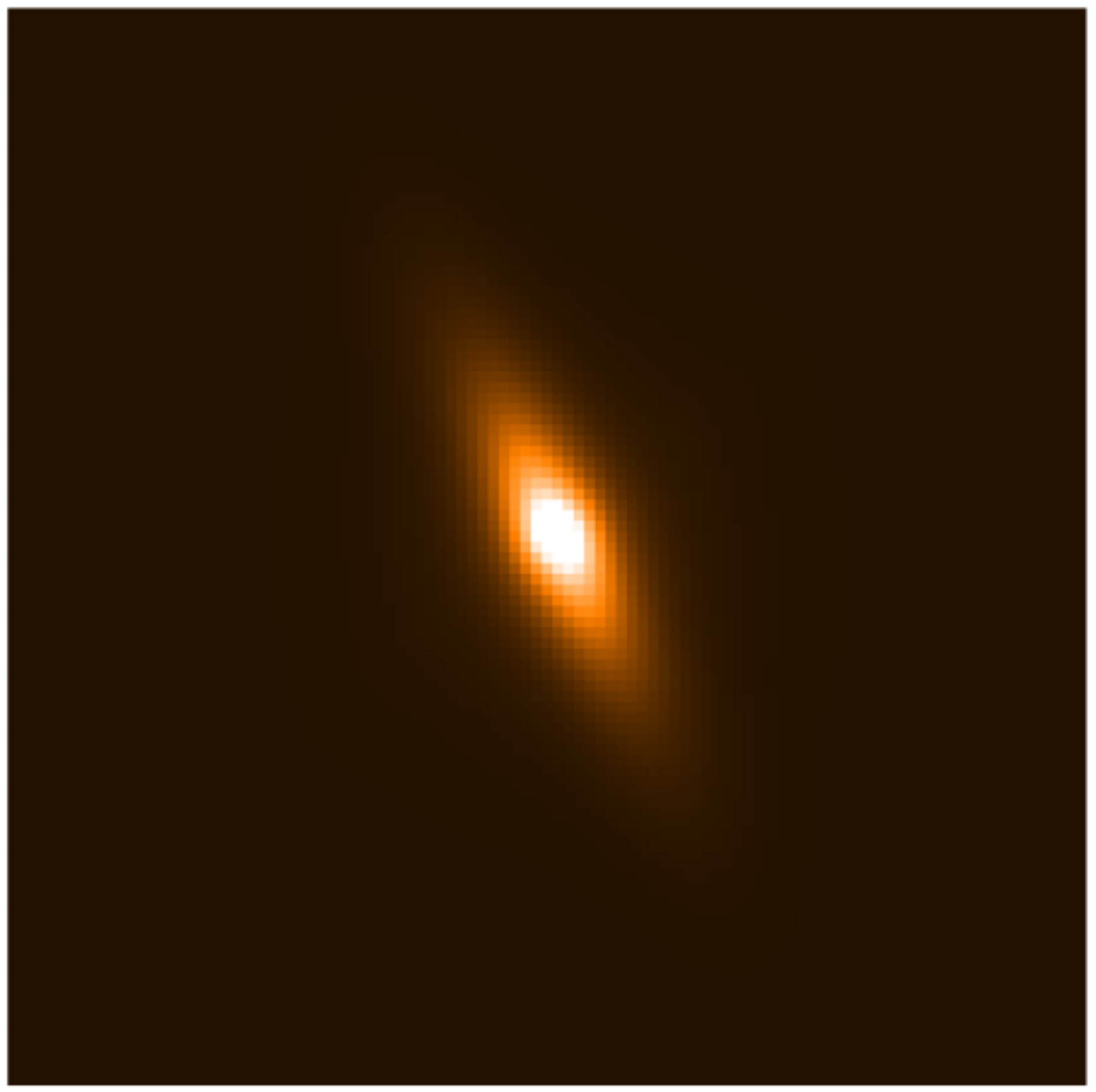}  &
\includegraphics[scale=0.15]{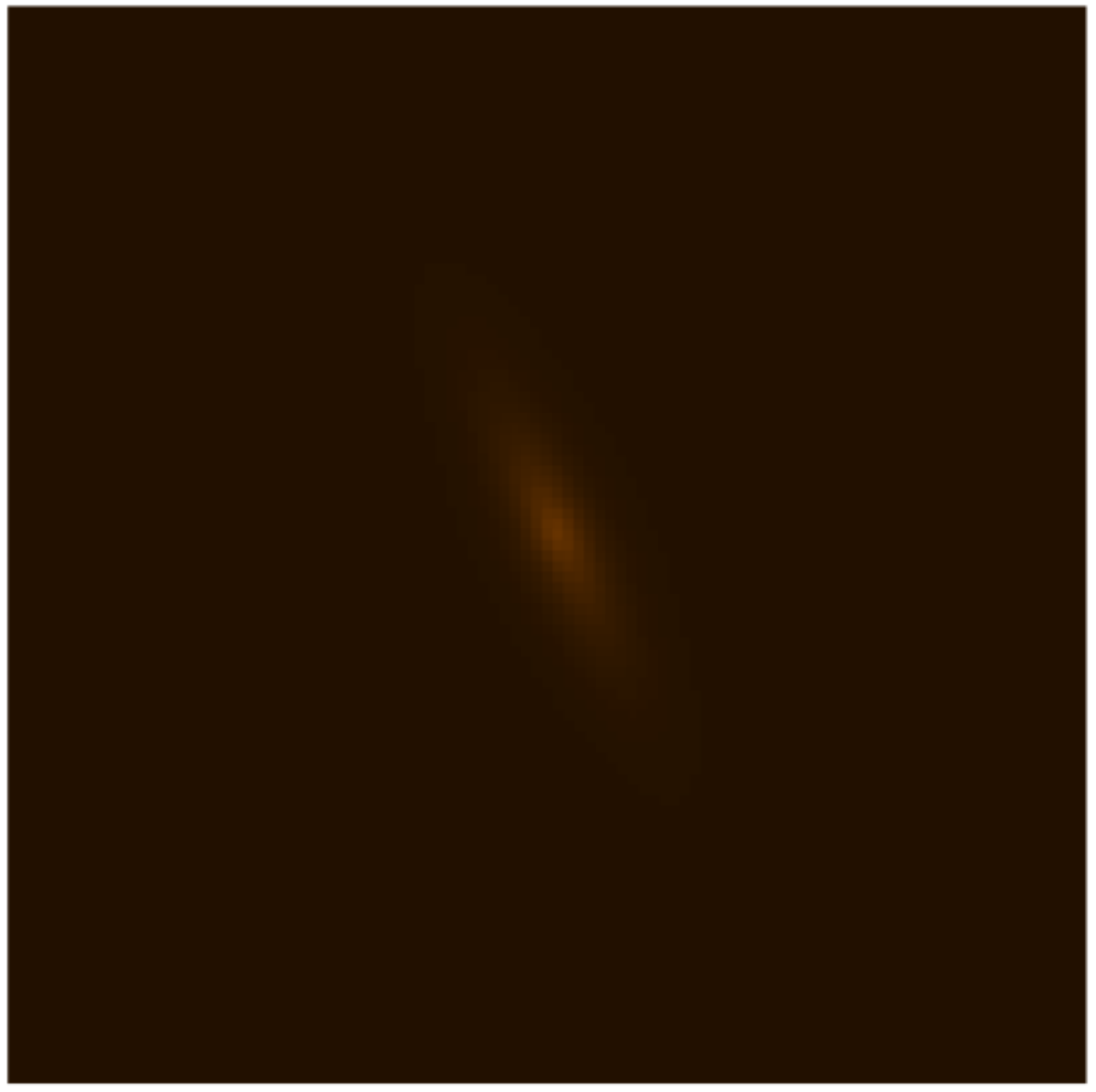}  &
\includegraphics[scale=0.15]{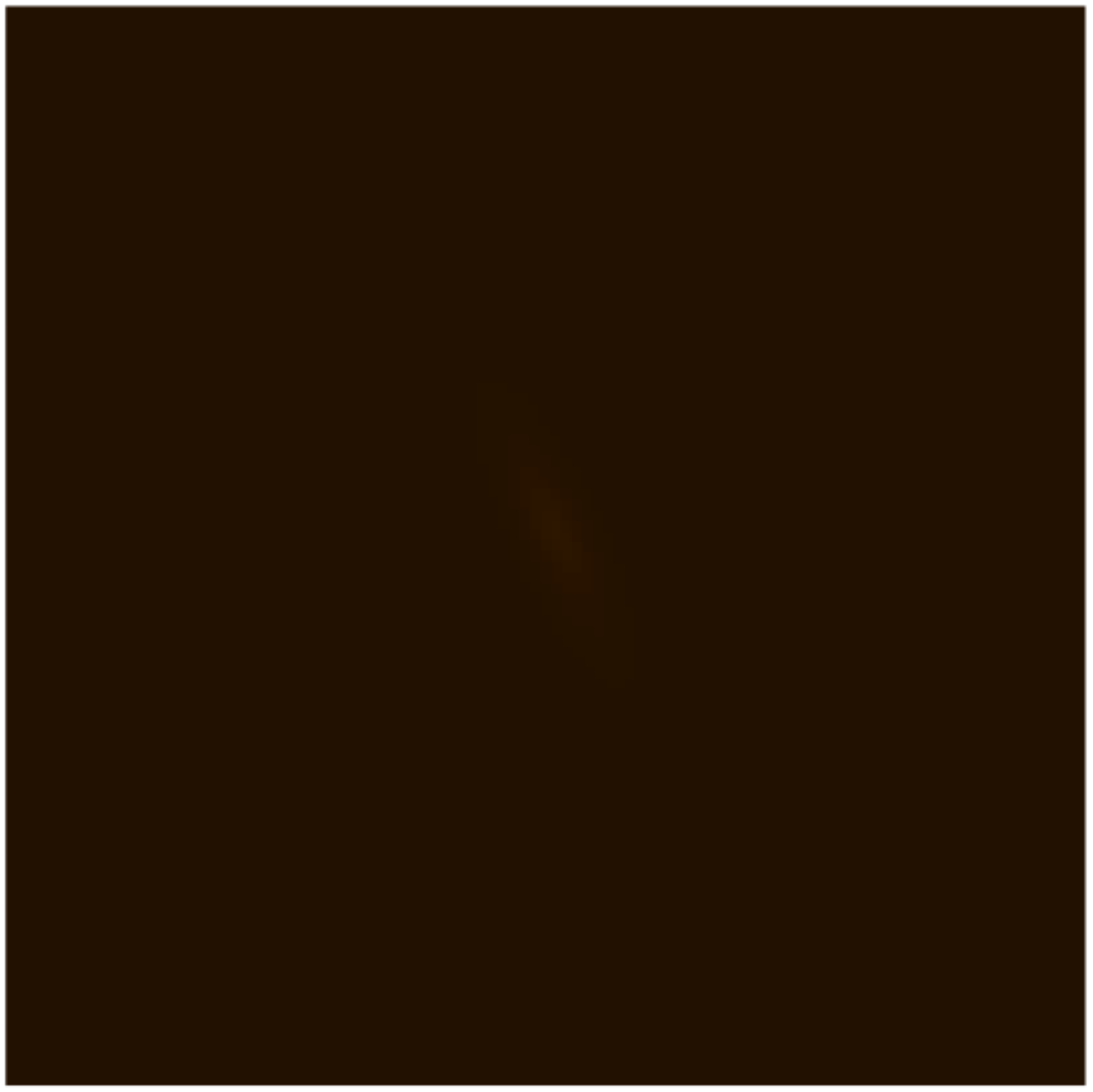}  \\
\includegraphics[scale=0.15]{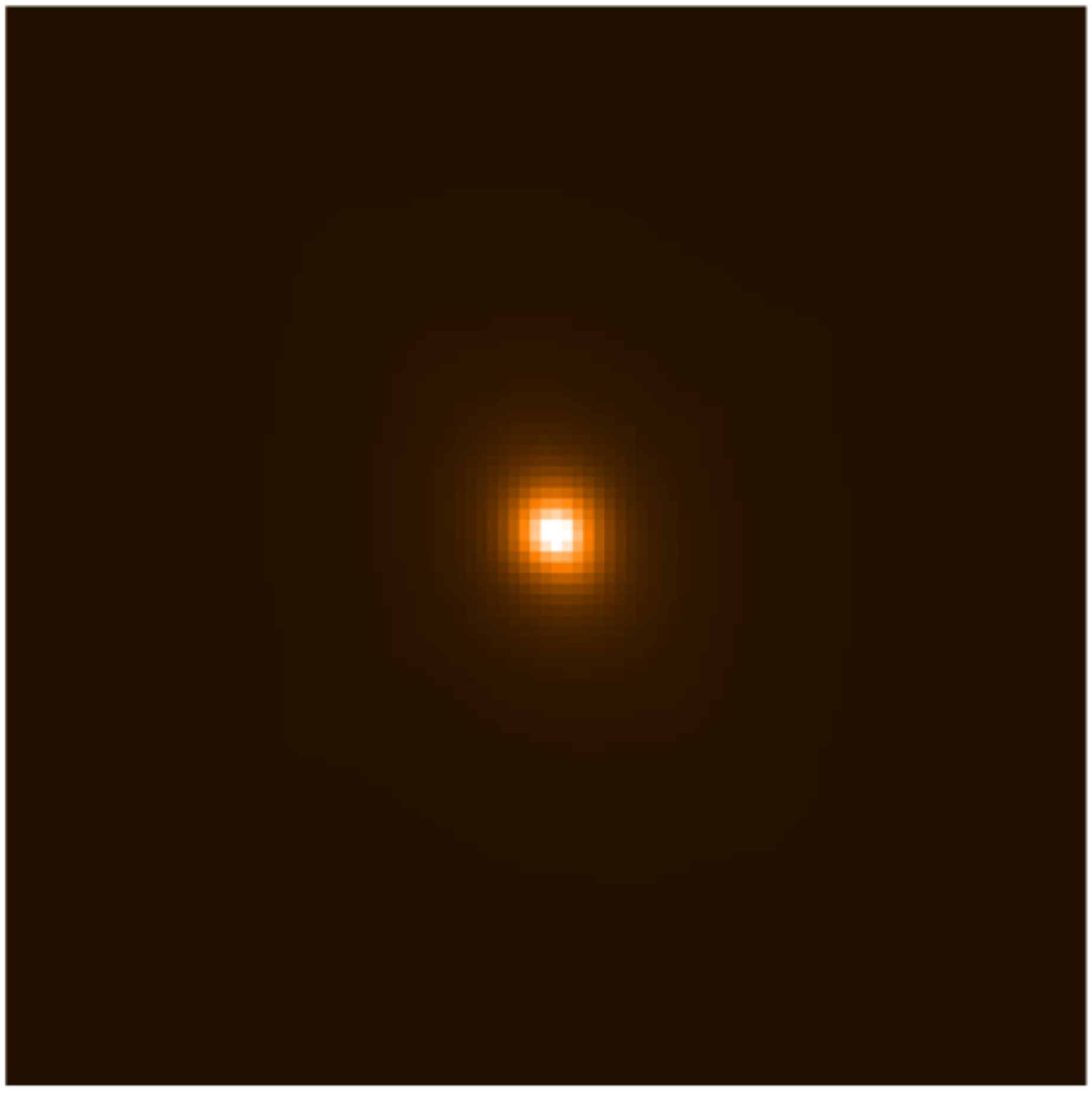}  &
\includegraphics[scale=0.15]{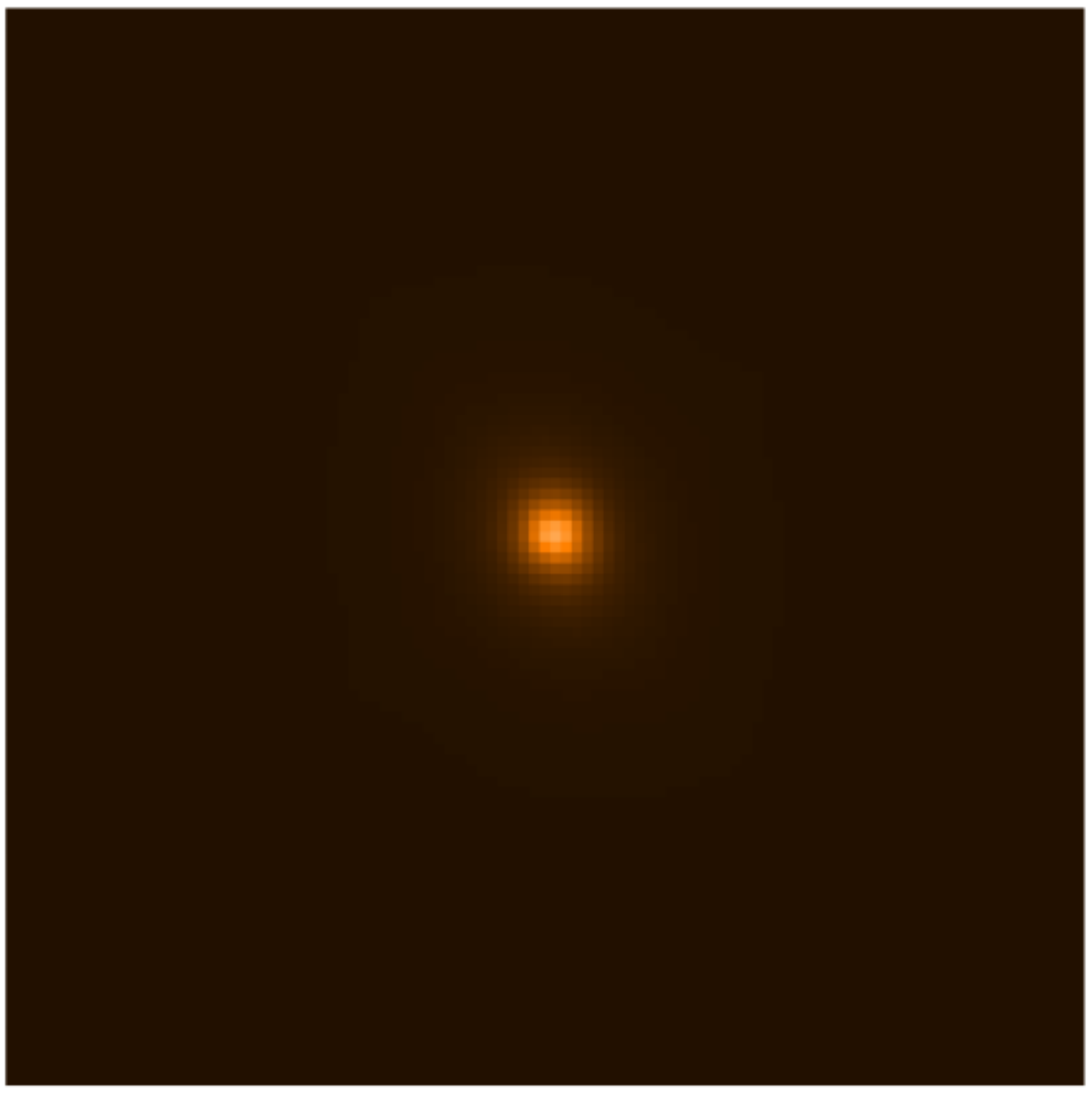}  &
\includegraphics[scale=0.15]{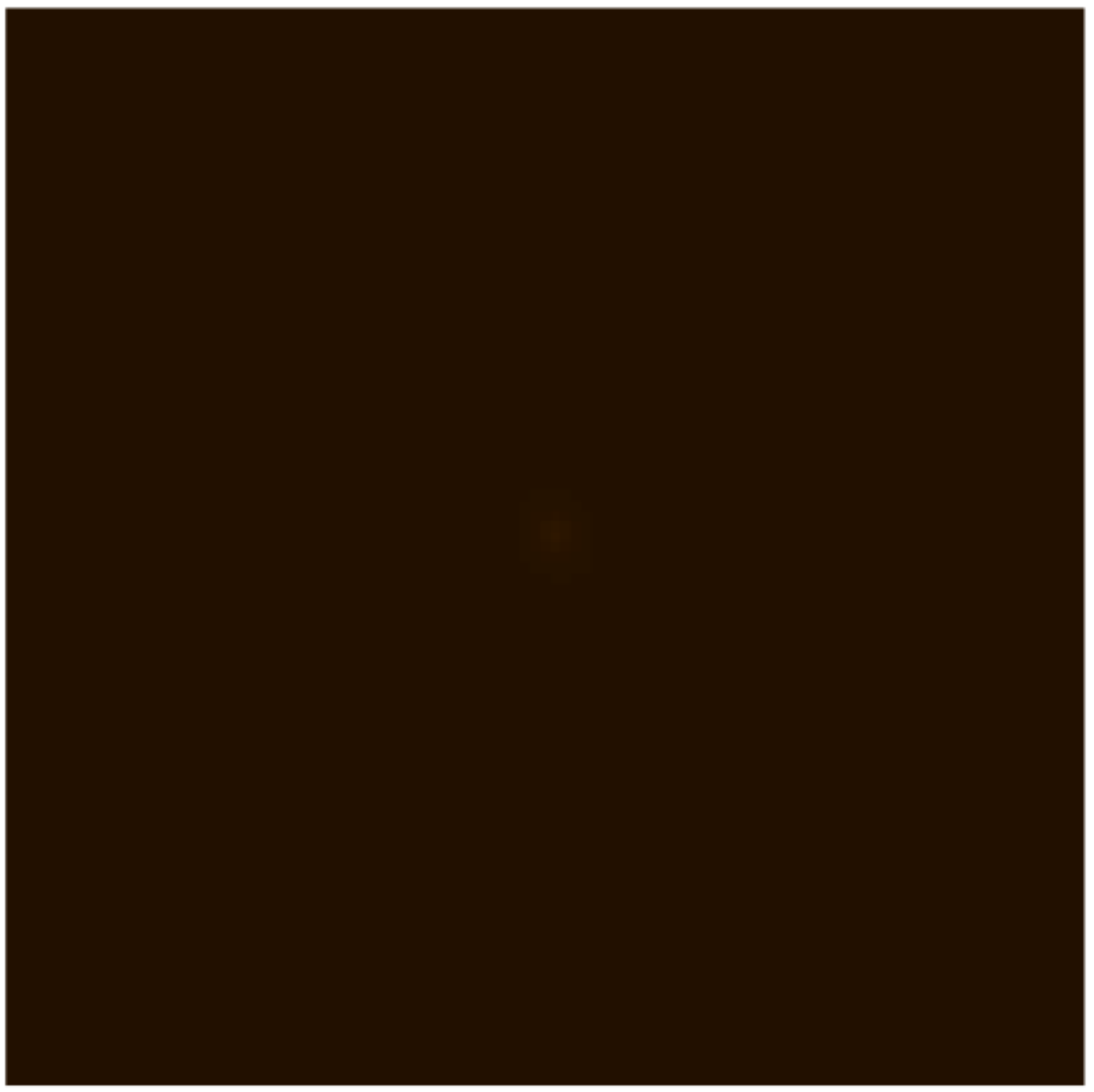}  &
\includegraphics[scale=0.15]{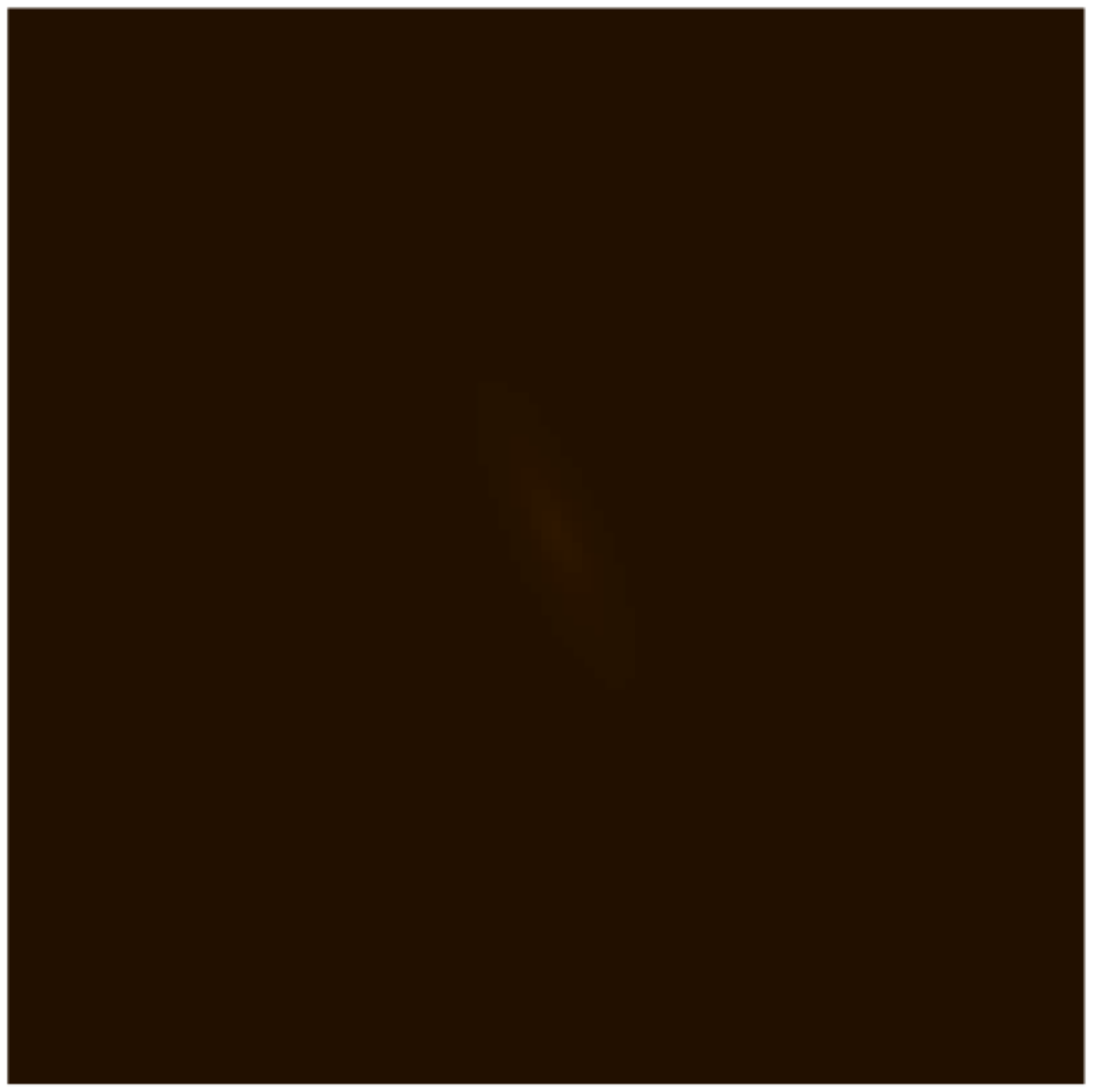}  \\
\includegraphics[scale=0.15]{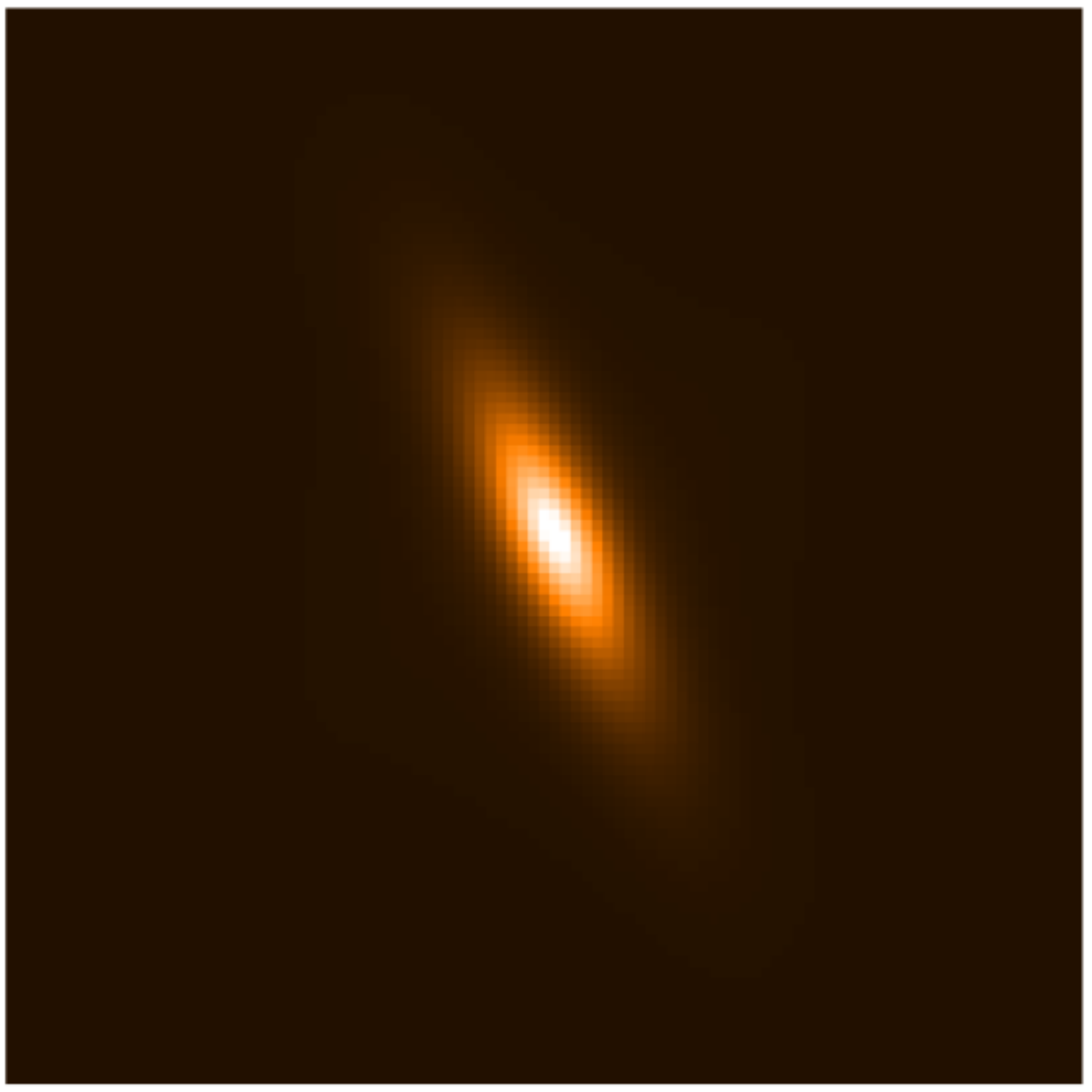}  &
\includegraphics[scale=0.15]{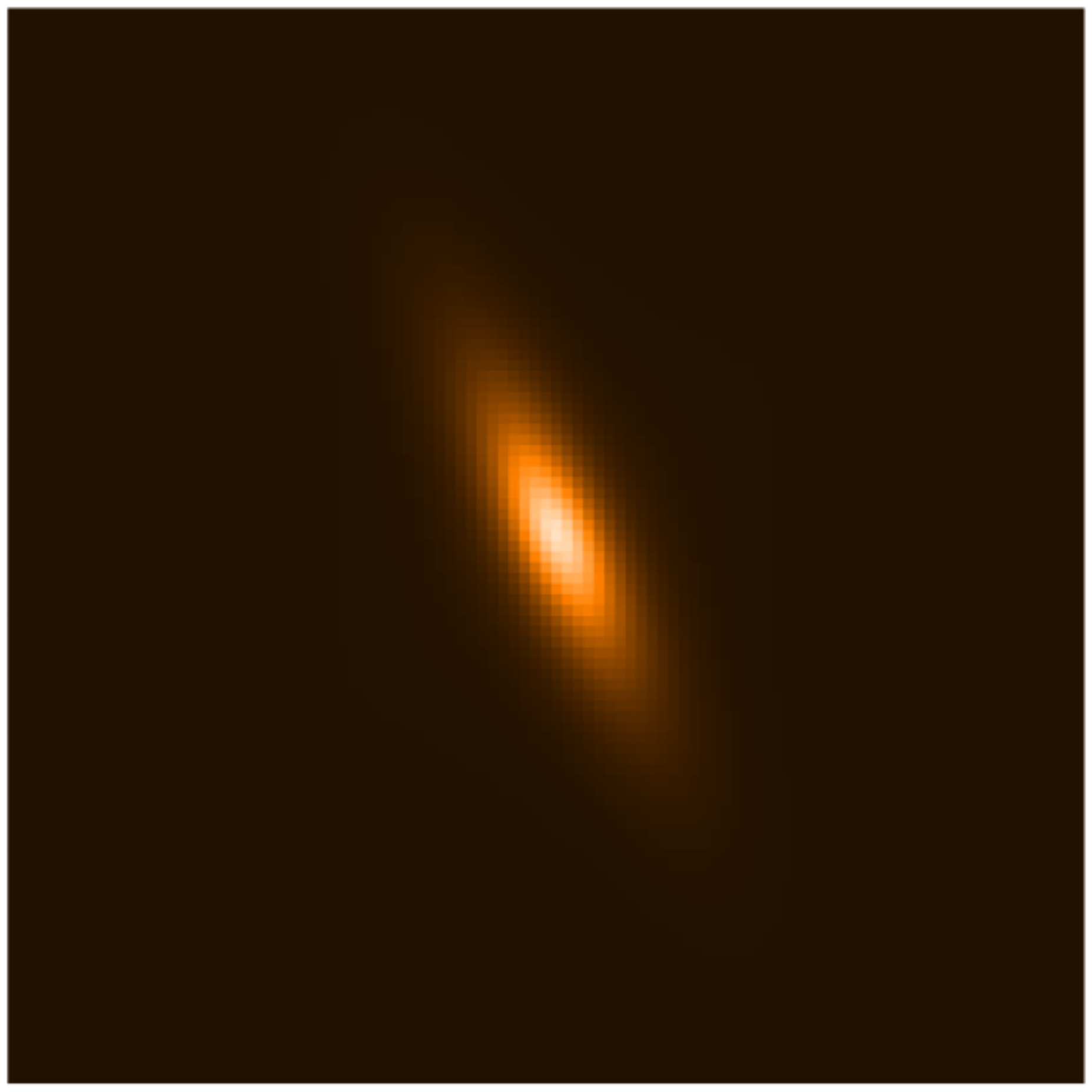}  &
\includegraphics[scale=0.15]{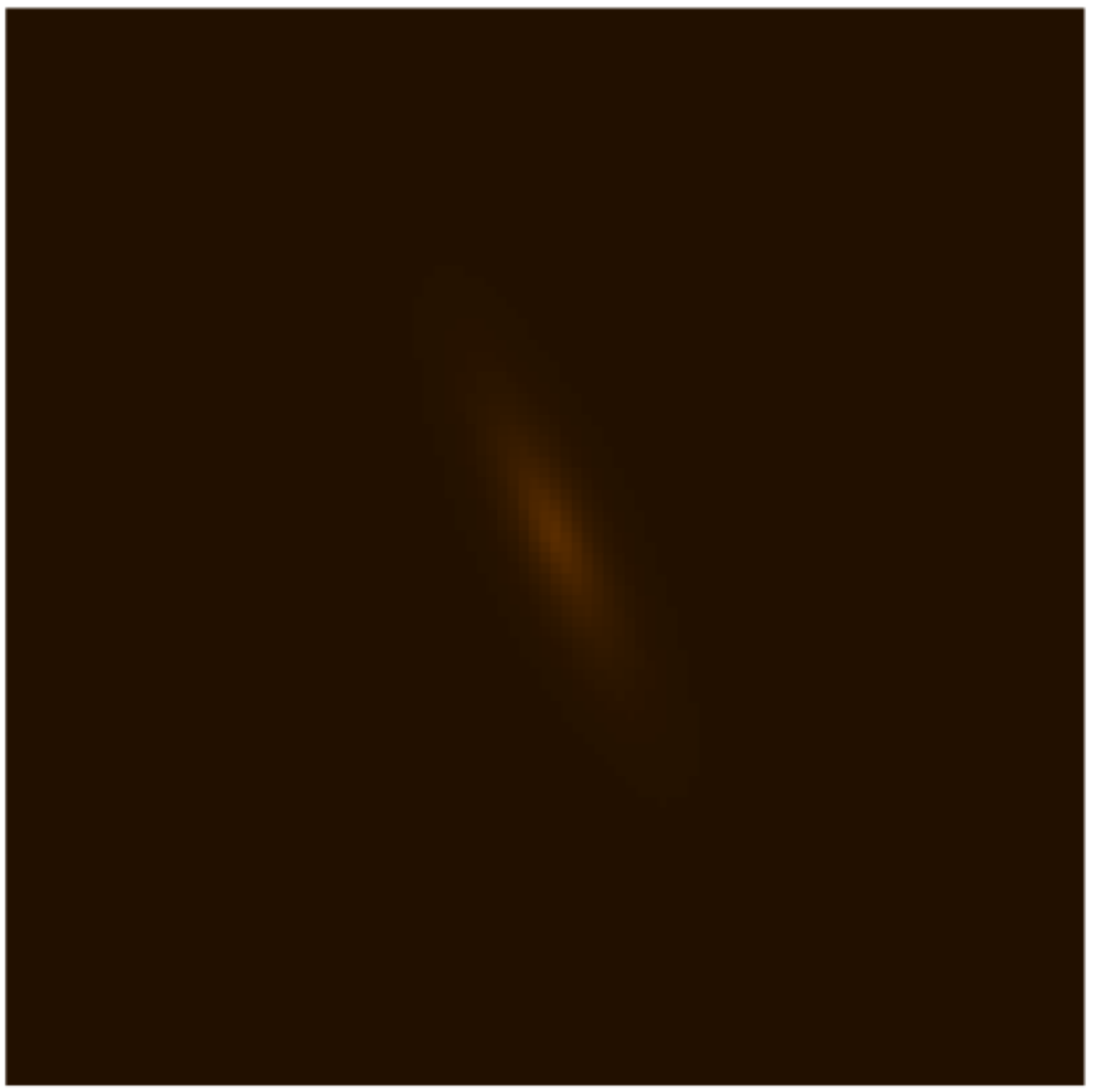}  &
\includegraphics[scale=0.15]{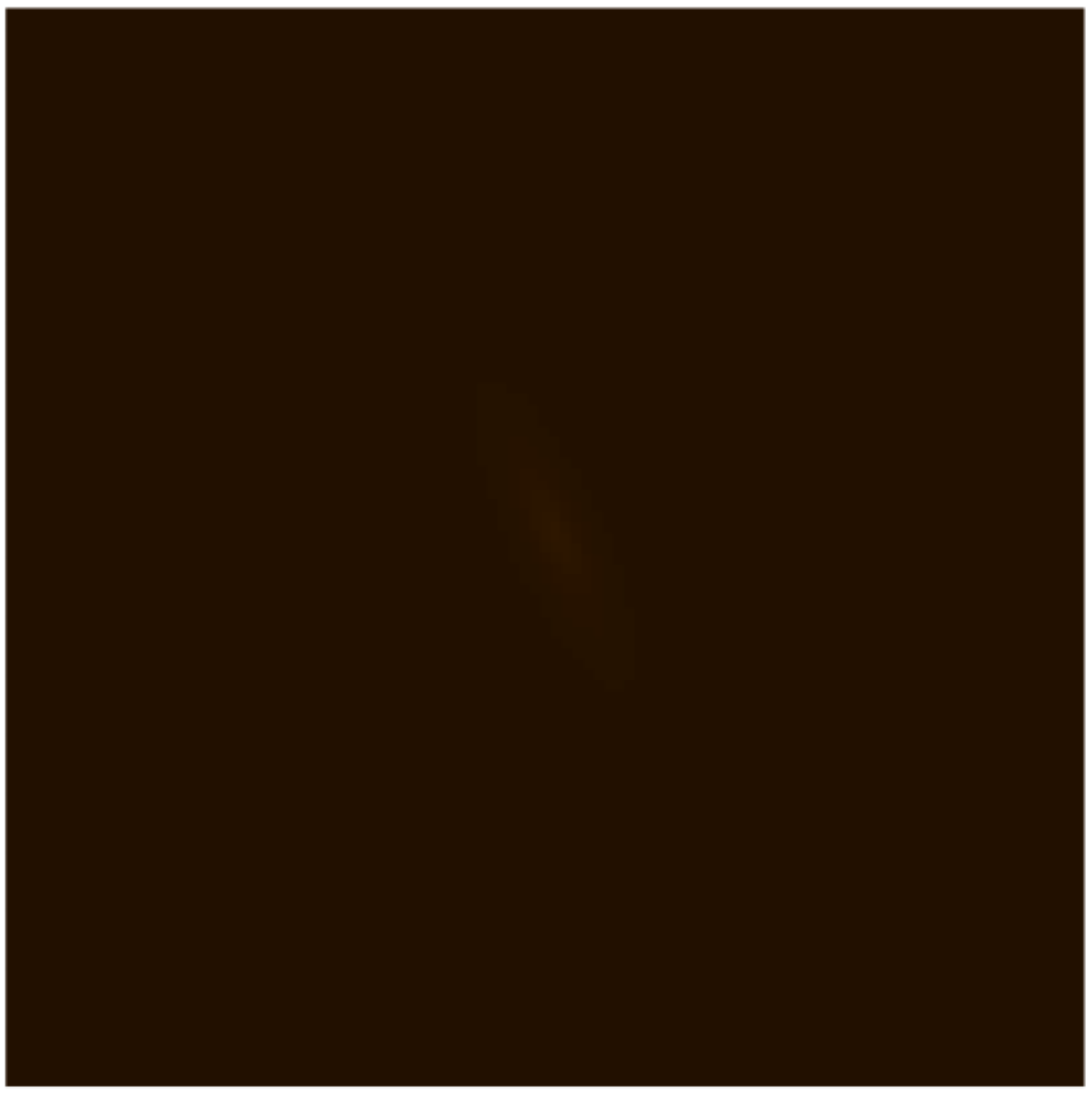}  \\
\includegraphics[scale=0.15]{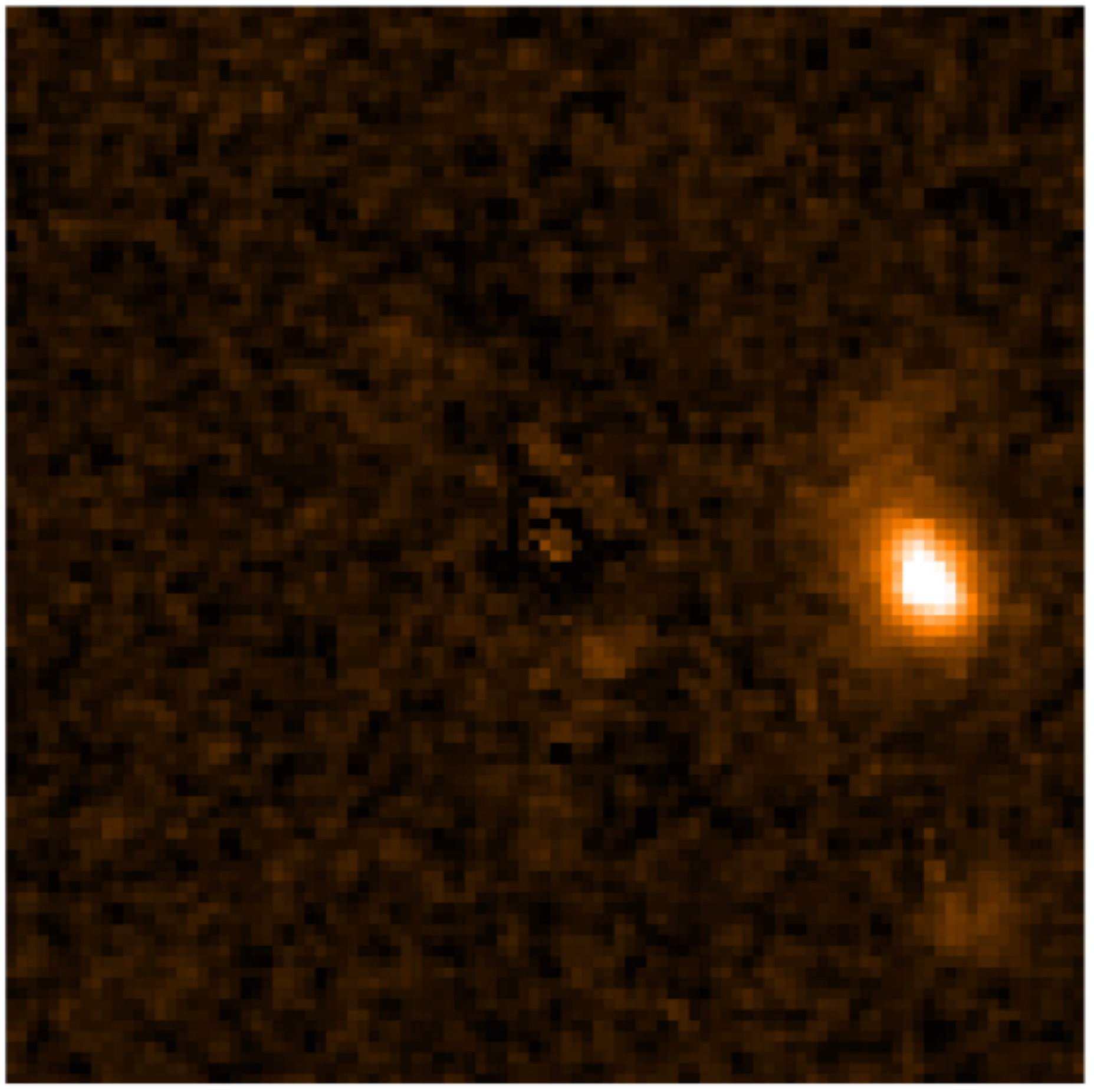}  &
\includegraphics[scale=0.15]{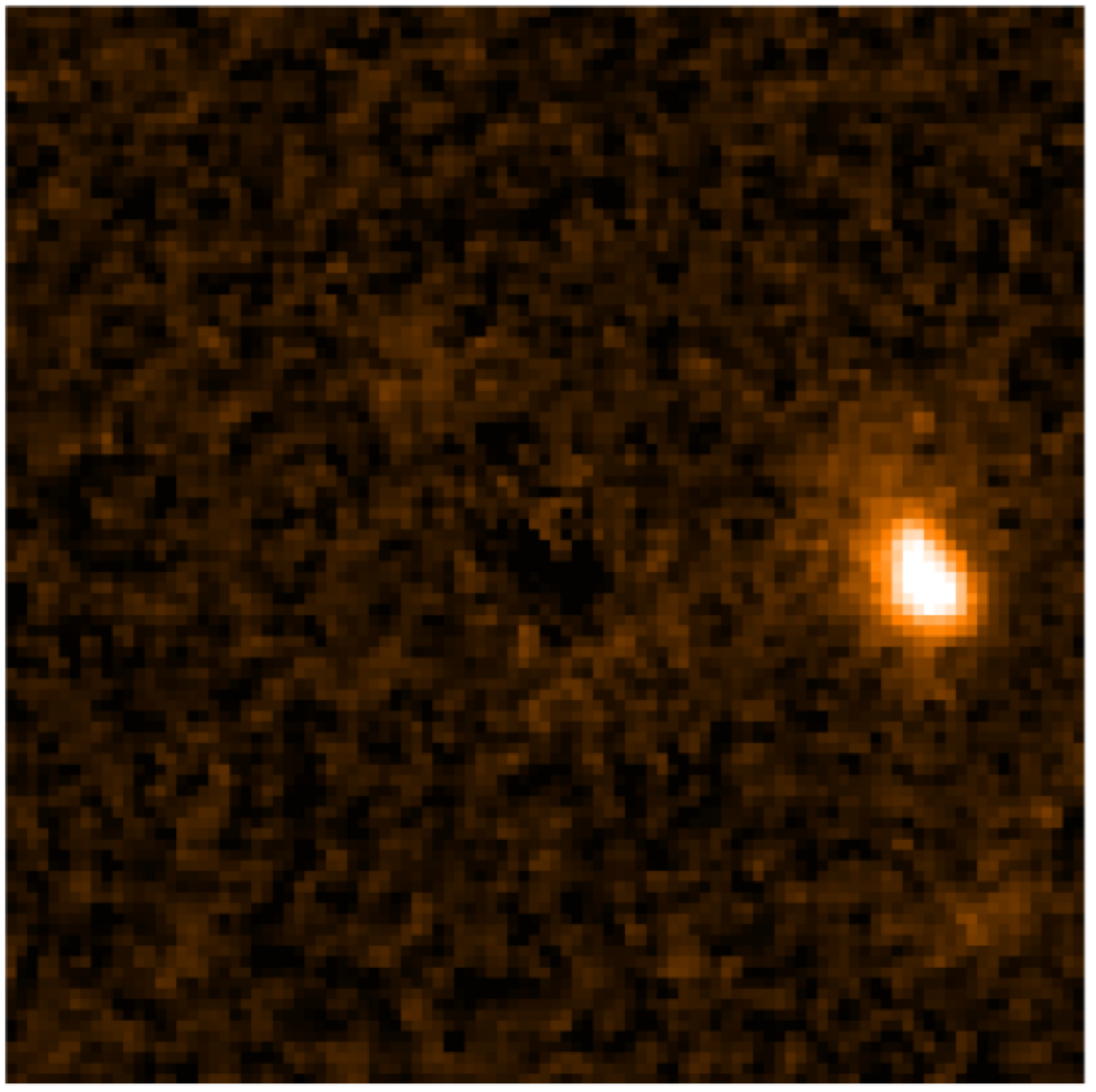}  &
\includegraphics[scale=0.15]{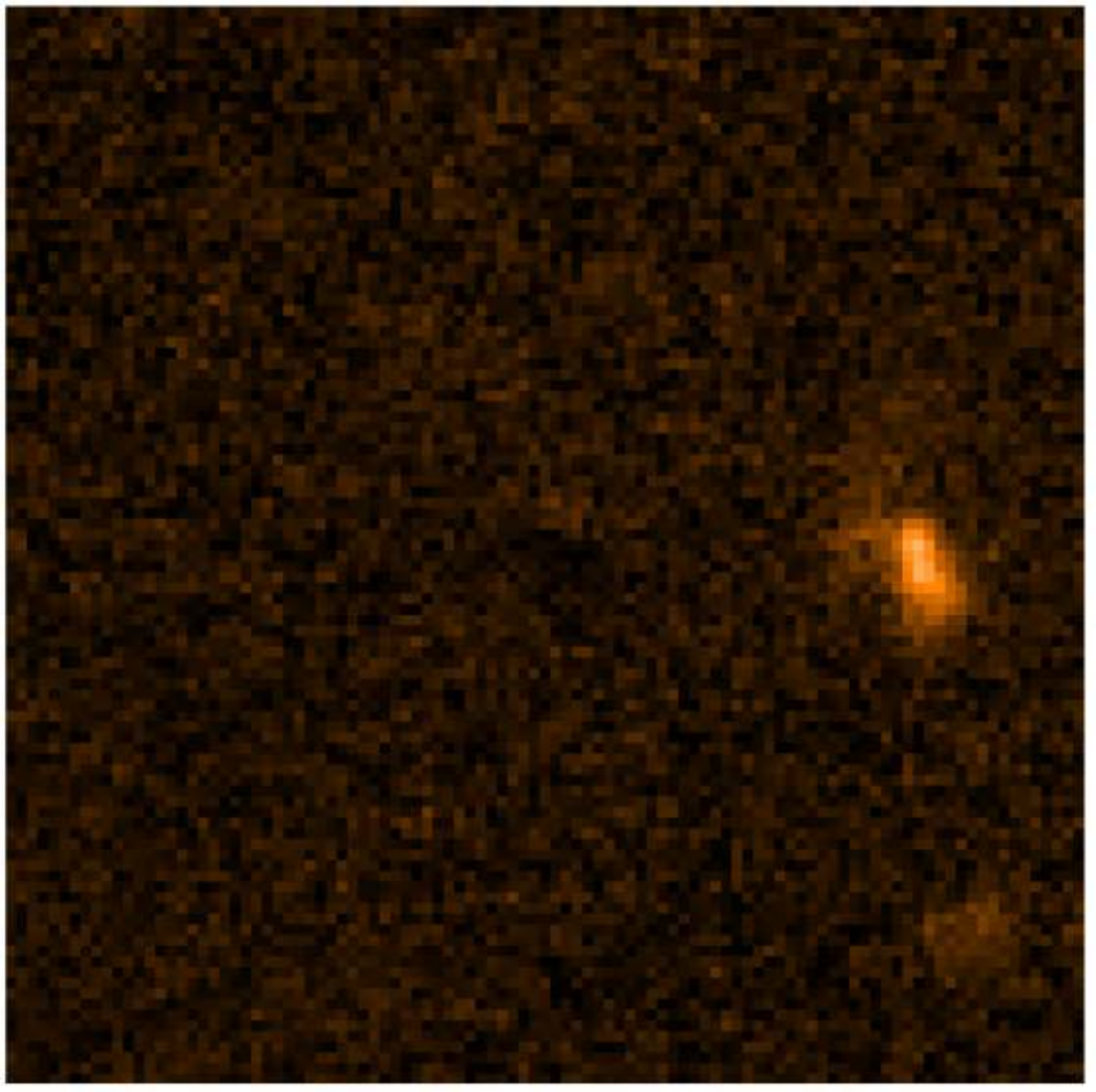}  &
\includegraphics[scale=0.15]{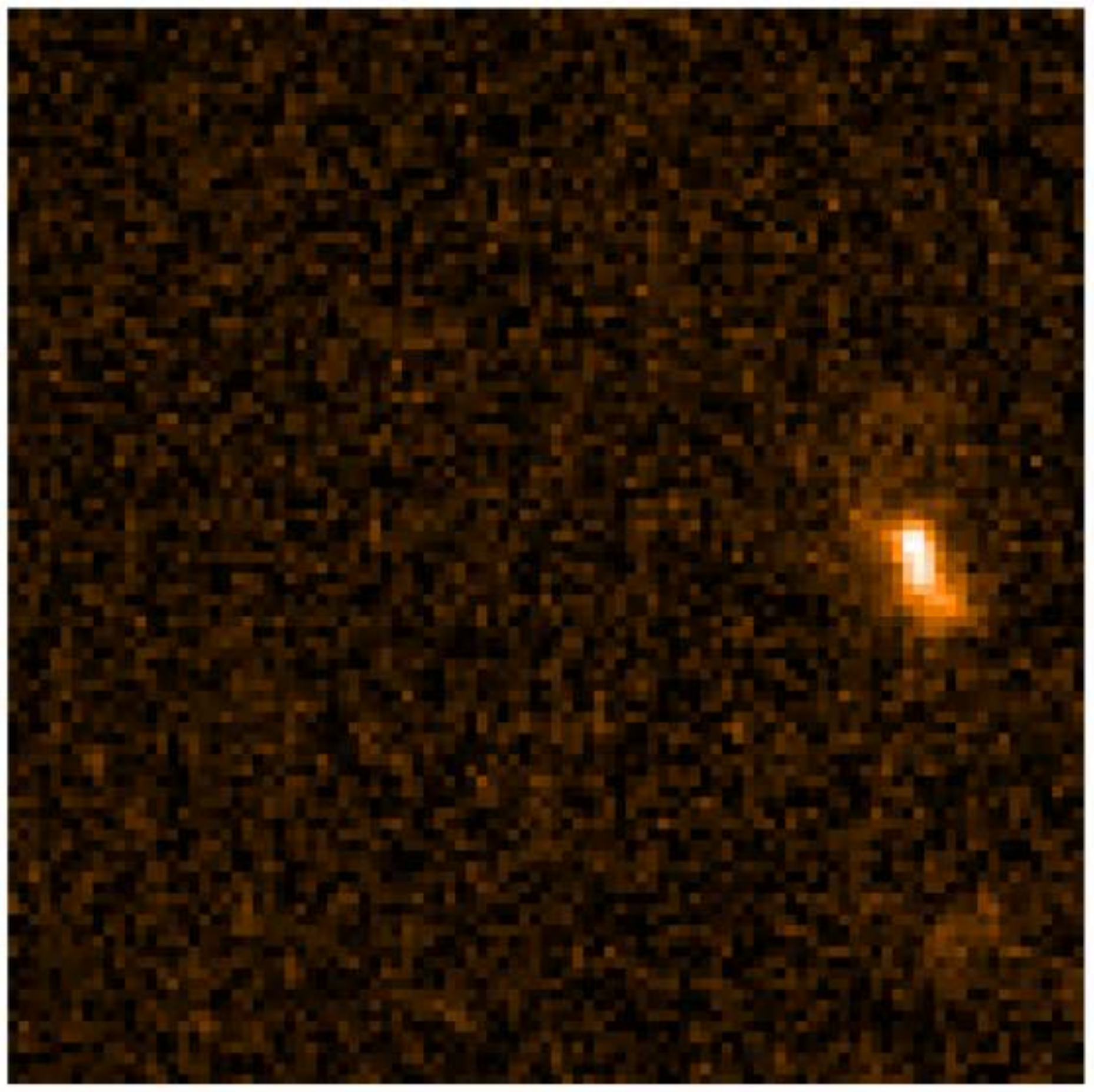}  \\
\end{tabular}
\caption[Image stamps of an example fit for a bulge+disk object with $B/T<0.5$.]{Image stamps for an example fit for a bulge+disk object with $B/T<0.5$. Here the stamps are again ranked from left to right by decreasing wavelength: $H_{160}$,$J_{125}$, $i_{814}$ and $v_{606}$, . The vertical placement is as follows, from top to bottom: images, best-fit combined bulge+disk models, individual best-fit bulge components, individual best-fit disk components, combined model residuals.}
\label{fig:bulge_less_50}
\end{center}
\end{figure*}

Thus, we have chosen to adopt fixed $r_{e}$ and $b/a$ parameter models for this multi-wavelength morphological analysis in order to avoid additional degrees of freedom, which are not required and may introduce an additional degree of bias. It should also be noted that the adoption of the fixed morphological parameter approach delivered magnitudes for each component over the four-band wavelength range available in this study, which transpire to yield realistic colours for the bulges and disks. This feature of our fits verifies the validity of this approach and further demonstrates the power in applying this simplified and well-constrained procedure.
A representative example fit across all four bands is shown in Fig.\,\ref{fig:bulge_less_50} for an object with an $H_{160}$-band best-fit model with $B/T<0.5$ to demonstrate how the contribution from the bulge component decreases at bluer wavelengths,while the disk-component remains more prominent in the blue bands, but is well fit by the fixed-parameter model. Further examples of the multi-band fitting of objects with best-fit  $H_{160}$-band single and multiple-component models are also given in Appendix A.

\section{Double-Component SED Fitting}
The photometry generated from the above multi-wavelength morphology fitting was then used to conduct separate bulge and disk SED fitting in order to provide more accurate masses for the two components. For this analysis we used only the 178 objects in the combined UDS and COSMOS sample which require both bulge and disk components. As detailed in \citet{Bruce2012} one of the conditions of the morphological decomposition is the criteria that each component in an acceptable model fit must contain $\geq10\%$ of the total flux of the object. This safeguards against selecting models which have a spurious additional component. However, for the purposes of individual component SED fitting, those objects from the original fits which chose to incorporate a ($\geq10\%$) PSF component in their best-fit model were modelled with only a bulge and disk component to provide photometry as although the addition of a PSF component can significantly influence the fitted parameters of a single-S\'{e}rsic model fit,  in the majority of cases the bulge and disk only model still provides a statistically acceptable fit. 

This decision to remove the PSF component was taken as several tests into the correlation between the presence and strength of a PSF component with x-ray, $24\mu$m and radio counterparts have provided no clear evidence that the adoption of a PSF component is motivated by the presence of either an AGN or a nuclear starburst. Moreover, as discussed in Section 3.1, from our mock galaxy simulations we find that there are a small number of cases ($\sim3\%$), where the best-fit multiple-S\'{e}rsic model will adopt a PSF component despite the fact that no component was included in the model galaxy. Therefore, despite this low level of degeneracy, it is not clear how any such fitted PSF component, or in fact any genuine PSF component, should be correctly physically modelled in an SED fit.

It is well known that the physical properties, most importantly the stellar masses, fitted by the template fitting SED approach can be strongly influenced by how well constrained the SEDs are by data across a broad wavelength range, particularly at the red end, where {\it Spitzer} IRAC data are important. In light of this, and given the limitations of using only the four-band $H_{160}$, $J_{125}$, $i_{814}$ and $v_{606}$ decomposed photometry available from our CANDELS analysis, we have adapted the SED fitting code of \citet{Cirasuolo2007} to additionally constrain the model fits to the four-band decomposed photometry at the extreme blue and red ends. This is done by fitting the sum of the bulge and disk photometry to the single-band photometry available at $\lambda<0.6\rm\mu$m and $\lambda>1.6\rm\mu$m which was used previously in the full single-component SED fitting.
 
In brief, the SED fitting code of \citet{Cirasuolo2007} generates a grid of different SED models from the input stellar population synthesis templates, dust correction steps, allowed ages, redshifts and star-formation histories. At each point in this grid a $\chi^{2}$ fit is performed between the data and the SED model using the uncertainty on each photometric data point. In order to allow the additional constraints to be applied at the red and blue ends,  we have adapted the SED fitting code to allow simultaneous fitting to both the bulge and disk component photometry separately. At every step of this model SED grid, we then imposed the condition that the $\chi^{2}$ of that fit must not only include the fit of the model bulge and disk SEDs to their respective photometry, but also that the sum of the bulge and disk components in the u' (and B-band for UDS), K/$K_{s}$, $3.6\mu$m and $4.5\mu$m bands are fitted to the overall photometry measured for the entire object in those bands. 

The errors on the input decomposed photometry data points were determined for each component by constructing a grid of different component magnitudes ranging from $-10$ to $+10$ magnitudes (in steps of 0.1 mags) from the component magnitude in the best-fit $H_{160}$ model (fixing all parameters) and re-running GALFIT to generate $\chi^{2}$ values for each of these different magnitude grid steps. This allowed the error on, for example, the bulge component magnitude to be estimated including also the (correlated) uncertainty on the disk component magnitude by exploring the two-dimensional parameter space. The error on each component magnitude was then taken to be the $1-\sigma$ level of the bulge-disk  $\chi^{2}$-contour for each object, unless this value fell below the minimum error limit imposed from the error associated with the direct measurement of the object magnitudes from the image, where this value is taken to be 0.1 mag, in which case this 0.1 mag error was adopted.

Upon conducting tests of this approach and comparing the final SED fits of the separate and combined bulge and disk components to the single-component original SED fit it was discovered that in the u' and B-bands often the flux of the total object measured from the {\sc sextractor} iso-magnitude was significantly brighter than the sum of the modelled bulge and disk components. This is due to the effect of close companions in the $H_{160}$-band image which, in that band, contribute a negligible amount to the total flux of the object and have not been de-blended by {\sc sextractor}. However, as the $H_{160}$ {\sc sextractor} segmentation map has been used in dual mode to measure the iso-magnitude fluxes of each of the objects in the accompanying bands, the flux of these close companions can begin to dominate the total flux at the shorter wavelengths (where the high-redshift galaxy is faint).

We have accounted for this effect by re-measuring the flux for our objects in the blue bands within a 2.5\,arcsec radius aperture, which is sufficient to include the full extent of the flux from the object, whilst minimising the contamination from the close companions. It is these values which are used to constrain the SED fits in the $u'$ and $B$-bands.

\begin{figure*}
\begin{center}
\includegraphics[scale=0.9]{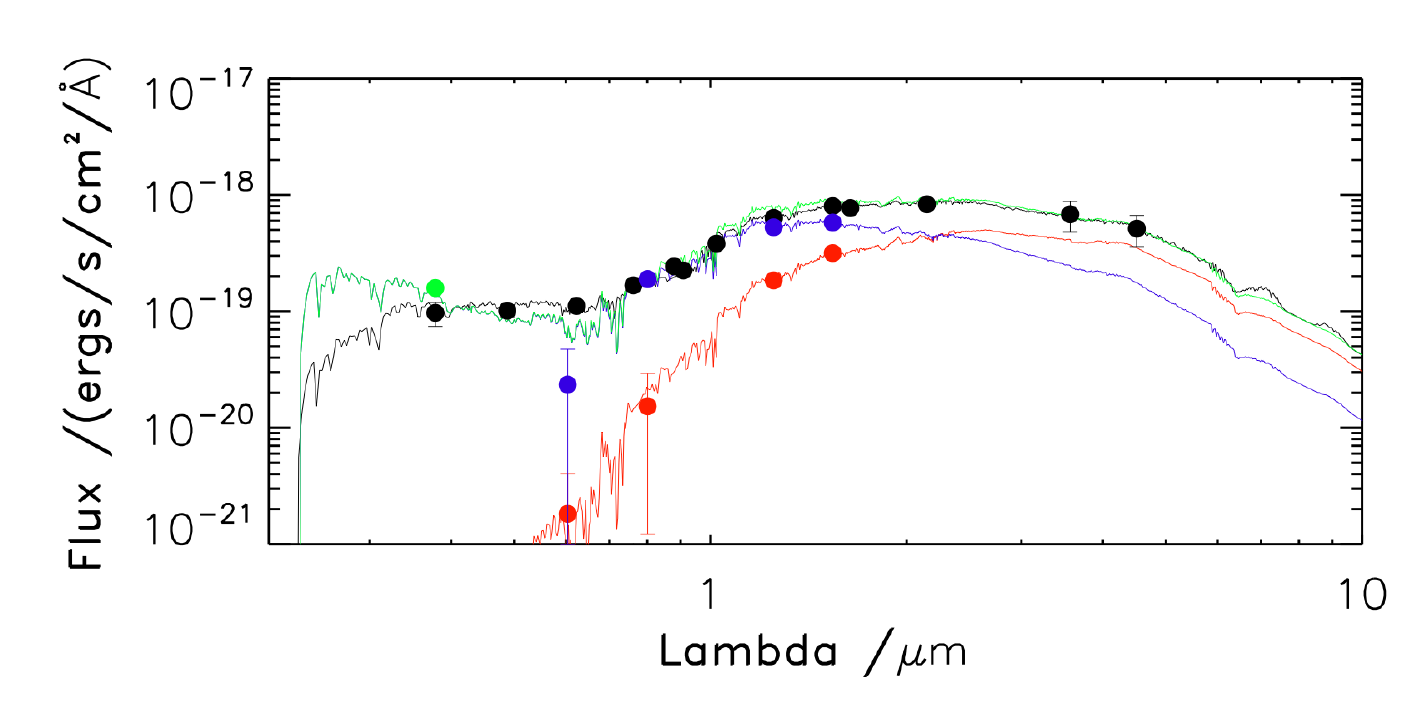} 
\end{center}
 \caption[SED models for example fits]{The SED fit for the example object shown in Fig.\,\ref{fig:bulge_less_50}. Plotted as black data-points and the solid black line is the total, overall galaxy photometry (with its associated error-bars) and the corresponding best-fit single-component SED. In blue is the modelled disk component photometry and the corresponding best-fit decomposed disk SED model, and in red is the modelled bulge photometry and the best-fit decomposed bulge SED model. Over-plotted in green is the sum of the best-fit bulge and disk SED models, which can be directly compared to the single-component fit in black and can be seen to be in good agreement with the overall galaxy photometry. Finally, the green point, and its error-bar, is the re-measured 2.5\,arcsec radius photometry for the $u'$-band.}
\label{fig:sedfit}
\end{figure*}

An example SED fit is given in Fig.\,\ref{fig:sedfit} for the same object shown in Fig.\,\ref{fig:bulge_less_50} with an $H_{160}$-band $B/T<0.5$ best-fit model. Here the single-component photometry points and best-fit SED are given in black (points and line, respectively), and have been over-plotted with the disk component photometry and the corresponding best-fit decomposed disk SED model in blue, and the best-fit bulge component photometry and decomposed bulge SED model in red. The sum of the best-fit bulge and disk SED models is shown in green, and can be directly compared to the single-component fit in black. Finally, the green points, and their error-bars, are the re-measured 2.5\,arcsec radius photometry for the blue bands. as this object is in the UDS field it has re-measured photometry for both the $u'$ and $B$ bands. Further example double-component SED fits are given in Appendix B.

\subsection{The effect of stellar template choice on stellar-mass estimates and ages}
During the simultaneous SED fitting of the separate bulge and disk components the dust attenuation, ages and star-formation histories of each component were allowed to vary freely and independently. This provides a clear distinction between our approach for multiple-component SED fitting, based purely on morphological decompositions, and the ``double-burst model'' fits of e.g. \citet{Michalowski2012}, who fit one set of photometry points with a composite of constrained old and young stellar components.

\begin{figure*}
\begin{center}
\includegraphics[scale=0.9]{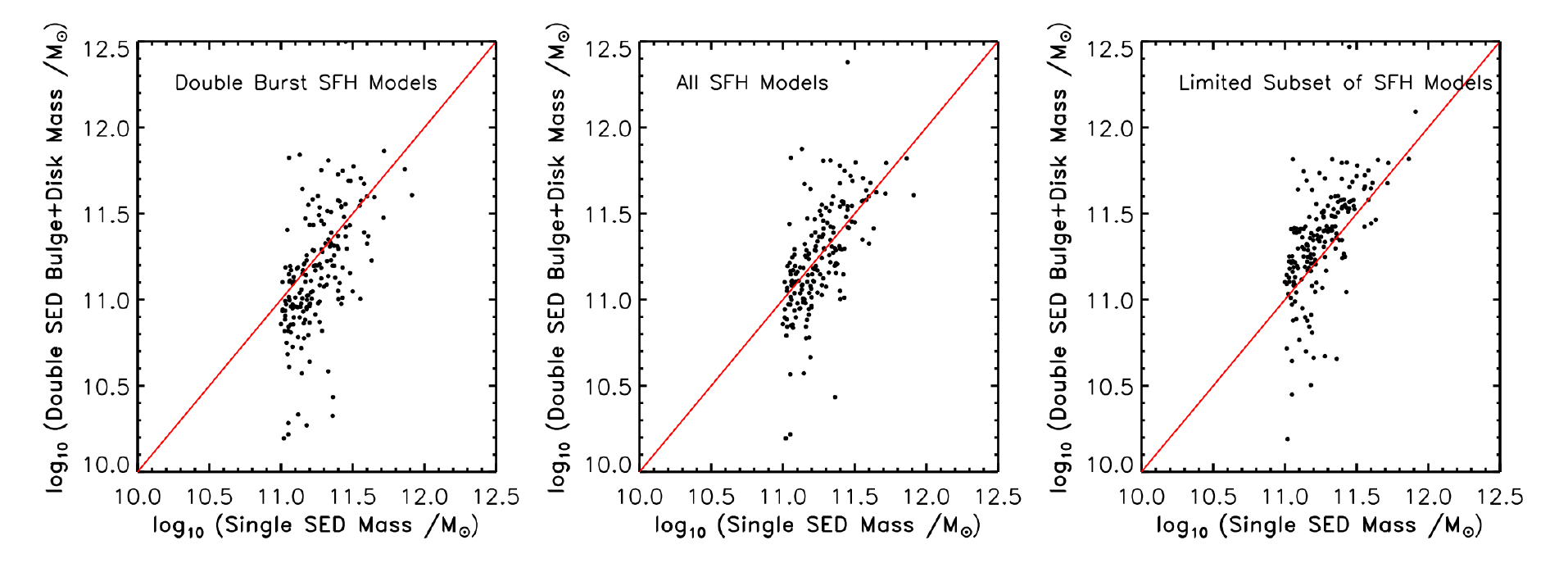} 
 \caption[Stellar-mass estimates from single and double-component models.]{Stellar mass estimates from double-burst models (left), double $0\leq \tau {\rm (Gyr)}\leq 5$ models (middle) and double $0.3\leq \tau {\rm (Gyr)}\leq 5$ models plotted against the mass estimates from the single-component SED fits. These panels clearly demonstrate that the double $0\leq \tau {\rm (Gyr)}\leq 5$ models provide stellar-mass estimates which are more comparable to the single-component masses. }
\label{fig:massmodels}
\end{center}
\end{figure*}

The adoption of multiple stellar population components can significantly influence the best-fit SED models so we have examined the impact of adopting different constraints and limitations on the input model parameters during our multiple-component fitting. For the single-component SED fitting, a minimum age limit of 50 Myr was necessary to ensure that, when fitting old objects with some on-going star-formation, the $\chi^{2}$ minimisation model parameter space did not become restricted to un-physically young ages with large amounts of dust extinction.
By adding the extra degrees of freedom to the models associated with the second component no such age restriction was needed for the multiple-component fitting.
We have also experimented with limiting the star-formation histories implemented in the models which are fitted (always adopting BC03 models). By running fits with both components limited to i) pure burst histories ($\tau$=0) ii) the components limited to $0.3\leq \tau {\rm (Gyr)}\leq 5$ iii) the full set of models ($0\leq \tau {\rm (Gyr)}\leq 5$), we have explored the effect on the mass determinations and the accompanying fitted ages of each component.

Having conducted this comparison we find that the full $0\leq \tau {\rm (Gyr)}\leq 5$ set of star-formation history models produce total bulge+disk component masses which are most comparable to the single-component SED fit masses (central panel of Fig.\,\ref{fig:massmodels}). Reassuringly, these models also represent the scenario in which we have applied the least constraints to the physical models fitted. By including the additional degrees of freedom from the second component there is no longer any physically-motivated reason to restrict either component in age or star-formation history, as a second younger burst, or exponentially decaying star-forming population can reasonably account for any continued or recent star-formation superimposed on an older, redder population.  A complete discussion of the ages and star-formation histories of the best-fit models is given in Appendix C.

The ability of the multi-wavelength photometry for the bulge and disk components (which has been decomposed based purely on their $H_{160}$-band morphologies) to produce colours which are well fit by physically-motivated SED templates demonstrates the validity of this technique, as it is clear that no fixed correlation or pre-determined trends with colour (such as the polynomial wavelength dependencies implemented in the single-S\'{e}rsic and multiple-component multi-wavelength fitting package of MegaMorph \citet{Haussler2013} and \citet{Vika2013} to fit lower $S/N$ objects) are required.

\subsection{Separate component star-formation rates}

In addition to the overall galaxy specific star-formation rate estimates discussed in Section 2.3, the decomposed specific star-formation rates of the individual components have also been determined from SED fitting of the individual-component photometry using models which include bursts and exponentially declining star-formation histories in the range $0.1\leq \tau {\rm (Gyr)}\leq 5$. Previous single component SED fitting studies by \citet{Wuyts2011} found, when comparing star-formation rates determined from dust-corrected SED fitting with those from combinations of non dust-corrected UV and infra-red contributions ($SFR_{UV+IR}$) that the adoption of very sharply declining star-formation e-folding timescales, $\tau<0.3$ Gyr, provided statistically improved SED fits, but that these fits had stellar population ages which were unrealistically young and had star-formation rates that were systematically lower than the estimates derived from $SFR_{UV+IR}$. As a result, they suggested removing such short e-folding timescales for SED fitting and it is this approach which we have adopted for the SED fitting of the single components. 

However, from the careful SED fitting comparisons conducted for the double-component models we have found that the extra degrees of freedom incorporated with the addition of the second component remove the bias of including these sharply declining star-formation histories in the SED fits. This has been found from examination of the ages of the old stellar component, which show that they are no longer biased towards implausibly young values and show an age distribution comparable to the ages of the single component fits which were limited to the $0.3\leq \tau {\rm (Gyr)}\leq 5$ subset (see Appendix C Fig.\,\ref{fig:ageagree}). Furthermore, by adding a second component we have also allowed the model fits to include the case of burst star-formation histories. The inclusion of a second component allows for the superposition of an old population, which experienced a burst of star-formation in the past, with an additional population exhibiting on-going or slowly-declining star-formation. However, while the incorporation of burst models in the double-component SED fitting has been validated by the above exploration, burst models obviously do not provide any estimates for even low levels of on-going star-formation in young components.

As a result, the star-formation rates derived for the separate-component modelling may be under-estimated in comparison with the star-formation rates derived from the single-component SED fitting, which was forced to have some level of on-going star-formation activity by adopting $0.3\leq \tau {\rm (Gyr)}\leq 5$. Thus, in order to better reconcile these two approaches, for the double-component model results we adopt different sSFR estimates for the several possible modelled scenarios, as follows.

\begin{enumerate}
\item For fits with double-$\tau$ ($0.1\leq \tau {\rm (Gyr)}\leq 5$) models, we use the UV dust-corrected star-formation rates (using the independent component best-fit $A_{v}$ extinction values) from the SED models of each component and divide through by the corresponding modelled mass of those components to provide the sSFR for each component separately. 
\item For fits with double-burst models where both components are older than $500$ Myr, we deem them both to be passive and allocate them a $sSFR= 10^{-12}{\rm yr}^{-1}$. Where one burst component is young ($<500$ Myr) we attribute the entire star-formation rate derived from the limited single-component SED fitting to that young component and calculate the sSFR by dividing through by the mass of that component modelled from the decomposed photometry, and assume the old component is passive and adopt a $sSFR= 10^{-12}{\rm yr}^{-1}$.
\item For a single burst $+\,\,\tau$ model we generate a sSFR for the $\tau$ model based on the UV dust-corrected SED fitted star-formation rate and mass of that component from the decomposed photometry and for the burst model we adopt the passive limit of $sSFR= 10^{-12}{\rm yr}^{-1}$.
\end{enumerate}

\begin{figure}
\begin{center}
\includegraphics[scale=0.4]{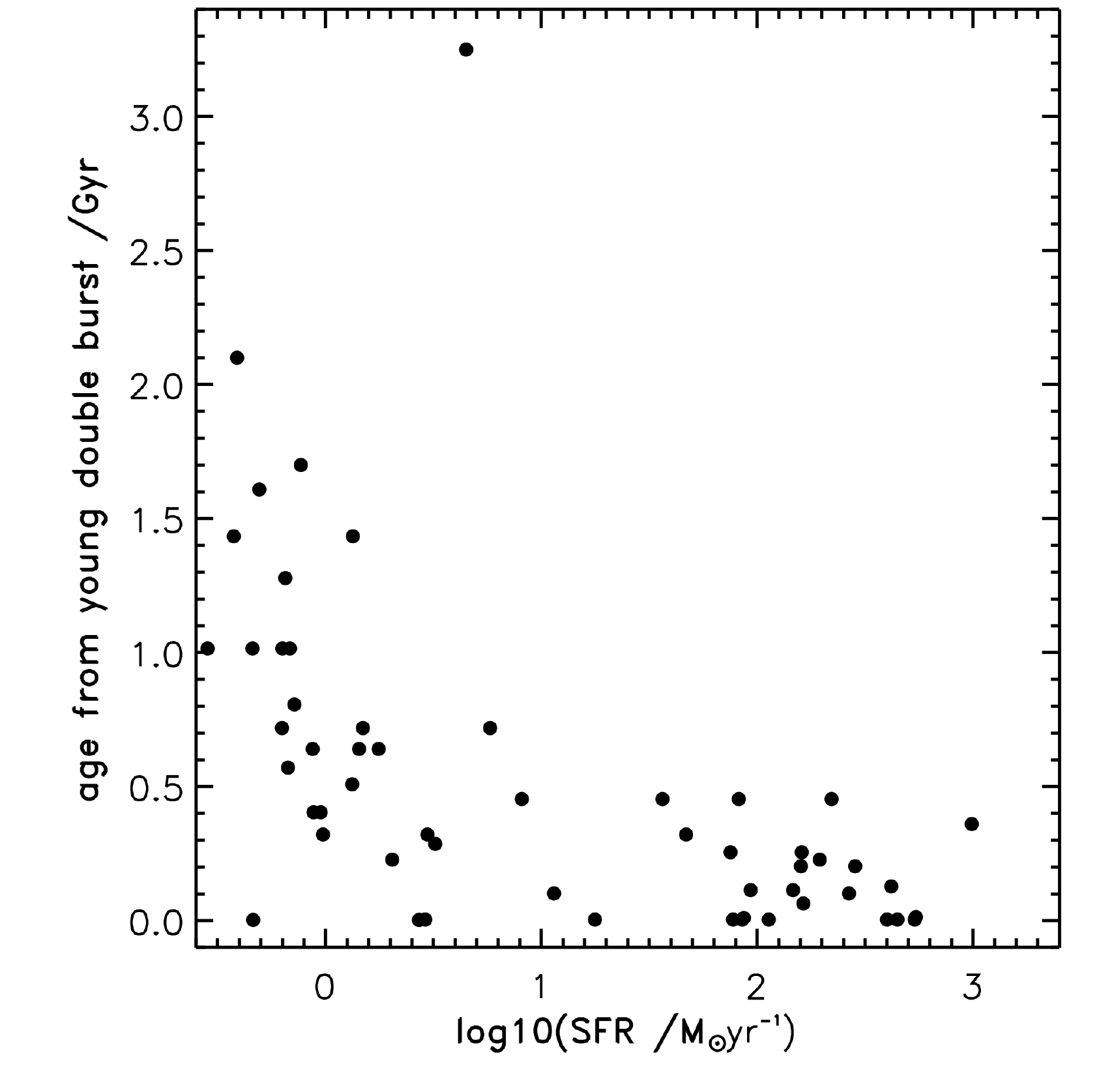} 
 \caption[Ages of the young double-burst models against the SFR from the single-component SED fit.]{Relation between the ages of the young component of the double-burst models and the star-formation rates from the single-component SED fit, illustrating good overall agreement between the ages of the components to which all star-formation activity in the galaxy has been attributed, and the overall star-formation rates for the galaxies.}
\label{fig:sfage}
 \end{center}
\end{figure}

In this way we have best accounted for the mass decomposition of the objects, by adopting a model which can incorporate both a burst and steeply declining star-formation rate for one or both of the components, and have provided an indication of the star-formation rates in each component, which otherwise would not have been achievable for model fits of burst star-formation histories. In fact, Fig.\,\ref{fig:sfage} demonstrates the general correlation between the ages of the young components from the double-burst models and the star-formation rates estimated from the single-component SED fitting.
Additional checks of this approach are presented in Appendix C.

\section{Morphological Evolution}

The evolution of the overall morphology of galaxies at $1<z<3$ can provide important insights into the physical processes which govern galaxy evolution, especially as there is growing evidence that it is within this epoch that massive galaxies are undergoing dramatic structural transformations (e.g. \citealt{Buitrago2008, Mortlock2013, ForsterSchreiber2009}). For completeness, we first present the separate CANDELS-UDS and COSMOS samples split by fractions into the different best-fit models, to allow comparison between the fields.

\begin{table*}
\begin{tabular}{m{3cm}m{2cm}m{2cm}m{2cm}m{2cm}m{2cm}m{2cm}}
\hline
&
bulge &
bulge
\newline +PSF &
disk &
disk
\newline +PSF &
bulge
\newline +disk &
bulge+disk
\newline +PSF \\
\hline
UDS &
$10\pm2\%$ &
$1\pm1\%$ &
$17\pm3\%$ &
$8\pm2\%$ &
$59\pm7\%$ &
$5\pm2\%$ \\
COSMOS &
$15\pm3\%$ &
$2\pm1\%$ &
$15\pm3\%$ &
$13\pm3\%$ &
$42\pm6\%$ &
$13\pm3\%$ \\
Combined&
$12\pm2\%$ &
$2\pm1\%$ &
$16\pm2\%$ &
$10\pm2\%$ &
$51\pm5\%$ &
$9\pm2\%$ \\
\end{tabular}
\caption{Top : the percentages of the re-selected UDS sample of 184 objects with multiple-component best-fits corresponding to each of the six fitted models. Middle: the percentages of the final COSMOS sample of 163 objects for each best-fit model. Bottom: the percentages of the combined UDS+COSMOS samples of 347 objects for each of the best-fit models.}
\label{table:tab4}
\end{table*}

From Table \ref{table:tab4}, it can be seen that, overall, there is good agreement between the fractions of objects best-fit by the different models between the CANDELS-UDS and COSMOS samples. However, this direct comparison also highlights that there are fewer objects fitted with a bulge+disk model in COSMOS compared to UDS, but more have a bulge+disk+PSF best-fit model. Nevertheless, this difference is small, and while only part of the discrepancy can be attributed to cosmic variance \citep{Newman2002}, the small number statistics involved do not provide significant evidence for any biases in the fitting of morphologies between these two fields. Moreover, it should also be noted, from this table, that overall the fractions of ``pure'' bulge and disk galaxies are statistically comparable in both fields. The fractions for the overall combined sample are given in the bottom line of Table \ref{table:tab4}, where they can be seen to be statistically consistent with those given for both the UDS and COSMOS fields.

\begin{figure*}
\begin{center}
\includegraphics[scale=0.7]{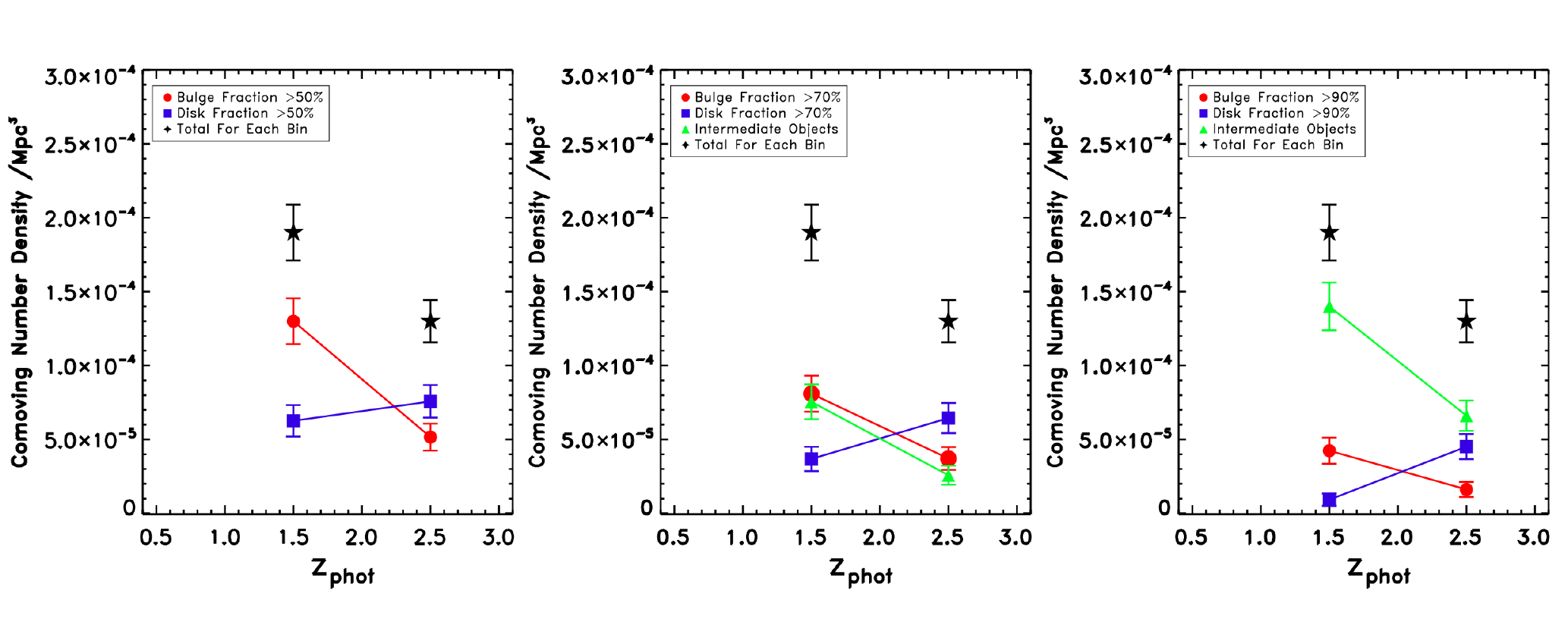}\\

\includegraphics[scale=0.7]{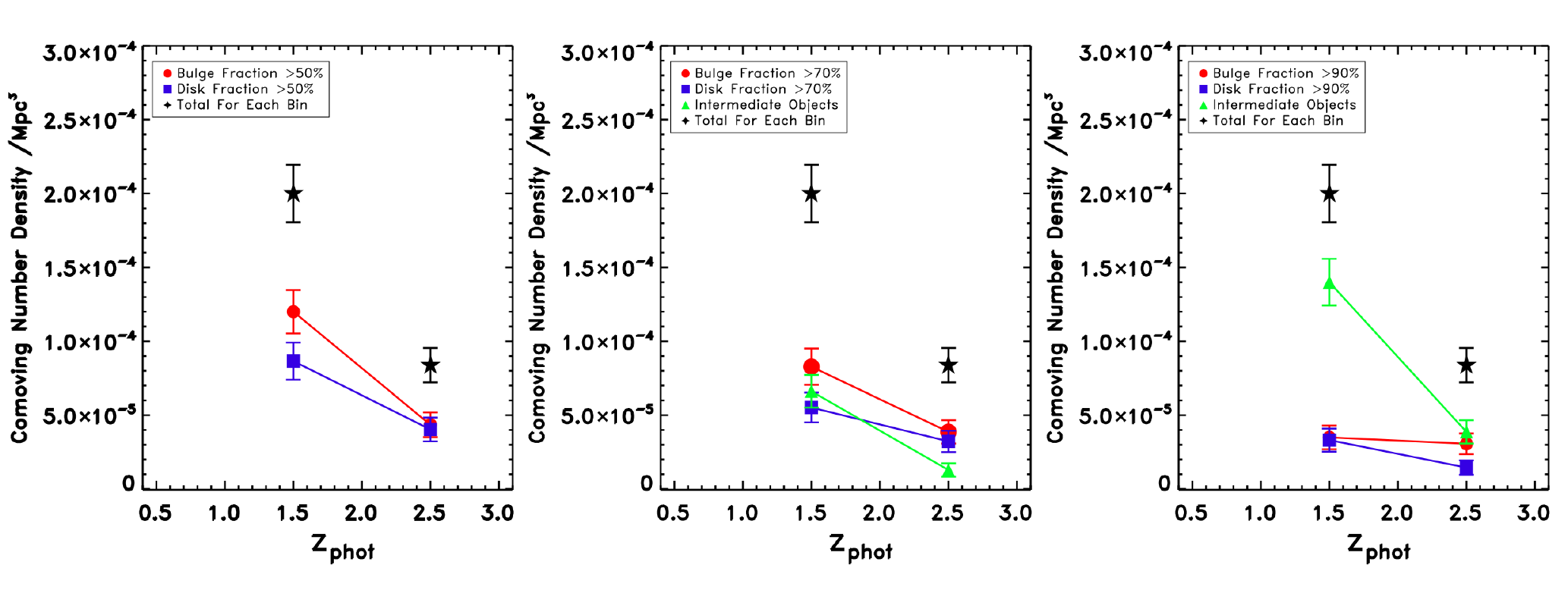}\\

\includegraphics[scale=0.7]{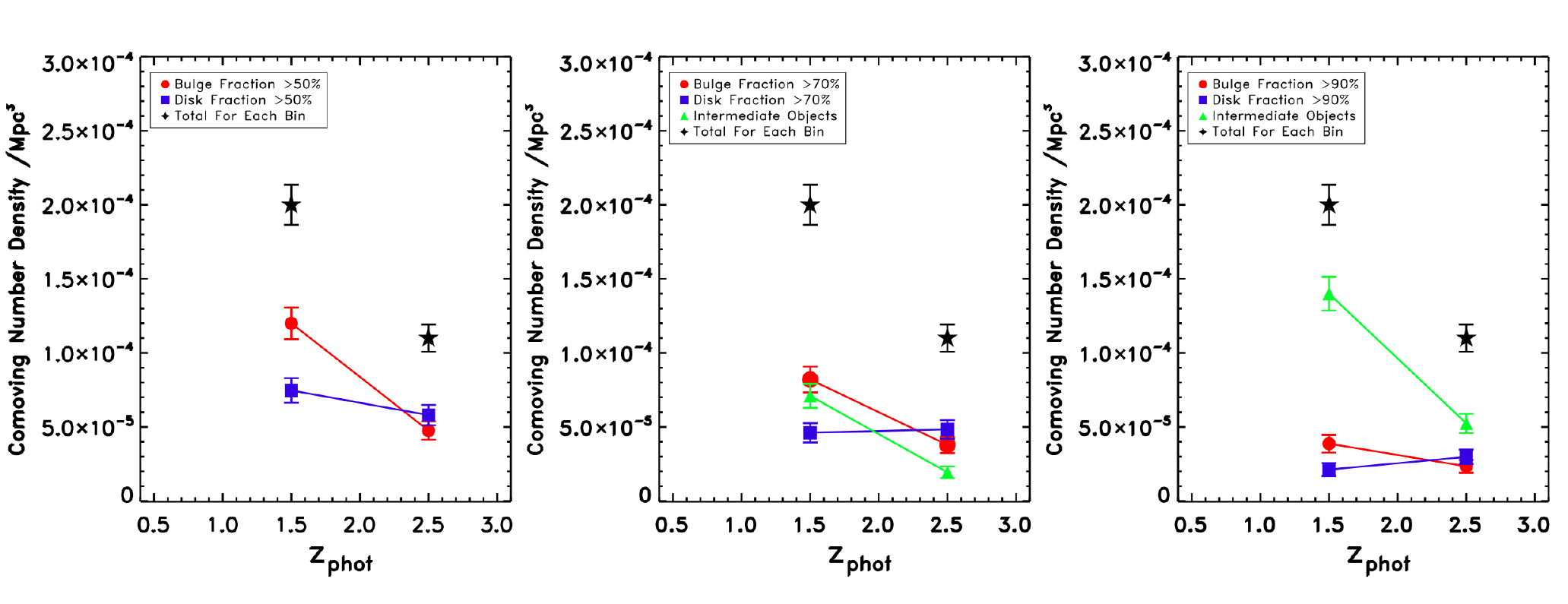}
\end{center}

 \caption[Co-moving number densities by SED mass fractions]{The co-moving number densities of the samples split by the different bulge and disk-dominated criteria according to their SED-fitted individual component stellar-mass fractions. The top panels show the UDS sample, the middle panels are the COSMOS sample, and the bottom panels are for the combined UDS+COSMOS total sample. The blue data-points and lines are for the disk-dominated galaxies, shown in red are the bulge-dominated galaxies and in green are galaxies classified as intermediate bulge+disk systems. The left-hand panels split the populations according to $B/T>0.5$ and $D/T>0.5$, the middle panels adopt $B/T>0.7$, $D/T>0.7$ and intermediate objects, and the right panels use $B/T>0.9$, $D/T>0.9$ and intermediate objects.}
\label{fig:2}
\end{figure*}

\begin{figure*}
\begin{center}
\includegraphics[scale=0.7]{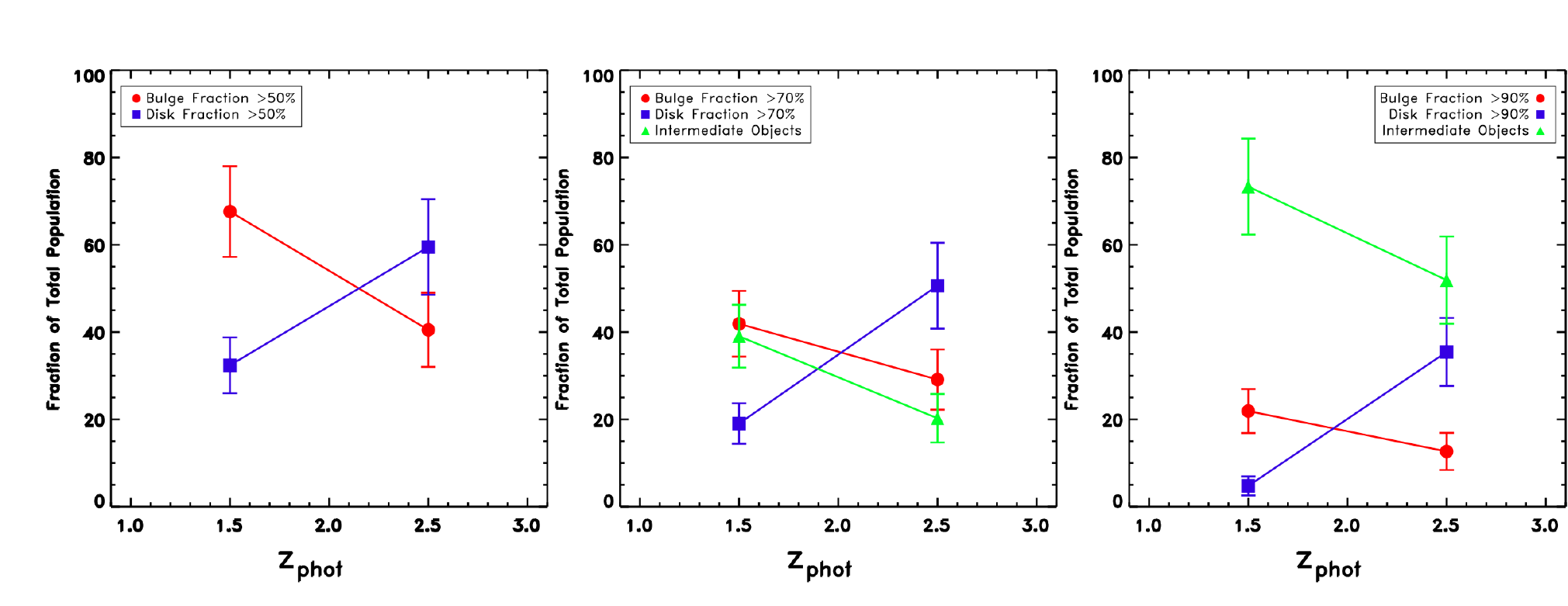}\\
\includegraphics[scale=0.7]{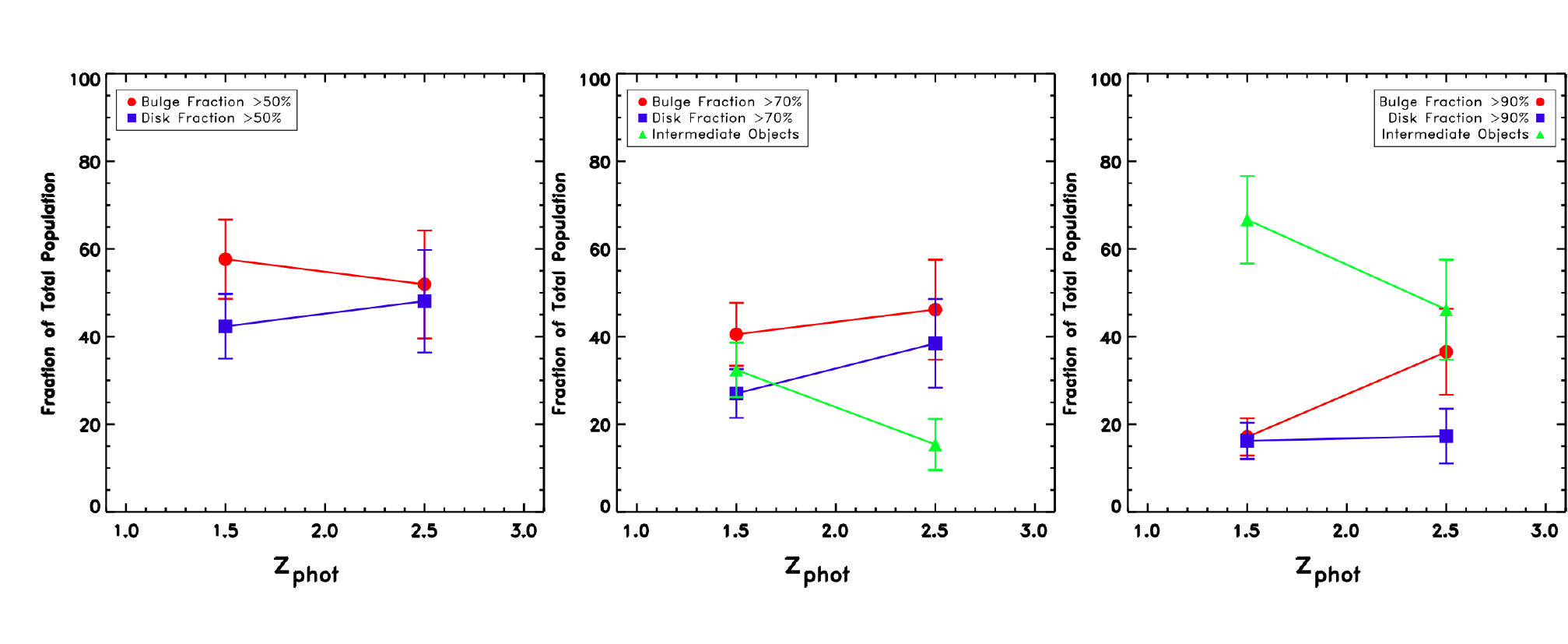}\\
\includegraphics[scale=0.7]{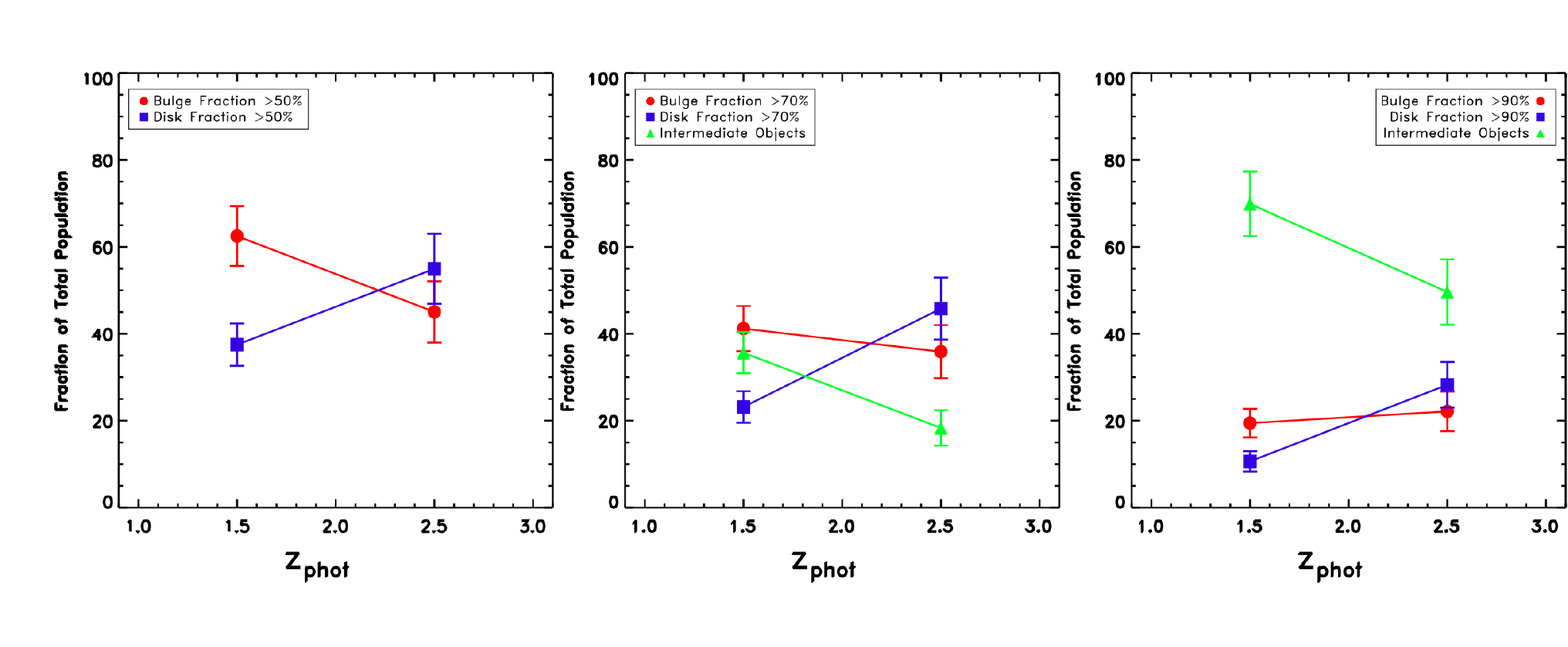}
\end{center}
 \caption[Morphological fractions determined from decomposed mass fractions.]{The fraction of the total number of objects in each sample classified by the different bulge and disk-dominated criteria according to their decomposed SED-fitted stellar-mass fractions. The top panels show the UDS sample, the middle panels are the COSMOS sample and the bottom panels are for the combined UDS+COSMOS total sample. The same configuration is used as in Fig.\,\ref{fig:2}. By adopting the fractions determined from the stellar-mass contributions and comparing them to those from the $H_{160}$ light fractions, it can be seen that the most massive galaxies become increasingly mixed bulge+disk systems morphologically from $z\sim3$ to $z\sim1$, but their masses become even more bulge-dominated. }
\label{fig:4}
\end{figure*}

\subsection{Trends with redshift}
In order to explore how the overall morphologies of these massive $1<z<3$ galaxies evolve with redshift, and keeping in mind that our fitted $B/T$ fractions are accurate to better than $\sim10\%$, we have binned the samples at $z<2$ and $z>2$ and have calculated the co-moving number densities of objects split according to the different morphology discriminators: $B/T>0.5$ and $D/T>0.5$ (where $D/T$ is used to avoid confusion arising from a PSF component contributing a significant fraction); $B/T>0.7$, $D/T>0.7$ and intermediate objects; and $B/T>0.9$ and $D/T>0.9$ and intermediate objects. The results from this binning, determined based on the multiple-component SED fitted masses, are shown in Fig.\,\ref{fig:2}, where the three rows are split further into the UDS, COSMOS and combined samples. These number densities have been over-plotted with the total number density of galaxies in each redshift bin to demonstrate how the overall number of galaxies falls above $z=2$. Compared to co-moving number density fractions obtained from using cuts based on the $H_{160}$ light fractions alone, these plots have a larger number of bulge-dominated systems, which is to be expected from the decomposed mass analysis due to the more evolved stellar populations and consequently higher mass-to-light ratios of bulge systems compared to disks. 
Nevertheless, to first order these plots reveal that, as previously found in \citet{Bruce2012}, in the UDS field the massive galaxy population is dominated by disk structures above $z=2$ and becomes an increasing mix of bulge+disk morphologies below this, with no evidence even by $z=1$ for the emergence of pure bulge systems. The rise in intermediate objects is similar in both the UDS and COSMOS fields. However, the COSMOS sample appears to show a less significant trend for the demise of dominant disks below $z=2$ as the redshift evolution for the $D/T>0.7$ and $D/T>0.9$ cuts is much flatter within the errors, and actually increases with redshift for the $D/T>0.5$ cut. These results appear to reveal significant evidence for a difference between the two fields.

However, when considering the evolution in the number of objects with disk and bulge-dominated morphologies in these plots, the overall evolution in the total number of galaxies in each redshift bin must also be taken into account, as it can clearly be seen that the total number of $z>2$ galaxies is significantly less than the number of objects at $z<2$. In order to better interpret these results, we have plotted the fraction of the total number of objects in each redshift bin which are split into bulge and disk-dominated according to the three different cuts in Fig.\,\ref{fig:4}, again using the decomposed stellar-mass estimates. From inspection of this plot, the variation between the disk-dominated trends with redshift between the fields is somewhat less prominent, although still present, and the evolution of the intermediate bulge+disk fractions remain consistent. This is also the case if the morphological cuts were based only upon the $H_{160}$-band light fractions.

\begin{figure}
\begin{center}
\includegraphics[scale=0.6]{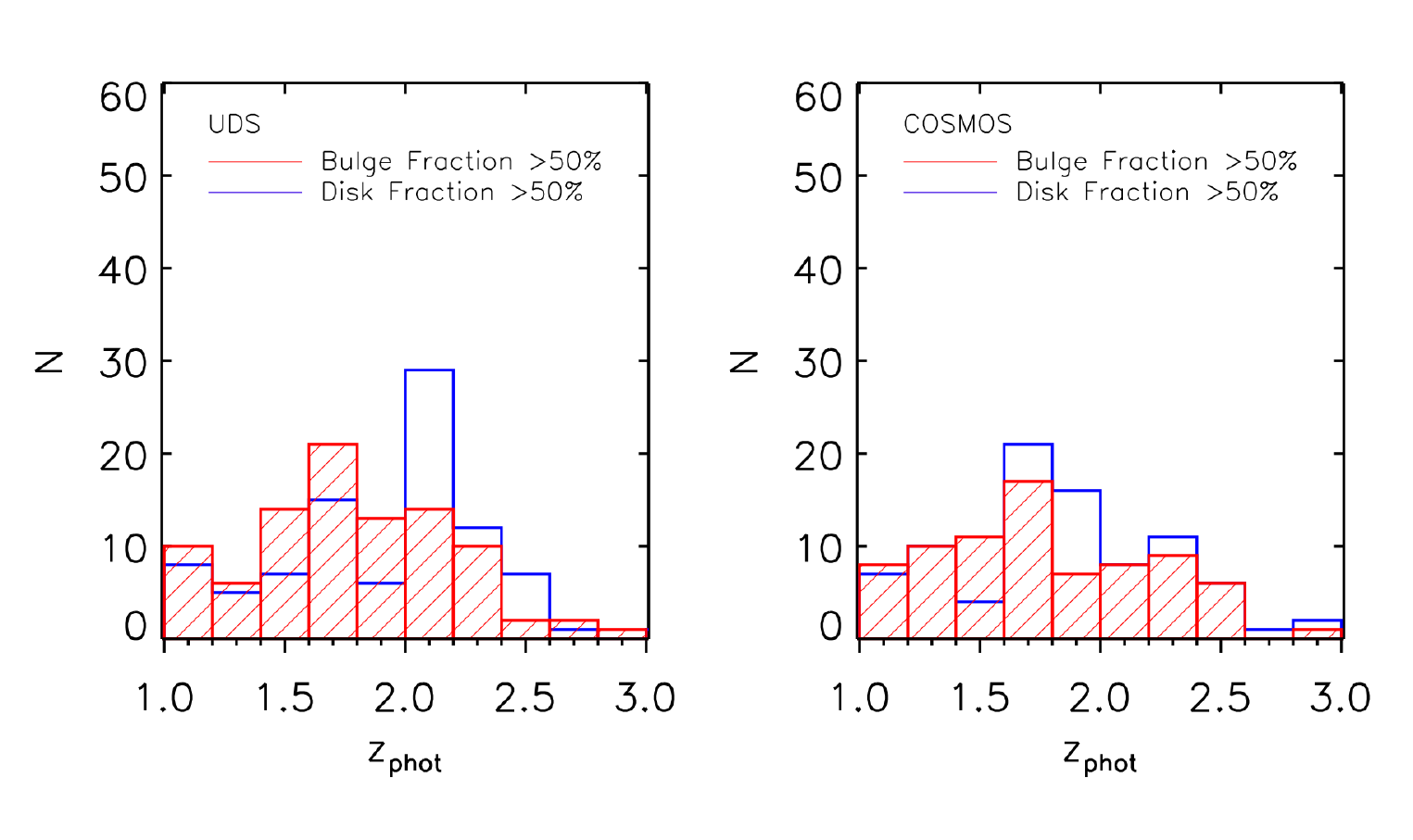} 
\end{center}
 \caption[Redshift distributions of the bulge and disk dominated components in the UDS and COSMOS.]{Redshift distributions of the bulge and disk dominated components using the $B/T>0.5$ and $D/T>0.5$ criteria for the UDS and COSMOS samples separately. Comparison of these distributions reveals a peak in the redshift distribution of the disk-dominated galaxies in the UDS at $z\sim2$, possibly indicative of redshift focussing in this sample. However, this comparison also shows that the number of bulge-dominated galaxies falls-off more steeply in the UDS than in COSMOS. This abundance of $z>2$ bulge-dominated galaxies in COSMOS will also contribute to the flatter evolution of the fraction of disk-dominated galaxies in COSMOS. }
\label{fig:5}
\end{figure}

This discrepancy between the two fields is somewhat surprising given the results from Table \ref{table:tab4}, which show that there are the same fraction of galaxies classified as ``pure'' bulges and disks in both fields. Further exploration of this issue revealed differences in the photometric redshift distributions of both the bulge and disk-dominated systems between the fields. These distributions are shown in Fig.\,\ref{fig:5} with the distributions for the UDS field on the left given in blue for the $D/T>0/5$ galaxies and red for the $B/T>0/5$, and the COSMOS fits on the right. It can be seen from these figures that there is a sharp peak in the redshift distribution of disk-dominated galaxies in the UDS at $z\approx2$, which is perhaps indicative of a photometric redshift focussing effect. There is also a steeper decline in the number of bulge-dominated systems above $z\sim2$ in the UDS field compared to COSMOS. These two effects act together to produce a flatter redshift evolution of disks compared to bulges in the COSMOS field.

The peaked redshift distribution of the disk-dominated objects in the UDS required further exploration to ensure that our results have not been biased by photometric redshift focussing effects. This issue is well-known for objects with relatively flat SEDs where there are no strong breaks for the SED template fitting approach to fit to and so may preferentially occur for the disk-dominated galaxies in our sample. Plausibly, this may also be more of an issue for the UDS sample due to the accompanying optical and near-IR photometry utilised for the SED fitting, as the UDS field makes use of much shallower Y-band data compared to the COSMOS field.

In order to test for this effect,  we have cross-matched our sample with the CANDELS-UDS photometric redshift catalogue of \citet{Dahlen2013}. The Dahlen et al. photometric redshift catalogue is constructed using a Bayesian approach which uses the redshift probability distributions of six different photometric redshift fits using different photometric-redshift fitting codes. Adopting the Dahlen et al. redshifts for our sources has a marginal effect on the redshift distribution of the disk-dominated galaxies in our UDS sample, but does not affect the trends for the combined UDS and COSMOS sample. Therefore, we have concluded for this work that our overall results are not biased by this issue and so present the trends displayed by the combined UDS and COSMOS samples, simply noting that the separate UDS and COSMOS samples display some differences which are at least partly due to cosmic variance \citep{Newman2002}.

In conclusion, the combined UDS and COSMOS samples reveal that the fraction of bulge-dominated galaxies remains relatively flat across the $1<z<3$ redshift range covered by this study, particularly when considering the ``pure'' bulges with $B/T>0.9$, where even at $z<2$ these most massive galaxies are predominantly mixed bulge+disk systems and a significant fraction of ``pure'' bulges, comparable to the giant elliptical systems which dominate the massive galaxy population locally, are yet to emerge.
Despite the above discussion, there is still some evidence for the demise of disk-dominated systems below $z=2$, both from the fractions determined from the $H_{160}$ light fractions and the decomposed stellar masses. This becomes particularly  evident when comparing the fraction of disk-dominated galaxies to not only the bulge-dominated, but also the intermediate objects, as it is clear that $z=2$ still marks a key phase below which these most massive galaxies gain an increasing contribution from bulge components. The strongest trend evident from this analysis is the increase in the intermediate classification with decreasing redshift, which reveals the rise of S0 type galaxies within this redshift regime. This trend is noticeably stronger when cutting by $H_{160}$ light fraction than decomposed mass. However, this is to be expected from the fact that bulge components are more dominant in terms of their contribution to the mass of the galaxy, so the decomposed mass trends show a weaker trend in the increase of intermediate systems but display a larger increase in the fraction of bulge-dominated systems.

\section{Star-Formation Evolution}
In the following sections we combine the information from the decomposed morphological analysis with star-formation activity estimates from both the overall galaxy, and the decomposed estimates for the separate bulge and disk components, to probe how the morphological transformations witnessed in this $1<z<3$ redshift regime are linked to the process, or processes, responsible for star-formation quenching. In order to facilitate better comparison with other studies, we have also compared our passivity cut at $sSFR<10^{-10}\,{\rm yr^{-1}}$ from the overall galaxy SED fitting to colour-colour cuts from UVJ diagrams following \citet{Williams2009}, and find good agreement between the two techniques for separating star-forming and passive galaxies.

\subsection{Star-formation and morphology}

\begin{figure*}
\begin{minipage}[b]{0.48\linewidth}
\begin{center}
\vspace{-1.2cm}
\includegraphics[scale=0.42]{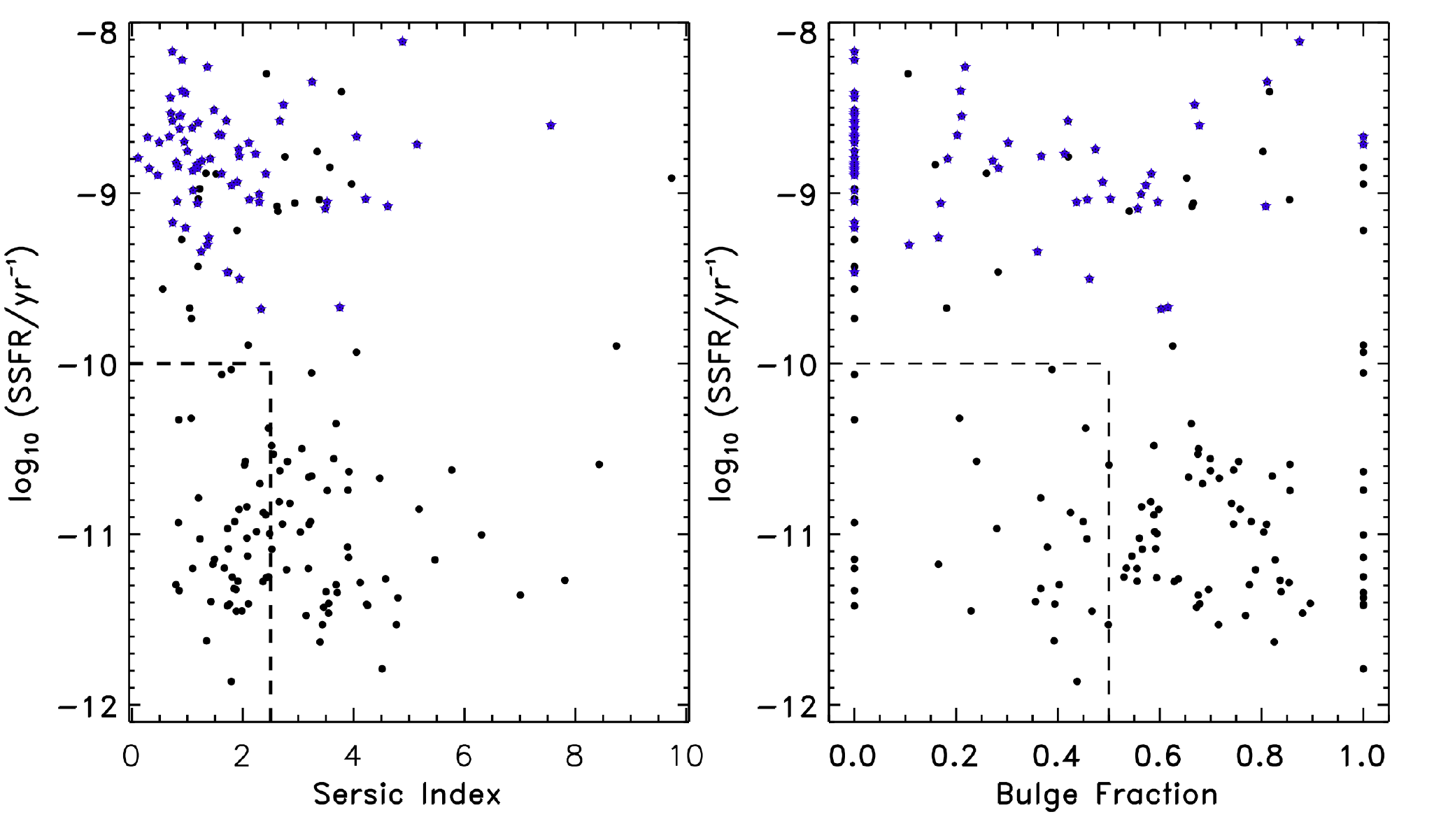} 
\end{center}
 \caption[ S\'{e}rsic index and bulge/total $H_{160}$ light fractions against sSFR for the UDS sample.]{sSFR for the galaxies in the UDS sample versus S\'{e}rsic index (left) and bulge/total $H_{160}$ light fractions (right). Galaxies with a $24\mu$m counter-part from either SpUDS or S-COSMOS have been highlighted by the blue stars, and a box has been placed around the passive ($sSFR<10^{-10}\,{\rm yr^{-1}}$) disk-dominated galaxies as judged by both $n<2.5$ and $B/T<0.5$. Comparison with the similar figure in \citet{Bruce2012} shows that the adoption of the re-evaluated star-formation rates for the UDS sample has removed objects which have a $24\mu$m counter-part from the passive region of the plot, but a significant number of passive disks and star-forming bulges remain. }
\label{fig:8}
\end{minipage}
\hspace{0.1cm}
\begin{minipage}[b]{0.48\linewidth}
\begin{center}
\vspace{0.2cm}
\includegraphics[scale=0.42]{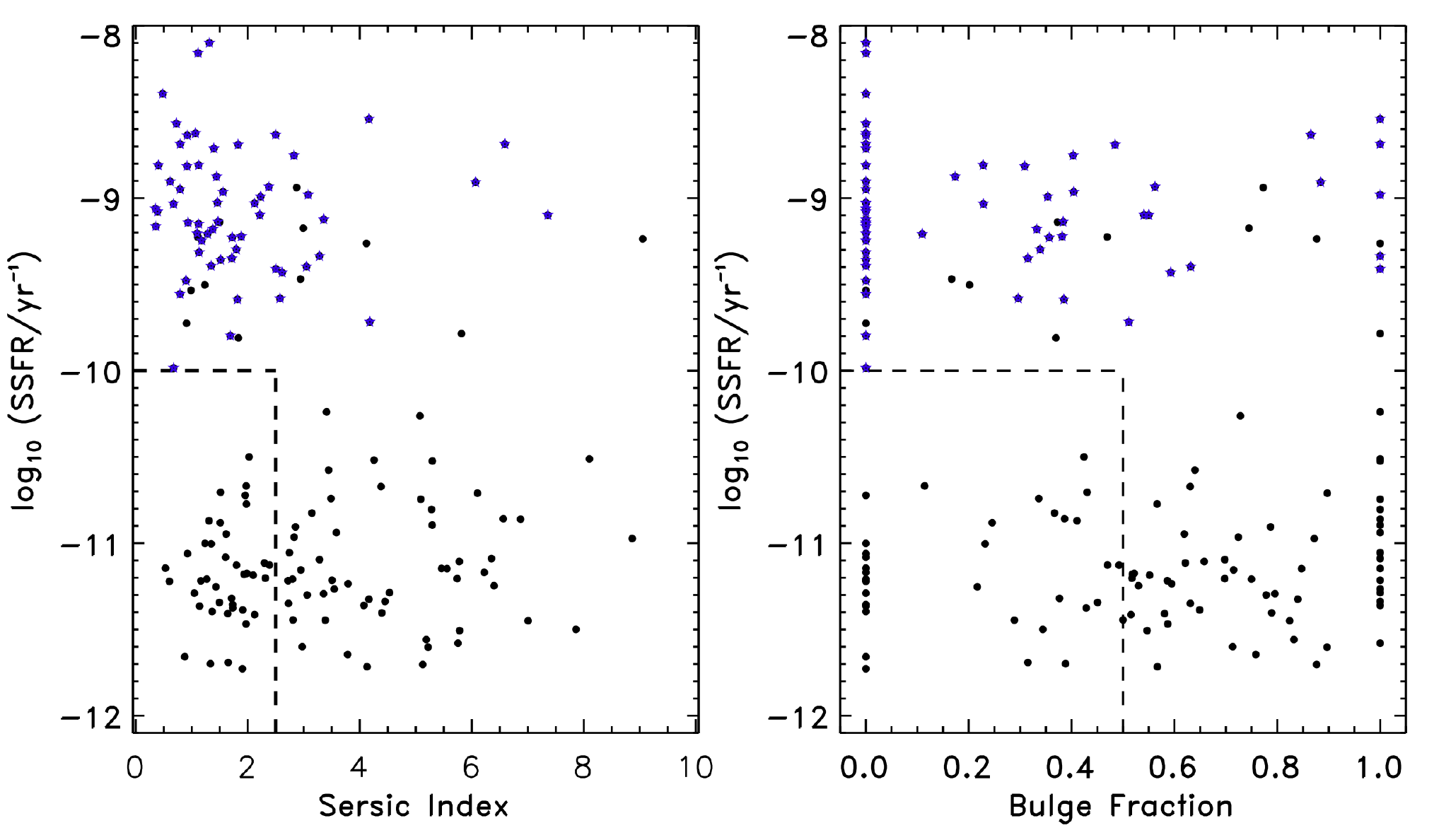} 
\end{center}
 \caption[S\'{e}rsic index and bulge/total $H_{160}$ light fractions against sSFR for the COSMOS sample.]{sSFR for the galaxies in the COSMOS sample versus S\'{e}rsic index (left) and bulge/total $H_{160}$ light fractions (right). Galaxies with a $24\mu$m counter-part from either SpUDS or S-COSMOS have been highlighted by the blue stars, and a box has been placed around the passive ($sSFR<10^{-10}\,{\rm yr^{-1}}$) disk-dominated galaxies as judged by both $n<2.5$ and $B/T<0.5$. In comparison to the UDS sample the COSMOS sample has the same overall number of objects in the star-forming bulge and disk-dominated and the passive bulge and disk-dominated sub-populations, however there do appear to be more passive $B/T=0$ disks in COSMOS, but these numbers are very small and may be affected by cosmic variance.}
\label{fig:9}
\end{minipage}
\end{figure*}

\begin{figure}
\includegraphics[scale=0.42]{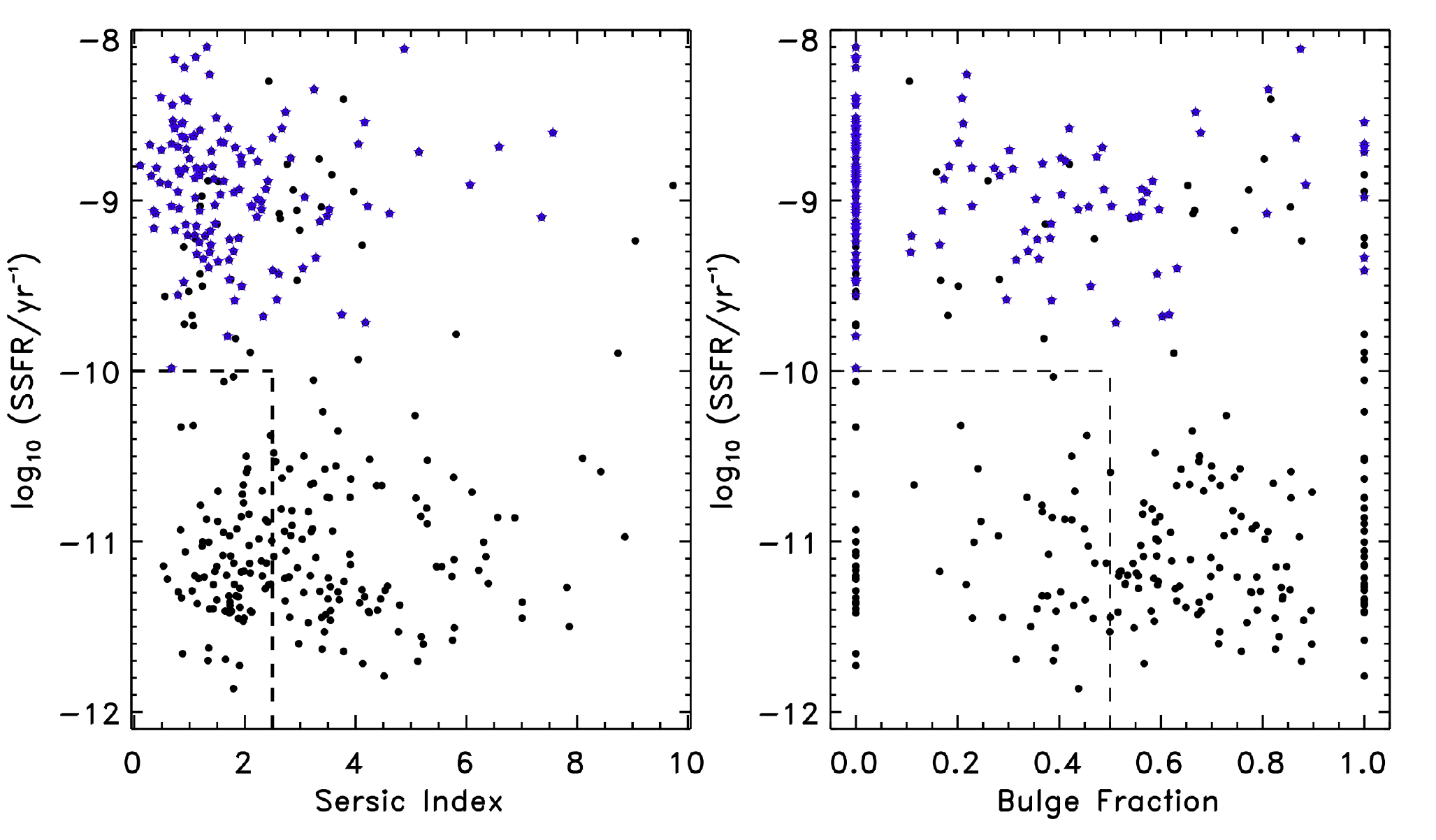} 
 \caption[S\'{e}rsic index and bulge/total $H_{160}$ light fractions against sSFR for the combined UDS+COSMOS sample.]{sSFR for the galaxies in the combined UDS+COSMOS sample versus S\'{e}rsic index (left) and bulge/total $H_{160}$ light fractions (right). Galaxies with a $24\mu$m counter-part from either SpUDS or S-COSMOS have been highlighted by the blue stars, and a box has been placed around the passive ($sSFR<10^{-10}\,{\rm yr^{-1}}$) disk-dominated galaxies as judged by both $n<2.5$ and $B/T<0.5$.}
\label{fig:10}
\end{figure}

 In Figs.\,\ref{fig:8},\ref{fig:9} and \ref{fig:10} we present plots of the total specific star-formation rate (sSFR) for each galaxy against both the single S\'{e}rsic index and the bulge/total light fraction for the UDS, COSMOS and combined fields respectively.

The fraction of passive galaxies (from the overall galaxy sSFR) which are disk-dominated and star-forming galaxies which are bulge-dominated in each field are given in Tables \ref{table:pd} and \ref{table:sfb}. There is also a comparable fraction of both passive disk-dominated and star-forming  bulge-dominated galaxies in both fields, judging morphology by either the single-S\'{e}rsic index cut at $n=2.5$ or by the decomposed $B/T=0.5$ measure, with $\sim30\%$ of all passive or star-forming galaxies being disk or bulge-dominated, respectively. 

However, one notable variation between the fields is the larger number of $B/T=0$ passive disks in COSMOS than in the UDS. For the COSMOS field there are 5 objects within this range that are best-fit by a ``pure'' disk, and another 10 objects which are best-fit with a disk+PSF model (it is worth noting that for the disk+PSF fits none of the PSF fractions exceed $30\%$, thus these objects would remain disk-dominated even in the extreme case where all of the PSF flux was attributed to the bulge component). The corresponding numbers for the UDS are 3 ``pure'' disks and 4 disk+PSF objects. Given the low number statistics involved, and that overall, as shown in Table \ref{table:tab4}, the same fraction of the population in both samples has been best-fit by disk and disk+PSF models, it is difficult to draw firm conclusions from these numbers. Moreover, given that the addition of a PSF component may indicate the presence of an AGN or nuclear starburst, we have tried to search for any evidence of AGN activity from the x-ray and radio catalogues available in both the UDS and COSMOS fields, but found no counter-parts within a 4\,arcsec matching radius for the x-ray catalogues and 2\,arcsec matching radius for the radio catalogues, for any of the ``pure'' disks or disk+PSF fits . Given the objects being fitted are from CANDELS images of the same $S/N$ there is also no reason to attribute this difference between fields to biases from the adoption of a spurious additional PSF in the fitting procedure, as explored with the simulations. In addition to this, from examining the spatial positioning of these objects in both fields we find no obvious evidence of clustering. Thus, we simply conclude that the classification of these objects is equally robust and unbiased in both fields, and attribute the offset in the small numbers to shot noise.

The verification of a significant population of both star-forming bulge-dominated galaxies and passive disk-dominated objects is particularly interesting as they may suggest that the processes which quench star-formation are distinct from those which drive morphological evolution. Despite previously limited evidence for these populations \citep{Stockton2008,McGrath2008,vanderWel2011,Cameron2011,Wang2012, Bruce2012}, more recently, the presence of a significant passive disk-dominated population has been corroborated by studies such as \citet{McLure2013,Chang2013, Fan2013, Lee2013,Talia2013} and \citet{Lang2014}. However, these studies use a mixture of single-S\'{e}rsic, non-parametric and visual morphological classifications, with the exception of \citet{vanderWel2011} and \citet{Lang2014} who perform double-component $n=1+$free fits and $n=1+n=4$ fits but adopt star-formation rates for the galaxy as a whole. This approach may be prone to mis-classification if, for example, the bulge component of the galaxy dominates in terms of stellar mass in which case the galaxy should arguably no longer be classified as disk-dominated and, moreover, the specific star-formation rate of the disk-only component may in fact be above the passivity threshold. This is clearly also a concern for the star-forming bulge-dominated galaxies, where one might expect the decomposition to reveal that, while the overall galaxy $sSFR>10^{-10}\,{\rm yr^{-1}}$, the bulge component itself is passive and the star-formation activity is limited to the disk component. In order to explore how many of the passive disks and star-forming bulges in our sample may be prone to mis-classification based on the adoption of the overall galaxy star-formation rate and morphological light-based bulge and disk fractions, we have examined the individual component stellar masses and star-formation rates for all objects within these populations.

\begin{table}
\begin{center}
\begin{tabular}{m{2.0cm}m{2.5cm}m{2.5cm}}
\hline
Field &
$n<2.5$ &
$B/T<0.5$\\
\hline
UDS &
$38\pm7\%$ &
$30\pm7\%$ \\
COSMOS &
$43\pm8\%$ &
$37\pm7\%$ \\
Combined &
$41\pm5\%$ &
$33\pm5\%$ \\
\end{tabular}
\caption{The fractions of passive galaxies which are disk-dominated, split by field and using both the single-S\'{e}rsic and multiple-component classifications.}
 \label{table:pd}
 \end{center}
\end{table}

\begin{table}
\begin{center}
\begin{tabular}{m{2.0cm}m{2.5cm}m{2.5cm}}
\hline
Field &
$n>2.5$ &
$B/T>0.5$\\
\hline
UDS &
$33\pm8\%$ &
$31\pm6\%$ \\
COSMOS &
$26\pm7\%$ &
$26\pm7\%$ \\
Combined &
$30\pm5\%$ &
$29\pm5\%$ \\
\end{tabular}
\caption{The fractions of star-forming galaxies which are bulge-dominated, split by field and using both the single-S\'{e}rsic and multiple-component classifications.}
\label{table:sfb}
\end{center}
\end{table}

\subsection{Passive disks}

Of the 184 passive galaxies in the full sample, only 146 are covered by both the ACS and WFC3 pointings and so have been modelled by the decomposed SED fitting. 

\begin{figure*}
\begin{center}
\begin{tabular}{m{3cm}m{3cm}m{3cm}}
\includegraphics[scale=0.15 ]{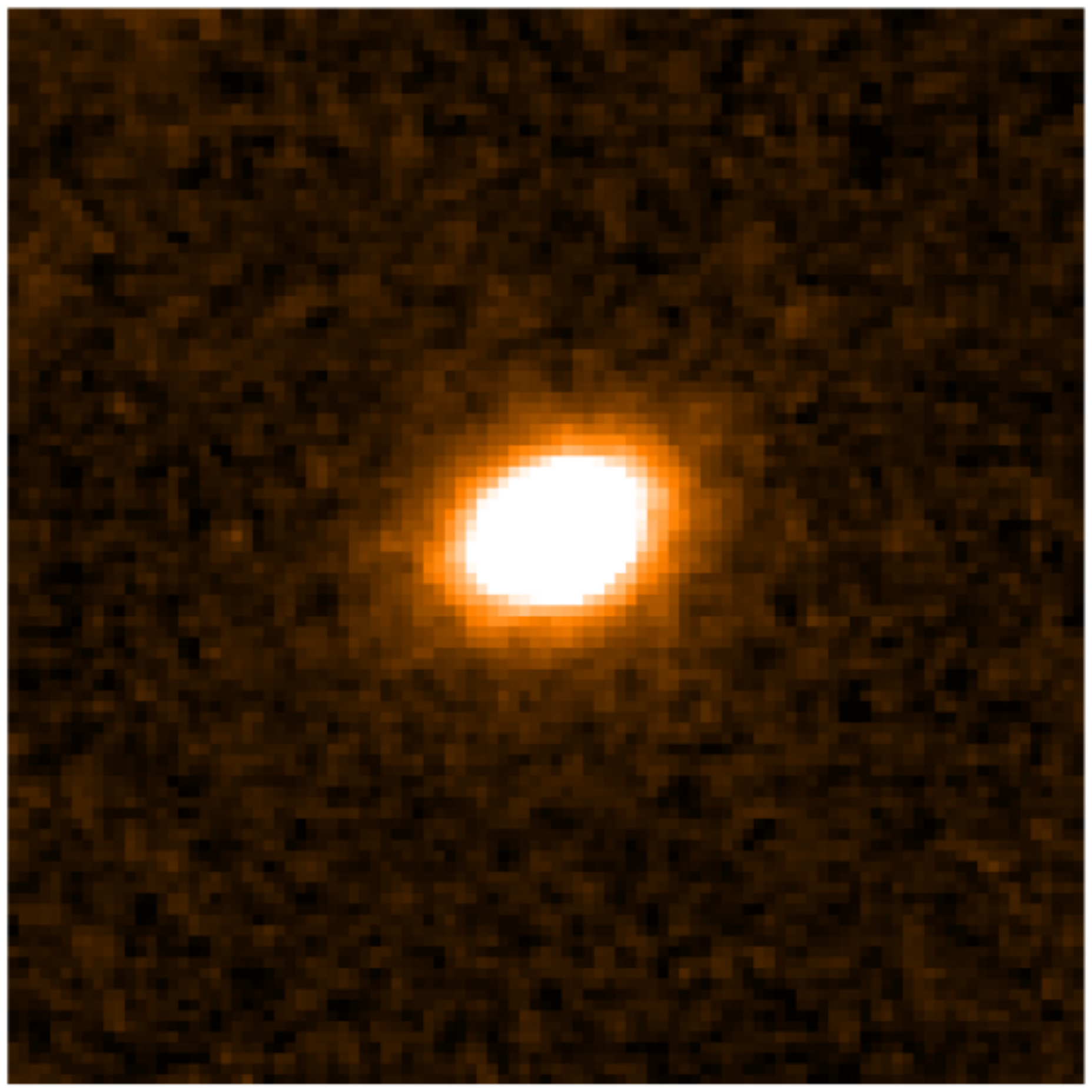}  &
\includegraphics[scale=0.15]{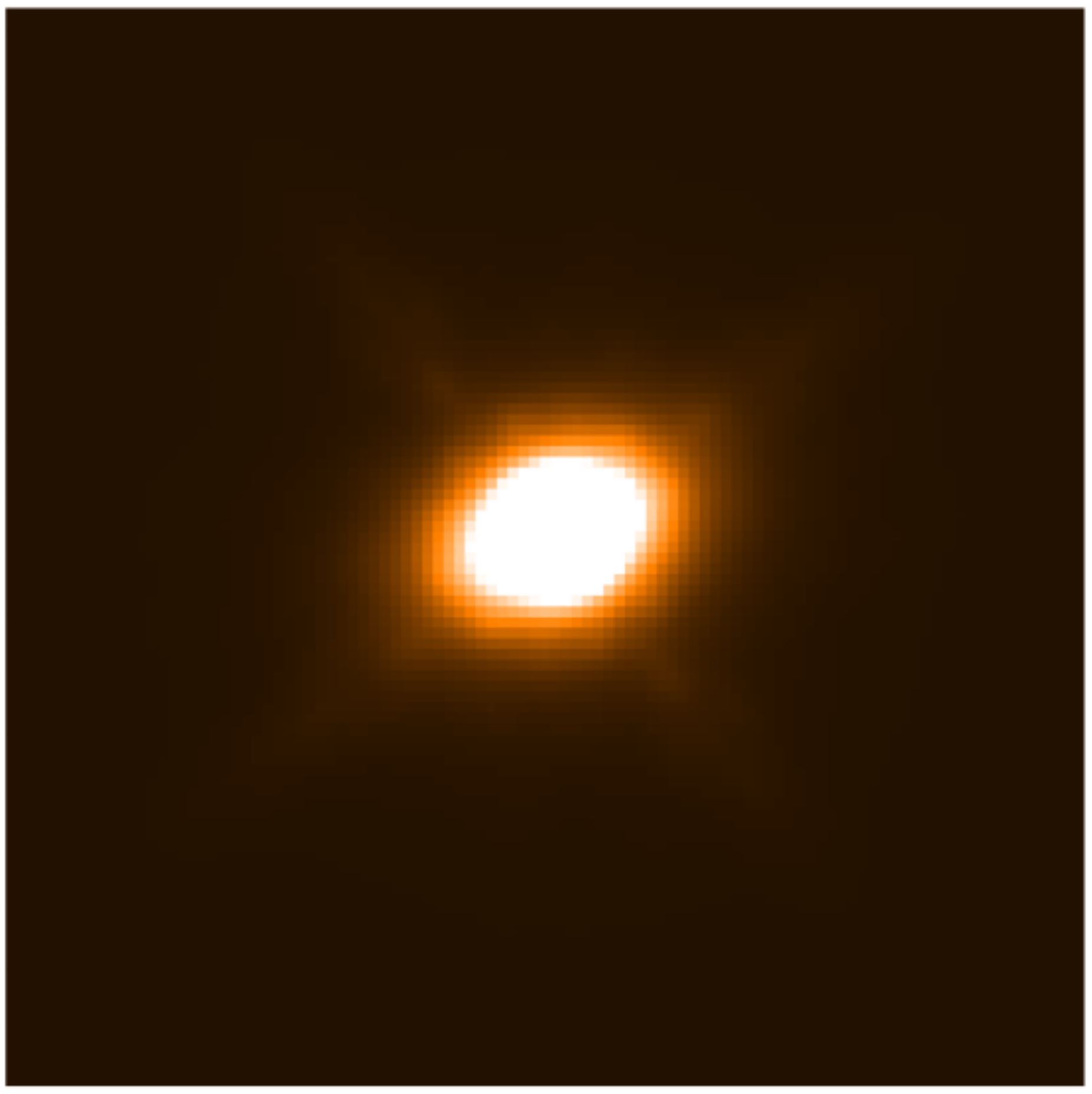}  &
\includegraphics[scale=0.15]{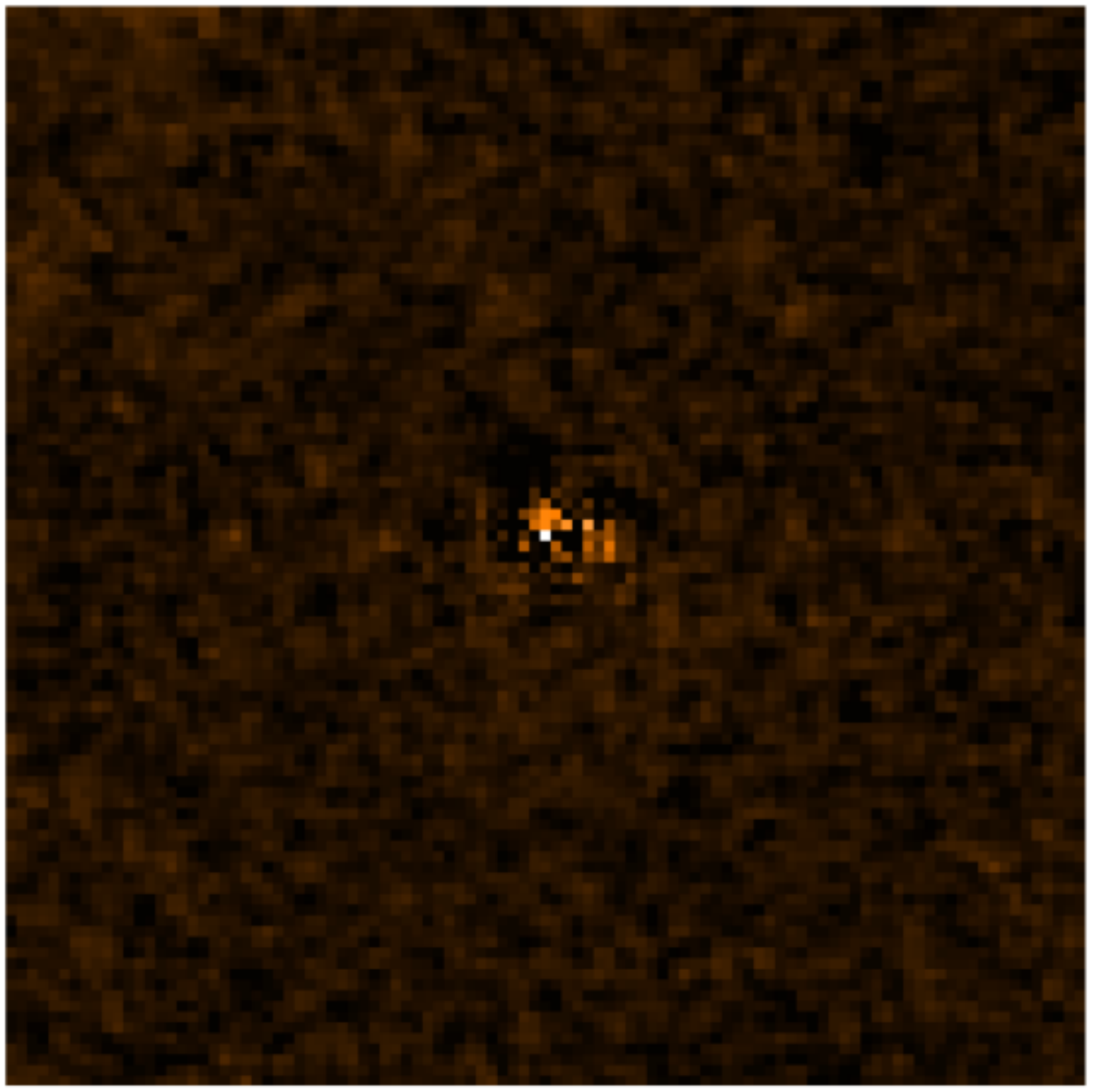}  
\end{tabular}
\begin{tabular}{m{3cm}m{3cm}}
\includegraphics[scale=0.15]{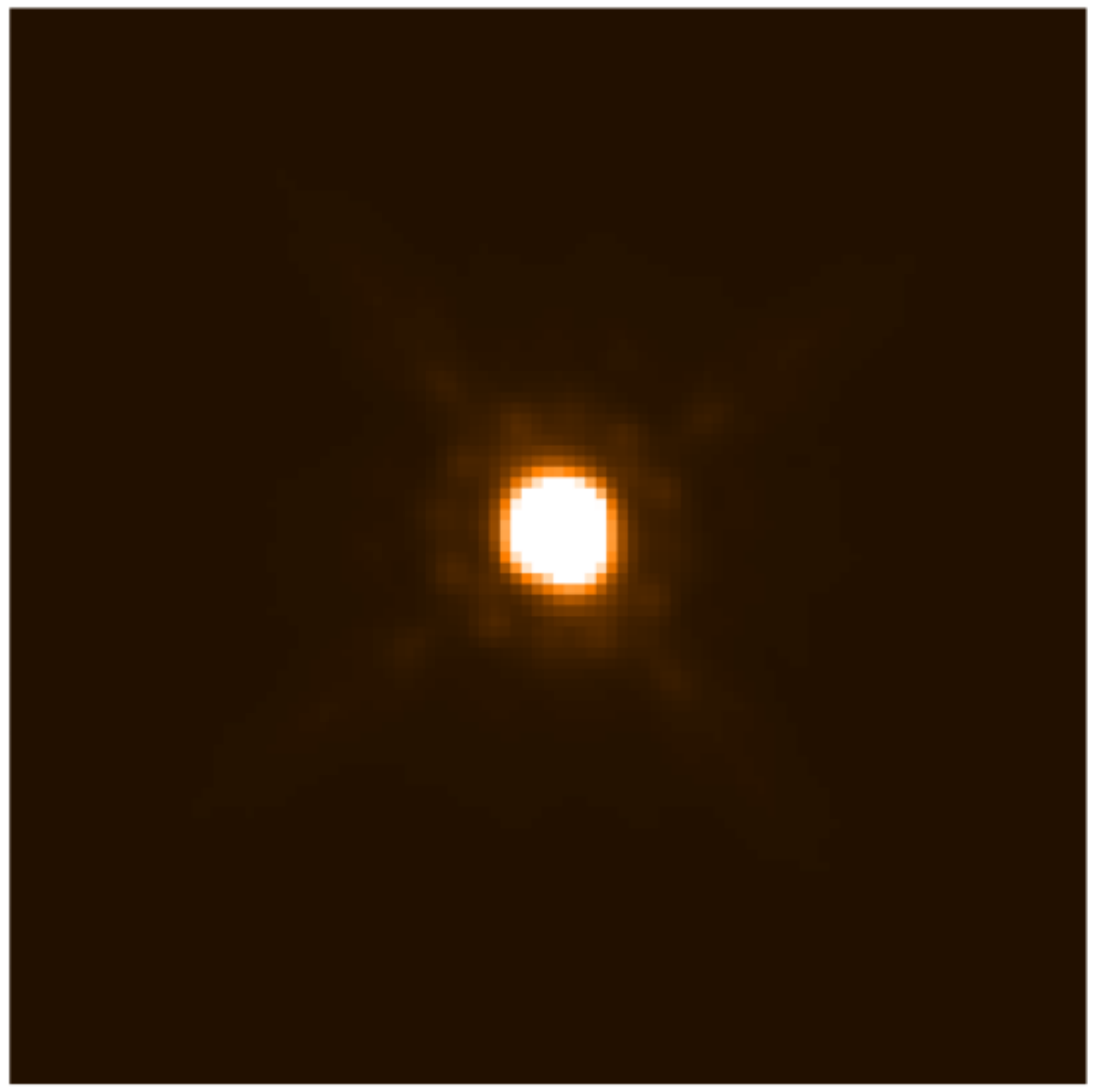}  &
\includegraphics[scale=0.15]{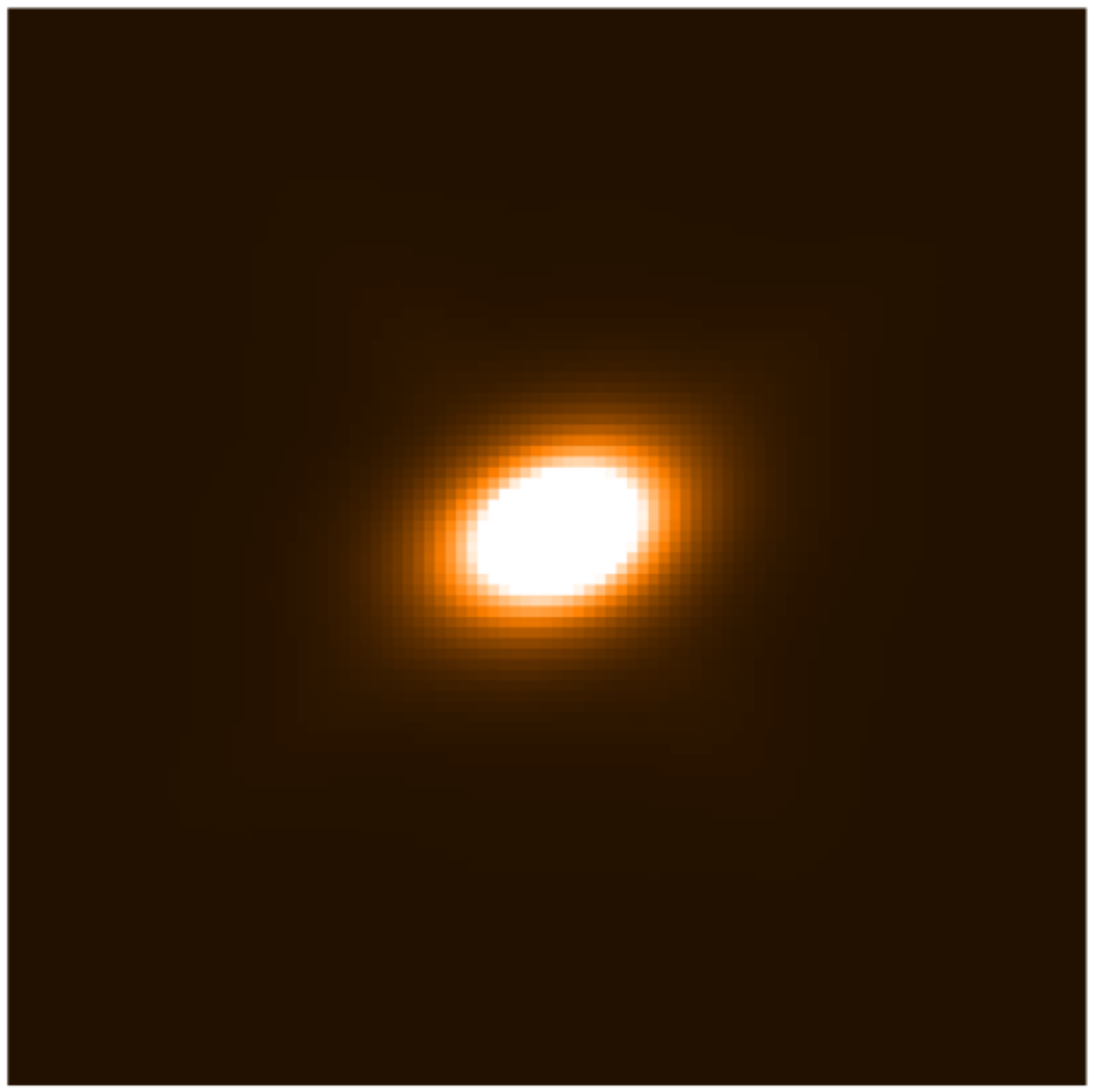}  
\end{tabular}
\begin{tabular}{m{9cm}}
\includegraphics[scale=0.7]{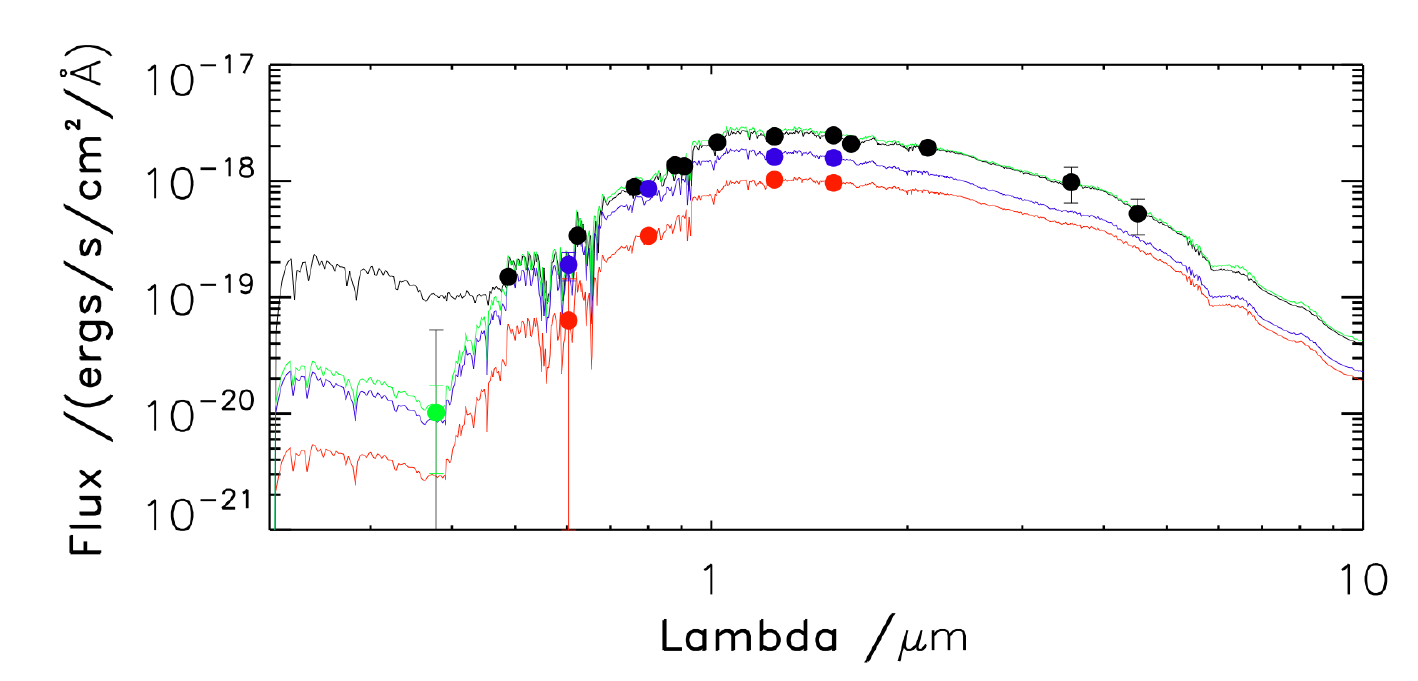}  \\
\includegraphics[scale=0.7]{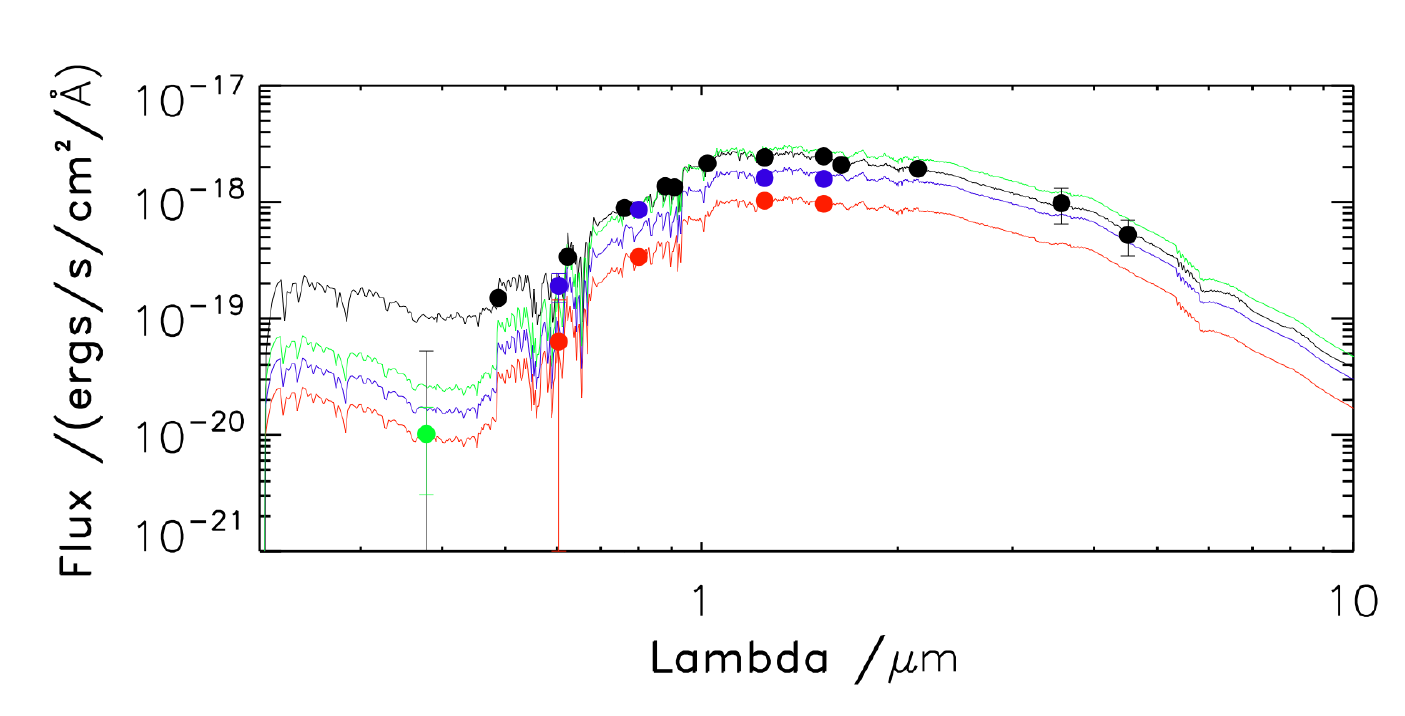}  
\end{tabular}
\caption[Example fit of one of the genuinely passive disk-dominated galaxies.]{One of the genuinely passive disk-dominated galaxies as judged by the most conservative criteria on the decomposed fits. The $H_{160}$ $6\times6$ arcsec stamps are shown. The first row, from left to right, contains the image stamp, the best-fit multiple-component model, the residual map, the best-fit bulge model component only and the disk only component. The second row displays the best-fit $0<\tau{\rm (Gyr)}<5$ decomposed SED models. For completeness, the bottom row contains the $0.3<\tau{\rm (Gyr)}<5$ SED fits which force the models to adopt a minimum low-level of on-going star formation.}
\label{fig:pd1}
\end{center}
\end{figure*}

Out of these 146 passive galaxies, 16 have been best-fit with either a ``pure'' disk or disk+PSF model (13 in COSMOS and 3 in the UDS). These objects have therefore not been subjected to the decomposed SED fitting, and given the above discussion on the lack of any $24\mu$m, x-ray or radio counter-parts for these systems, we have no reason to assume they have any obscured star-formation and/or AGN activity. Thus, we have adopted the overall galaxy star-formation rate for these galaxies and report them as genuine pure passive disks. For  completeness, for the case of only the ``pure'' disks, there are 2 in the UDS and 4 in COSMOS.

In addition to the ``pure'' disk and disk+PSF fits, there are 30 objects which are passive and disk-dominated which have been fit with a multiple-component model. In order to report the most conservative and robust fraction of  passive disk-dominated galaxies in our sample using the decomposed stellar-mass and star-formation rate estimates, we have adopted the criteria that these objects must be classified as passive with a total bulge+disk decomposed $sSFR<10^{-10}\,{\rm yr^{-1}}$ {\it and} a decomposed disk $sSFR<10^{-10}\,{\rm yr^{-1}}$ (where we have used the decomposed disk stellar-mass to calculate the specific star-formation rate). This leaves 26/30 candidates as passive, where these objects all additionally had decomposed bulge $sSFR<10^{-10}\,{\rm yr^{-1}}$. Finally, to remain in the passive disk-dominated population, we required that the objects must be classified as disk-dominated by their decomposed disk/bulge+disk stellar masses. Imposing this criterion removed 15 galaxies and left 11 galaxies which are genuinely passive and disk-dominated by even our strictest definitions. 

These 11 passive disk-dominated galaxies were then combined with the 16 ``pure'' disk and disk+PSF objects and are taken as a fraction of the 146 passive galaxies. This  provides the estimate that ($27/146$) $18\pm5\%$ of all passive galaxies are disk-dominated. This fraction becomes $12\pm3\%$ if the objects with a best model fit with a disk+PSF are removed. However, given that the results from our simulations reveal that in $\geq90\%$ of models which have a compact bulge component our fitting procedure is able to correctly attribute the flux from the model component to a bulge rather than a PSF, it is evident that these disk+PSF fits are not significantly contaminated by disk-dominated objects which host a centrally compact bulge, and in any case the PSF fractions of these fits do not exceed $30\%$. Therefore, these objects are still disk-dominated and there is no obvious reason to remove them.

An example of one of the passive disk-dominated galaxies which is best-fit with a bulge+disk model and has been subjected to decomposed SED fitting is shown in Fig\,\ref{fig:pd1}, where we have displayed the master $H_{160}$ $6\times6$\,arcsec image, the best-fit bulge+disk model, the bulge component of the best-fit model , the disk component of the best-fit model, and the residual stamp. These images clearly illustrate that this is a genuinely morphologically disk-dominated galaxy, the disk component is not some remnant of poor fitting,  and that the model is a good fit to the galaxy. We have also included the best-fit model SED output for this galaxy. As before, the blue data-points and line represent the disk component, the red data-points and line represent the bulge component, and the sum of the best-fit bulge and disk SED model is given in green. This SED demonstrates the quality of the fit to the photometry and the genuine passivity of the disk component. In order to provide the most conservative estimates of the passive disk-dominated fraction, we have also ensured that even by adopting the limited sub-set of SED models with $0.3<\tau{\rm (Gyr)}<5$, which forces the galaxy to have at least some low-level of measurable on-going star-formation, the disk components of this population remain passive. The best-fit SED using this limited $\tau$ model sub-set is given in the bottom panels of Fig\,\ref{fig:pd1} and serves to further confirm the classification of this object.

\begin{figure*}
\begin{center}

\begin{tabular}{m{2.8cm}m{2.8cm}m{2.8cm}m{2.8cm}m{2.8cm}}
\includegraphics[scale=0.15 ]{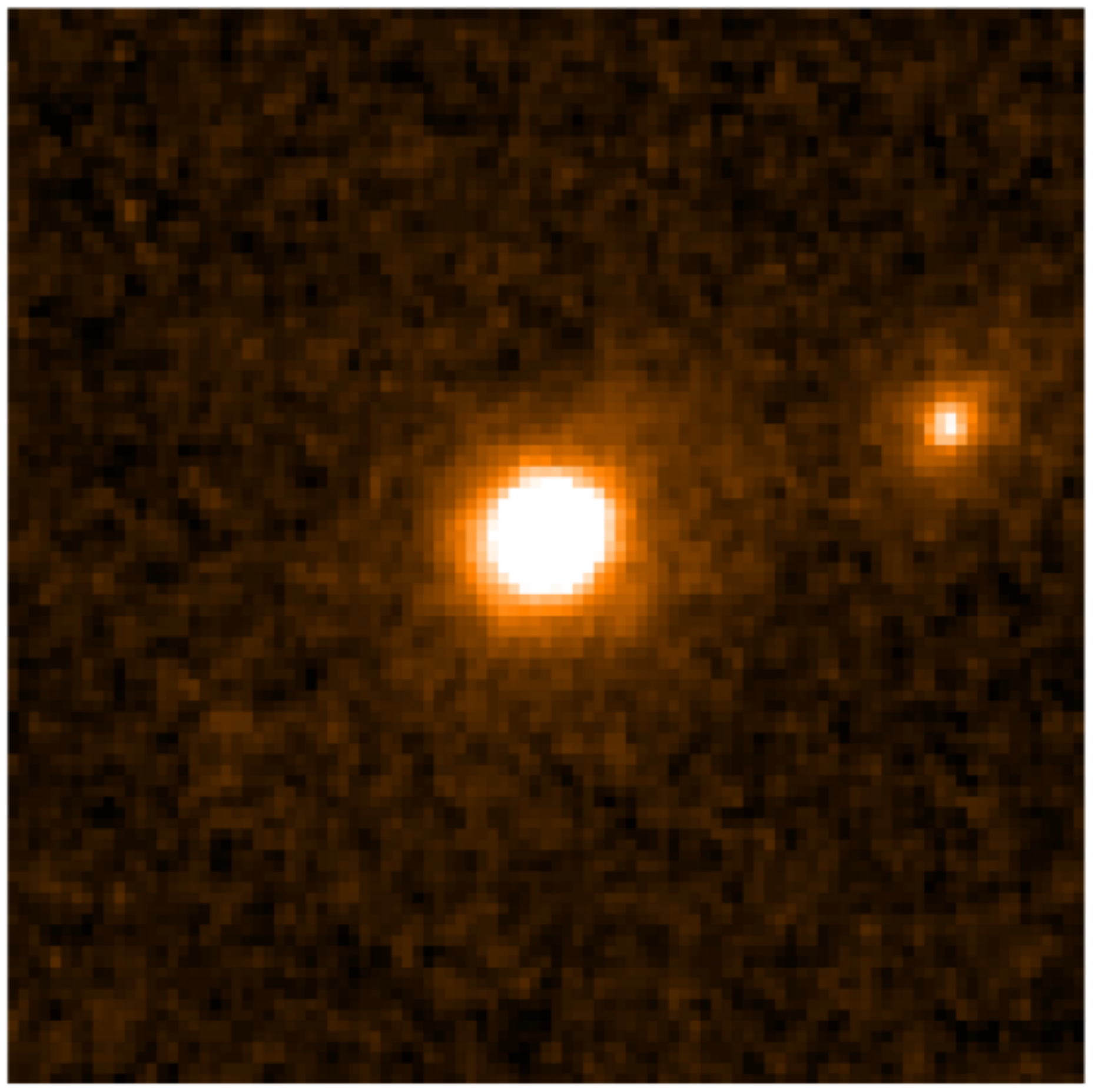}  &
\includegraphics[scale=0.15]{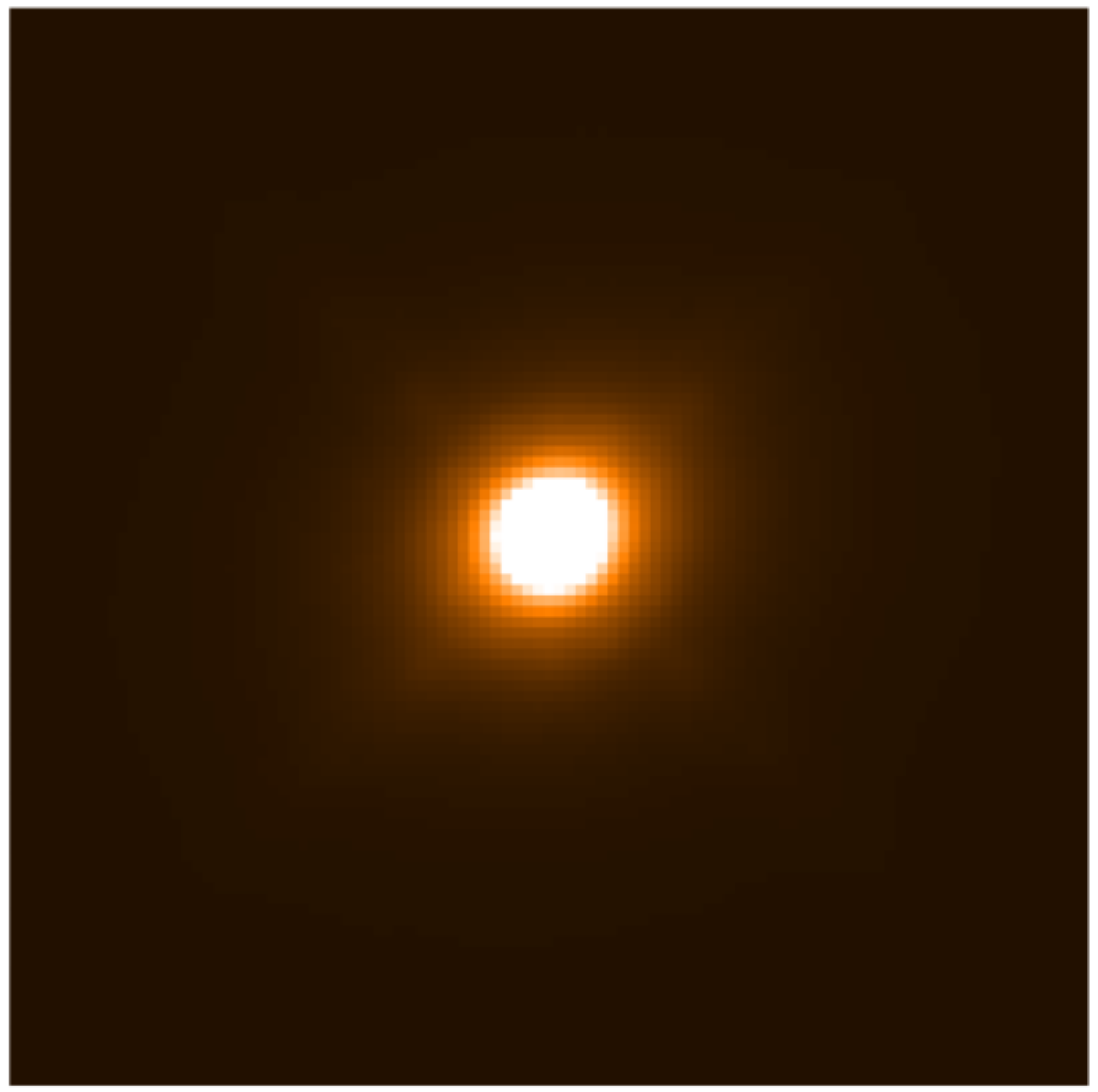}  &
\includegraphics[scale=0.15]{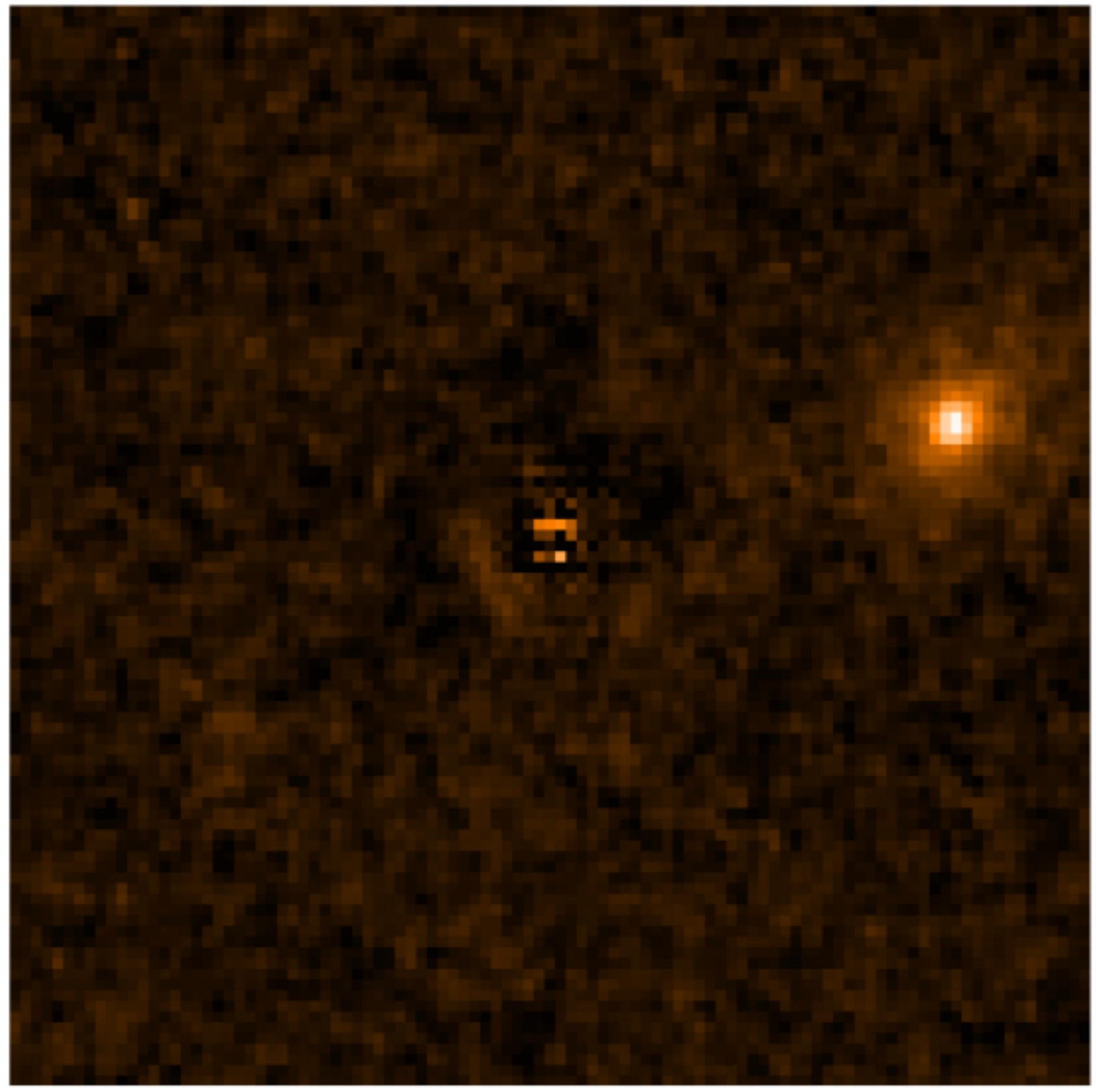}  &
\includegraphics[scale=0.15]{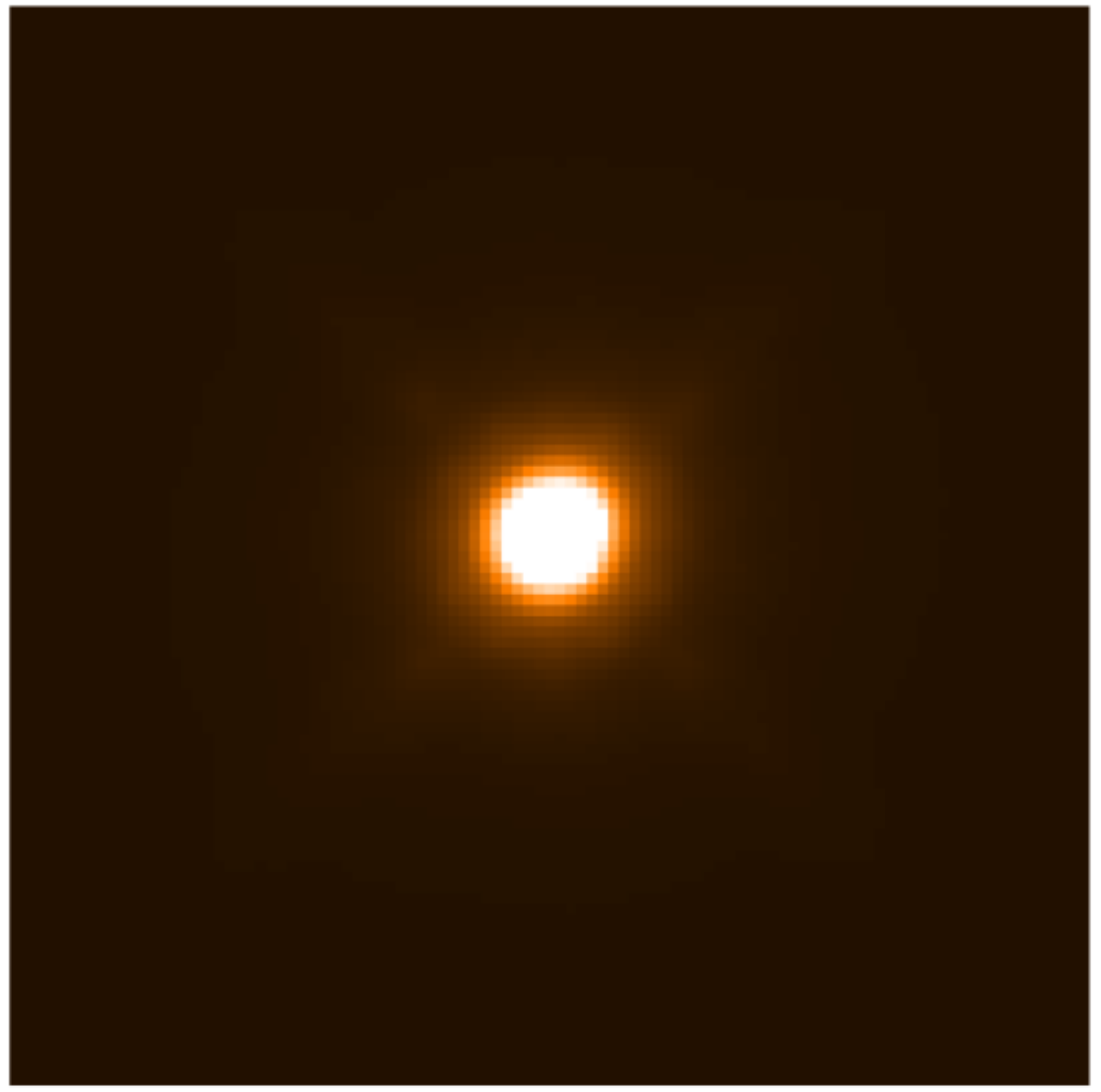}  &
\includegraphics[scale=0.15]{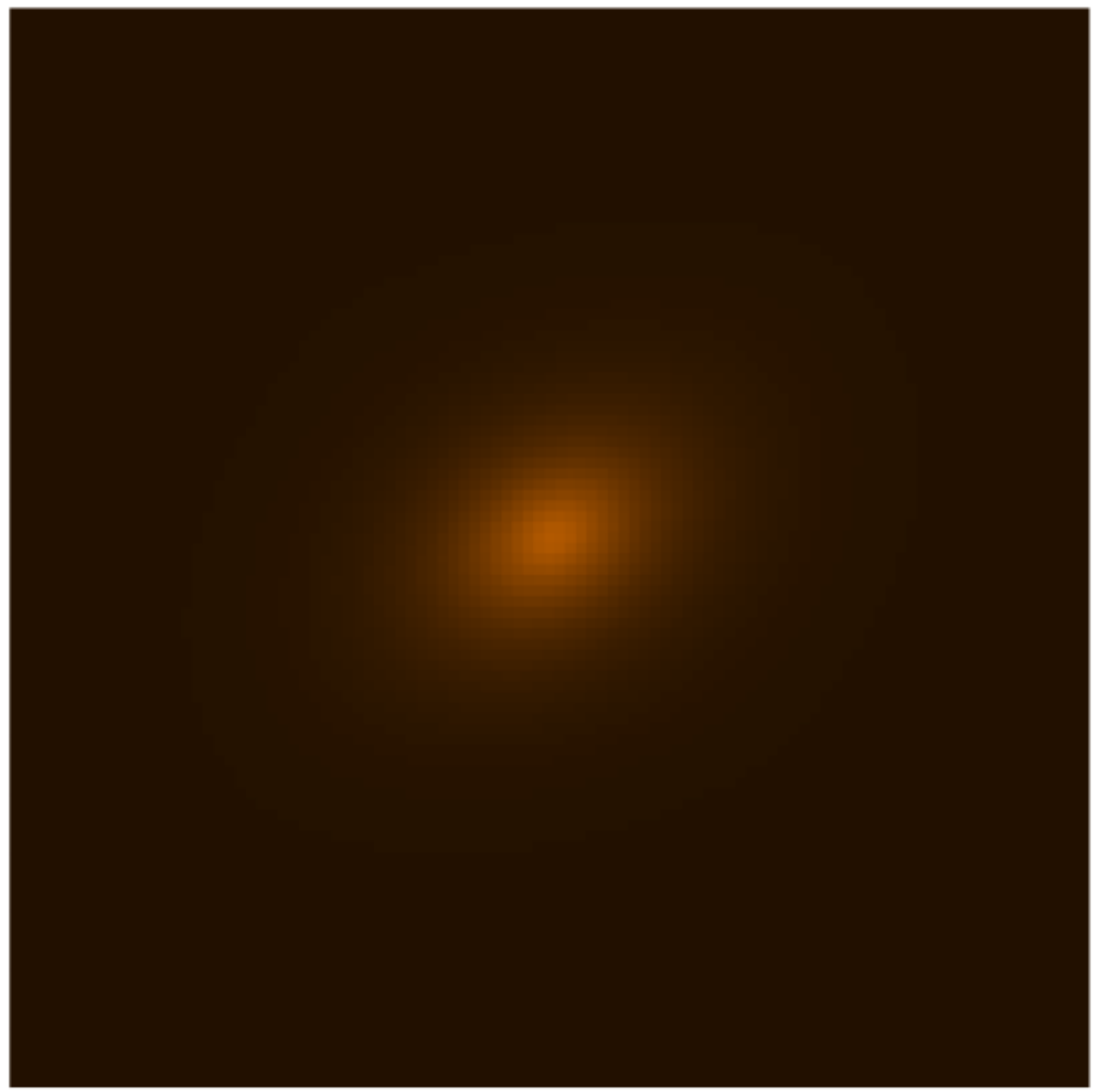}  
\end{tabular}

\begin{tabular}{m{2.8cm}m{2.8cm}m{2.8cm}m{2.8cm}m{2.8cm}}
\includegraphics[scale=0.15 ]{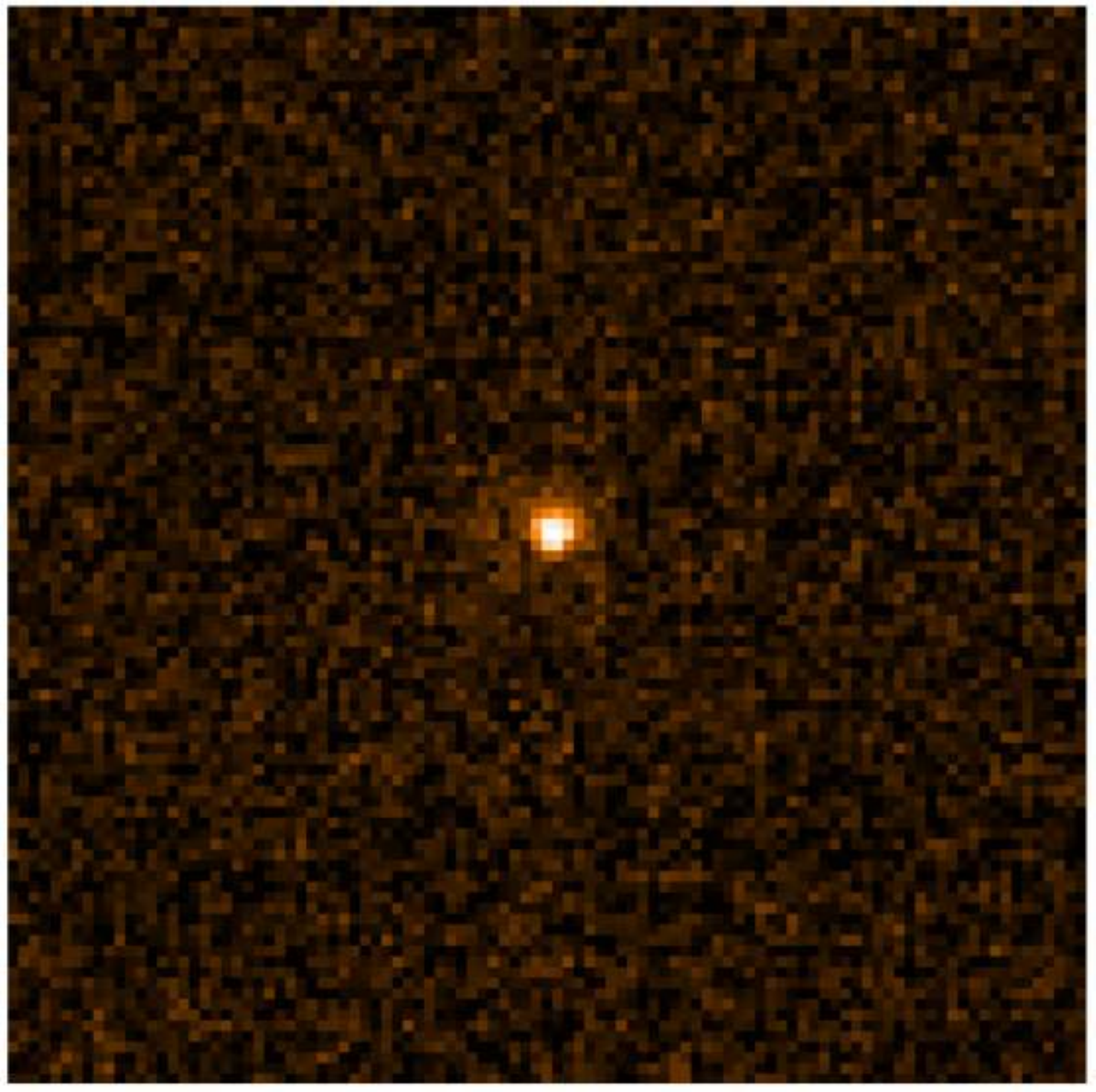}  &
\includegraphics[scale=0.15]{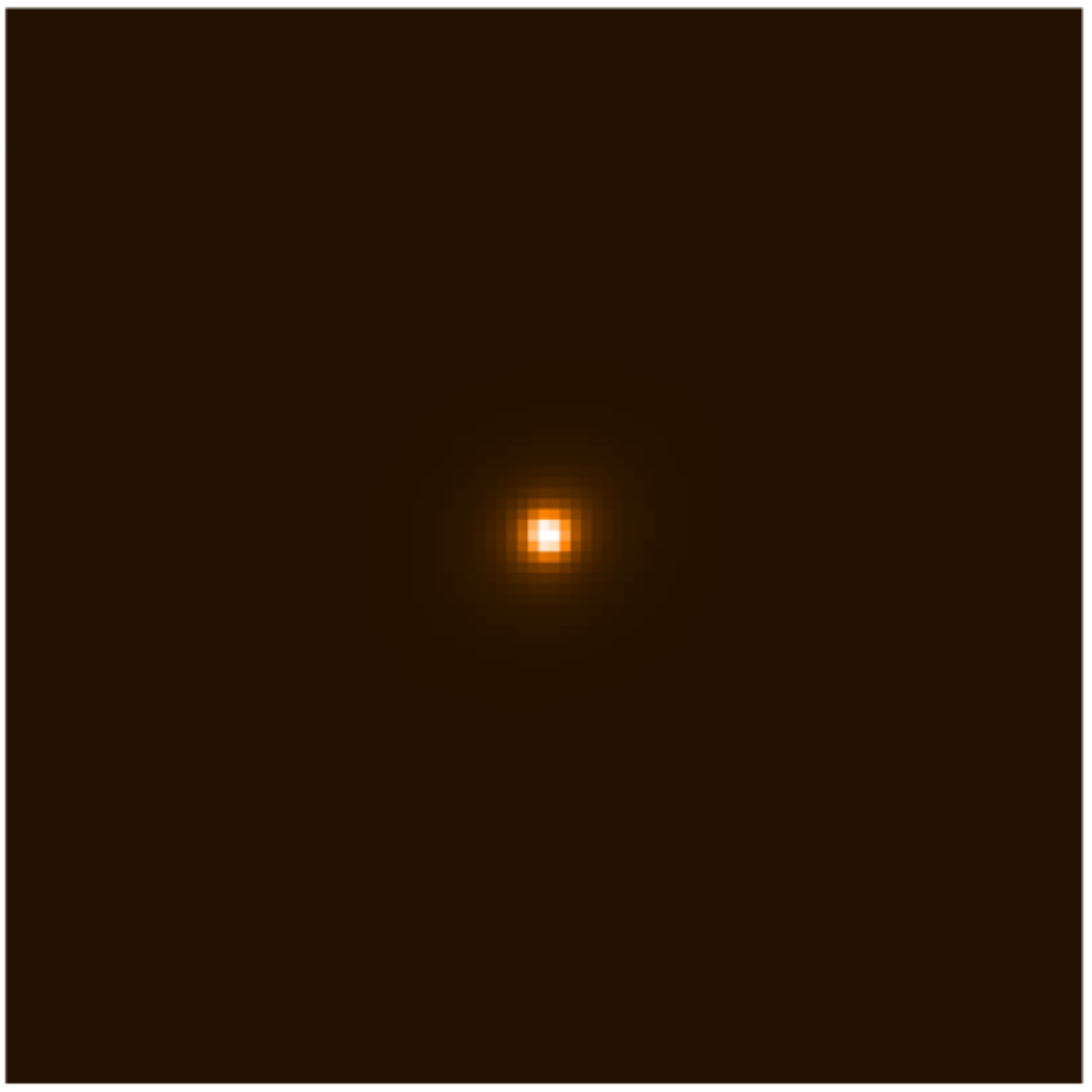}  &
\includegraphics[scale=0.15]{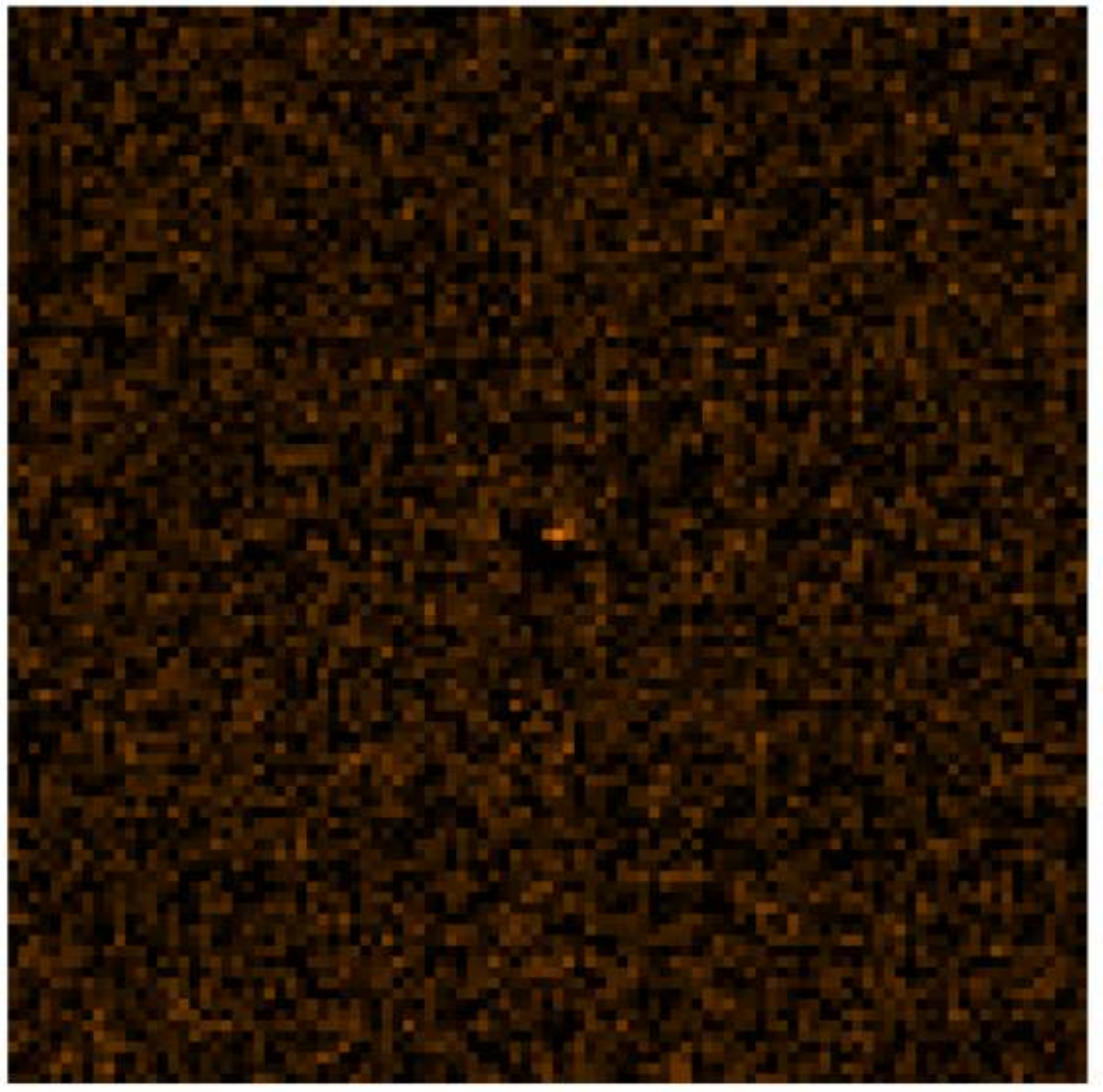}  &
\includegraphics[scale=0.15]{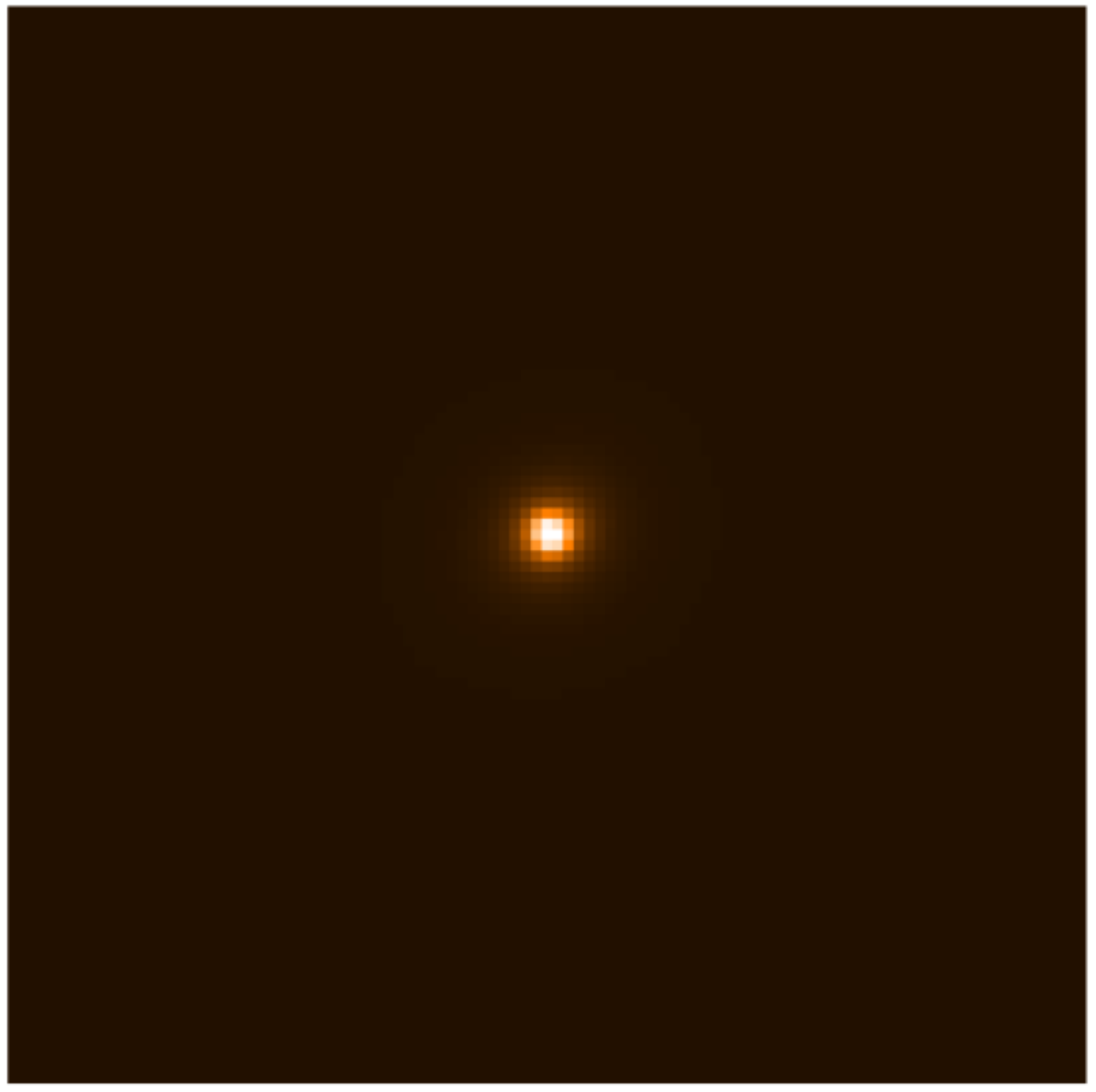}  &
\includegraphics[scale=0.15]{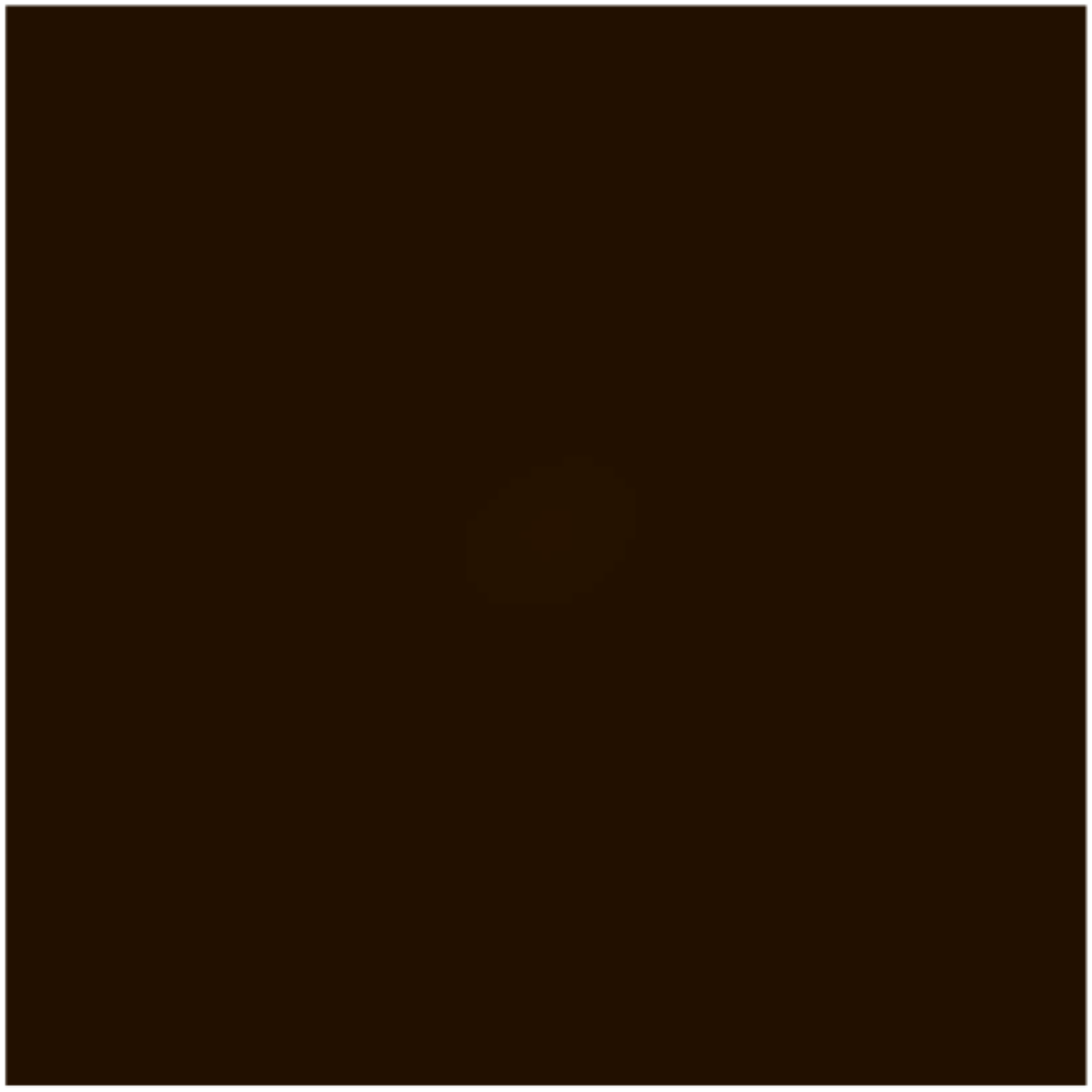}  
\end{tabular}
\begin{tabular}{m{9cm}}
\includegraphics[scale=0.7]{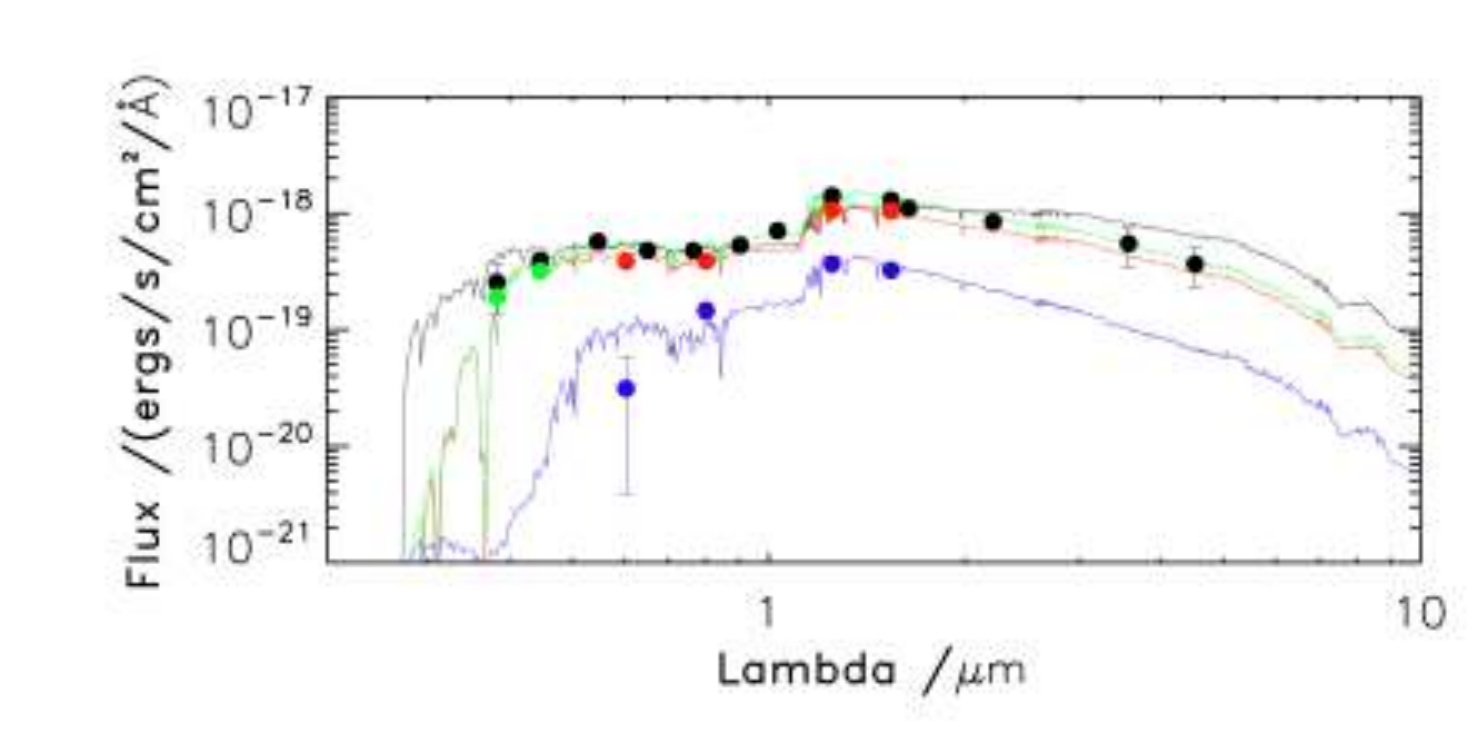}  \\
\includegraphics[scale=0.7]{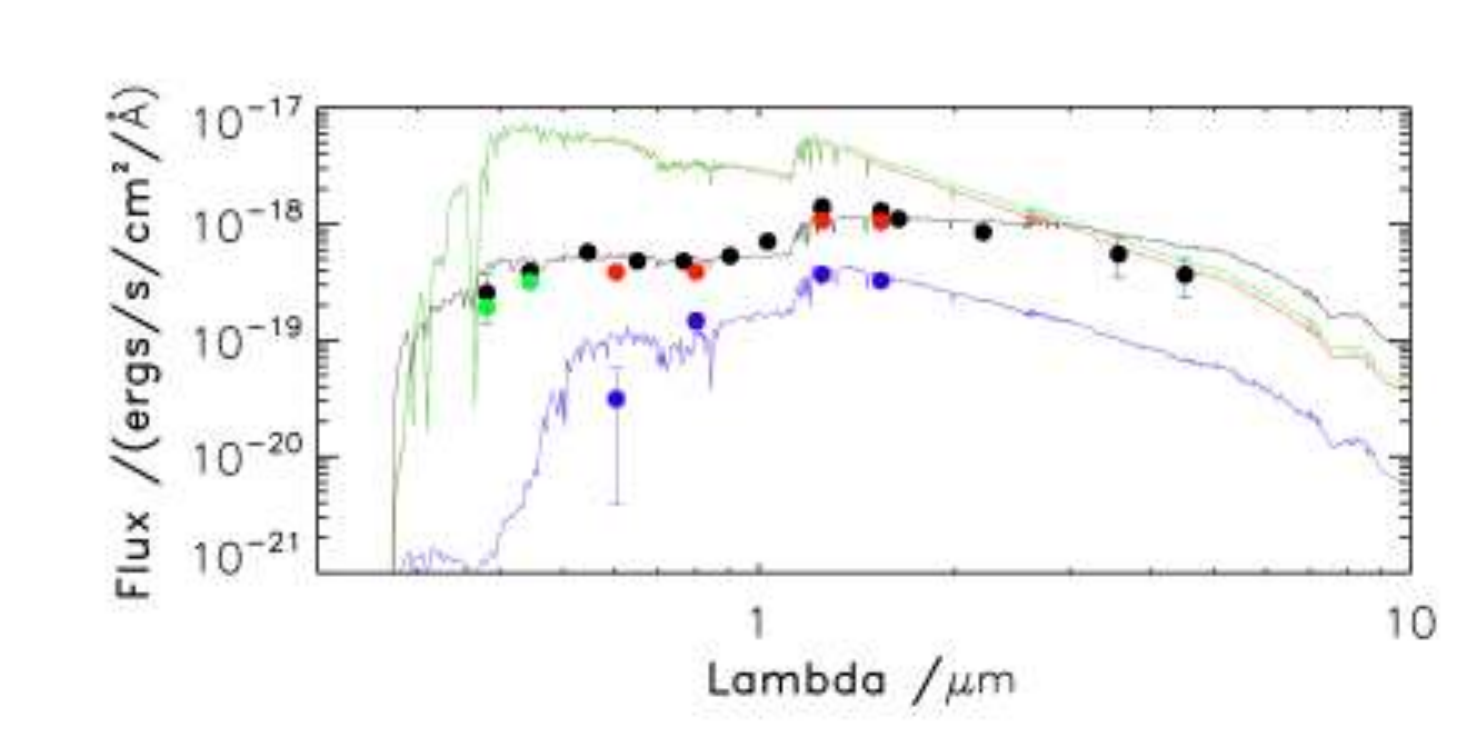}
\end{tabular}
\caption[The first genuinely star-forming bulge-dominated galaxy.]{The first genuinely star-forming bulge-dominated galaxy in our sample. The first row displays the $6\times6$ stamps for the $H_{160}$ images. From left to right they are the image stamp, the best-fit multiple-component model, the residual, the bulge component of the best-fit model and the disk component of the best-fit model. The second row follows the same configuration for the blue $v_{606}$ stamps. The third row show the best-fitting decomposed SED. The bottom panel shows the dust-corrected best-fit SED models to allow a direct comparison of the contribution from the  bulge and disk components.}
\label{fig:sfb1}
\end{center}
\end{figure*}

\subsection{Star-forming bulges}
We have also examined the decomposed stellar-mass and star-formation rate estimates for the star-forming bulge-dominated systems. Again, only a limited sub-set of these objects have been covered by both WFC3 and ACS pointings, which gives a total of 136 star-forming galaxies. Out of these, 11 are best-fit by ``pure'' bulges (comprising 6 objects in COSMOS and 5 objects in the UDS) and there are an additional 2 objects best-fit by bulge+PSF models. There are also 24 candidate star-forming bulge-dominated galaxies with best-fit multiple component models.

For these 24 candidate star-forming bulge-dominated systems we then insist that in order to remain in this sample they must be bulge dominated in terms of their bulge/bulge+disk decomposed stellar mass fractions, and that the bulge $sSFR>10^{-10}\,{\rm yr^{-1}}$.  Only 4 objects meet these criteria, as in the vast majority of cases the star-formation rate decomposition reveals that it is the disk components which are active. For all four of these objects the disk component has a $sSFR<10^{-10}\,{\rm yr^{-1}}$. As above, we also further restrict the sample to the most conservative fraction by requiring that, when the limited $0.3<\tau{\rm (Gyr)}<5$ SED models are adopted, the bulge component remains star-forming. In one of the four cases the limited $\tau$ SED models fit both the disk and bulge components with $sSFR<10^{-10}\,{\rm yr^{-1}}$. This suggests that this best-fit model is particularly degenerate so this object is removed form the sample and we have retained only 3 robust star-forming bulge-dominated galaxies.

\begin{figure*}
\begin{center}

\begin{tabular}{m{2.8cm}m{2.8cm}m{2.8cm}m{2.8cm}m{2.8cm}}
\includegraphics[scale=0.15 ]{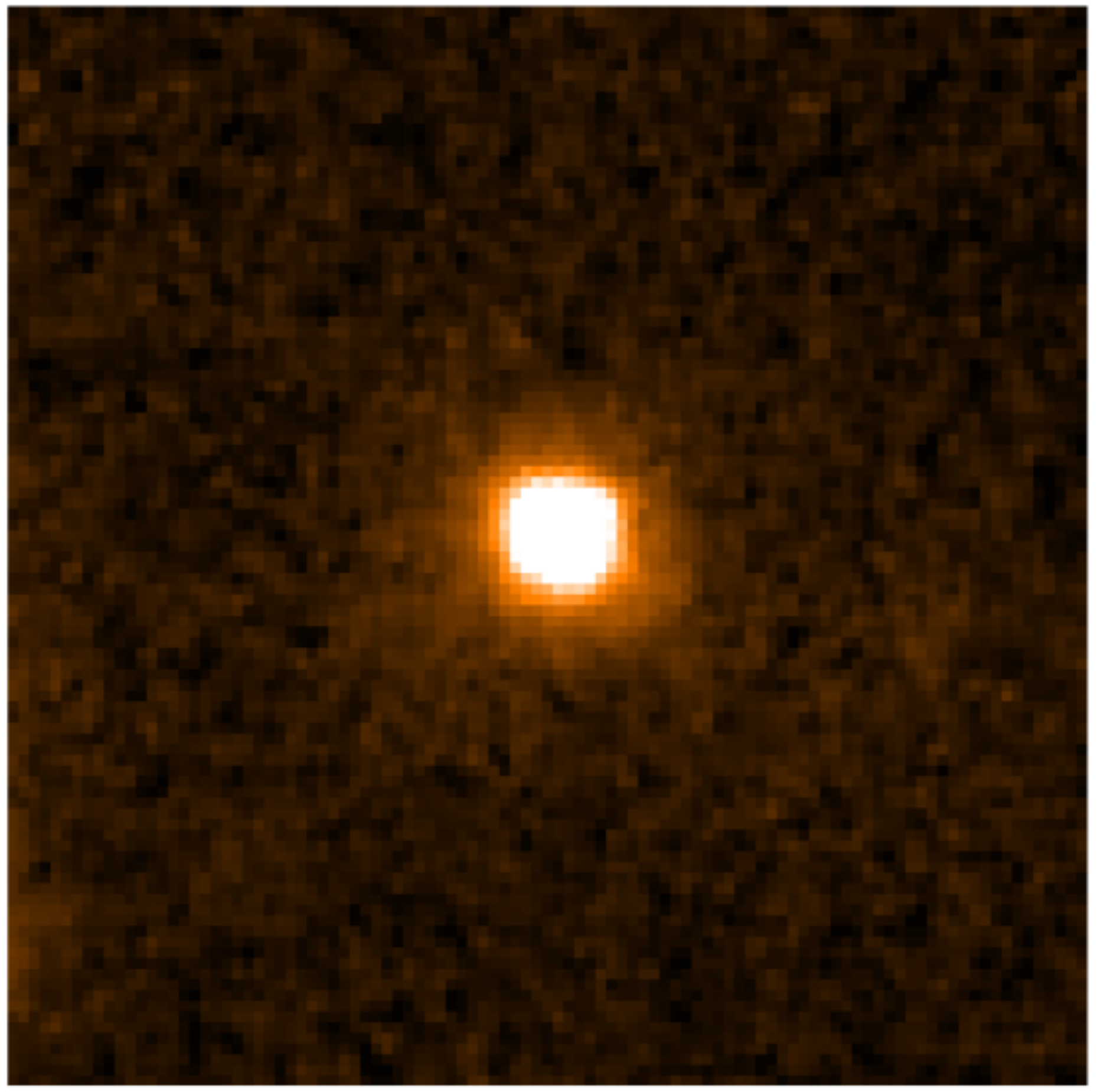}  &
\includegraphics[scale=0.15]{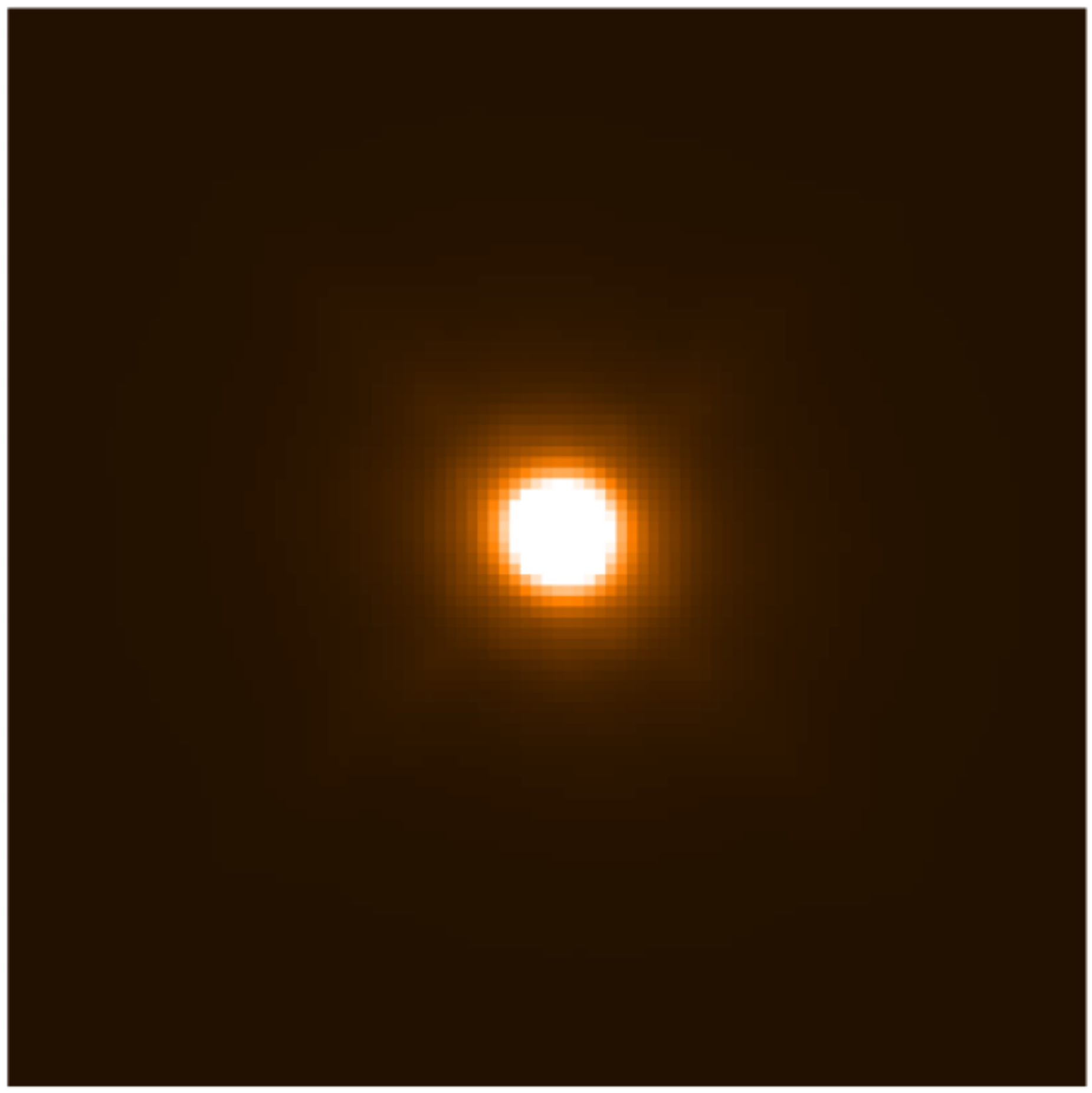}  &
\includegraphics[scale=0.15]{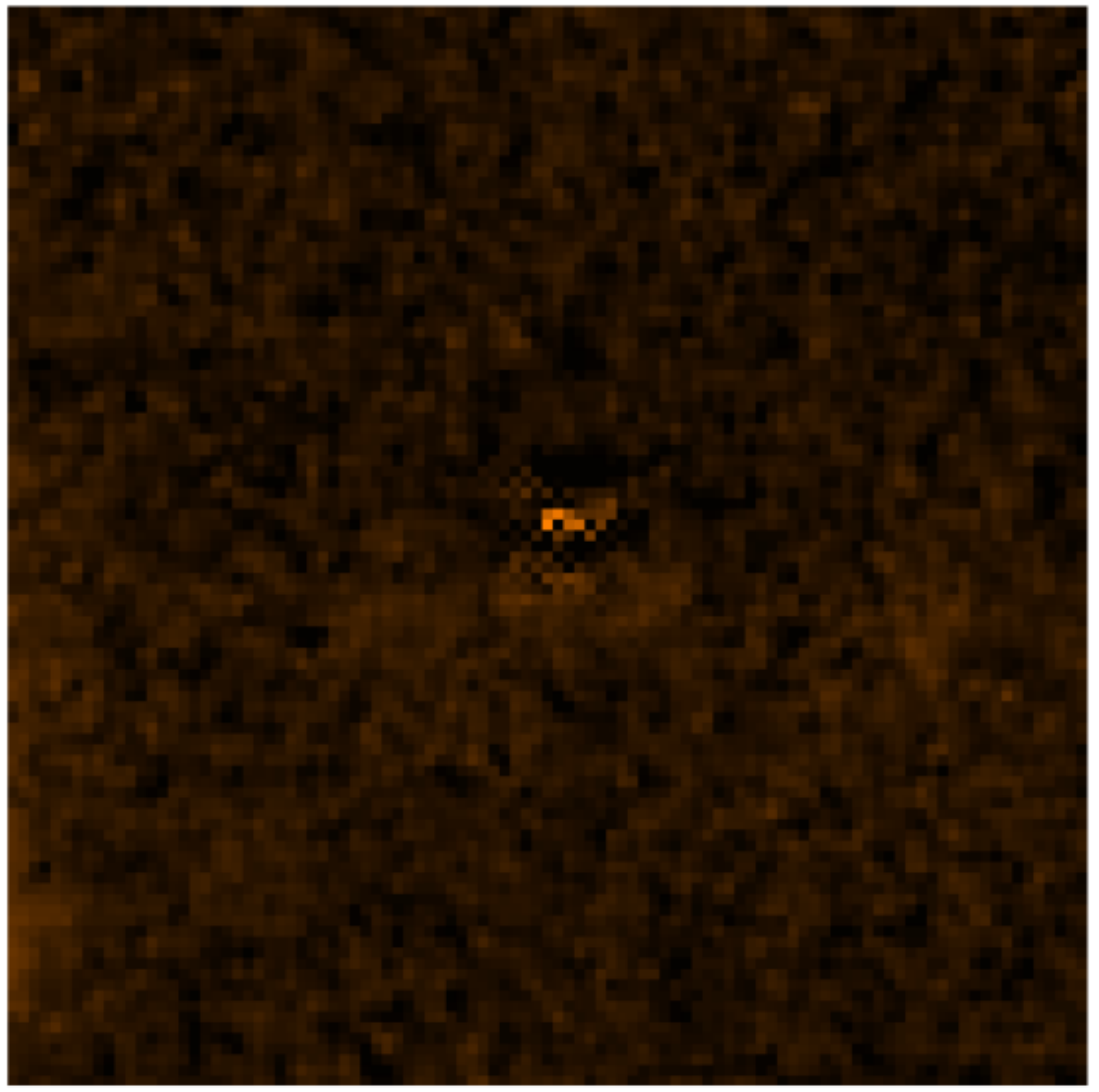}  &
\includegraphics[scale=0.15]{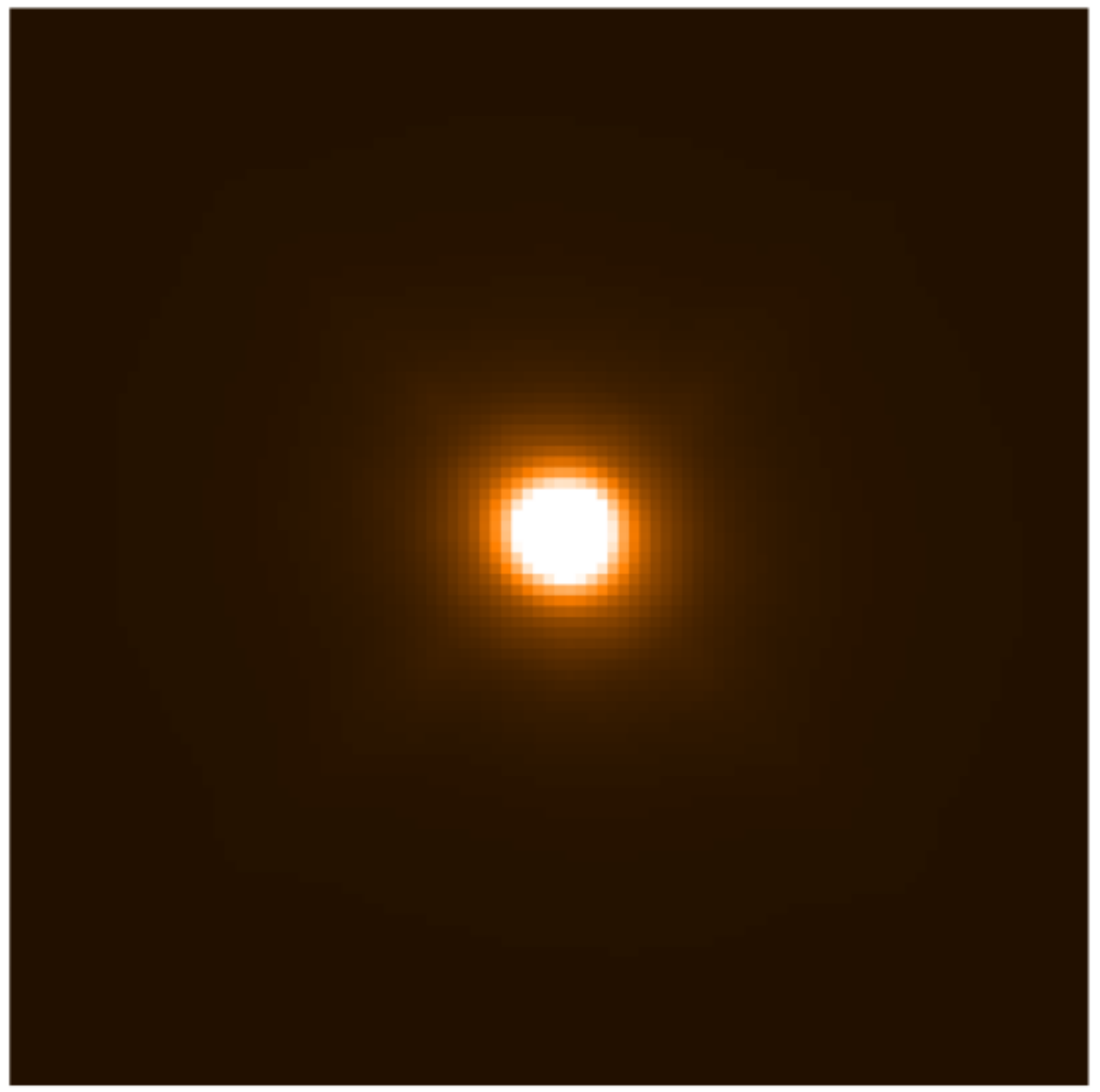}  &
\includegraphics[scale=0.15]{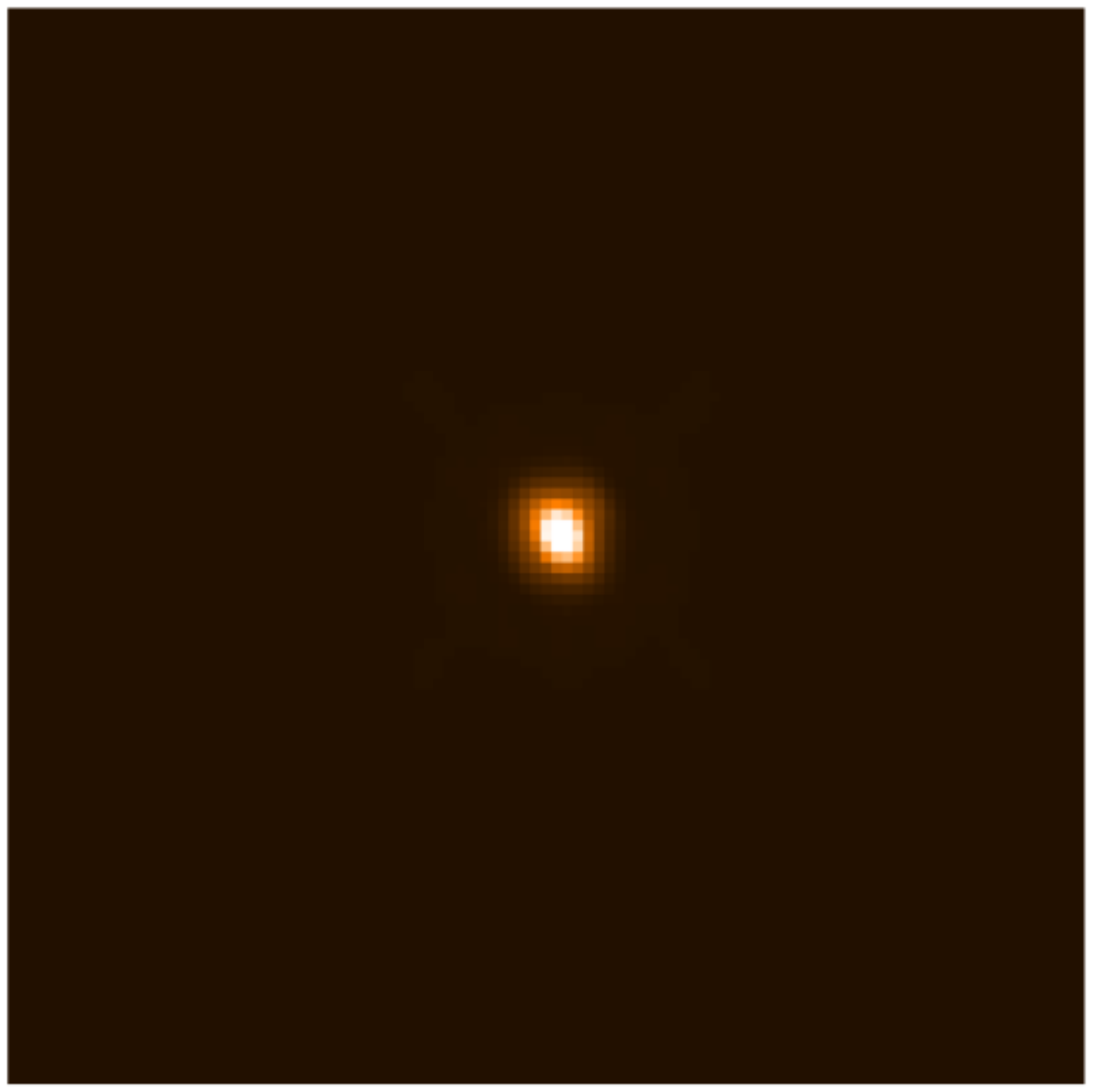}  
\end{tabular}

\begin{tabular}{m{2.8cm}m{2.8cm}m{2.8cm}m{2.8cm}m{2.8cm}}
\includegraphics[scale=0.15 ]{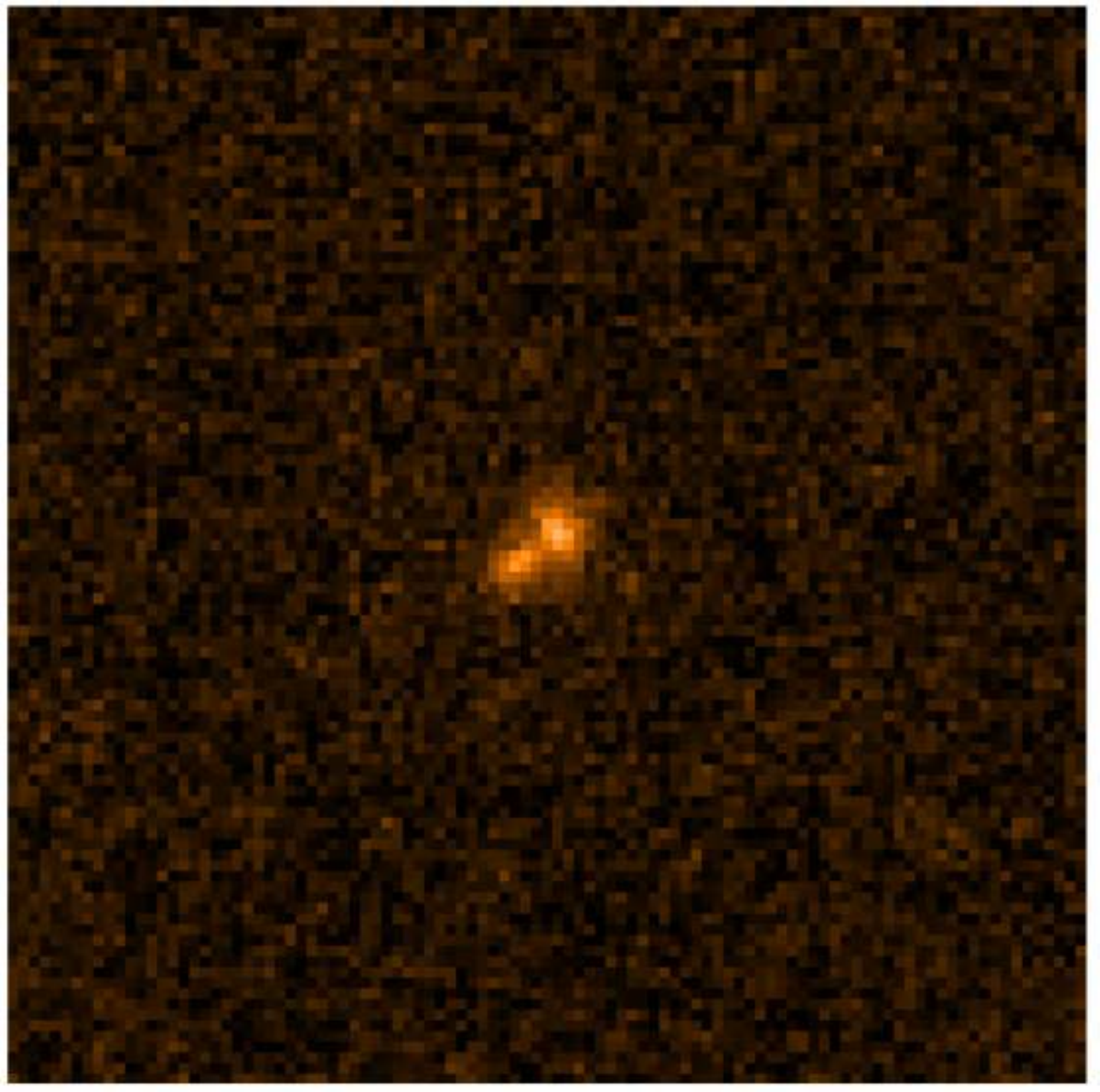}  &
\includegraphics[scale=0.15]{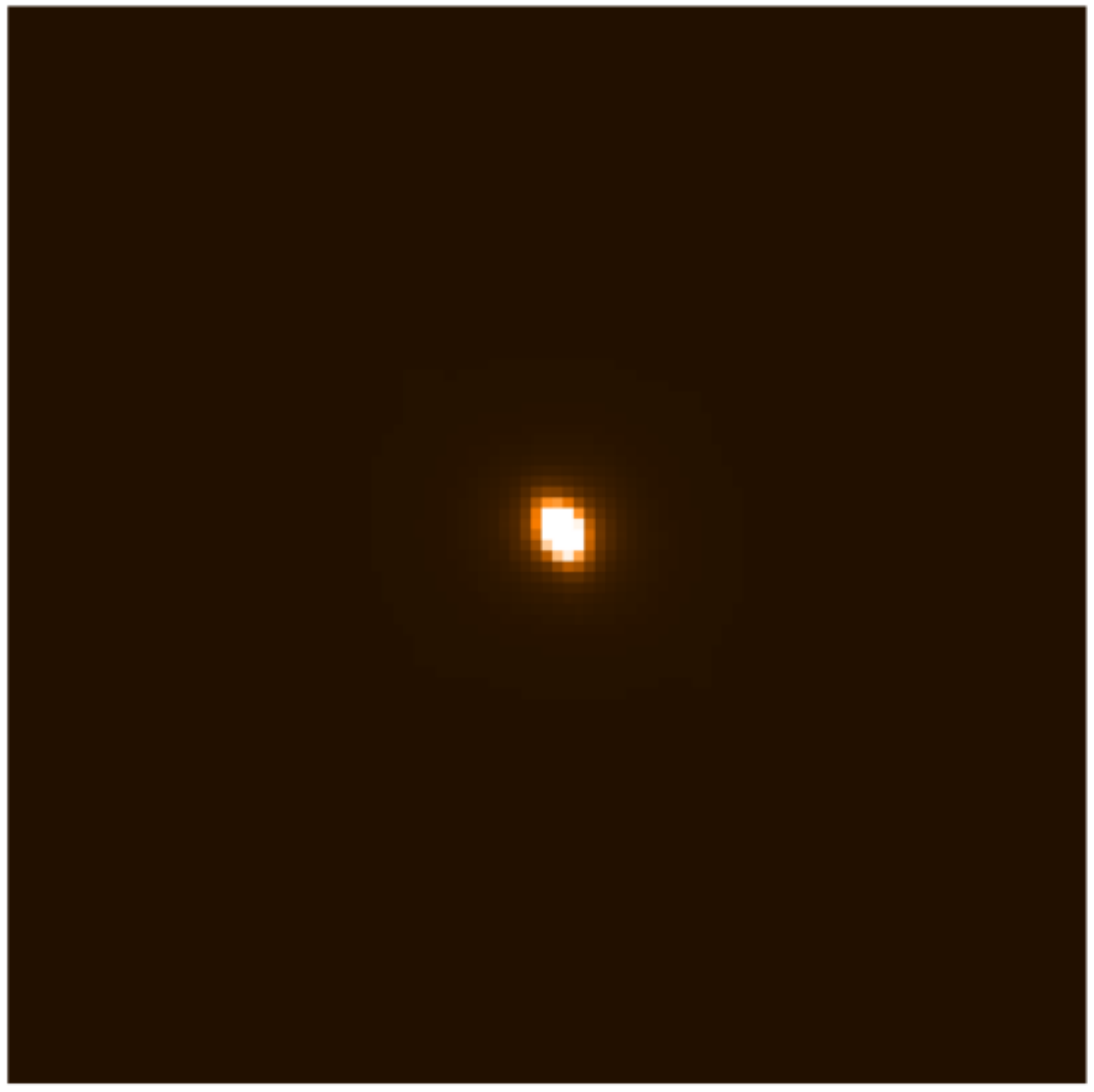}  &
\includegraphics[scale=0.15]{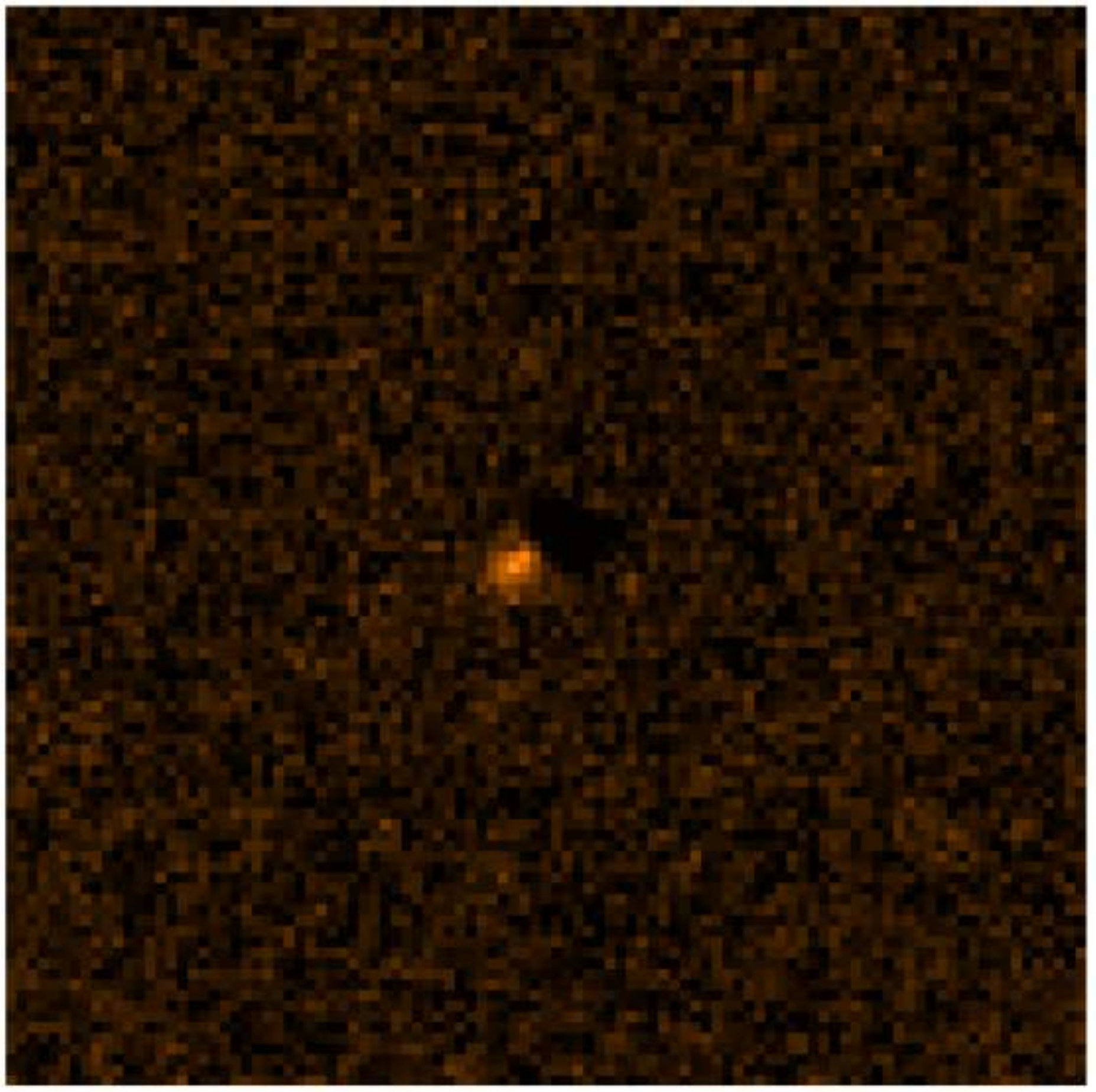}  &
\includegraphics[scale=0.15]{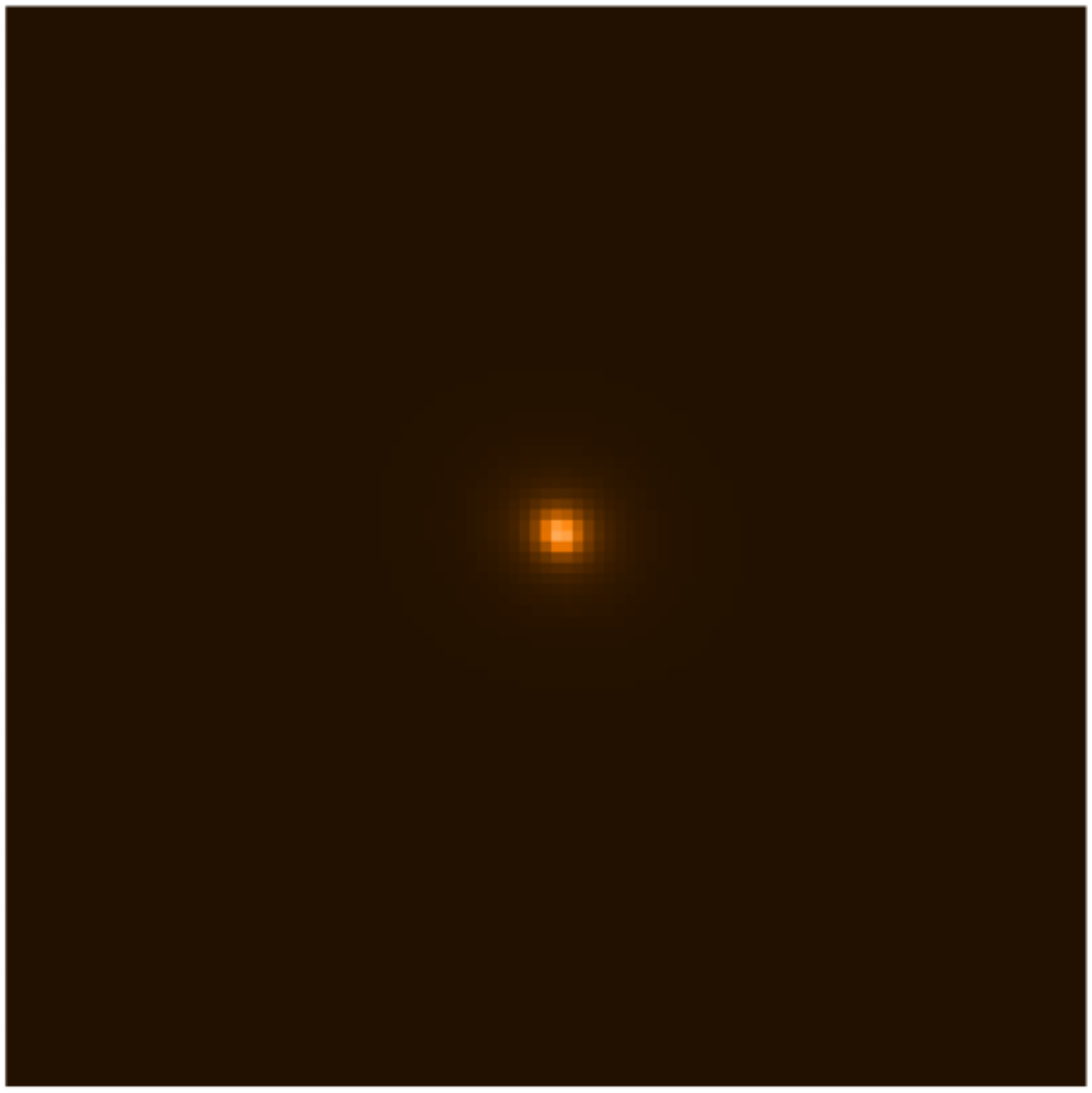}  &
\includegraphics[scale=0.15]{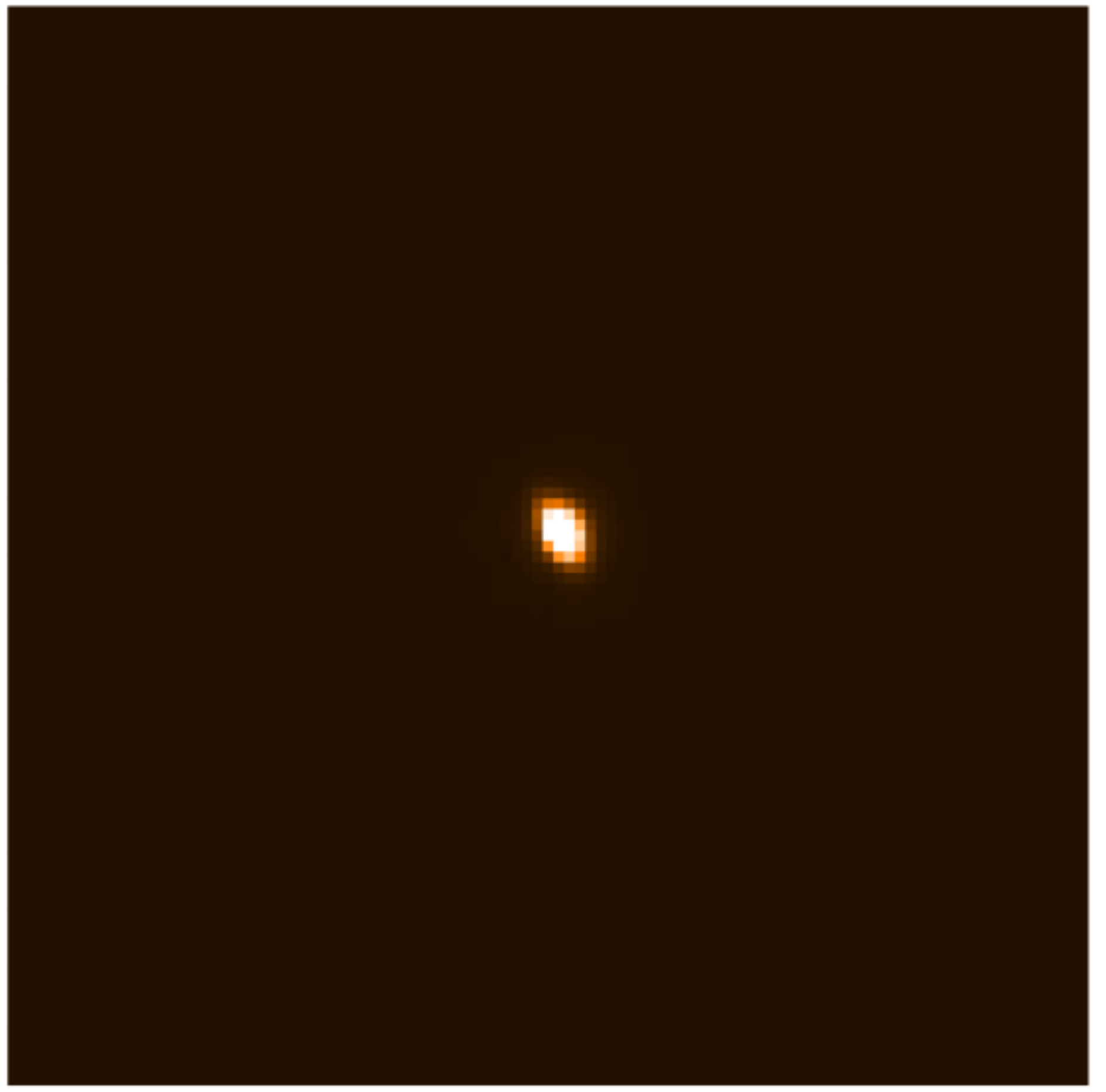}  
\end{tabular}
\begin{tabular}{m{9cm}}
\includegraphics[scale=0.7]{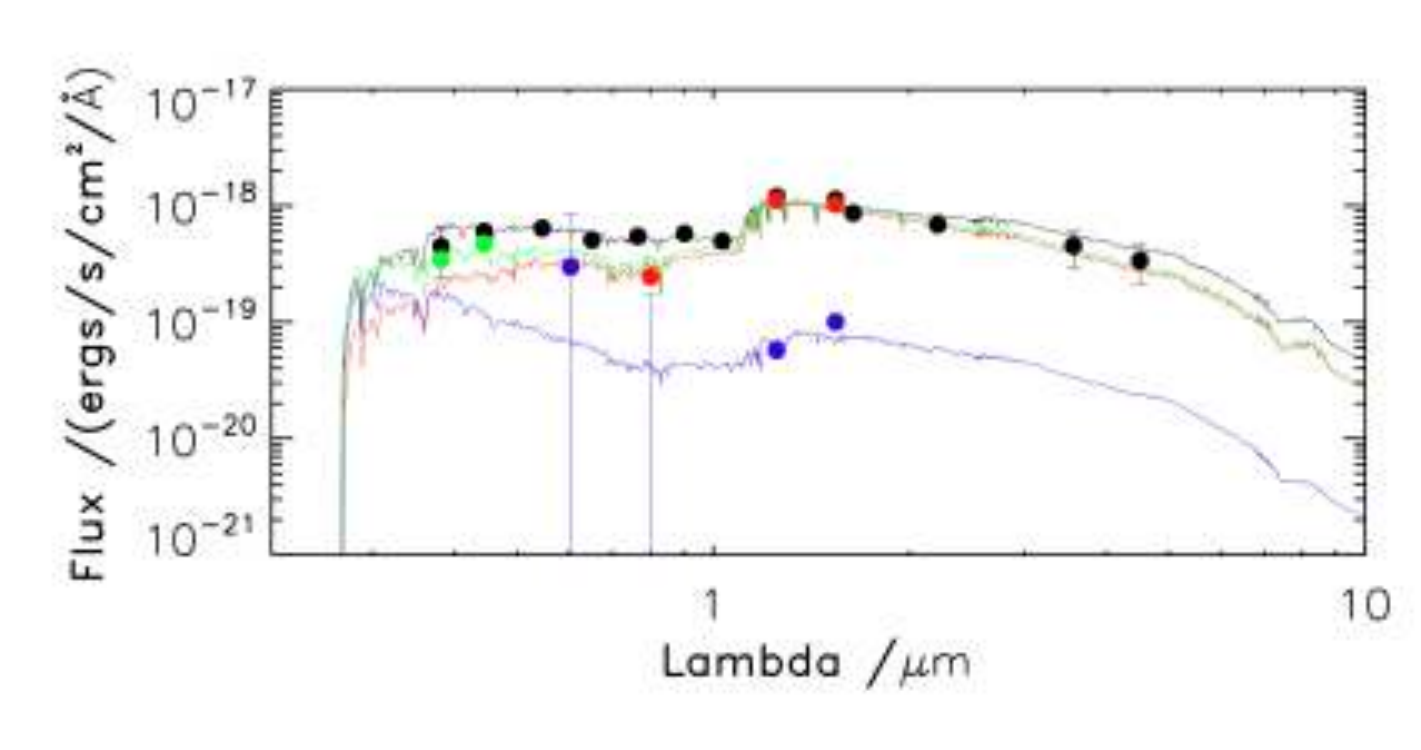}  \\
\includegraphics[scale=0.7]{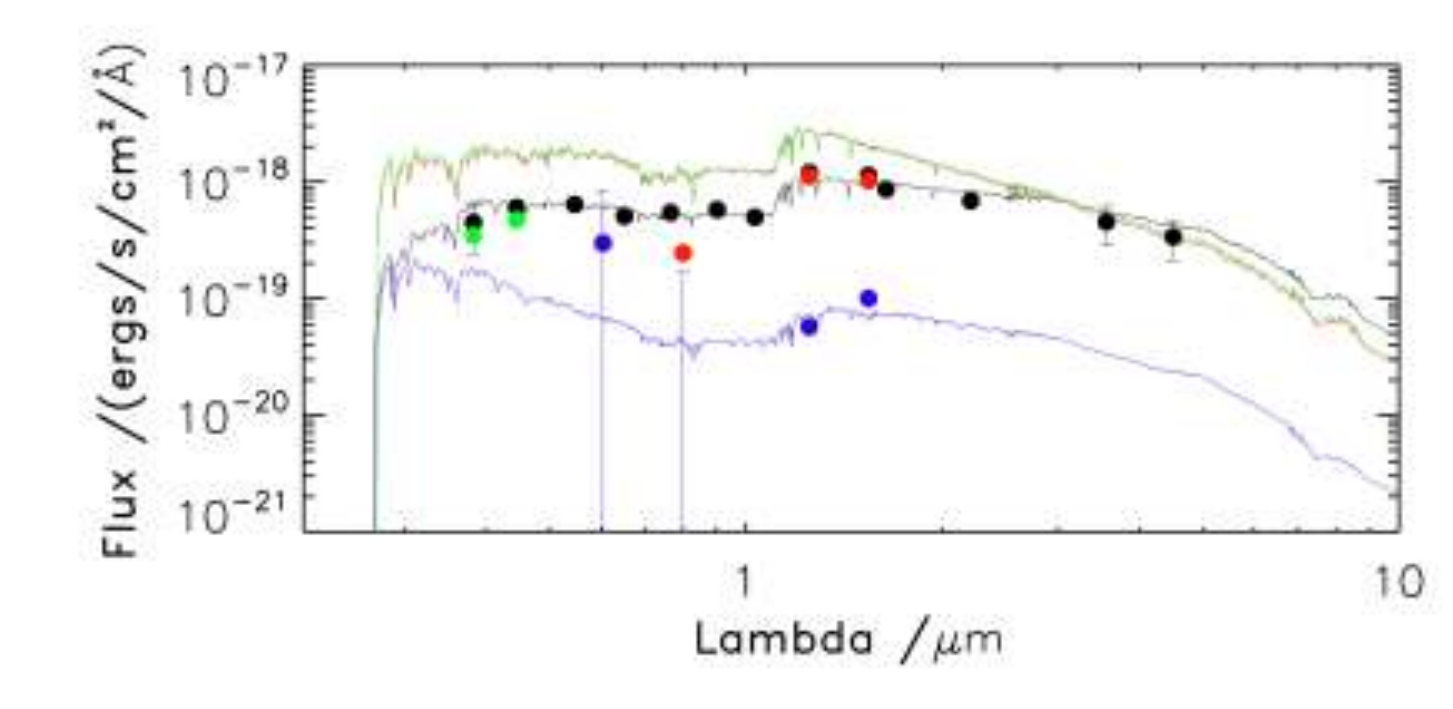}
\end{tabular}
\caption[The second genuinely star-forming bulge-dominated galaxy.]{The second genuinely star-forming bulge-dominated galaxy in our sample. This figure follows the same configuration as Fig.\,\ref{fig:sfb1}, where in the bottom panels, which illustrate the SED fits, the green data-points are the re-measured 2.5\,arcsec radius aperture photometry in the u' and B bands. }
\label{fig:sfb2}
\end{center}
\end{figure*}

Visual examination of one of these object revealed that it was a mis-classified AGN and has in fact been identified as a Type1, unobscured, AGN by \citet{Trump2009}. Therefore, we have removed it from the star-forming bulge population and combine the two remaining bulge-dominated systems with the 11 ``pure'' bulges and the 2 bulge+PSF fits, and comparing them to the 136 star-forming galaxies also with ACS coverage, provides an estimate for ($15/135$) $11\pm3\%$ of all star-forming galaxies to be bulge-dominated systems.

These 2 star-forming bulge-dominated bulge+disk systems are shown in  Figs.\,\ref{fig:sfb1} and \ref{fig:sfb2}. These plots show the master $H_{160}$ and blue $v_{606}$ stamps for the images, best-fit models, separate model bulge and disk components and residuals, and justify that for the first two objects in  Figs.\,\ref{fig:sfb1} and \,\ref{fig:sfb2} these bulge-dominated morphologies are good fits to the objects. For these star-forming bulges we have also shown the best-fit SED models and below them the same best-fit models this time corrected for the modelled dust obscuration. These dust-corrected SEDs allow a direct comparison between the contribution to the flux of the galaxy at the blue end from the bulge and disk components and have been included as they support the classification of the bulge components as star-forming.

\begin{figure*}
\begin{center}
\includegraphics[scale=0.7]{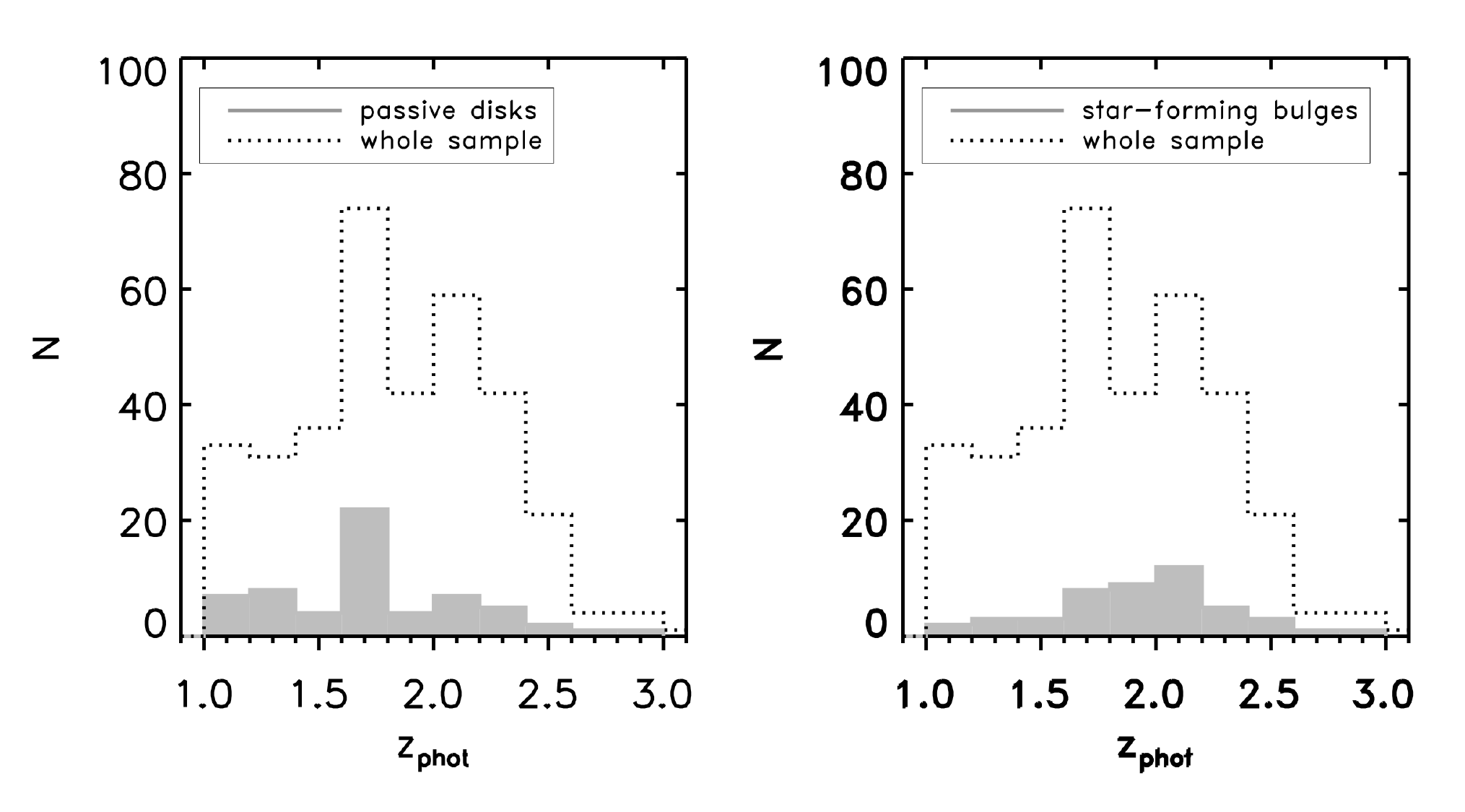} 
 \caption[Redshift distribution of the passive disks and star-forming bulges.]{The redshift distribution of all the $B/T<0.5$ passive disks (left) and $B/T>0.5$ star-forming bulges (right), compared to the whole sample of objects. This appears to show that the star-forming bulges preferentially lie at higher redshifts. This result is not statistically significant, which may be due to the small number of objects in the star-forming bulge sample, but a K-S test does reveal that the passive disks and star-forming bulges are not consistent with being drawn from the same distribution at the $3-\sigma$ level.}
\label{fig:}
\end{center}
\end{figure*}

Finally, we have also explored the mass and redshift distributions for the entire star-forming bulge-dominated galaxy population, using the overall $sSFR<10^{-10}\,{\rm yr^{-1}}$ and $B/T>0.5$ $H_{160}$ light-fraction criteria to obtain larger numbers for comparison and find that the star-forming bulge-dominated galaxies appear to preferentially lie at higher redshifts compared to both the full $M_*>10^{11}\,{\rm M_{\odot}}$ sample and all bulge-dominated galaxies. This result is not statistically significant, which may be due to the small number of objects in the star-forming bulge sample, but a K-S test does reveal that the passive disks and star-forming bulges are not consistent with being drawn from the same distribution at the $3-\sigma$ level. In fact, given the mass selection imposed for this study, this result is to be expected in the context of downsizing, where the most massive galaxies are observed to be more active at higher redshift and experience an accelerated evolution, quenching before less massive systems. This is further supported by the similarity in the redshift distributions of the star-forming bulges and all star-forming galaxies.

\section{Axial Ratios}

In addition to providing new insight into how the overall morphologies of the star-forming and passive components evolve, the detailed morphological analysis employed in this work has also allowed us to explore the axial ratios of the passive and star-forming populations. Axial ratio measurements provide valuable additional information about the structure of these components and any trends with redshift can offer further indicators of the physical processes which govern galaxy evolution within this epoch.

\begin{figure*}
\begin{center}
\includegraphics[scale=0.7]{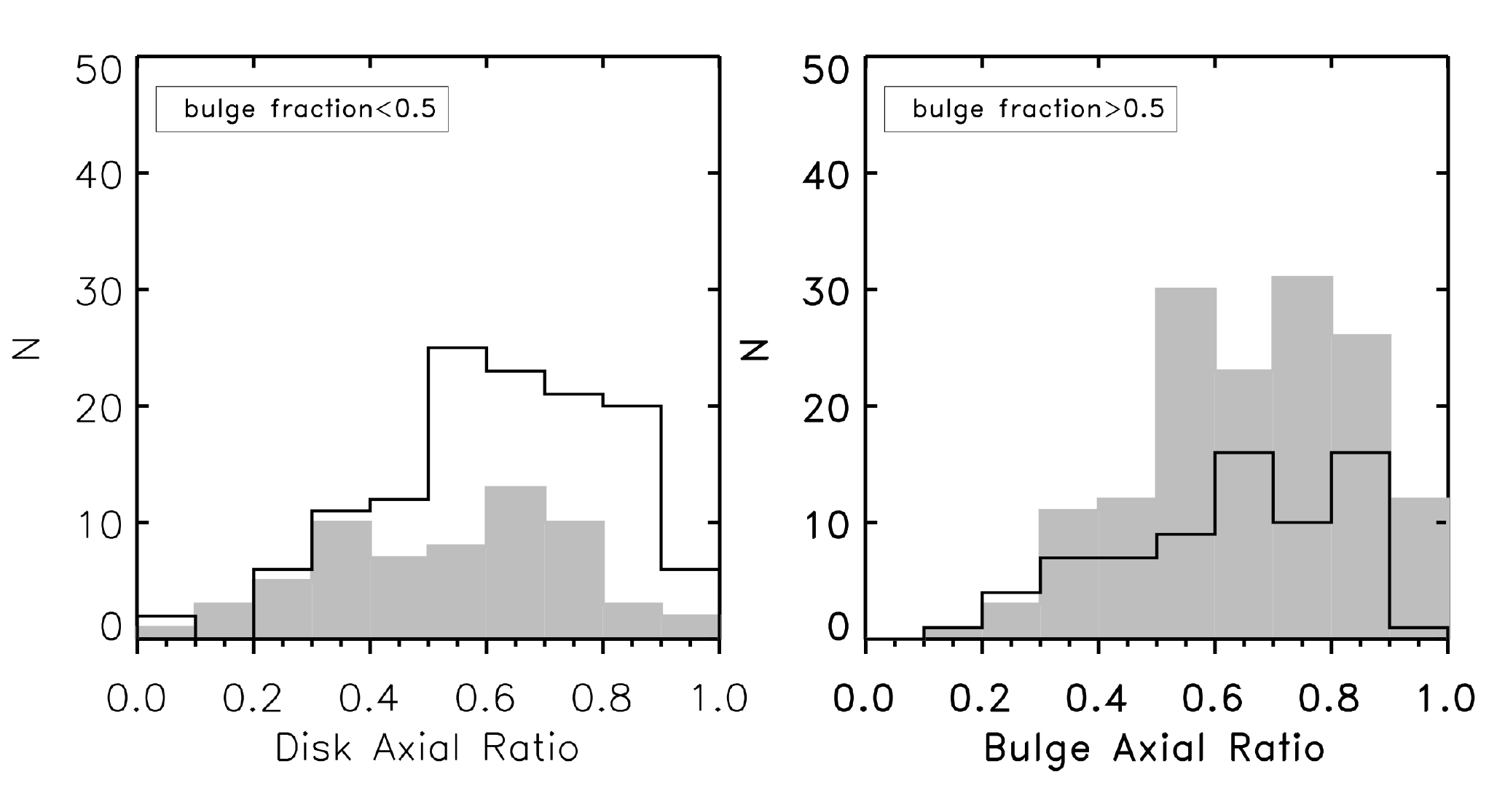} 
\end{center}
 \caption[Axial ratio distributions for the star-forming and passive bulges and disks.]{Axial ratio distributions for the star-forming and passive disk (left) and bulge (right) components of the disk and bulge-dominated galaxies. In both plots the star-forming components are plotted in the black solid line, while the passive components are represented by the filled grey regions. }
\label{fig:axis1}
\end{figure*}

 The axial ratio distributions for the disk components of disk-dominated galaxies (as judged by $H_{160}$ light fractions) and bulge components of the bulge-dominated galaxies are shown in Fig.\,\ref{fig:axis1}, where the samples have been split into passive and star-forming sub-populations using the total galaxy specific star-formation rate at $sSFR=10^{-10}\,{\rm yr^{-1}}$, for simplicity. Similar to the results in \citet{Bruce2012}, these distributions reveal that the axial ratios of both the star-forming and passive bulge components are peaked around $b/a\approx0.7$, consistent with bulges in the local Universe (e.g. \citealt{Padilla2008}). However, visually the passive disks display a markedly flatter distribution to the star-forming disks, which look to be more consistent with the bulge components. Even so, based on the results from K-S tests, all four distributions shown in Fig.\,\ref{fig:axis1} are actually consistent with being drawn from the same underlying distribution. This includes the passive and star-forming disks which, despite appearing to be significantly different, are not statistically distinguishable at the $2-\sigma$ level ($p=0.11$).

The peaked distribution of the bulge components is consistent with a population of tri-axial objects, and similar distributions have been found for bulge-dominated systems at both low (e.g. \citealt{Padilla2008}) and high redshifts (e.g. \citealt{Ravindranath2006}). Similarly, the apparently flatter distribution of the passive disks is in agreement with expectations from a population of randomly oriented thin disks.

\subsection{Passive disks}
The apparently flat distribution for the passive disks provides further corroboration that the passive disk-dominated galaxies are genuine, and agrees well with other axial ratio studies at $z>1$ such as \citet{Chang2013b}. This study used CANDELS photometry to explore the evolution of the axial ratios of early-type galaxies defined by low star-formation rates from rest-frame colours, where no additional morphological distinction was made. \citet{Chang2013b} found, from de-projecting the observed axial ratios of the galaxies in their sample, that both the local SDSS distribution and their $1<z<2.5$ distribution were not consistent with a single population of structures which are randomly oriented, but can be accurately modelled by two-components: a round tri-axial (bulge-like) population and a flatter oblate (disk-like) population. In their high-mass $M_*>10^{10.8}\,{\rm M_{\odot}}$ bin, they also found evidence that this oblate population increases as a fraction of the total number of objects from $20\pm2\%$ at $z=0$ to $60\pm1\%$ at $1<z<2.5$. These results not only support the flat distribution that we find for the passive disks but also serve to independently substantiate the findings that, at higher redshifts, these most massive galaxies are increasingly disky, and that pure bulge galaxies emerge slowly within the $0<z<3$ epoch from a population of mixed systems.
 
\begin{figure*}
\begin{center}
\includegraphics[scale=0.7]{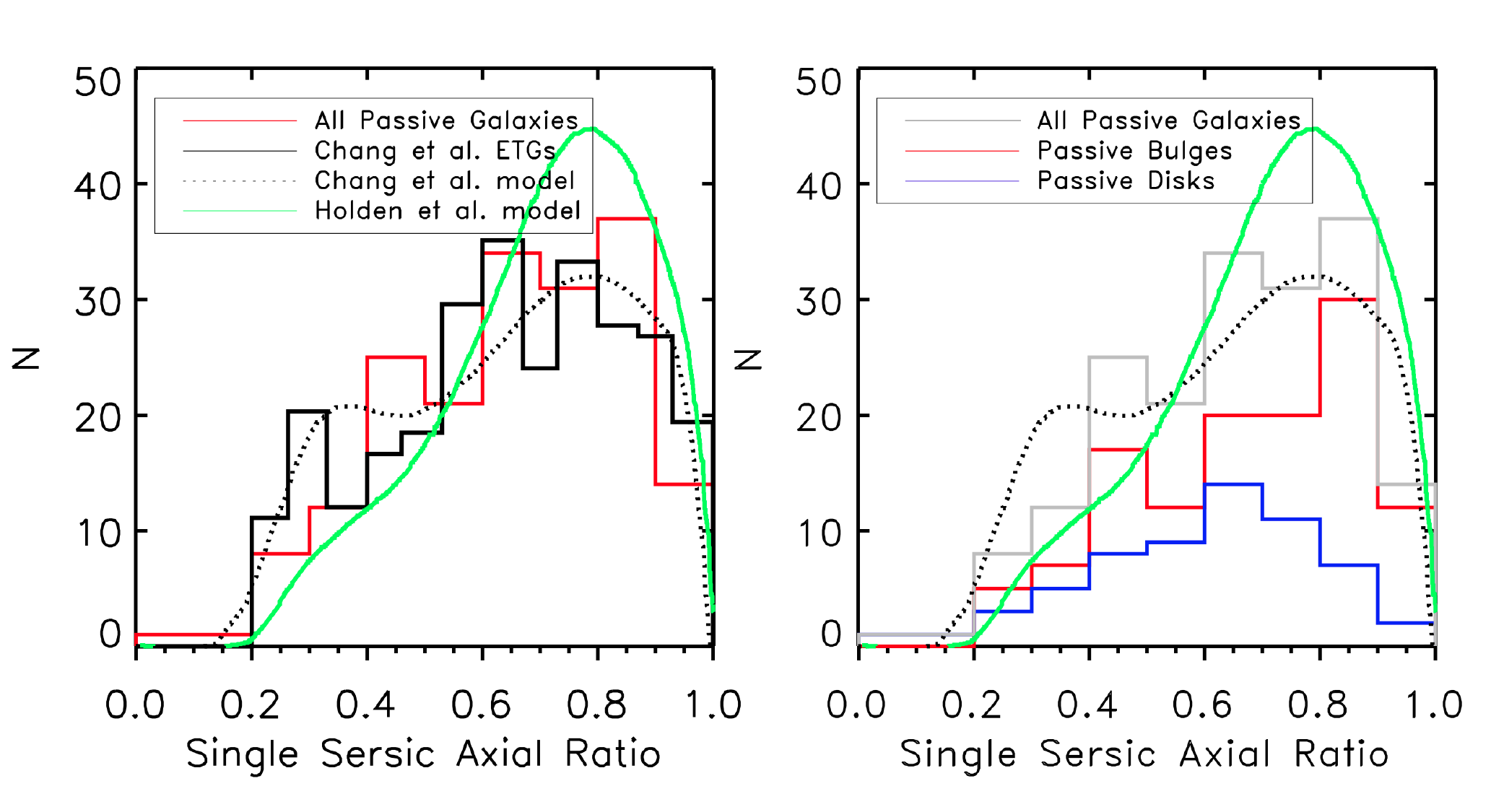} 
\end{center}
 \caption[The single-S\'{e}rsic axial ratio distributions for the passive and star-forming galaxies.]{Left: The single-S\'{e}rsic axial ratio distributions for the passive (red) galaxies in our sample over-plotted with the \citet{Chang2013b} observed distribution in black and their model in the dotted black line, where these distributions have been re-normalised to match the area under our histograms. Also included in this plot is the local SDSS model distribution from \citet{Holden2012}. The passive galaxies in our mass-selected sample display a distribution which appears to be intermediate between the $1<z<2.5$ \citet{Chang2013b} ETGs, and the local SDSS sample. Right: the axial ratio distributions from the single-S\'{e}rsic fitting for the passive galaxies (grey) also split into the bulge (red) and disk-dominated (blue) populations. By splitting the passive galaxies into their morphological classifications we find that the single-S\'{e}rsic axial ratios also separate into a flatter distribution for the oblate disk-dominated objects and a peaked distribution for the tri-axial bulge-dominated galaxies. This is roughly consistent with the \citet{Chang2013b} results, although the contribution from the oblate population in the models of \citet{Chang2013b} to the flat tail of the distribution appears to be significantly larger in comparison with our sample. }
\label{fig:axis3}
\end{figure*}

Given that  \citet{Chang2013b} used single-S\'{e}rsic morphological fits, to allow a more direct comparison we have included the single-S\'{e}rsic axial ratio fits from our passive disks in the left panel of Fig.\,\ref{fig:axis3}. Also included in the comparison in Fig.\,\ref{fig:axis3} is the model distribution for the local SDSS galaxies from \citet{Holden2012}. From this comparison our passive disks appear to display a distribution which is somewhat intermediate between the peaked local distribution and the $1<z<2.5$ ETG sample, which has a more extended flat tail.  By explicitly splitting the passive population into the bulge and disk-dominated samples there is a more discernible difference between the contribution of the oblate population to the overall axial ratio distribution in the \citet{Chang2013b} models, compared to the distribution of the passive disks in our mass-selected sample. From this comparison the extended flat tail from disks appears more populated in the \citet{Chang2013b} models than with respect to our observed distributions, especially at $b/a<0.35$. However, there are still visual indications that the passive disks have a flatter distribution than the star-forming disks,  although these are not statistically significant. Thus, both our single-S\'{e}rsic and morphologically decomposed results broadly agree with the results of \citet{Chang2013b}, especially given the difference in the mass ranges adopted.  

\citet{Chevance2012} have also explored the axial ratios of 31 compact ($R_{e}<2\rm{kpc}$), passive galaxies at  $z>1.5$ and find that while the single-S\'{e}rsic index distributions of their high-redshift sample are more consistent with local disks, the axial ratio distribution has only a $5\%$ probability of being drawn from the same population and is more consistent with the distribution of local ETGs. These results suggest that despite the passive galaxies having ``disk-like'' light profiles, their axial ratios are rounder. \citeauthor{Chevance2012} propose that their high-redshift sample is either a mixture of bulges and disks or that the high-redshift galaxies are a genuinely distinct population with no real local analogues. By cutting our full sample at $z>1.5$ and plotting the single-S\'{e}rsic index and axial ratio distributions of all the passive galaxies and just the most compact systems with $R_{e}<2\rm{kpc}$, we find agreement with the  \citet{Chevance2012} results. The single-S\'{e}rsic index distributions are more similar to the distributions of local disks than ETGS, peaking at a value of $n\sim 1.5-2$, while the axial ratio distributions are more peaked than the flat distributions observed from local disks. In fact, examining the bulge-to-total light fractions from our decompositions reveals that these passive, compact, objects are indeed a mixture of bulge plus disk systems with a median $B/T\sim 0.59$. These results are in line with our previous findings that within the $1<z<3$ era all massive galaxies become increasingly mixed systems, which gives rise to measured single-S\'{e}rsic indices which have intermediate values. 

\subsection{Star-forming disks}

While the flat distribution of the passive disk components is supported by previous studies, the relatively peaked distribution of the star-forming disk components has been less well explored in the literature, and remains less understood. Nevertheless, this peaked distribution does agree with the results of \citet{Law2012}, who conducted a single-S\'{e}rsic and non-parametric morphological analysis of  $M_*=10^{8}-10^{11}\,{\rm M_{\odot}}$ star-forming galaxies at $1.5<z<3$ using {\it HST}/WFC3 imaging and reported an axial ratio distribution for $n<2.5$ galaxies peaked at $b/a\approx0.6$.

\begin{figure*}
\begin{center}
\includegraphics[scale=0.7]{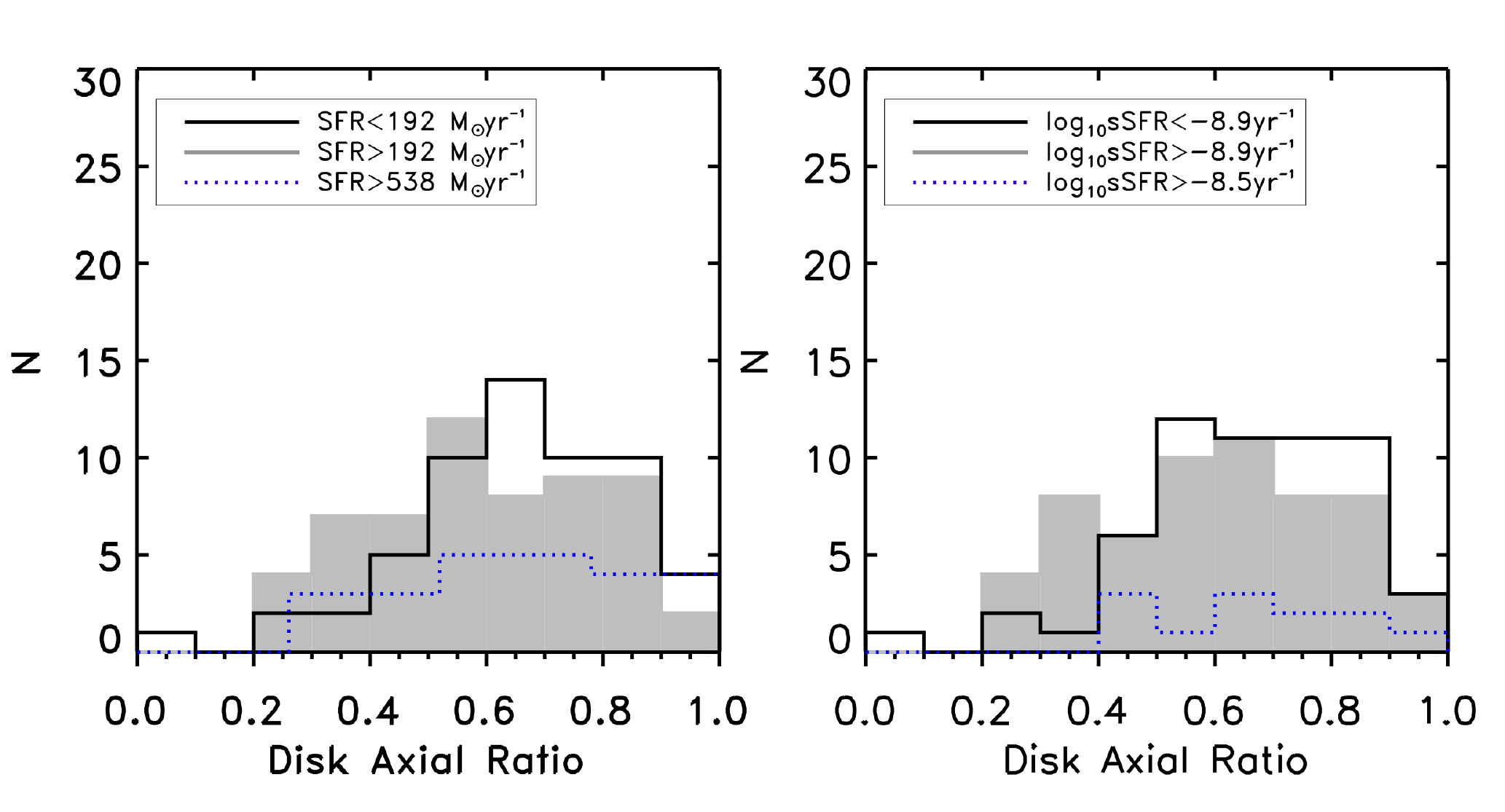} 
\end{center}
 \caption[Axial ratio distributions for the star-forming disks, split by specific and absolute star-formation rate.]{Axial ratio distributions for the star-forming disks split by different measures of specific and absolute star-formation rate. On the left the star-forming disks have been split at the median absolute star-formation for the galaxies in this sub-population and the two distributions have been plotted with the less active star-formers in black and the more active disks in grey. We have also constructed an additional sub-sample of objects which have SFRs above the $90$th percentile and have over-plotted the axial ratios of their disk components in blue. This comparison appears to show tentative evidence that the most active disks have a flatter axial ratio distribution.
 On the right the star-forming disks have been split at the median specific star-formation for the galaxies in this sub-population and the two distributions have been plotted with the less active star-formers in black and the more active disks in grey. We have also constructed an additional sub-sample of objects which have sSFRs above the $90$th percentile and have over-plotted the axial ratios of their disk components in blue. This comparison appears to also show some weak evidence that the most active disks have a flatter axial ratio distribution.}
\label{fig:axis4}
\end{figure*}

To better explore the origin of this peaked distribution, we have cut our star-forming, disk-dominated, disk component sample according to several different criteria. In the first case, we have cut the sample according to different indicators and measures of star-formation. We first compared the axial ratio distributions of the whole star-forming disk-dominated population with only those objects which also had a $24\mu$m detection from either SpUDS or S-COSMOS. However, this was not informative as the majority of star-forming disk-dominated galaxies have a $24\mu$m counter-part. We next compared the distributions for the sample split by their absolute and specific star-formation rates, as shown in Fig.\,\ref{fig:axis4}. The left-hand panel of Fig.\,\ref{fig:axis4} displays the cuts made at increasingly higher absolute star-formation rates for the overall galaxy. The $SFR>190\,{\rm M_{\odot}yr^{-1}}$ cut corresponds to the median absolute star-formation rate for the full star-forming disk-dominated galaxy sample, and the $SFR>540\,{\rm M_{\odot}yr^{-1}}$ cut corresponds to the $90$th percentile value. There is perhaps tentative evidence from this comparison that by imposing the extreme SFR cut, the distribution begins to flatten out, however the number of galaxies within this bin becomes very small, so this result is by no means robust. The right-hand panel of Fig.\,\ref{fig:axis3} displays similar cuts made at increasingly higher specific star-formation rates for the overall galaxy. The $sSFR>10^{-8.9}\,{\rm yr^{-1}}$ cut now corresponds to the median specific star-formation rate for the full star-forming disk-dominated galaxy sample, and the $sSFR>10^{-8.5}\,{\rm yr^{-1}}$ cut corresponds to the $90\%$ quartile value. These distributions agree with the results from splitting by absolute star-formation rate as they also display a weak trend for the most active disks defined by specific star-formation rate to have flatter axial ratios.

\begin{figure*}
\begin{center}
\includegraphics[scale=0.7]{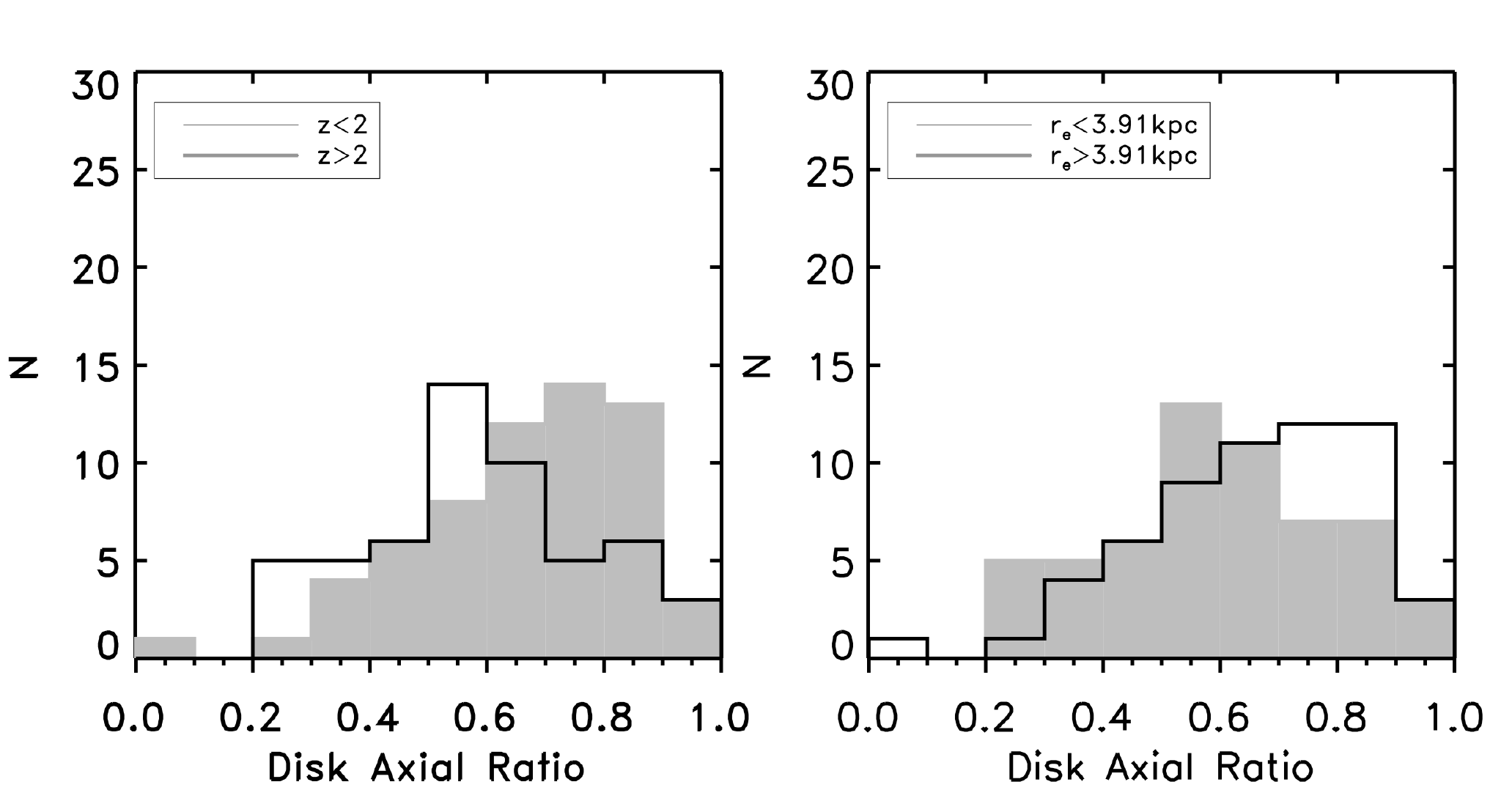} 
\end{center}
 \caption[Axial ratio distributions for the star-forming disks, split by redshift and disk size.]{The axial ratio distributions for the star-forming disks split by photometric redshift (left) and disk size (right), where the disks have been split according to the median disk size of $3.9$\,kpc. These distributions reveal a trend for lower redshift galaxies to have flatter axial ratios, although this can be explained by the redshift size evolution of the disks. In fact, these distributions demonstrate that the smallest disks are the most tri-axial, which may be due to an absolute increase in disk scale-height due to star-formation at high redshift which would have a greater affect on the axial ratios of disks with smaller scale-lengths.}
\label{fig:axis5}
\end{figure*}

In addition to studying how the axial ratio distribution of the disk components varies with the star-formation rate of the galaxy, we have also examined how it varies with other properties of the galaxy. To do this, we first split the star-forming, disk-dominated, disk components above and below $z=2$ to investigate if there was any redshift evolution evident. These distributions are given in the left-hand panel of Fig\,\ref{fig:axis5}, and reveal that the star-forming disk components at $z>2$ are peaked towards a higher $b/a$ value compared to the $z<2$ components. This is interesting as, naively the $z>2$ disks would be expected to be more active than those at  $z<2$. Given that for the previous figure we reported that there might be a potential trend for the distribution to flatten out for the most active disks, this result appears to be in direct conflict with this assertion. Moreover, we have also conducted a K-S test for these two redshift binned distributions and found that at the $2-\sigma$ level ($p=0.02$) they are inconsistent with being drawn from the same distribution. However, in addition to a star-formation rate evolution with redshift, we have also reported on the size evolution of the bulge and disk components with redshift in Bruce et al. (2013b), so in the right-hand panel of Fig\,\ref{fig:axis5} we have split the disks according to their median size. This clearly reveals that  the largest disks have a flatter distribution and the smaller disks have a distribution peaked towards higher values. In this case the K-S test provides  a value of $p=0.03$, again making them formally different at the $2-\sigma$ level.

The trend for smaller star-forming disks to be rounder and larger disks to be flatter can be explained if at these redshifts star formation in disks depends only on the self-gravity of the disk, not on scale-length, and this intense star-formation phase induces feedback which ``puffs up'' the disk scale-height. In which case, the ``puffiness'' of the star-forming disk is independent of the disk size, and smaller disks will appear more tri-axial in structure than larger disks. Alternatively, if these high redshift disks are clumpy then it possible that an increase in scale height may be driven by asymmetries, but the star-forming clumps in high-redshift disks have been found to not contribute significantly to the near-infrared flux \citep{Wuyts2012}, where our morphological decompositions are based.

For completeness, we have also ensured that these trends are not biased by also selecting only the disk components of disk-dominated galaxies as classified by their decomposed disk/bulge+disk masses and by splitting the sample by disk decomposed mass above and below the median value of $M_{disk*}$. 

It is worth noting that previous studies and GALFIT simulations have expressed some concern about the robustness of GALFIT results for small, flat objects due to the resolution limit, where $b/a\leq0.1$ values of small objects result in scale-heights less than a single pixel (\citealt{vanderWel2012} and private communication). Having explored this with our own simulation results we also find evidence that the single pixel effective radius disk models with axial ratios 0.1 are not well recovered and axial ratio distribution of the best-fits is centred on $b/a=0.05$ with a secondary, small, peak in the distribution at $b/a=0.2$. Such offsets are not witnessed for small disk models with $b/a=0.3,0.6$ or $1.0$, nor for our disk models with effective radii of 5 or more pixels, but in which cases there is a slight tail in the fitted $b/a$ distributions towards higher $b/a$ values such that the recovered fits for the $b/a=0.1$ models are not symmetric. (For completeness, examination of the fitted axial ratios with $b/a>0.1$ for larger model disks also show no offsets but have slightly less spread in the fitted $b/a$ distributions, which shows that for $b/a>0.1$ 1 pixel disk models the fitting procedure is not biased but does carry a larger random uncertainty.)

However, this is clearly not the reason for the dearth of $b/a<0.3$ values for the star-forming disks within our sample, as not only are the sizes of our objects too large to force the scale-heights of the majority of $b/a$ fits to be less than a pixel, but if this were the case then we would have found an over-abundance of $b/a<0.1$ values, which is not seen.

\section{Comparison with Sub-mm Galaxies}

One of the main insights from this work is that the disk components of the star-forming disk-dominated galaxies have surprisingly peaked axial ratio distributions but that we find tentative evidence that the most active star-forming galaxies have the flattest distributions. However, in order to isolate the most active systems we restricted this cut to a very small number of objects, which prevented us from drawing any robust conclusions. As a result, this issue is clearly worthy of further investigation. 
To this end, in this section we have considered the mm/sub-mm AzTEC selected sample of \citet{Targett2013} in the GOODS-South field, which has been selected from sub-mm imaging at a uniform depth over a similar area to the CANDELS-UDS and COSMOS fields. Sub-mm selected galaxies are widely agreed to be extreme star-forming galaxies, although there is debate over whether this activity is triggered by major mergers or whether it is just the high-end tail of the normal star-formation main-sequence. One of the main conclusions of \citet{Targett2013} is that the (sub-)mm selected galaxies represent the extreme star-forming end of the morphologically disk-dominated population, thus they provide the ideal sample for extension of this axial ratio study.

\begin{figure*}
\begin{center}
\includegraphics[scale=0.7]{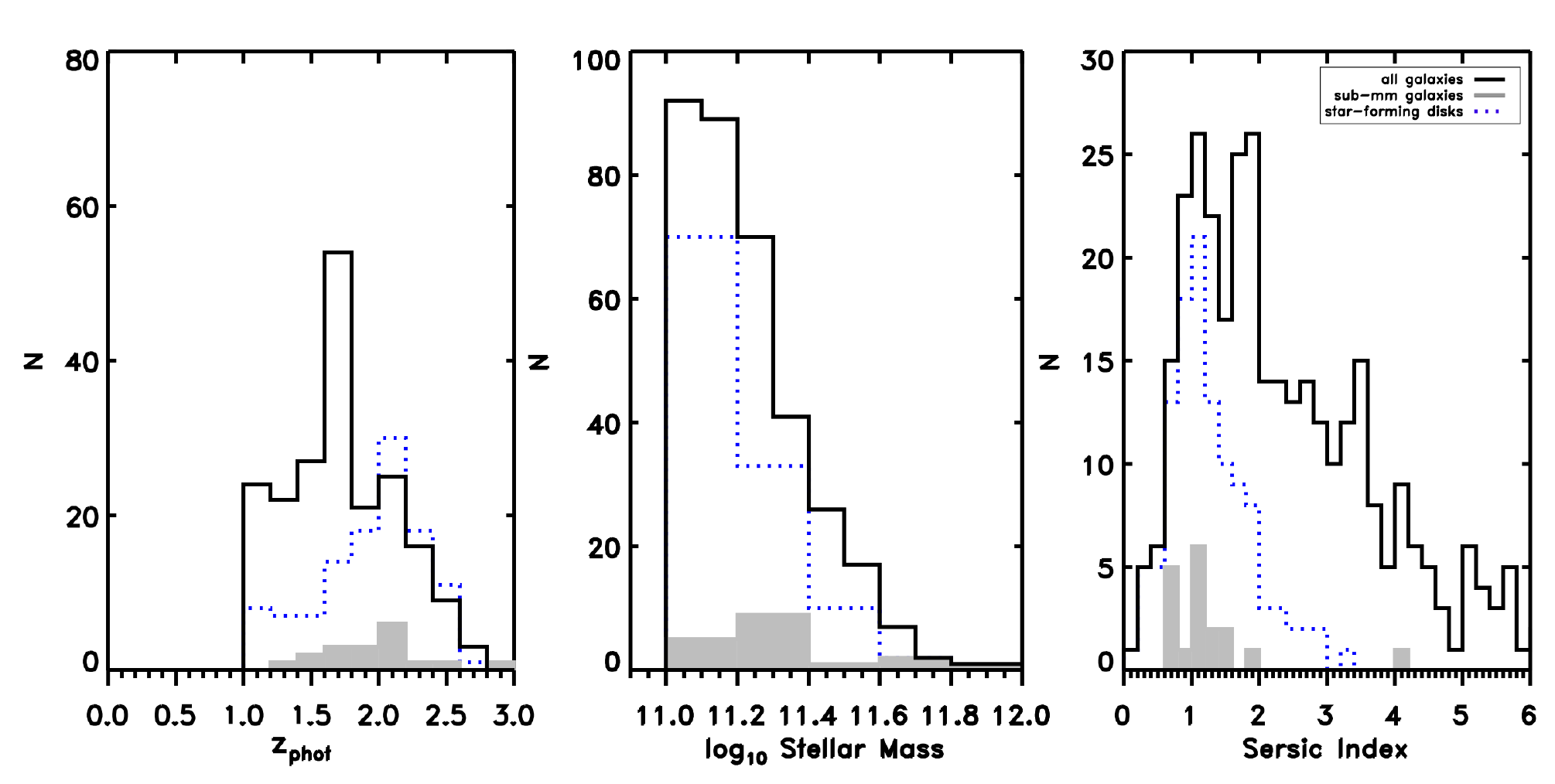} 
\end{center}
 \caption[Properties of the (sub-)mm galaxies.]{Comparison of the properties of our sample and the mass-selected sample of \citet{Targett2013}. The combined UDS and COSMOS CANDELS samples are plotted in the black solid histograms and the star-forming disk-dominated galaxies in the blue dashed histograms, where here the properties for the overall galaxies rather than for the disk components are used to allow a fairer comparison to the (sub-)mm selected samples. The \citet{Targett2013} (sub-)mm sample is shown in the light grey filled histogram. The left panel displays the redshift distributions, the middle panel shows the total galaxy stellar-mass distributions and the right panel shows the single-S\'{e}rsic index distributions. The comparison between these samples verifies that the GOODS-S and AzTEC (sub-)mm sample contains comparably massive, disk-dominated galaxies at similar redshifts to our CANDELS galaxies. }
\label{fig:submm}
\end{figure*}

First of all, in order to ascertain how valid a comparison is between our star-forming disks and the (sub-)mm galaxy sample, we have compared the properties of the samples in Fig.\,\ref{fig:submm}. Our entire mass-selected galaxy sample is plotted in black, our star-forming disks are plotted in blue, and the (sub-)mm galaxies are plotted in light grey. This colour scheme is adopted for all subsequent figures. The far-left panel of  Fig.\,\ref{fig:submm} shows the photometric redshift distributions of the samples, and confirms that the (sub-)mm galaxies span the full extent of the redshift regime covered by our study. The mass distributions of the samples are plotted in the middle panel of Fig.\,\ref{fig:submm}, and reveal that the (sub-)mm selected sample provide a comparable sample for an extreme population with similar stellar masses to our purely mass-selected sample. The final panel of Fig.\,\ref{fig:submm} shows the distribution for the single-S\'{e}rsic indices of the samples. It should be noted here that the study of \citet{Targett2013} was conducted using both a single-S\'{e}rsic index morphological fit and a detailed decomposition of the individual clumps in each (sub-)mm galaxy, but it is the overall single-S\'{e}rsic index results which are used here and compared with our single-S\'{e}rsic index fits.

Finally, as we have previously found that the axial ratios of the star-forming disks depends on the size of the disk components, we have also compared the size distributions for the overall galaxies, again to provide the most direct comparison of the whole galaxy we use sizes estimated from the single-S\'{e}rsic fits. This shows that the (sub-)mm galaxies span roughly the same size range as our star-forming disk-dominated galaxies.

\begin{figure}
\begin{center}
\includegraphics[scale=1.2]{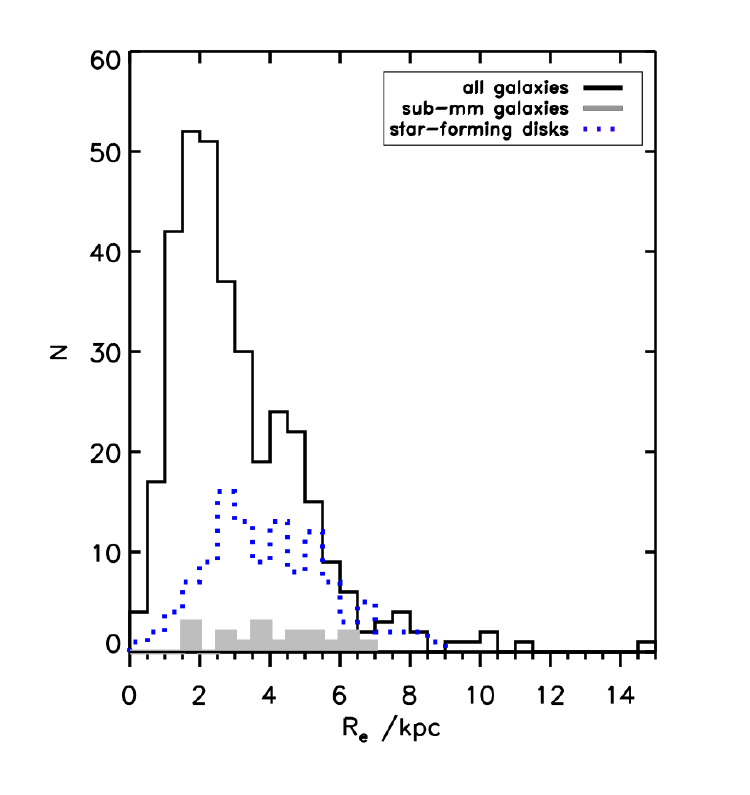} 
\end{center}
 \caption[Size distributions of the (sub-)mm galaxies.]{Size distributions of the CANDELS galaxies (black= all galaxies, blue= star-forming disks), over-plotted with the size distributions of the (sub-)mm galaxies (grey). This shows that the (sub-)mm selected galaxies span roughly the same size range as our star-forming disk-dominated galaxies.}
\label{fig:submm3}
\end{figure}

\begin{figure*}
\begin{minipage}[b]{0.49\linewidth}
\begin{center}
\includegraphics[scale=0.95]{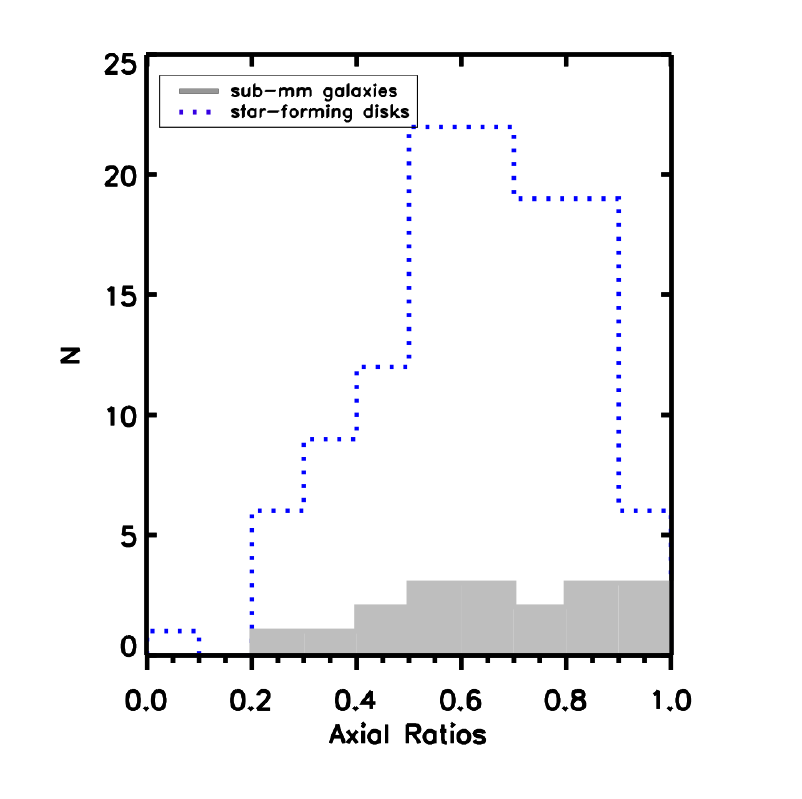} 
\end{center}
\end{minipage}
\hspace{-2.4cm}
\begin{minipage}[b]{0.49\linewidth}
\begin{center}
\includegraphics[scale=0.95]{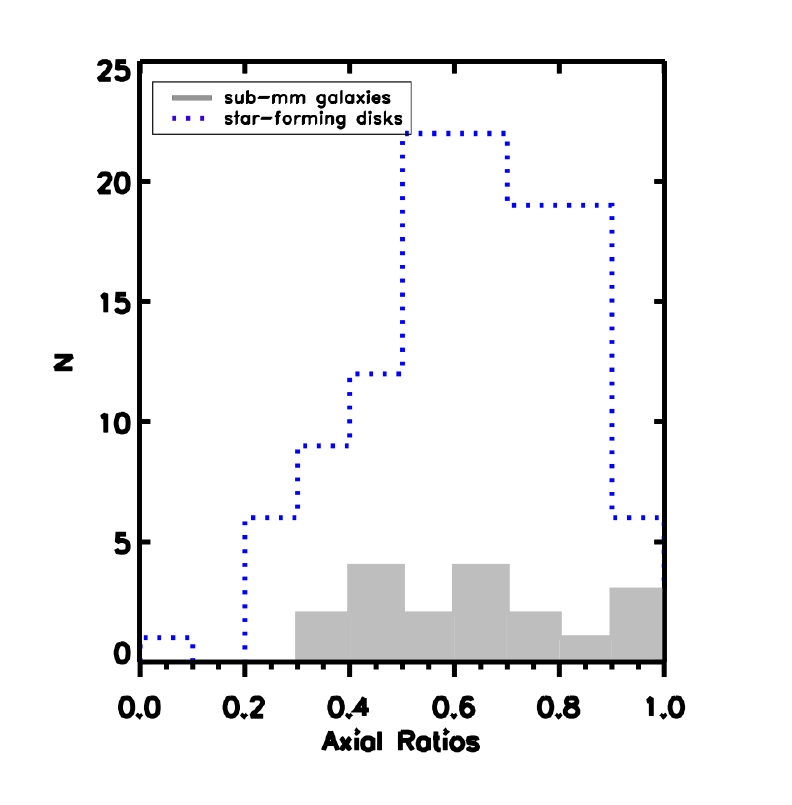} 
\end{center}
\end{minipage}
 \caption[Axial ratios of the (sub-)mm galaxies.]{The axial ratio distributions of the (sub-)mm galaxies, where again the (sub-)mm selected sample is in light grey. The (sub-)mm selected sample was modelled in \citet{Targett2013} using both a single S\'{e}rsic fit and multiple component fits to each clump. Therefore, on the left we have plotted the axial ratio distributions from the single S\'{e}rsic fits and in the right-hand plot the distributions for the axial ratios of the flux-dominant component/clump, even in the case where that component was best fit with $n>2.5$. In both plots the axial ratio distirbutions for the (sub-)mm selected galaxies are considerably flatter than the distributions for the CANDELS star-forming disks. Although the numbers within the sample are too small for a robust statistical comparison, it does appear that the trend for the axial ratios of the most actively star-forming galaxies to flatten-out extends to the extreme star-forming (sub-)mm selected population.}
\label{fig:submm4}
\end{figure*}

Having verified that the (sub-)mm selected galaxies provide a legitimate sample with which to conduct a direct morphological comparison between the ``normal'' star-forming galaxies in our CANDELS study and an extreme star-forming population, we have examined the axial ratio distributions for the different populations and have plotted them in Fig.\,\ref{fig:submm4}. For the above comparisons, which were included to establish the validity of the morphological comparison between the near-IR and (sub-)mm selected samples, we have used the properties of the overall galaxies. In most cases these properties correspond to genuinely disk-dominated star-forming objects and so a direct comparison between the overall galaxy axial ratios for the (sub-)mm selected sample and the distributions for the disk components of our star-forming disk-dominated objects is justifiable. However, \citet{Targett2013} conducted a decomposition of the individual clumps, so in  Fig.\,\ref{fig:submm4} we have over-plotted the axial ratio distributions from the disk components of the star-forming disk-dominated CANDELS sample with the distributions from the single-S\'{e}rsic model of the (sub-)mm galaxies in the left panel, and with the primary (flux dominant) component of the decomposed model of the (sub-)mm galaxies in the right panel, even in the case where the primary clump was best-fit with an $n>2.5$ profile (which will only result in a more peaked distribution). 

In fact, comparison of the axial ratio distributions reveals that the distributions for the (sub-)mm galaxies are flatter than for our star-forming disk components, which lends further support to our previous, tentative,  findings that the axial ratio distribution of the most active star-forming galaxies appears to flatten out. The statistics for this comparison are weak, but this apparent trend allows speculation that if there is a maximum surface density of star-formation at high-redshift, then the most star-forming disks, will also be expected to be the largest, which agrees with the other observed trend between larger sizes and flatter axial ratio distributions and so goes some way to delivering at least a self-consistent picture of star-forming high-redshift disks.

\begin{figure*}
\begin{center}
\includegraphics[scale=0.7]{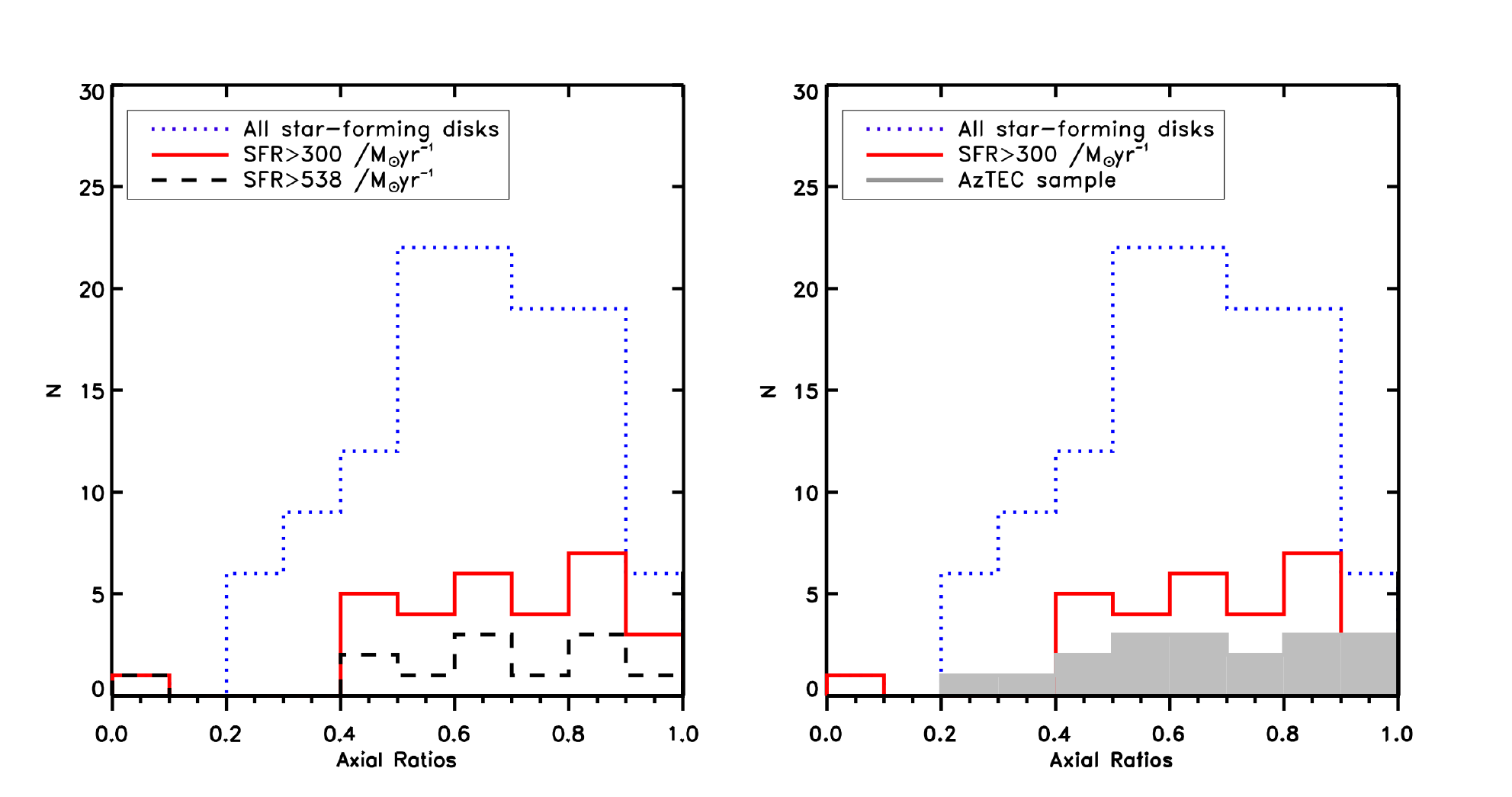} 
\end{center}
 \caption[Axial ratio distributions of the extreme star-forming disks in comparison with the (sub-)mm selected sample.]{Left: the axial ratio distributions of all the star-forming disks in our sample (blue dotted histogram), the sub-set cut at a similar star-formation rate ($SFR>300\,{\rm M_{\odot}yr^{-1}}$) as the (sub-)mm  AzTEC selected sample (red solid histogram), and the extreme $90$th percentile cut imposed previously (black dashed histogram). Right: the axial ratio distributions of all the star-forming disks (blue) and the $SFR>300\,{\rm M_{\odot}yr^{-1}}$ sub-set (red), over-plotted with the distributions for the (sub-)mm  AzTEC selected sample (grey filled histogram). This further demonstrates that by selecting the most  active star-forming disks, comparable to the extreme star-forming (sub-)mm selected sample, the axial ratio distributions appear comparably flat.}
\label{fig:submm6}
\end{figure*}

We have also explored cutting our sample at a star-formation rate that roughly matches the surface number density of the (sub-)mm galaxy sample. Allowing for the survey areas this match in surface number density is achieved at $SFR>300\,{\rm M_{\odot}yr^{-1}}$. A comparison between the axial ratios of our full star-forming disk sample, the $SFR>300\,{\rm M_{\odot}yr^{-1}}$ sub-set and the previously flat distribution for the extreme  $SFR>538\,{\rm M_{\odot}yr^{-1}}$ ($90\%$ quartile) cut is given in the left panel of Fig.\,\ref{fig:submm6}. The whole star-forming disk sample and the $SFR>300\,{\rm M_{\odot}yr^{-1}}$ sub-set are also directly over-plotted with the axial ratio distributions for the (sub-)mm selected sample in the right-hand panel of Fig.\,\ref{fig:submm6}. This further demonstrates that, by selecting the most active star-forming disks, comparable to the extreme star-forming (sub-)mm selected sample, the axial ratio distributions appear, comparably flat (although statistics from a K-S test are inconclusive).

\section{Discussion}

By exploiting our decomposed stellar-mass and star-formation rate estimates, we have been able to better explore the evolution of the full range of morphological properties exhibited by our massive, high-redshift galaxy sample.

The most fundamental morphological measurement is the relative dominance of the bulge and disk components, as measured in terms of the galaxy morphology from the light fractions and by their contribution to the total galaxy mass. Here, although we continue to find a decrease in the fraction of disk-dominated objects from $z=3$ to $z=1$, we also see that the evolution of bulge-dominated, and essentially ``pure'' bulges ($B/T>0.9$, where our $B/T$ ratios are accurate to within $\sim10\%$), is relatively slow. The decrease in disk-dominated objects is instead accompanied by an increase in mixed bulge+disk systems with decreasing redshift, where, by incorporating the decomposed stellar-mass results, we find that this morphological evolution is accompanied by an increase in the bulge component mass of the objects. 
 
 These results suggest that, instead of a clear transition at $z=2$ when the most massive galaxies transform straight from disk to bulge-dominated systems, at $1<z<3$ there is a more gradual emergence of a mass-dominant bulge component while the galaxies still display a mixed bulge+disk morphology. Even by $z=1$ these massive galaxies do not display the ``pure'' bulge morphologies exhibited by the ETGs in the local Universe, and in fact are more comparable to local S0 galaxies.

At high redshifts an increased fraction of massive disk-dominated galaxies is in agreement with previous studies, covering similar mass and redshift regimes, such as \citet{Buitrago2008}, \citet{Conselice2011} and \citet{Buitrago2013}. In addition to these studies, \citet{Mortlock2013} have conducted visual morphological classifications of $M_*>10^{10}\,{\rm M_{\odot}}$ galaxies at $1<z<3$ using CANDELS WFC3/IR imaging. For comparison, \citet{Mortlock2013} find that classifying bulges and disks by the $n=2.5$ cut returns a morphological evolution for their highest mass bin ($M_*>10^{10.5}\,{\rm M_{\odot}}$) in agreement with our findings, whereby there is a transition from bulge to disk-dominated morphologies at $z\sim2$. However, their visual classification suggests that, even for this high mass bin, the fraction of bulge and disk-dominated systems increases with decreasing redshift, and is accompanied by a fall in the fraction of ``peculiar'' morphologies. This may be explained if, at high redshift, star-forming disks have not yet relaxed into stable disks and display disturbed and/or interacting morphologies. We note that although $\sim 80\%$ of the disk-dominated objects within our sample were well-fit by the symmetric profiles, the remaining fraction required re-fitting with additional masking to achieve acceptable fits. Despite the fact that the underlying profiles of these re-fitted objects were disk-like exponentials, the additional structure (whether clumps, spiral features or other asymmetries) may influence visual classifications.

In addition to the morphological evolution with redshift, our analysis of the star-formation activity in the bulge and disk components has also delivered new insights into the evolution of the massive galaxies within our sample. One of the most interesting results from recent morphological studies has been the discovery that a significant fraction of passive galaxies are in fact disk-dominated, and star-forming galaxies are bulge-dominated. Star-forming bulges and passive disks have been previously observed at both high and low redshift, however the large fractions found within the sample of \citet{Bruce2012} were unexpected. By effectively doubling the sample size (by combining the CANDELS-UDS and COSMOS samples) the reported fractions from \citet{Bruce2012} have been further substantiated and we find that using the {\it overall} galaxy star-formation rates from SED fitting as discriminators (which correlate extremely well with UVJ classifications of passive and star-forming) $\sim30\%$ of all star-forming and passive galaxies are bulge and disk-dominated, respectively.

Several other recent studies have also found higher fractions of passive disk-dominated galaxies (the majority of which can potentially contain a bulge component at some level), more comparable to the $\sim30\%$ quoted here. In particular: \citet{Wang2012} report a passive, visually-classified, disk-dominated fraction of $\sim30\%$ for  $M_*>10^{11}\,{\rm M_{\odot}}$ at $1.5<z<2.5$;  \citet{Lee2013} have published broadly consistent results from non-parametric and single-S\'{e}rsic fitting for their 35 $1.4<z<2.5$ passive galaxies (although note the low $S/N$ of these sources); \citet{McLure2013} find a passive, single-S\'{e}rsic determined, disk-dominated fraction of $44\pm12\%$ for their $M_*>6\times10^{10}\,{\rm M_{\odot}}$, $1.3<z<1.5$ sample; \citet{vanderWel2011} extrapolate from the 14 galaxies with $M_*>10^{10.8}\,{\rm M_{\odot}}$, $1.5<z<2.5$ in their sample, based on visual, S\'{e}rsic decomposition and axial ratio results, that $64\pm15\%$ of all high-redshift passive galaxies are disk-dominated; and \citet{Fan2013} find that $68\pm24\%$ of the 19 passive $z\approx3$ galaxies in their sample are disk-dominated with $n<2$ and with consistently flat axial ratios.

However, we have also discussed how these statistics may be biased due to the use of star-formation rate estimates for the whole galaxy rather than for the individual bulge and disk components. To this end we re-evaluated the fractions of galaxies in our sample in each population using the strictest criteria possible in terms of both star-formation rate and component dominance. This conservative approach revealed that only $\sim18\pm5\%$ of passive galaxies are disk-dominated and $\sim11\pm3\%$ of star-forming galaxies are bulge-dominated by $H_{160}$-band morphology and by mass.

The gradual emergence of a mixed bulge+disk (but increasingly bulge dominated) population at $1<z<3$, coupled with the discovery of a significant population of both star-forming bulges and passive disks, reveals that while in most cases bulge growth and quiessence are correlated, there is significant scatter, suggesting that the physical processes which quench these massive galaxies are not simply connected to the mechanisms which bring about the morphological transitions witnessed within this era.

While major gas-poor mergers between $1<z<3$ can account for the transition of disk-dominated to bulge-dominated systems \citep{Robertson2006a}, this process is fairly rapid and appears at odds with the gradual emergence of increasingly bulge-dominated revealed by our study. The model of violent disk instabilities (VDI) is more consistent with the morphological evolution displayed by the galaxies within our $1<z<3$ sample. The VDI scenario has gained increasing support from both recent simulations and observations. This model proposes that, in high-redshift star-forming disks, massive clumps form from gravitational instabilities within the disk. These clumps then migrate towards the centre of the disk transferring gas and stars to the centre of the system and build a bulge component. There is significant observational evidence of star-forming clumps within high-redshift disks from, for example, \citet{Elmegreen2005}, \citet{Elmegreen2009}, \citet{Guo2011},  \citet{Wuyts2012} and Mozena et al. (2014, in preparation). These studies agree that $\sim70\%$ of massive $1<z<3$ star-forming galaxies contain at least a few kpc-sized star-forming clumps. In fact \citet{Elmegreen2009} claim that, within these disks, the contribution to the total star-formation rate is split equally between the clumps and the diffuse disk. The formation and migration of giant clumps within a gas-rich disk can be well modelled (e.g. \citealt{Ceverino2010, Martig2012, Hopkins2012}), as can the formation of a central bulge and the reshaping of the remaining disk to an exponential density profile (e.g. \citealt{Bournaud2007b, Krumholz2010, Bournaud2011}). However, there is some debate over the extent to which these clumps can survive or reform after disruption from feedback, and whether they can survive long enough to migrate to the centre of the disks (e.g. \citealt{Genzel2011,Hopkins2012}). 
 
 The latest simulations (for example \citealt{Bournaud2013, Dekel2013}) address this issue by modelling the effects of photo-ionisation, radiation pressure and supernova feedback on the clumps and find that, in their models, the clumps can account for any momentum-driven mass outflows and tidal stripping of gas and stars by re-accreting from the gas-rich disk, so that they maintain constant  masses and star-formation rates for a few hundred Myr, which is long enough to complete their migration to the centre to merge to form a bulge. However, there remains observational debate (e.g. \citealt{Wuyts2012,Elmegreen2009}) over whether the masses in these clumps are actually sufficient to build a central bulge.

The VDI model could plausibly account for the morphological trends observed from our analysis. However, the clump migration mechanism acts to build a central star-forming bulge. It is unclear what becomes of these star-forming central bulges, as this work has found evidence for only a small sub-set of star-forming galaxies which are bulge-dominated, where these systems also tend to be at higher redshifts. 

Despite the growing evidence for a correlation between star-formation quenching and the build-up of a massive bulge displayed by the majority of objects from this and other studies, and debate over whether the witnessed trends can be attributed to major mergers, the presence of passive disks and star-forming bulges is inconsistent with the idea that bulge growth and quenching are simply linked by single process. Thus, these sub-populations suggest that the gas-poor major mergers at $1<z<3$ are not solely responsible for both morphological transformations and quenching.
There are several competing quenching mechanisms which would not necessarily alter the morphologies of the systems or do so through a less stochastic processes than gas-poor mergers. These include quenching through AGN feedback, halo quenching (e.g. \citealt{Birnboim2003,Keres2005, Dekel2009a}) and morphological quenching (e.g. \citealt{Martig2009}). The VDI model is also consistent with the presence of passive disks if the clump migration which builds a central bulge, but leaves a massive disk in place, is accompanied by morphology quenching in the disk. However, given the reduction in the fraction of genuinely passive disks and star-forming bulges from our full decomposition analysis, it remains to be seen if these models can account for the population statistics.

Finally, in addition to the bulge and disk dominance, and star-formation activity of the individual components, we have also explored the difference between the axial ratio distributions of the sub-populations comprising our sample. We found that the axial ratios of the passive disks of disk-dominated galaxies display a flat distribution, consistent with a population of randomly oriented thin disks, whereas the distribution for the star-forming disks is peaked at $b/a\sim0.7$, consistent with a more tri-axial population. 

The flat distribution for the passive disks agrees with both the distributions of local disks and the results from \citet{vanderWel2011} and \citet{Chang2013b}. \citet{Chang2013b} explored the axial ratio distributions of local and $z\sim2.5$ ETGs identified on the basis of low star-formation rates from rest-frame colours. By de-projecting the observed axial ratios \citet{Chang2013b} found that the low and high redshift ETG samples comprise two populations: tri-axial objects, and flatter disk-like galaxies. The identification of disk-like galaxies within the \citet{Chang2013b} ETG sample by axial ratio distributions (which are similar to the distribution for our passive disks) independent of S\'{e}rsic index or bulge/disk ratios, helps to confirm that there is a significant population of passive, genuinely disk-dominated galaxies at $z>1$.

As well as the flat distribution for the passive disks, our two-component analysis has also revealed that the star-forming disks have a distribution more similar to that of the bulges, peaked at  $b/a\sim0.7$ and with a relative dearth of objects with $b/a<0.3$. Having explored how the axial ratio distribution varies with size and star-formation rate, we found that the smaller galaxies tend to be rounder, and the most star-forming galaxies flatter. Due to the small number of objects in our most active star-formation rate cut we also compared our axial ratio distributions to those for (sub-)mm galaxies which are expected to be extreme star-formers, and found that this trend also holds for the more extreme (sub-)mm population. These results lead us to speculate that, at high redshift, feedback from star-formation in disks acts to ``puff-up'' their scale-heights, which accounts for the peaked axial ratio distribution of the disks with smaller scale-lengths and the flatter axial ratios of larger disks. However, we note that, given in the local Universe systems such as pseudo-bulges or dwarf spheroidals which are not flattened rotating disks can display low S\'{e}rsic indices, the true structure of these high-redshift galaxies may also be similarly complex.

\section{Conclusions}

We have decomposed the most massive ($M_*>10^{11}\,{\rm M_{\odot}}$) galaxies at $1<z<3$ in the CANDELS UDS and COSMOS fields into their separate bulge and disk components across multiple bands based on their {\it HST} WFC3 $H_{160}$-band morphologies. By extending this analysis to conduct individual component SED fitting we have been able to estimate decomposed bulge and disk stellar-mass and star-formation rate estimates. This decomposition analysis has provided us with new insight into the morphological and star-formation evolution of these systems.

By comparing the dominance of the bulge and disk fractions in terms of both the $H_{160}$ light fractions and the stellar-mass contributions, we find that, within the $1<z<3$ redshift era, these most massive galaxies are more disk-dominated at higher redshifts and become increasingly mixed bulge+disk systems with decreasing redshift, where they are more bulge-dominated by mass than by light. However, even at $z=1$ the ``pure'' bulges comparable to the giant ellipticals have yet to emerge.

Despite confirmation that the colour-morphology bimodality is already well established at $1<z<3$, our simple {\it overall} galaxy $sSFR$ and $H_{160}$-band light fraction classifications have identified a significant fraction of passive galaxies which are disk-dominated ($33\pm5\%$) and star-forming galaxies which are bulge-dominated ($29\pm5\%$). These results challenge the idea that major mergers are the main mechanism for galaxy quenching as the gas-poor major mergers at $1<z<3$ would both quench these systems and transform their morphologies from disk to bulge dominated. In order to better probe this population we have examined the decomposed stellar-mass and star-formation rate estimates for the passive disks and star-forming bulges adopting the strictest possible criteria. We conclude that, in fact, only $18\pm5\%$ of passive galaxies are genuinely disk-dominated and  $11\pm3\%$ of star-forming galaxies are bulge-dominated. Both these fractions are significantly lower than the fraction determined from the overall galaxy specific star-formation rates and stellar-masses reported by previous studies at similar redshifts and masses, which clearly illustrates the advantage of the full SED decompositions over and above single model fits for this type of analysis. 

Nevertheless, the confirmation of a significant population of passive disks and star-forming bulges, coupled with the observed gradual emergence of increasingly mixed bulge+disk systems, with larger bulge/total mass fractions, from $z=3$ to $z=1$ suggests that, whilst some of these most massive galaxies may undergo major mergers which both quench their star formation and transform their morphologies, there must also be other physical processes which quench star-formation but leave a massive disk intact. This evolutionary scenario is more consistent with the models of AGN quenching, halo quenching \citep{Dekel2006,Dekel2009a} or possibly violent disk instabilities \citep{Dekel2009b,Ceverino2010}.

Finally, in addition to studying the overall morphologies and star-formation rates of the most massive galaxies at $1<z<3$, our morphological decomposition has also allowed us to investigate how the axial ratio distributions for the passive and star-forming sub-populations of the bulge and disk components differ. The axial ratios of these components have provided another key indicator of the structure of these systems as they reveal that the passive disks have flattened axial ratio distributions consistent with a population of randomly oriented thin disks, similar to disks in the local Universe. This further verifies that these are genuine disk-dominated passive galaxies. However, the star-forming disks have a distribution peaked at higher values of $b/a\sim0.7$, more consistent with the distributions for the passive and star-forming bulges in this sample, and those for lower redshift bulge-dominated systems. 

By exploring the trends within the star-forming disk population we find evidence that the smallest galaxies are the roundest, and that the most actively star-forming disk galaxies are the flattest. This star-forming - axial ratio trend has also been supported by comparison with a (sub-)mm selected sample. These results are thus consistent with a scenario in which, at high redshift high star-formation rates correlate with increased scale-height of disks \citep{Dekel2014}, which affects the axial ratios of the smaller galaxies but has a lesser impact on the larger disks. If there is also a maximal surface-density of star formation in high-redshift disks, then the largest, flattest galaxies would also be expected to be the most star-forming, thereby reconciling both of these observations.

\section{Acknowledgments}
VAB acknowledges the support of the Science and Technology Facilities
Council (STFC) via the award of an STFC Studentship.
VAB and JSD acknowledge the support of the EC FP7 Space project ASTRODEEP (Ref. No: 312725).
JSD, RAAB, FB, TAT acknowledge the support of the 
European Research Council via the award of an Advanced Grant. 
JSD and RJM acknowledge the support 
of the Royal Society via a Wolfson Research Merit Award and a University 
Research Fellowship respectively.
RJM acknowledges the support of the Leverhulme Trust via the award of a Philip Leverhulme Research Prize.
MC acknowledges the support of the Science and Technology Facilities
Council (STFC) via the award of an  STFC Advanced 
Fellowship.
US authors acknowledge  support from NASA grants for {\it HST} Program GO-12060.
SMF, DCK, DDK, and EJM also acknowledge support from NSF grant AST-08-08133.

This work is based in part on observations made with the NASA/ESA {\it Hubble Space Telescope}, which is operated by the Association 
of Universities for Research in Astronomy, Inc, under NASA contract NAS5-26555.
This work is based in part on observations made with the {\it Spitzer Space Telescope}, which is operated by the Jet Propulsion Laboratory, 
California Institute of Technology under NASA contract 1407.

\bibliographystyle{mn2e}

\bibliography{bibtex}

\appendix 

\section{Multiple-Component Fitting}

\begin{figure*}
\begin{center}
\includegraphics[scale=0.8]{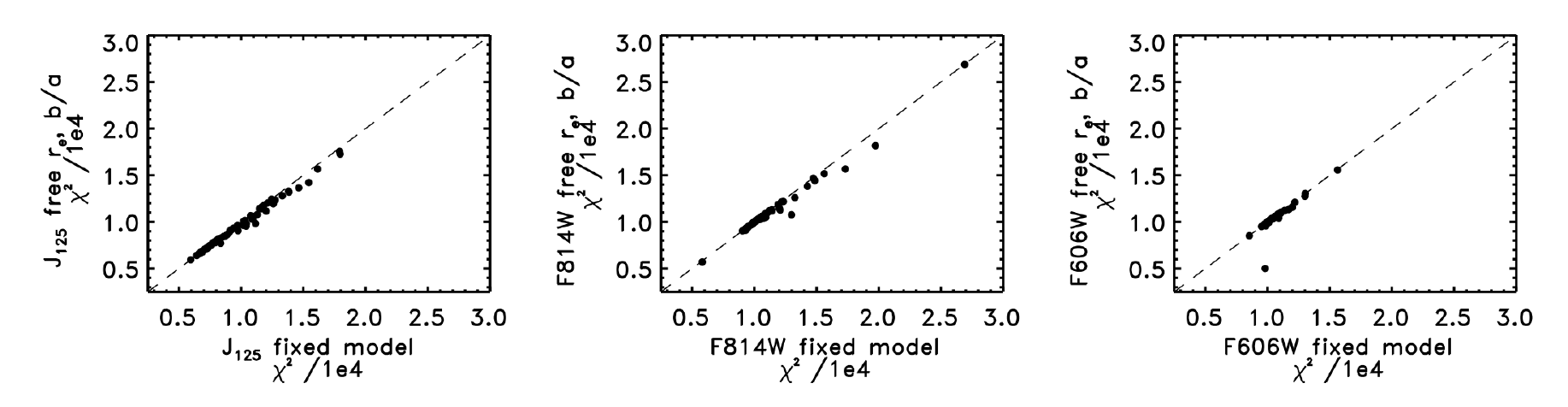} 
 \caption[Comparison of the $\chi^{2}$ statistics of models with and without fixed parameters.]{Comparison of the minimum $\chi^{2}$ between the fixed $r_{e}$ and $b/a$, and the free $r_{e}$ and $b/a$ model fits. Left: comparison in the $J_{125}$-band, middle: for the $i_{814}$-band, right: for the $v_{606}$-band. These comparisons illustrate that the additional degrees of freedom introduced by allowing the $r_{e}$ and $b/a$ parameters to be freely fitted significantly improve the fits for only a small subset of the galaxies in our sample.}
\label{fig:chis}
 \end{center}
\end{figure*}

For this analysis we have adopted multi-wavelength decompositions which are based on the best-fit model parameters of the $H_{160}$-band fits, with all model parameters fixed at other wavelengths except for the magnitude of each component. Several tests of the validity of this approach have been conducted by also allowing the effective radii and axis ratios of the components to be included as free parameters in the fitting and we have explored how these affect not only the $\chi^{2}$ statistics of the fits, but also the output bulge and disk magnitude contributions and fitted sizes.

\begin{figure*}
\begin{center}
\includegraphics[scale=0.8]{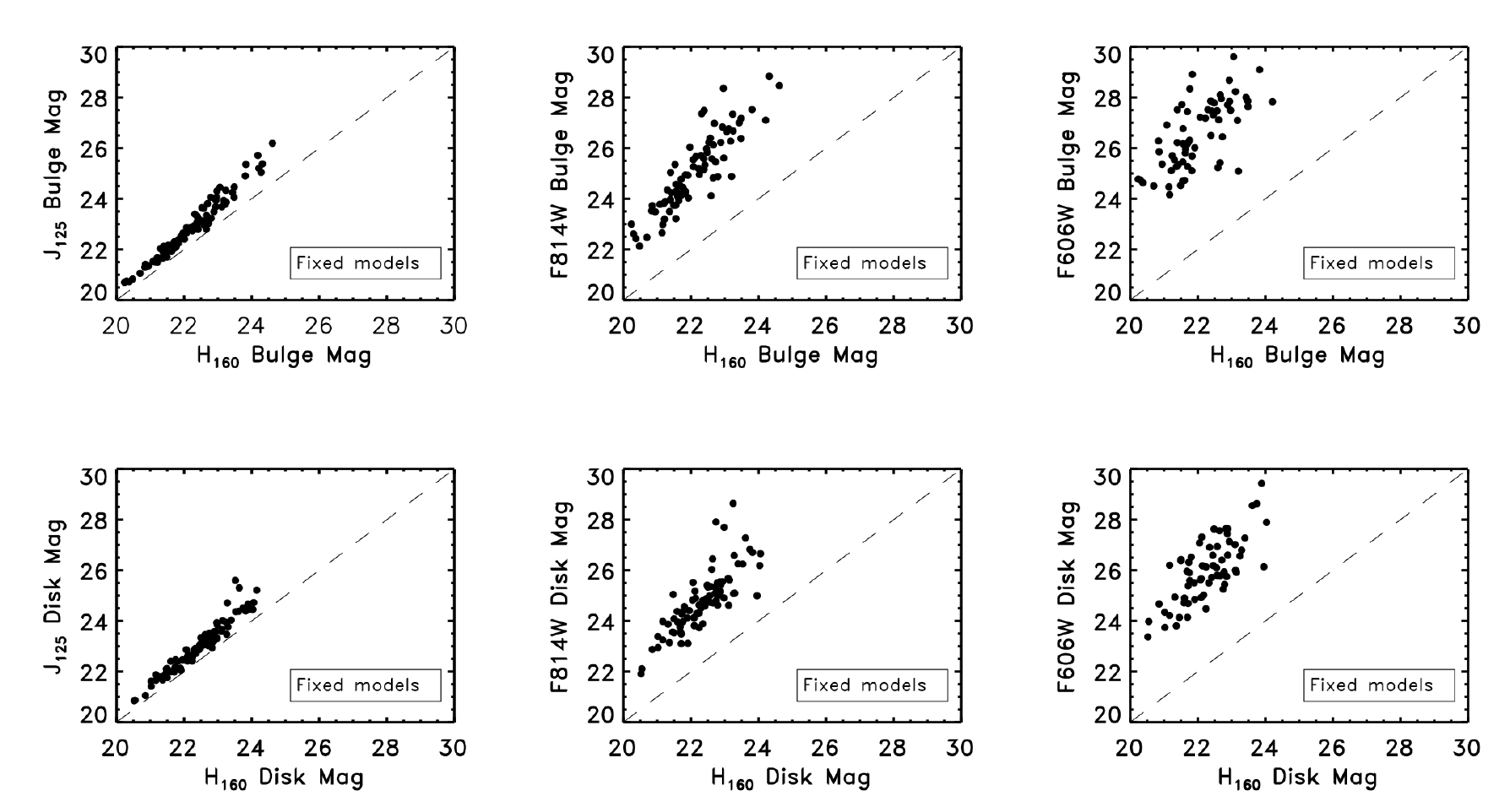} 
 \caption[Comparison of the fitted magnitudes of the fixed-parameter models for both components in relation to the $H_{160}$ model fits.]{Comparison of the fitted magnitudes of the fixed-parameter multi-wavelength models for the separate bulge and disk components in relation to the $H_{160}$ model fits. The top panels show the comparison between the magnitudes of the bulge components across the additional three bands and the original $H_{160}$ model fits, and the bottom panels illustrate this comparison for the disk components.}
\label{fig:magcomp}
 \end{center}
\end{figure*}

\begin{figure*}
\begin{center}
\includegraphics[scale=0.8]{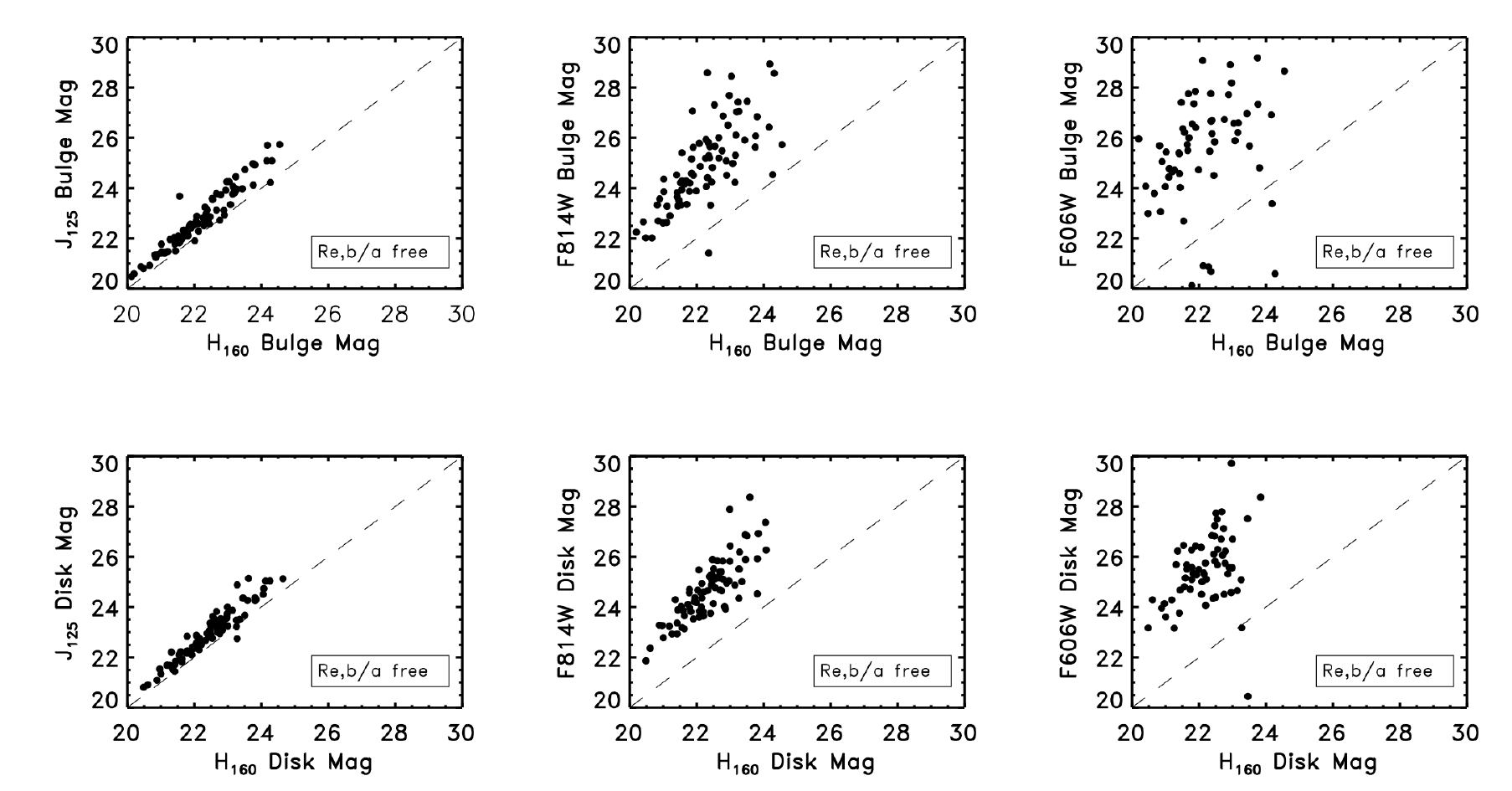} 
 \caption[Comparison of the fitted magnitudes of the free-parameter models for both components in relation to the $H_{160}$ model fits.]{The fitted magnitudes of the free-parameter multi-wavelength models for the separate bulge and disk components in relation to the $H_{160}$ model fits. The panels follow the same configuration as in Fig.\,\ref{fig:magcomp}, with the comparison for the bulge components in the top panels and the disks in the bottom panels.}
\label{fig:magcomp2}
 \end{center}
\end{figure*}

The first tests conducted were to study how the introduction of the additional free parameters influenced the acceptability of the best-fit models. While allowing the effective radii ($r_{e}$) and axis ratios ($b/a$) parameters to be freely fitted in the modelling did, unsurprisingly, provide appreciable improvement in the $\chi^2$ fits for a few of the objects in the sample (Fig.\,\ref{fig:chis}), we found evidence that there is the potential for significant biases to be introduced as a consequence of adopting the increased degrees of freedom. Moreover, when comparing the bulge and disk component magnitudes in all bands it is clear that in both cases, where $r_{e}$ and $b/a$ are held fixed (Fig.\,\ref{fig:magcomp}) and allowed to fit freely (Fig.\,\ref{fig:magcomp2}), the overall trend for the magnitude of both components to become fainter than the $H_{160}$ estimates remains, with the introduction of only additional scatter to this relation for the case where the $r_{e}$ and $b/a$ parameters are fitted freely.

\begin{figure*}
\begin{center}
\includegraphics[scale=0.8]{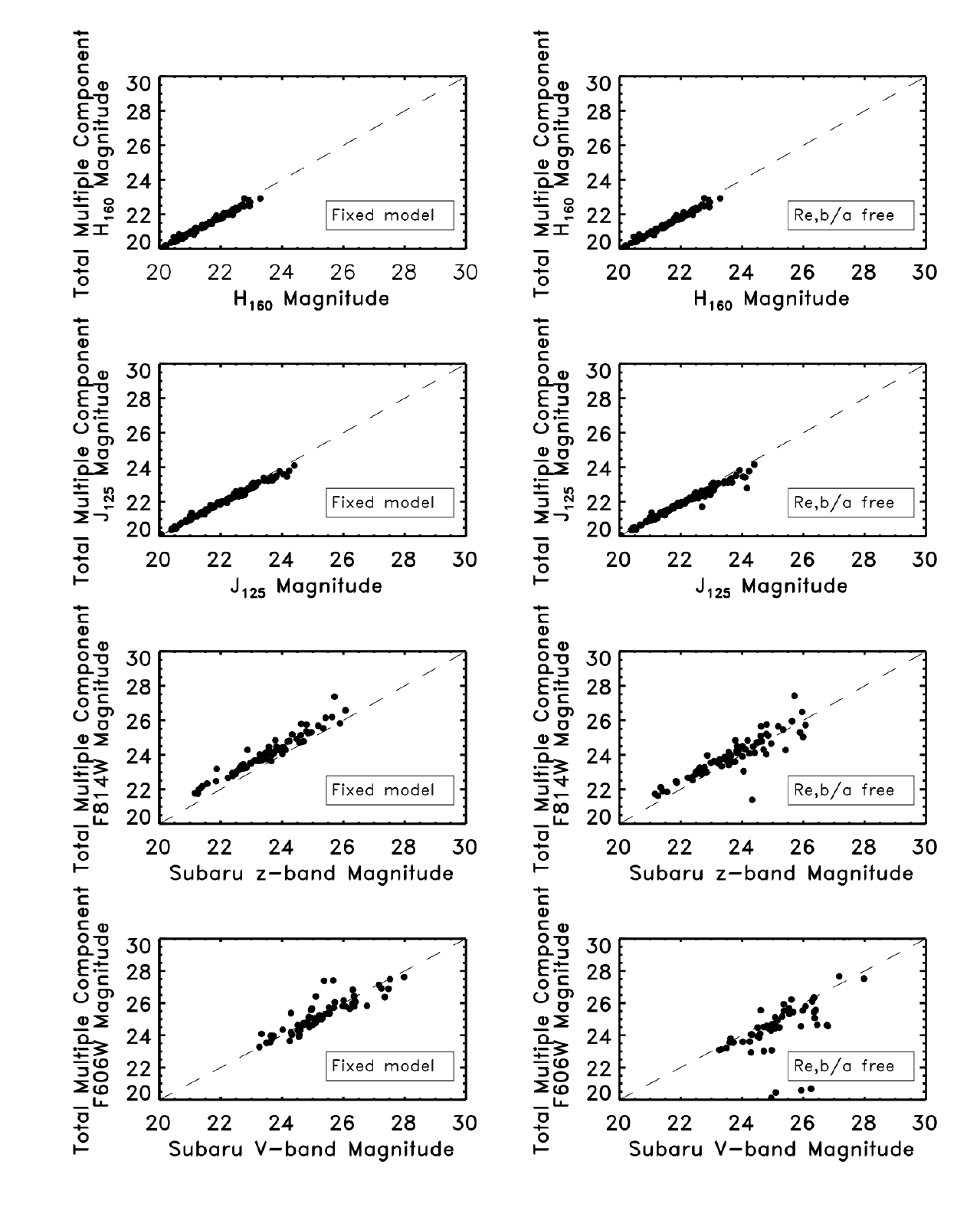} 
 \caption[Relation between the sum of the best-fit multiple-component model magnitudes and the measured iso-magnitude of each object for the different models.]{Relation between the sum of the best-fit multiple-component model magnitudes and the measured iso-magnitude of each object for the fixed and free $r_{e}$ and $b/a$ parameter models. The fixed parameter models are shown on the left, with the free parameter models on the right, and the plots are ranked from top to bottom by decreasing wavelength.The constant offset between the best-fit modelled $z_{814}$-band magnitudes and the measured Subaru z-band magnitudes is due to the mis-match in the central wavelengths of these filters. These plots illustrate how the adoption of the additional free parameters results in some case where the sum of the best-fit bulge+disk model magnitudes exceeds the measured iso-magnitude photometry and so demonstrates that this extra freedom can result in biases in the fitting which are hard to understand.}
\label{fig:total}
 \end{center}
\end{figure*}

Furthermore, from examining the total multiple-component (bulge+disk and bulge+disk+PSF) magnitudes of each component from both the fixed and free $r_{e}$ and $b/a$ parameter models, we found that towards the bluer wavelengths, the scatter in the relation between this integrated measure of the total magnitude of the object and the iso-magnitude estimate, measured directly from the image, increases in the case of the free $r_{e}$ and $b/a$ fits. This is demonstrated in Fig.\,\ref{fig:total}, where it can also be seen that this increase in scatter is preferentially in the direction of the total multiple-component magnitude being brighter than the measured iso-magnitude. 
 
\begin{figure*}
\begin{center}
\includegraphics[scale=0.8]{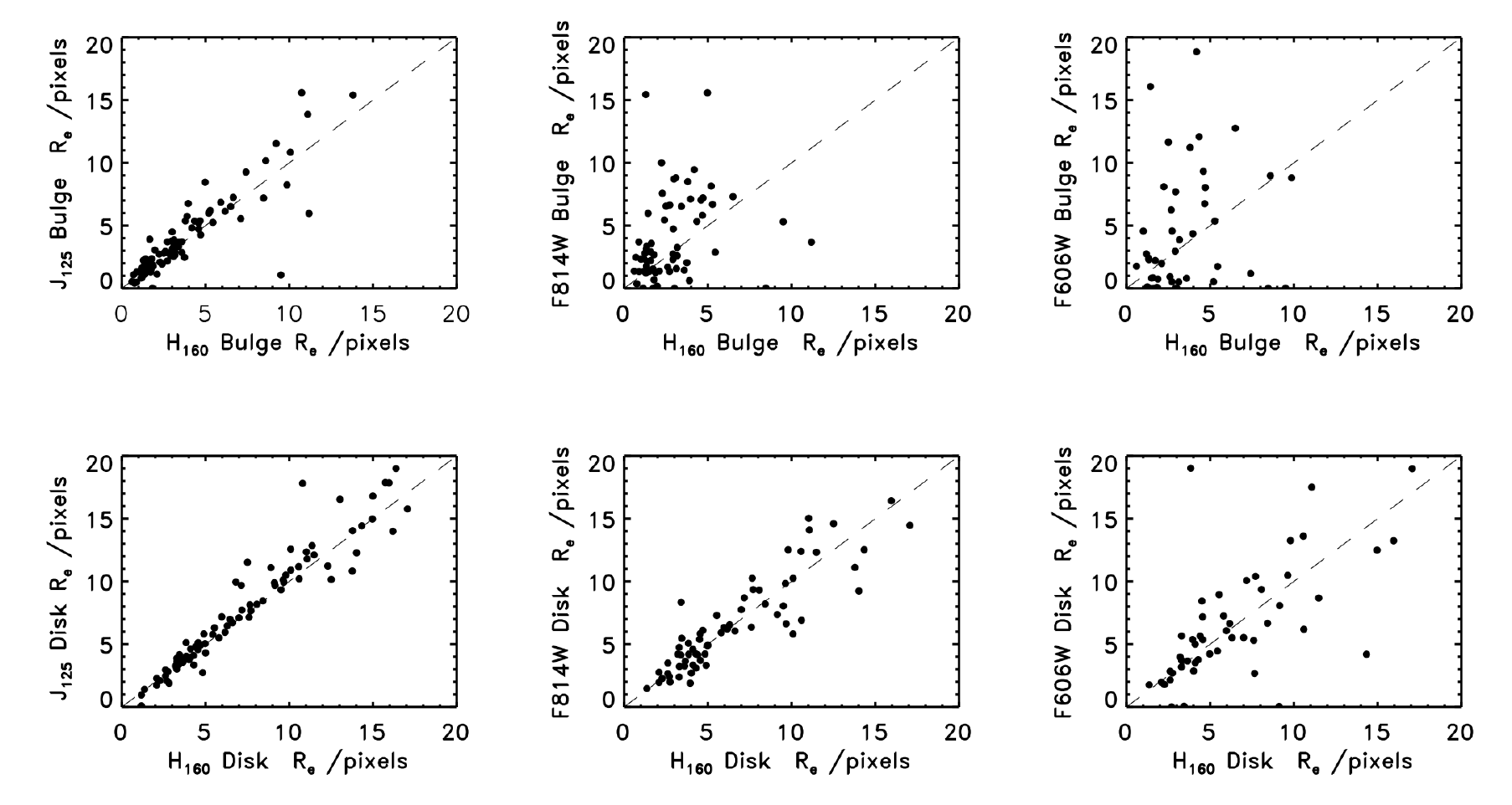} 
 \caption[Comparison of the fitted sizes of individual components between the fixed and free-parameter models.]{Comparison of the fitted sizes of the individual components between the fixed and free-parameter models for the four bands. These plots illustrate that allowing the $r_{e}$ and $b/a$ parameters to be fitted freely, even in the bluest band ($v_{606}$), results in cases where the disk component becomes both larger and smaller compared to the master $H_{160}$-band fit, and, moreover, highlights cases where in the $v_{606}$-band the bulge size exceeds the fitted sizes in the redder bands.}
\label{fig:sizecomp}
 \end{center}
\end{figure*}

In addition to this, a comparison of the bulge and disk sizes of each component in the 4 different bands revealed that by allowing the effective radius and axis ratio to be free parameters, the scatter in the relation between the effective radius of each component in the given bands compared to the master fit from the $H_{160}$-band (Fig.\,\ref{fig:sizecomp}) also increases, and highlights clear cases where in the $v_{606}$-band, the bluest available, the bulge component becomes significantly larger than in the $H_{160}$-band.  These trends suggest that introducing the effective radius and axis ratio as free parameters, and removing the $H_{160}$-band constraints on these parameters, introduces new biases in the fitting procedure which, while hard to correctly interpret, do not significantly improve the ability of the models to reproduce the underlying morphologies of the galaxies in our sample. Moreover, this also holds true for the case of objects which were modelled as pure bulges in the $H_{160}$-band, and where we have added an additional n=1 disk component in the bluer bands (where for this additional disk model all other parameters except for the centroid position and S\'{e}rsic index were allowed to vary).

For these objects, the addition of the second component to account for any faint disks becoming dominant in bluer bands still did not provide a significant improvement in the ability of the model to fit the data, and due to low number statistics, it is hard to argue from examining the fraction of light which becomes attributed to the disk in the $v_{606}$-band compared to the $H_{160}$-band that the addition of such a disk component is motivated. 

In conclusion we decided to adopt fixed $r_{e}$ and $b/a$ parameter models for the multi-wavelength morphological analysis in order to avoid adopting additional degrees of freedom, which are not required and may introduce an additional degree of bias.

\begin{figure*}
\begin{center}
\begin{tabular}{m{3cm}m{3cm}m{3cm}m{3cm}}
\includegraphics[scale=0.15 ]{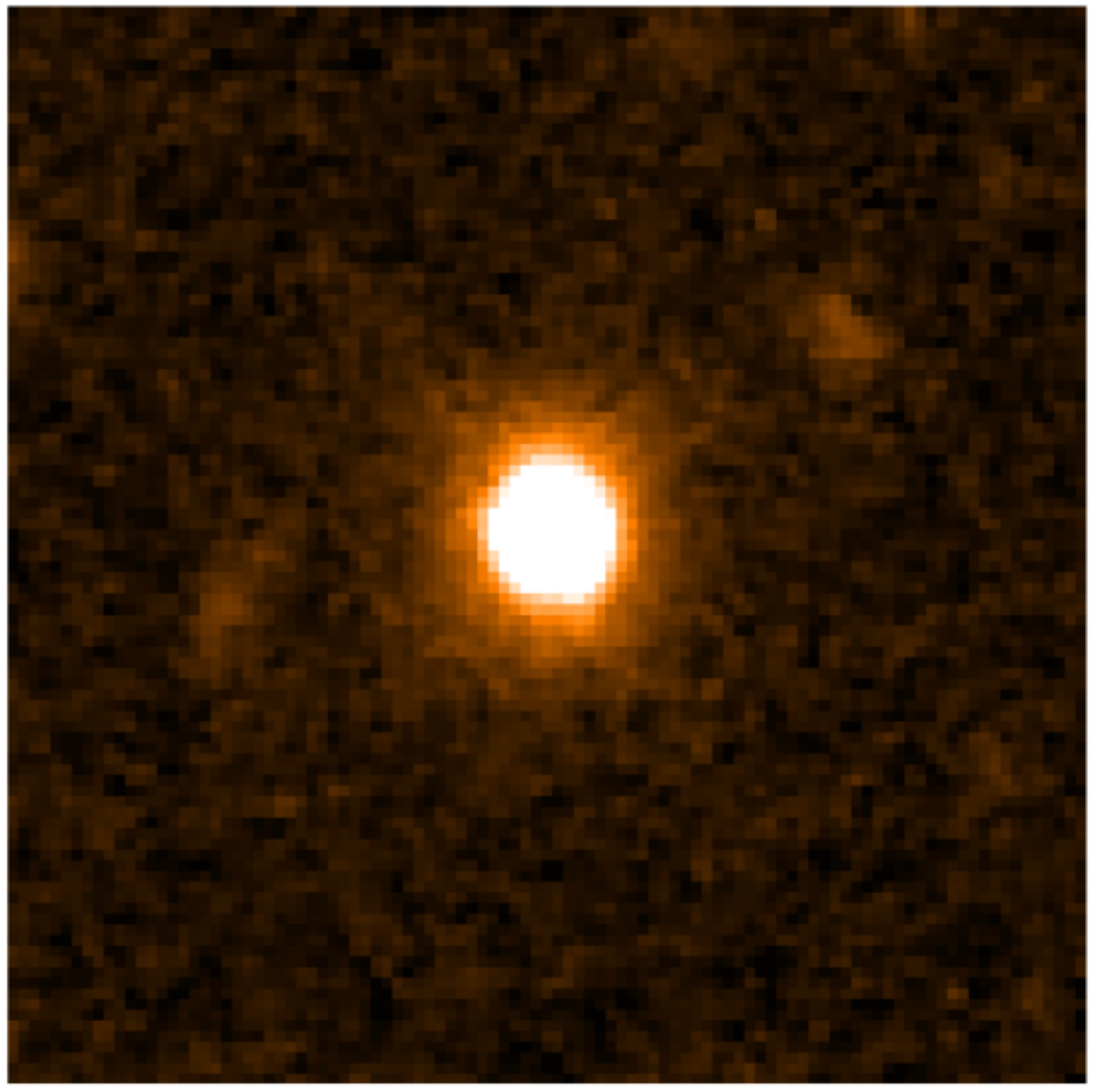}  &
\includegraphics[scale=0.15]{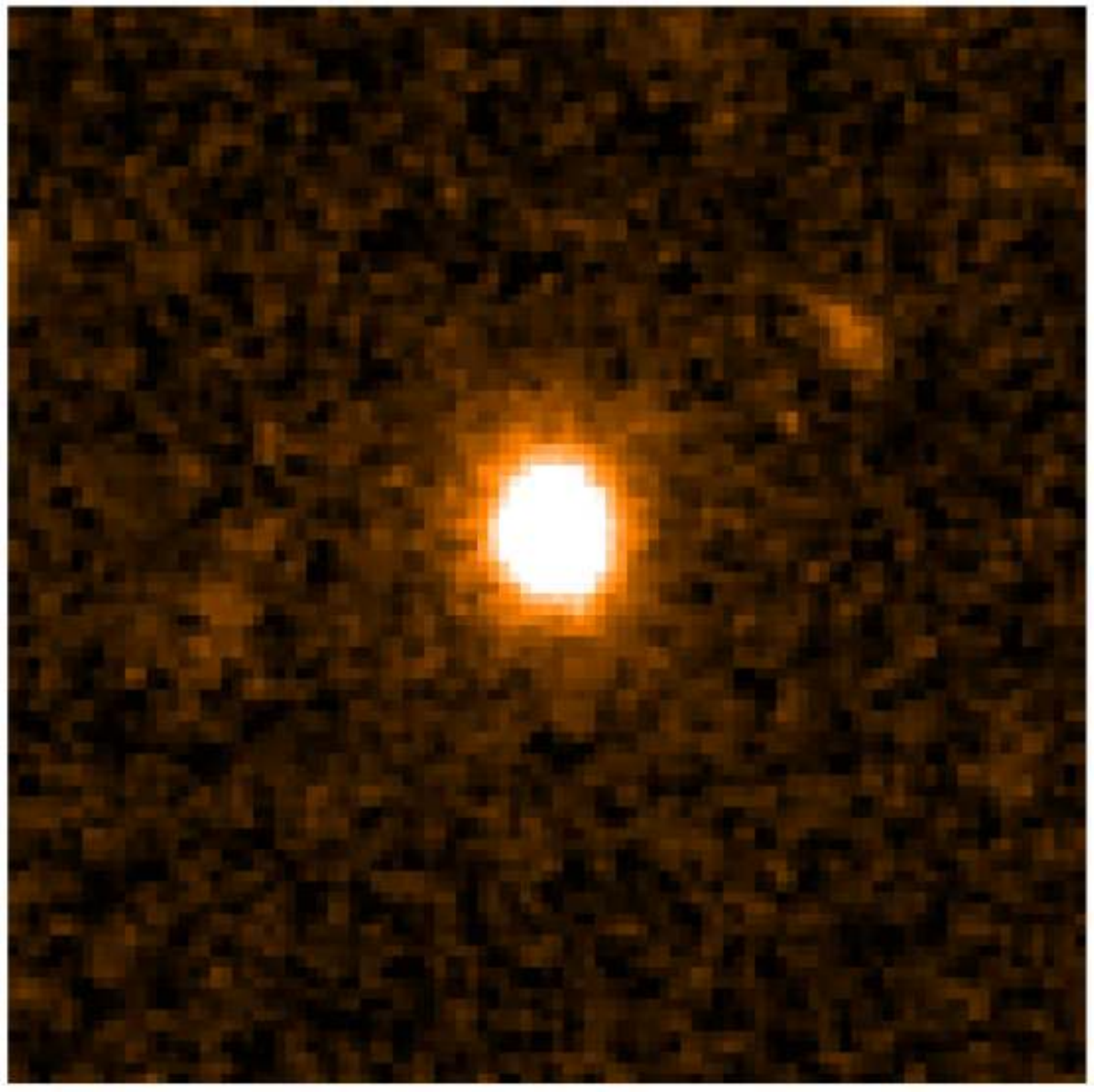}  &
\includegraphics[scale=0.15]{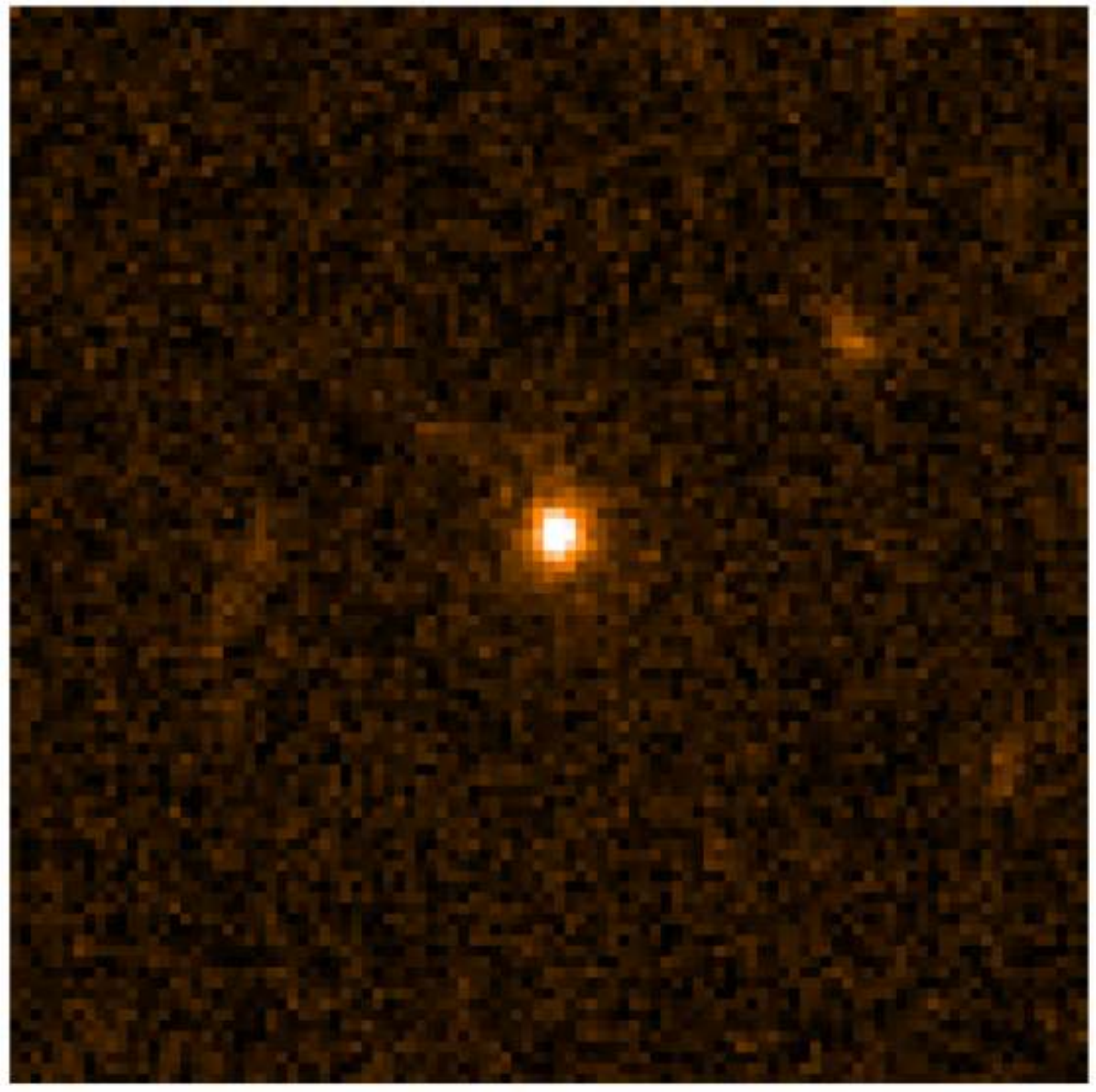}  &
\includegraphics[scale=0.15]{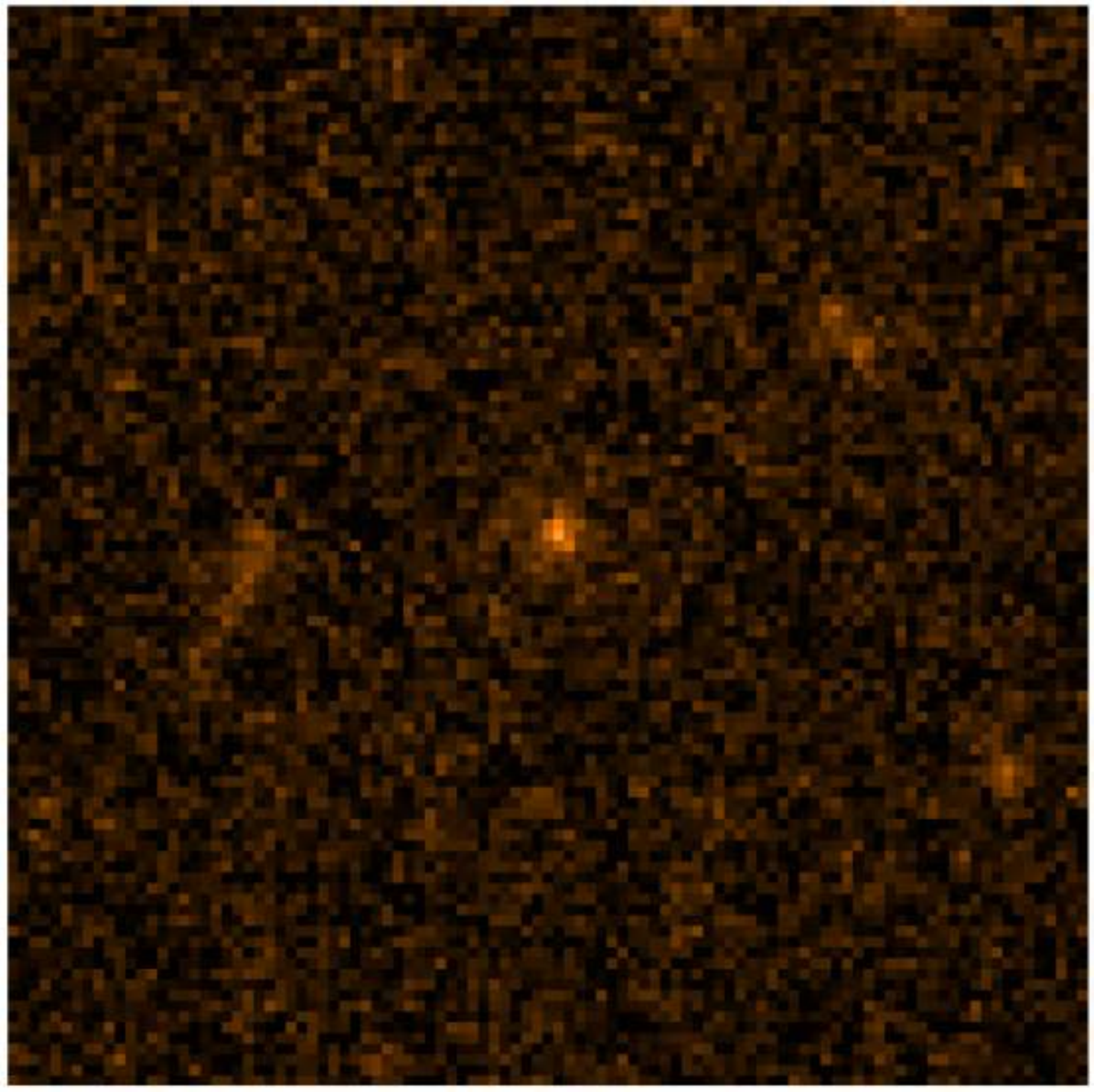}  \\
\includegraphics[scale=0.15 ]{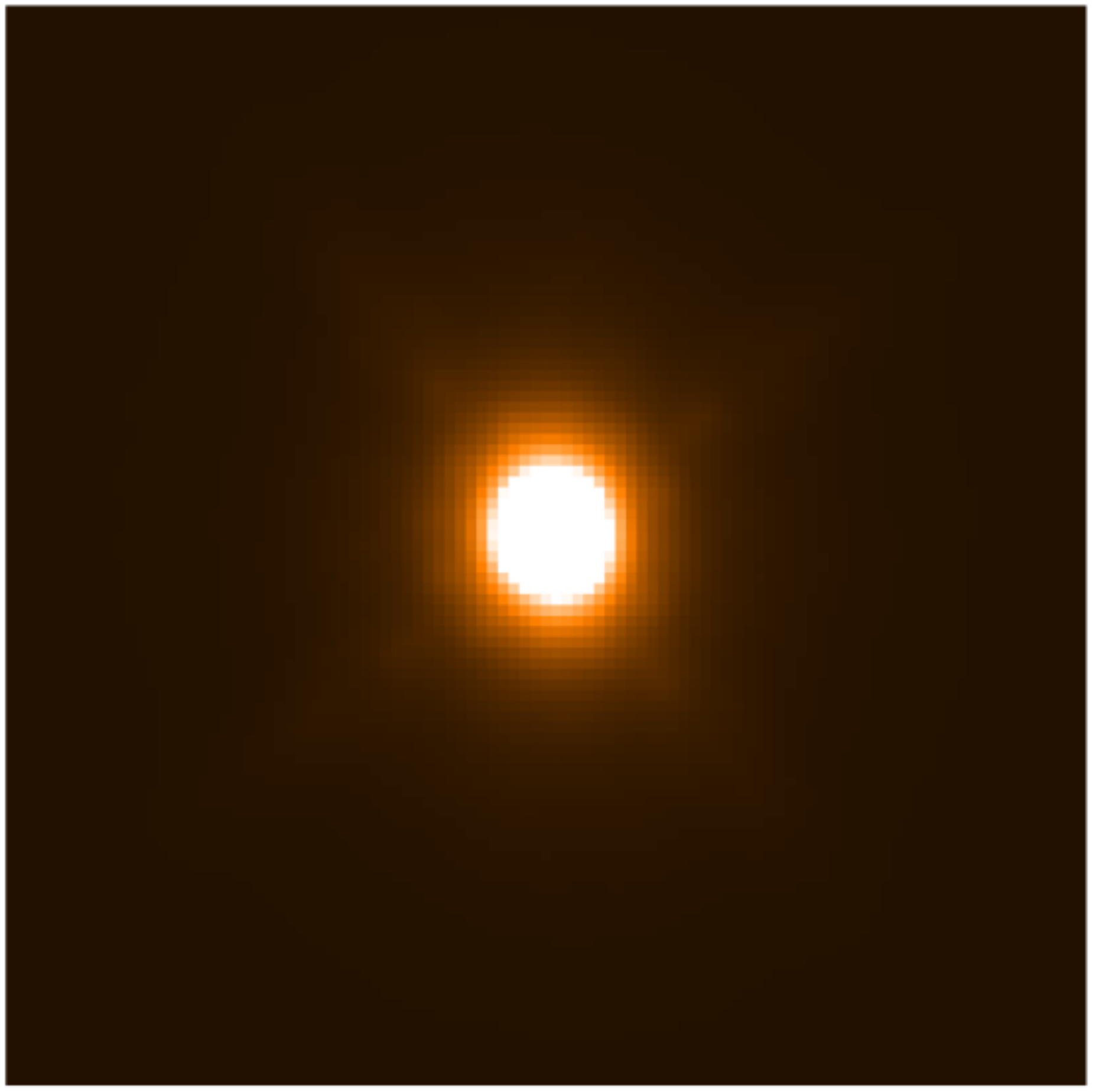}  &
\includegraphics[scale=0.15]{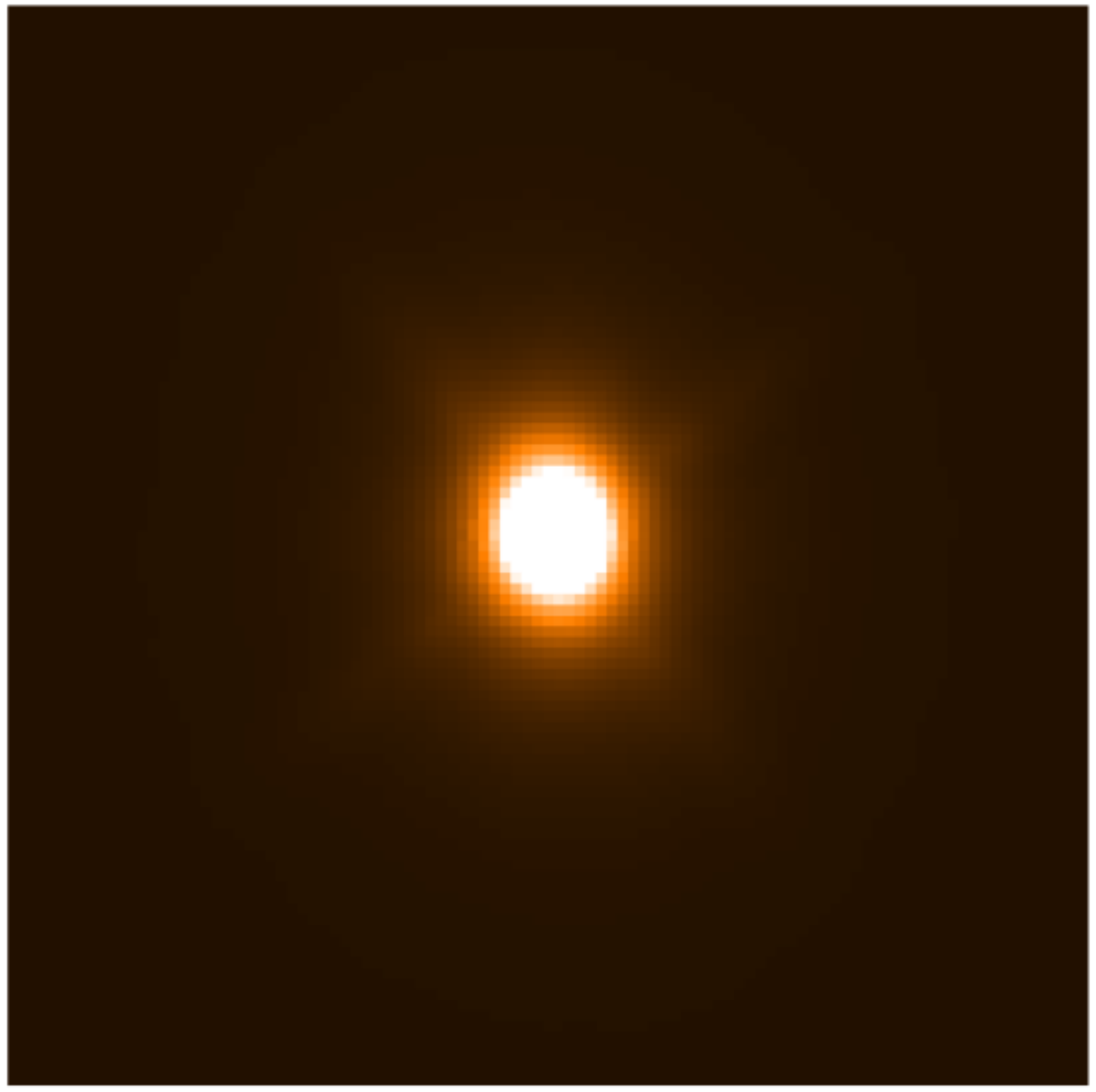}  &
\includegraphics[scale=0.15]{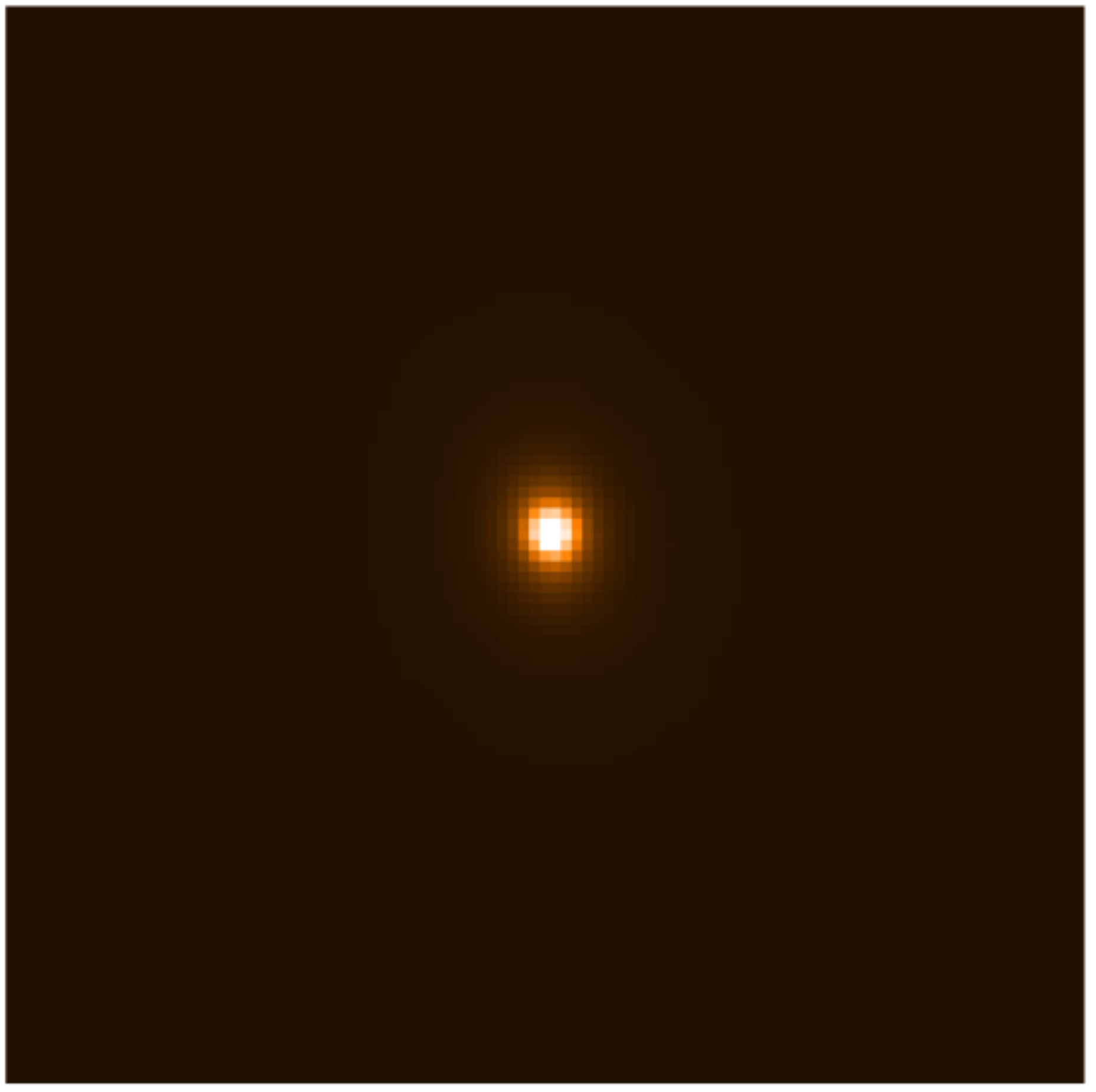}  &
\includegraphics[scale=0.15]{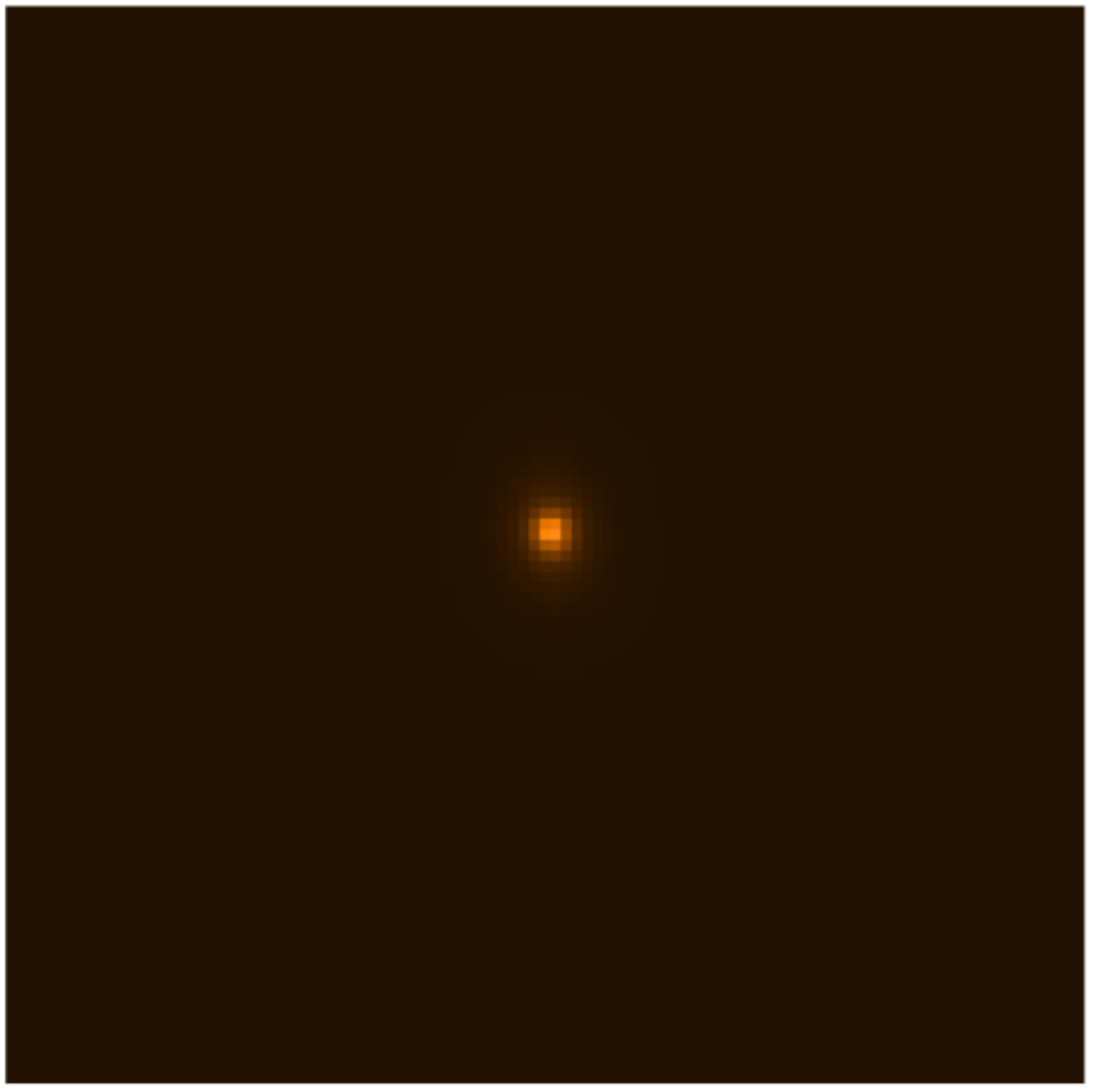}  \\
\includegraphics[scale=0.15]{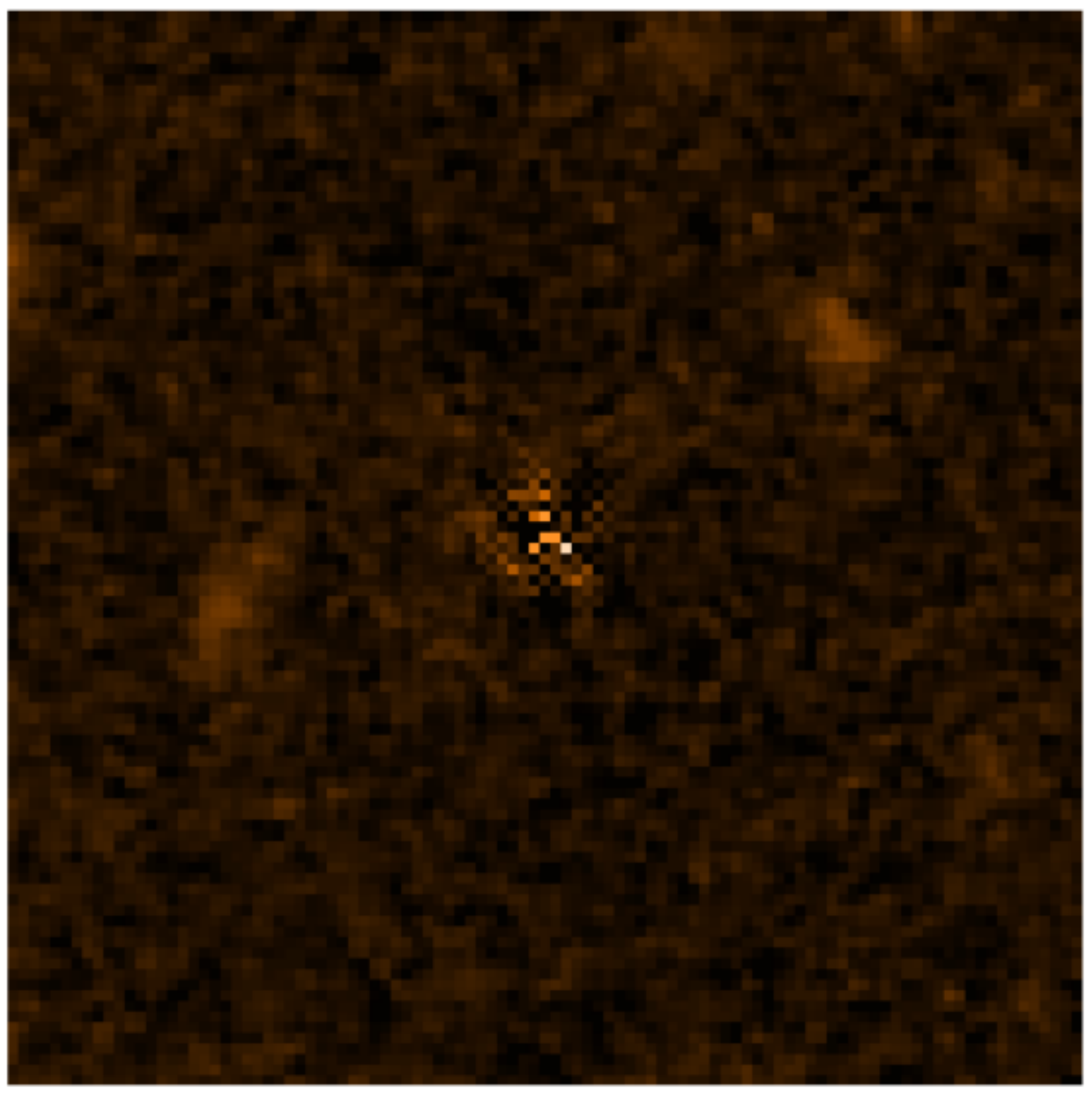}  &
\includegraphics[scale=0.15]{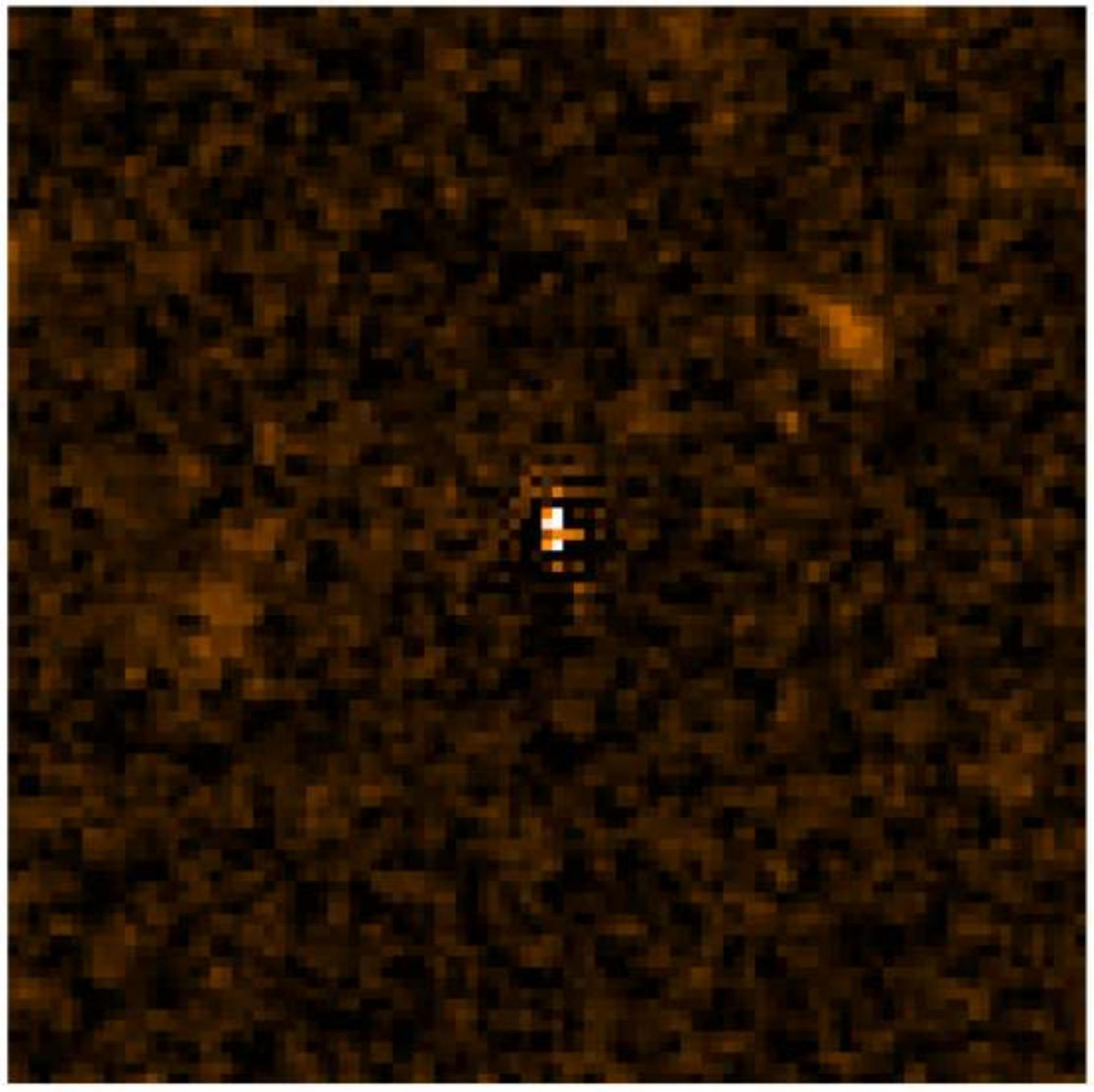}  &
\includegraphics[scale=0.15]{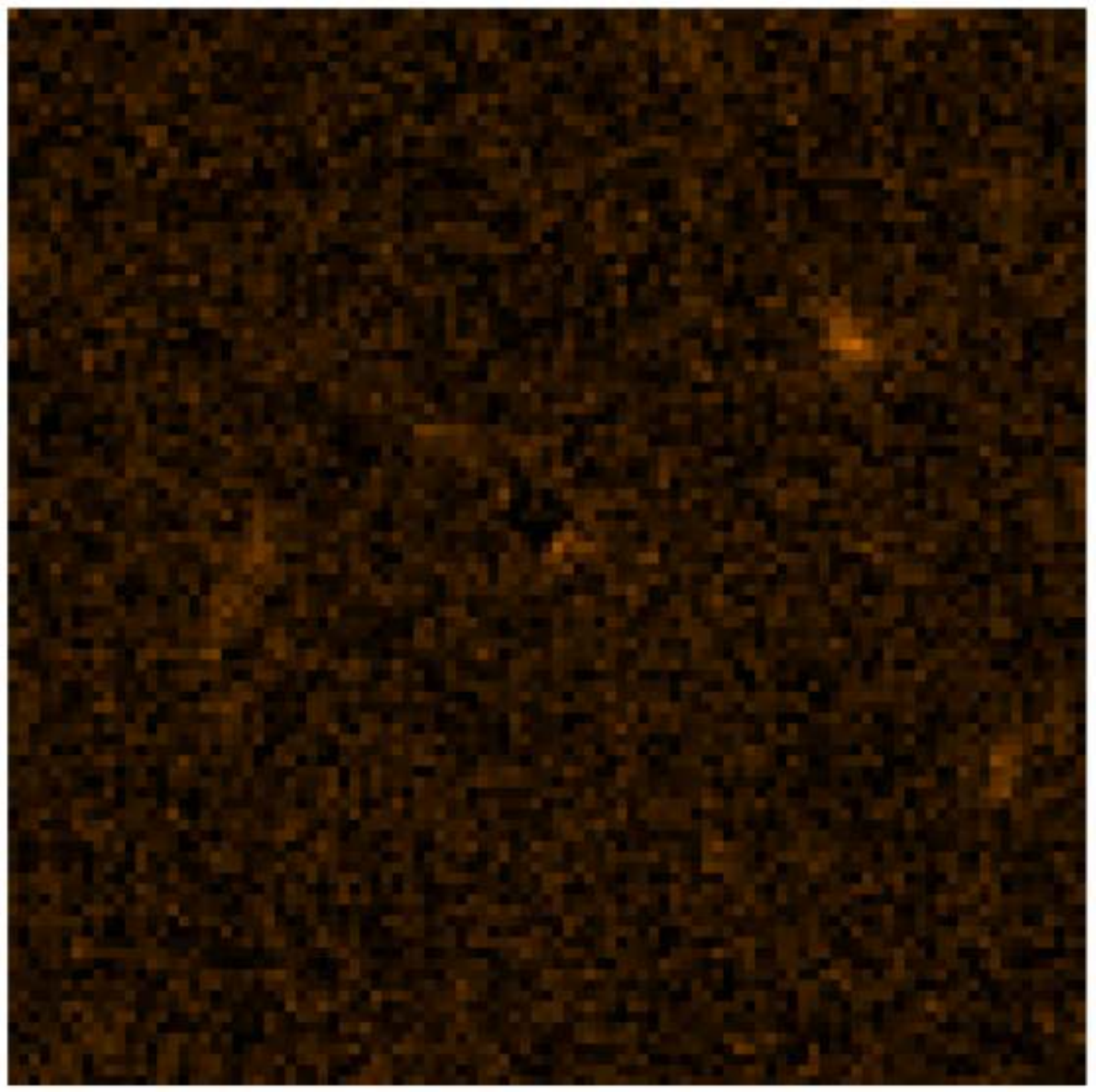}  &
\includegraphics[scale=0.15]{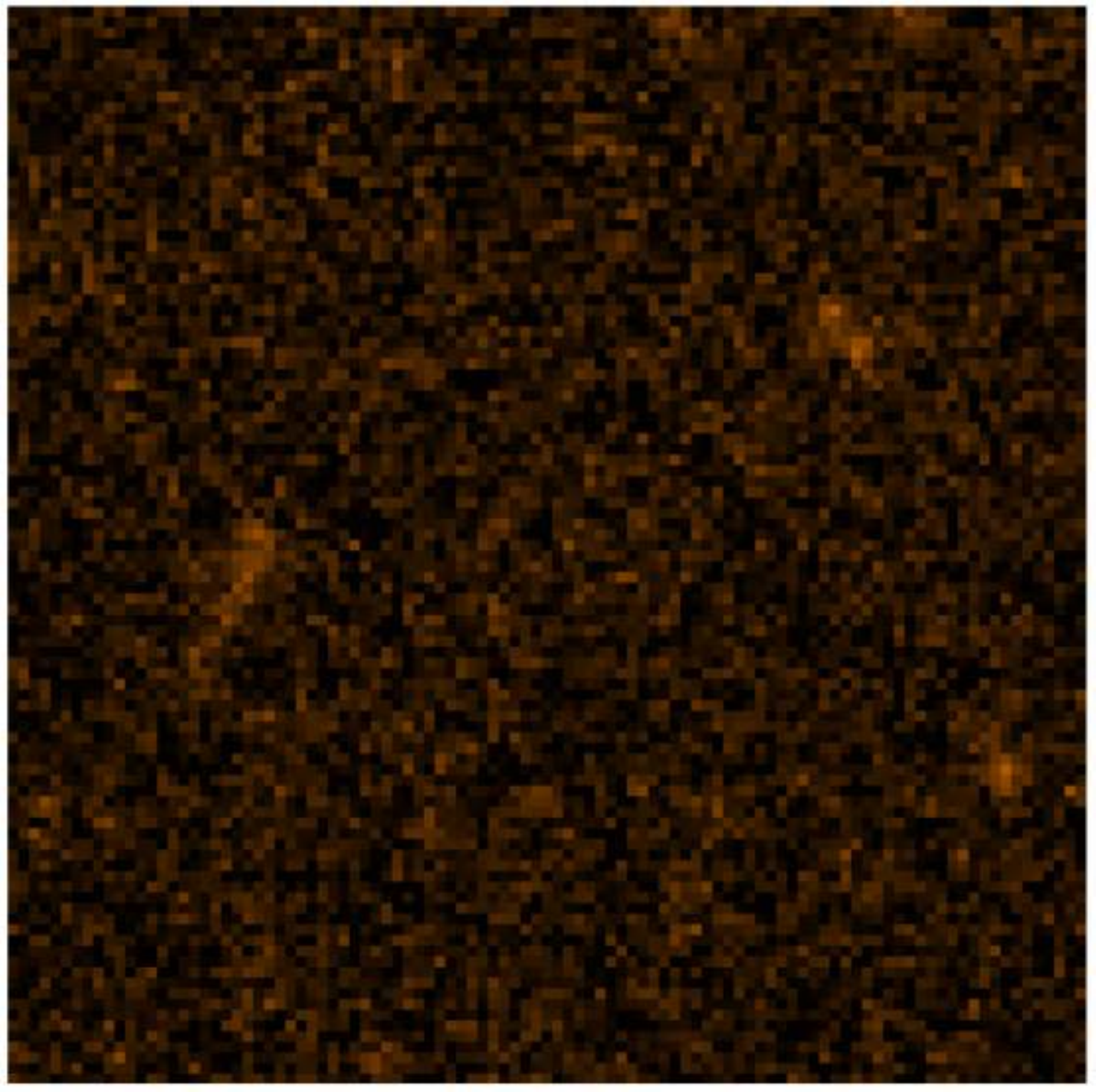}  \\
\end{tabular}
\caption[Image stamps of an example fit for a pure bulge object.]{Image stamps of an example fit for a pure-bulge object. The images displayed are $6\times6$ arcsec stamps with the images in the top panels, the best-fit models in the middle panels and the residual stamps in the bottom panels. The columns are ranked from left to right by decreasing wavelength, with the $H_{160}$-band stamps on the far-left, then the $J_{125}$-band stamps, the $i_{814}$-band stamps and the $v_{606}$-band stamps on the far-right. This is a representative fit for an object which is best fit with a pure bulge model, and illustrates the level of fits achieved across the full wavelength range in this study by fixing all model parameters, except magnitudes, to the $H_{160}$-band fits.}
\label{fig:pure_bulge}
\end{center}
\end{figure*}

\begin{figure*}
\begin{center}
\begin{tabular}{m{3cm}m{3cm}m{3cm}m{3cm}}
\includegraphics[scale=0.15 ]{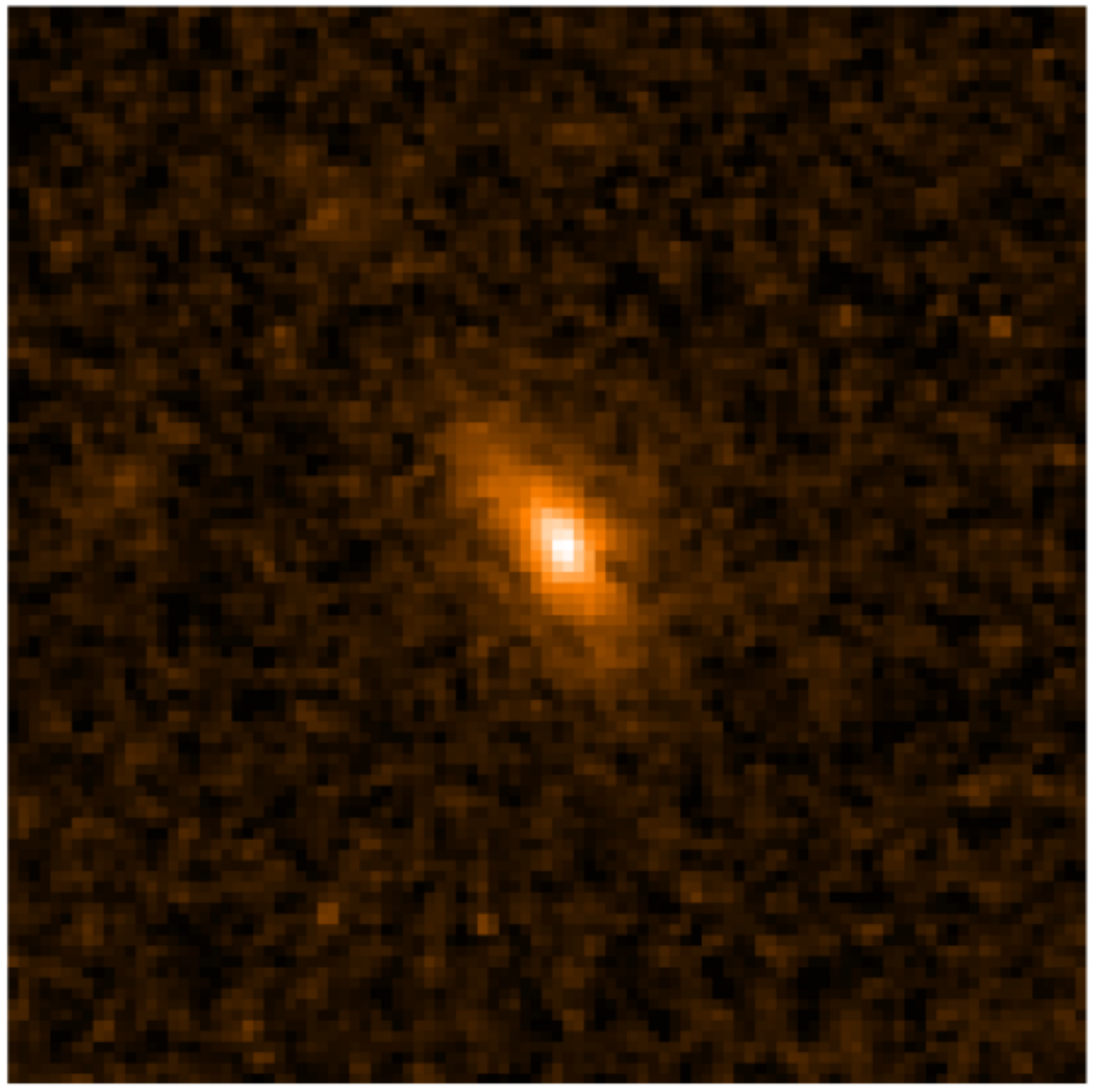}  &
\includegraphics[scale=0.15]{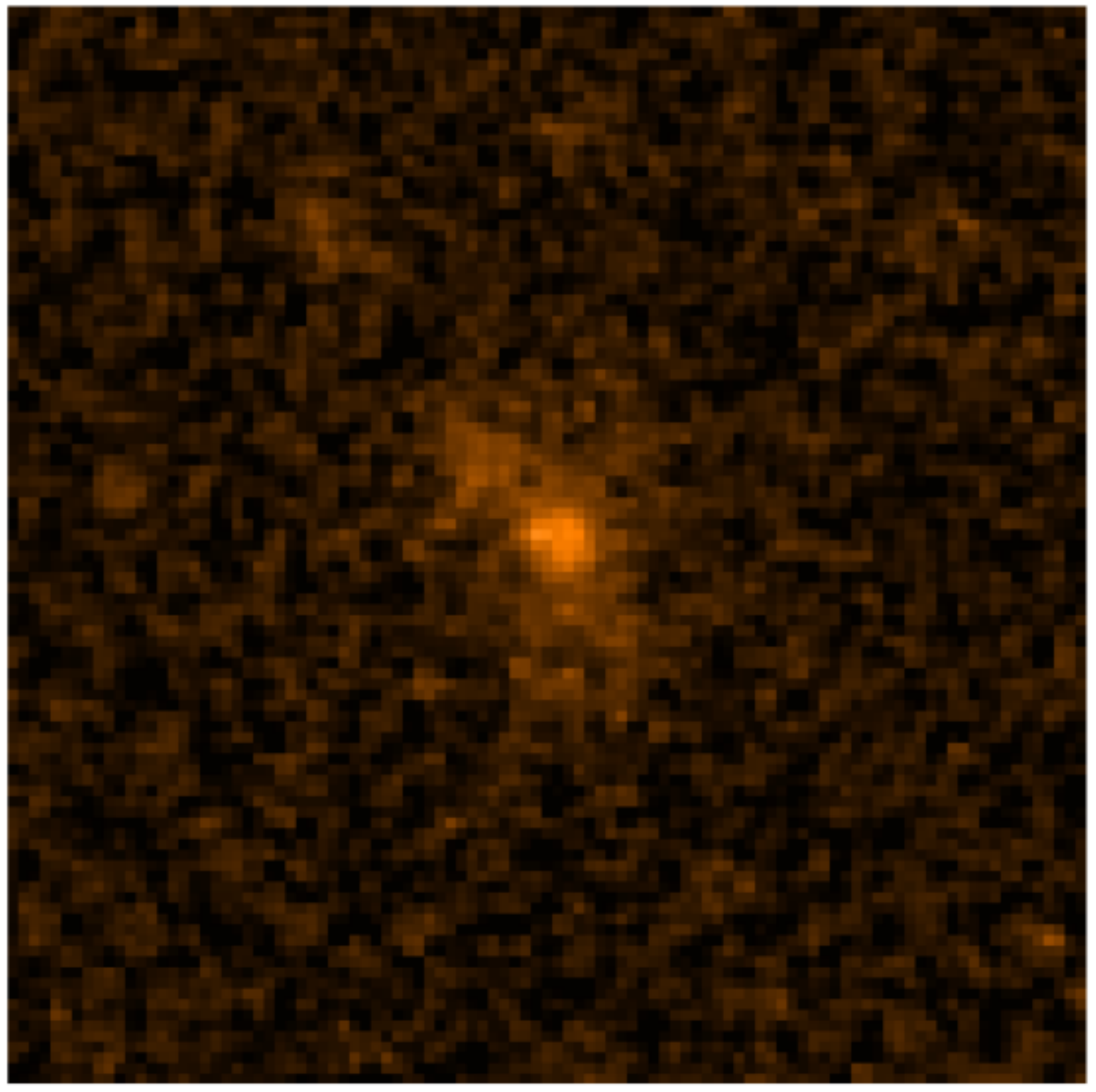}  &
\includegraphics[scale=0.15]{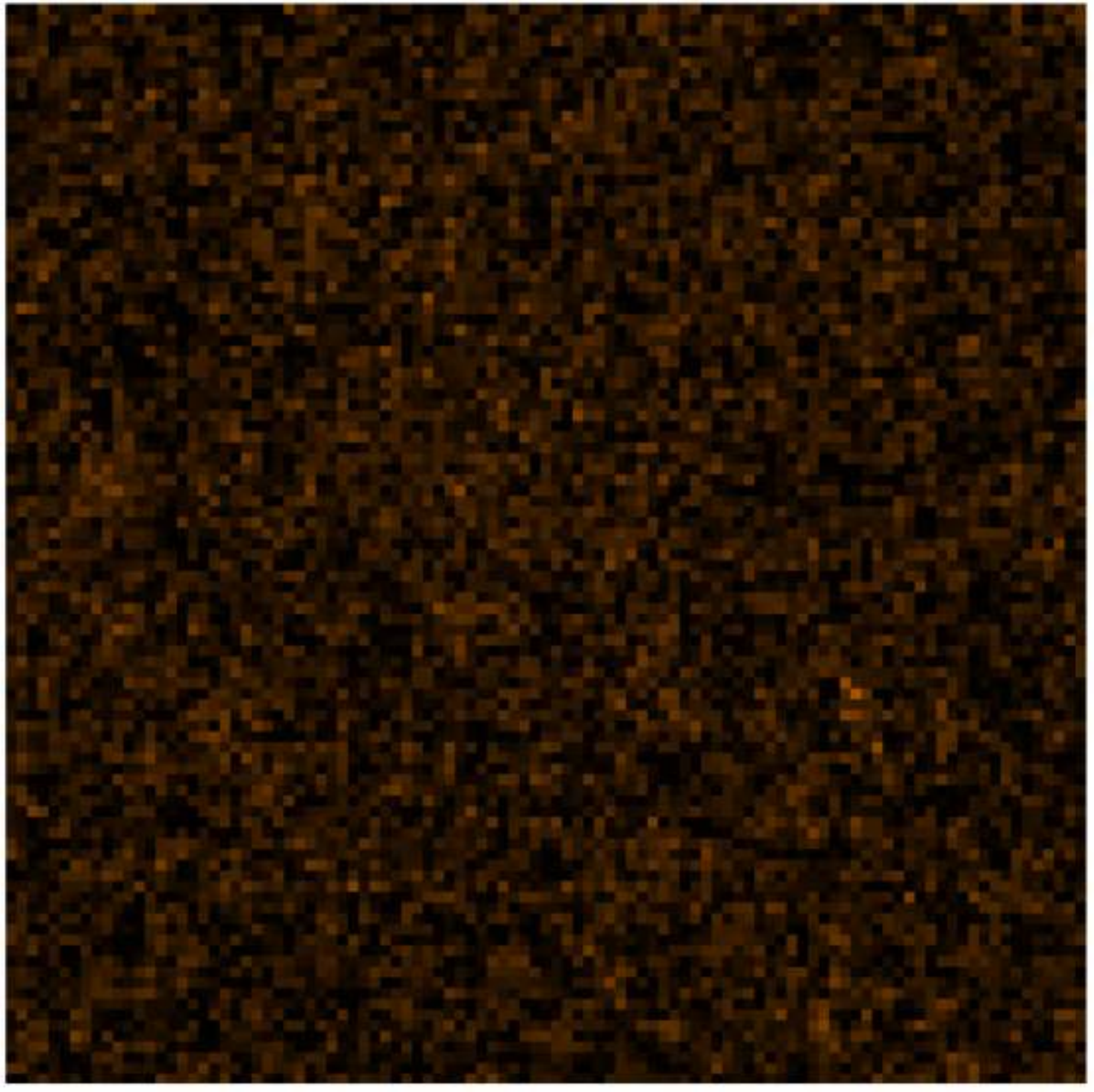}  &
\includegraphics[scale=0.15]{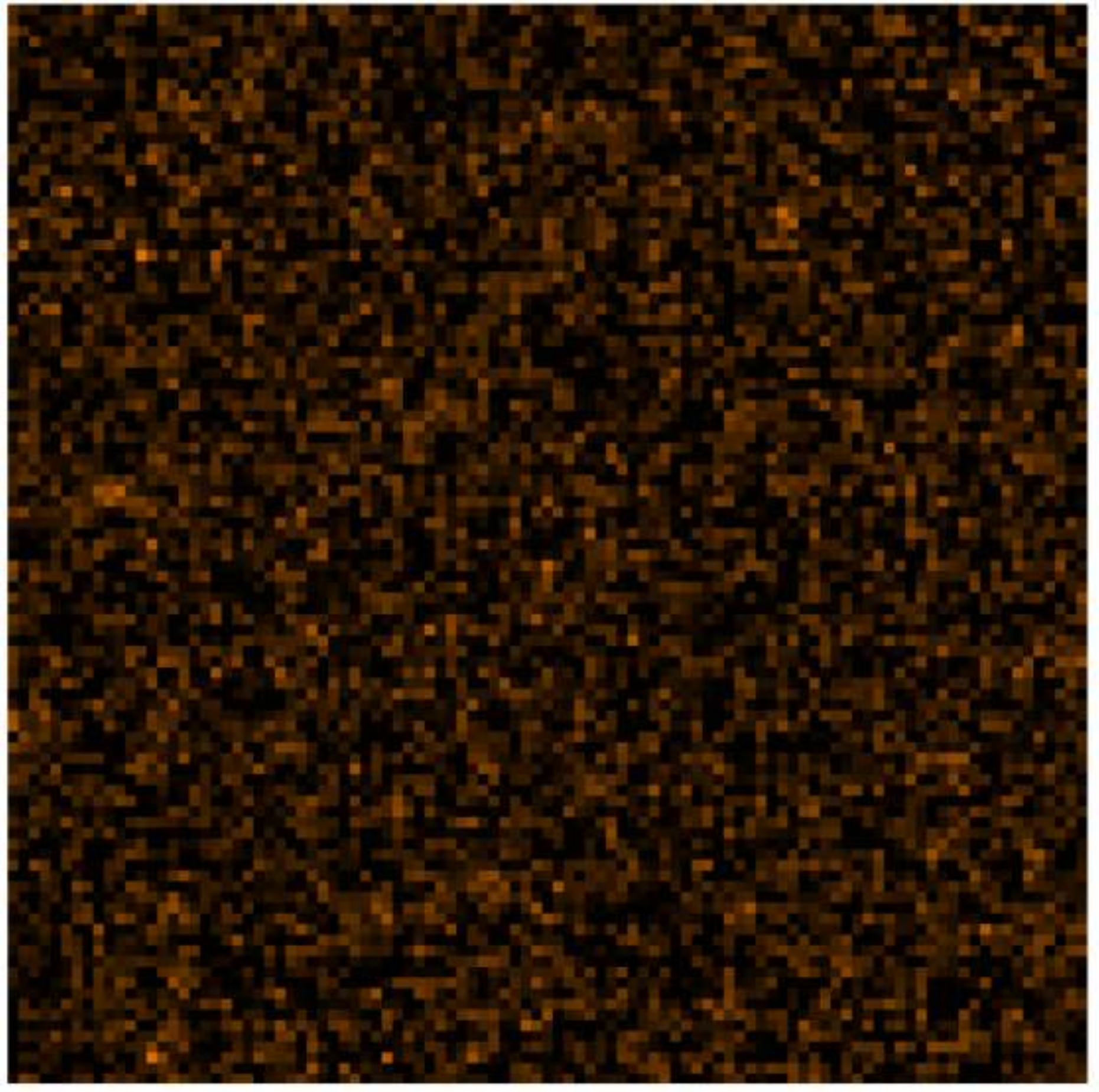}  \\
\includegraphics[scale=0.15]{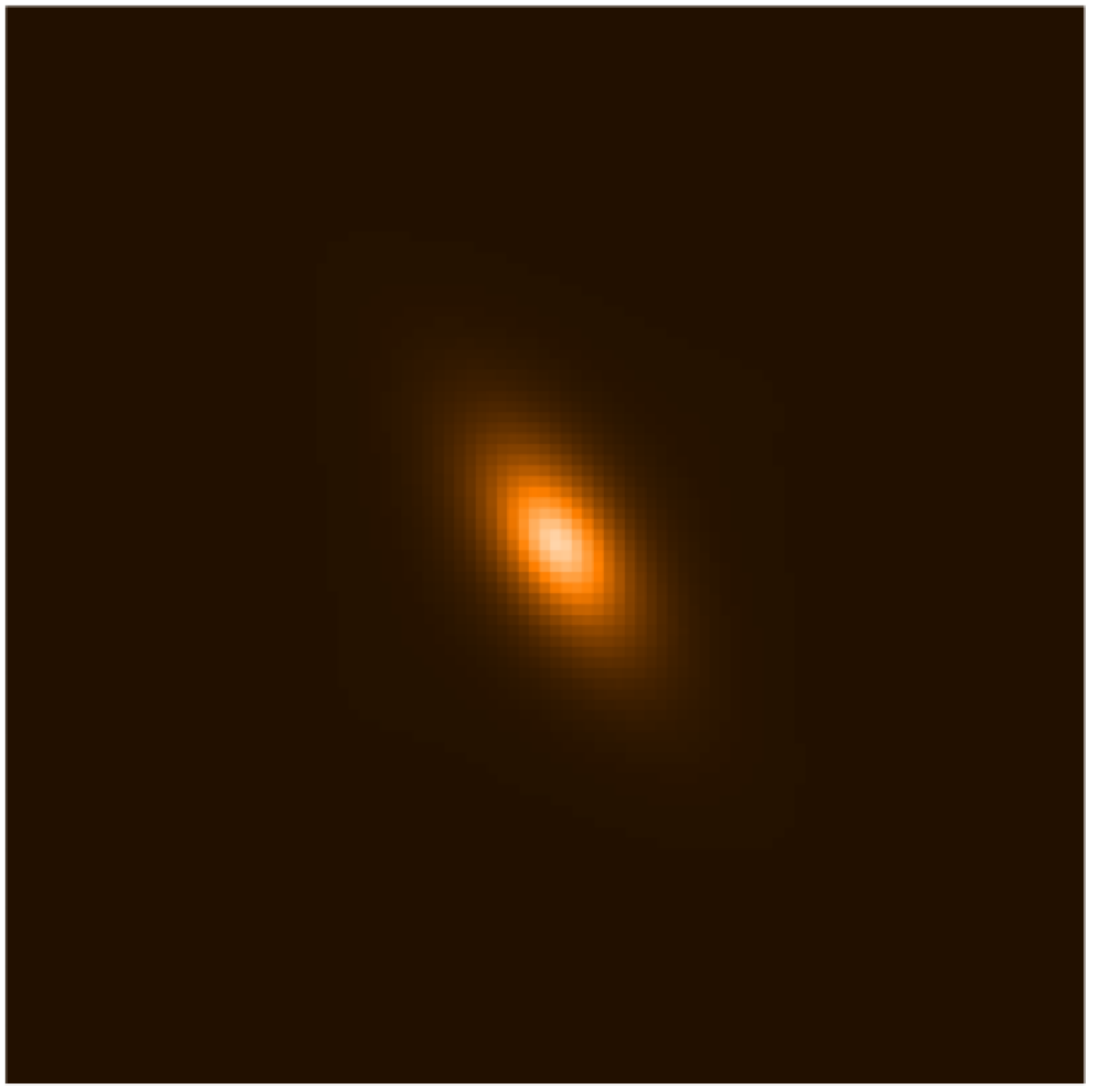}  &
\includegraphics[scale=0.15]{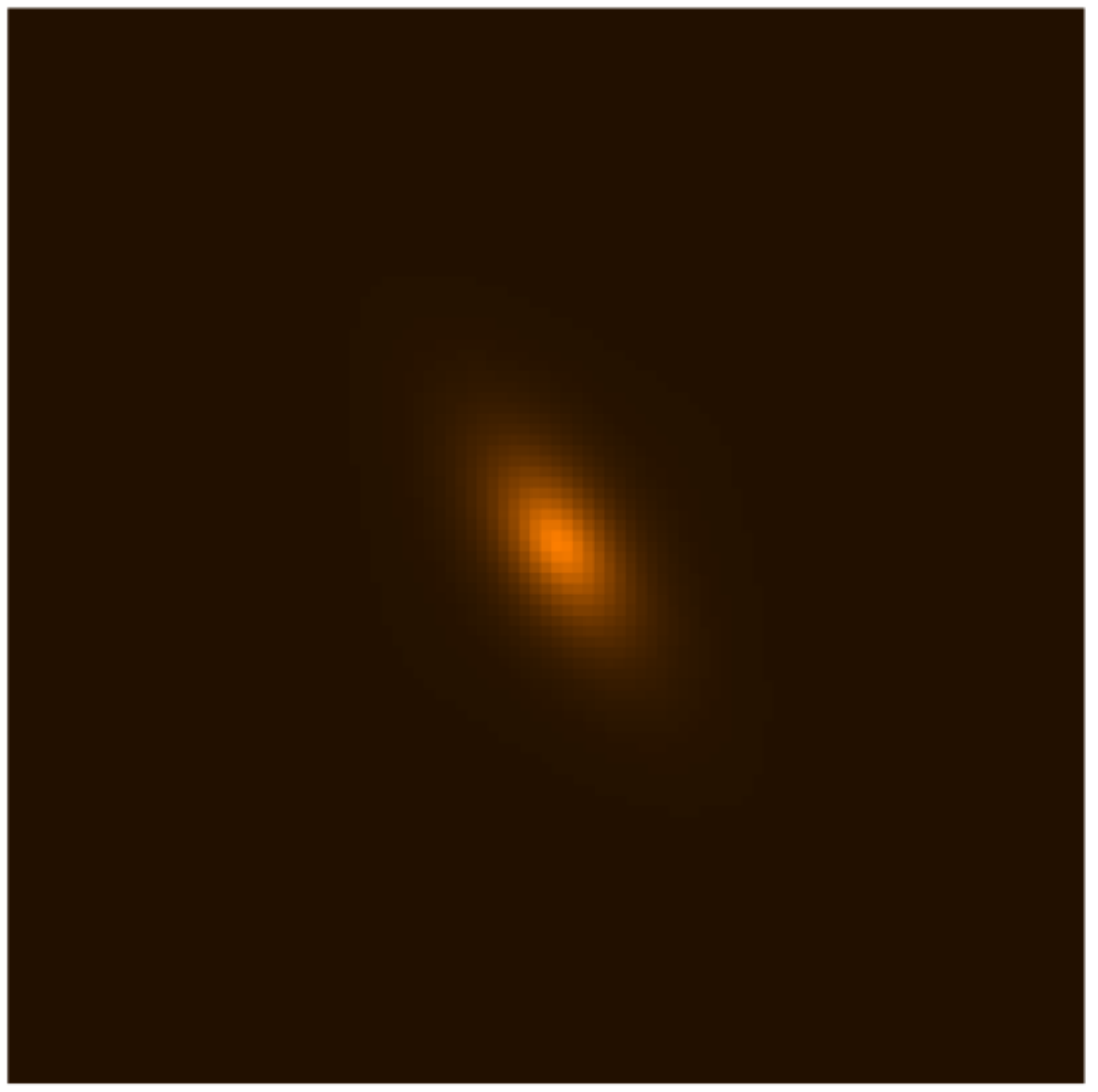}  &
\includegraphics[scale=0.15]{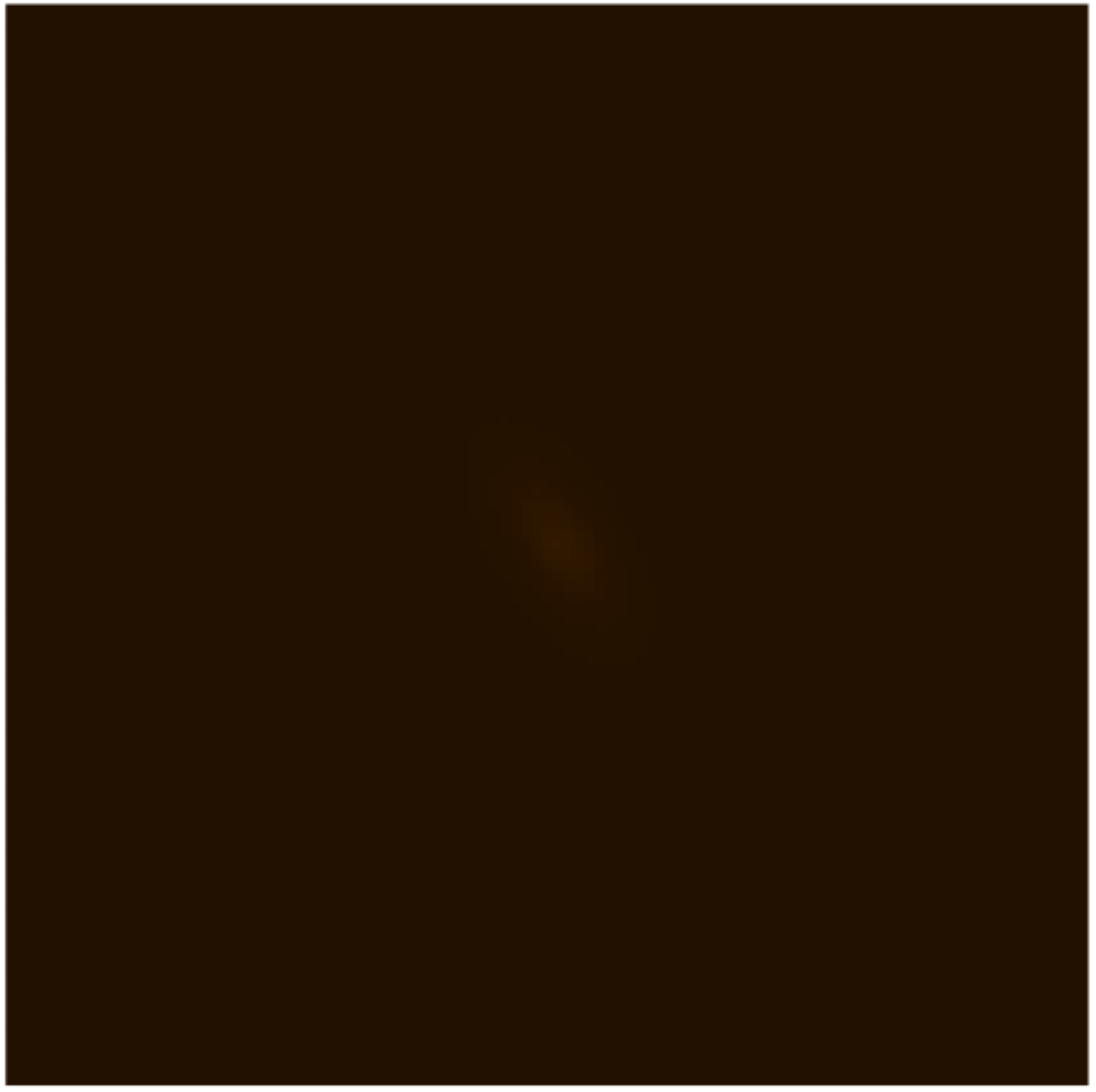}  &
\includegraphics[scale=0.15]{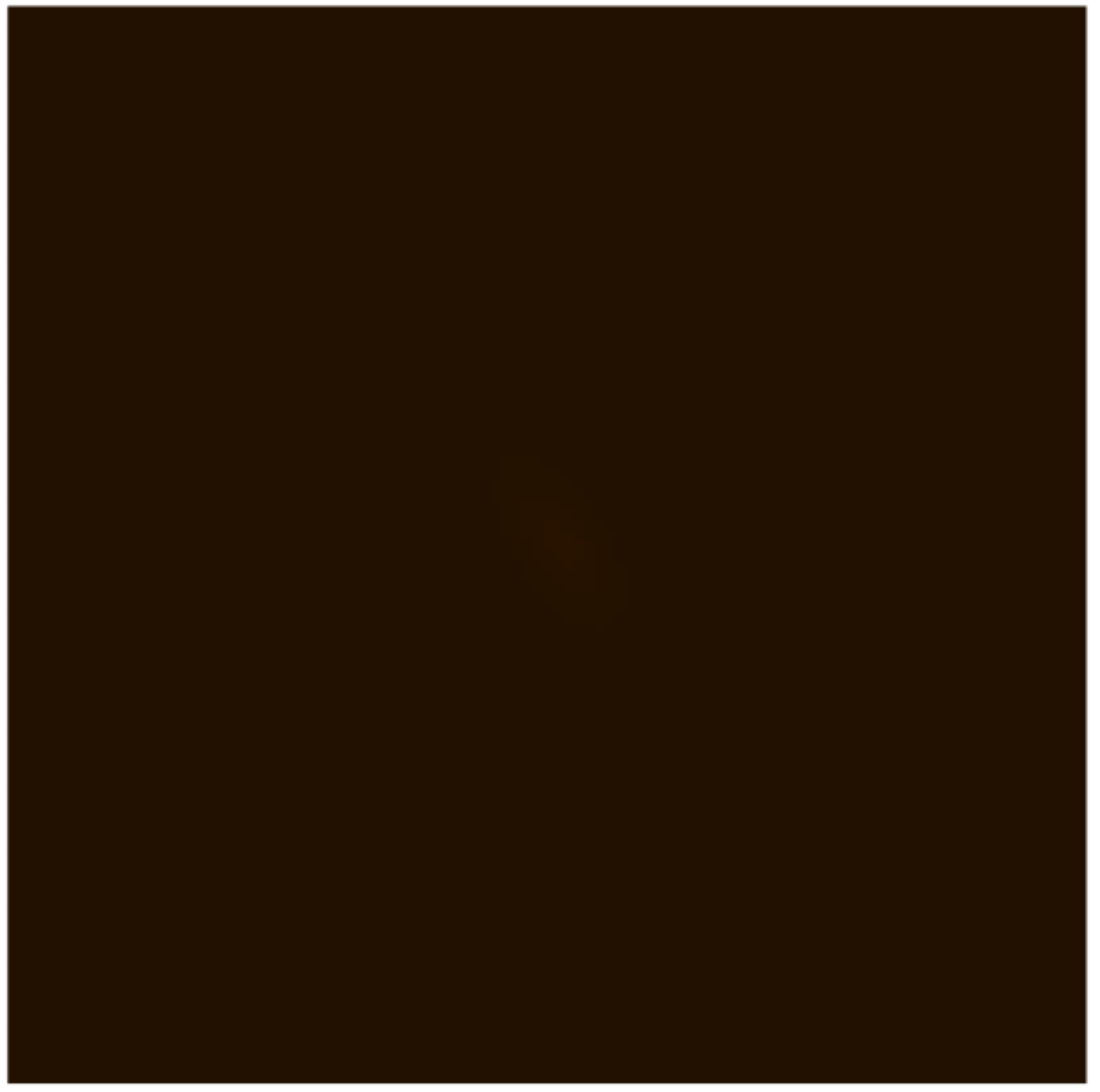}  \\
\includegraphics[scale=0.15]{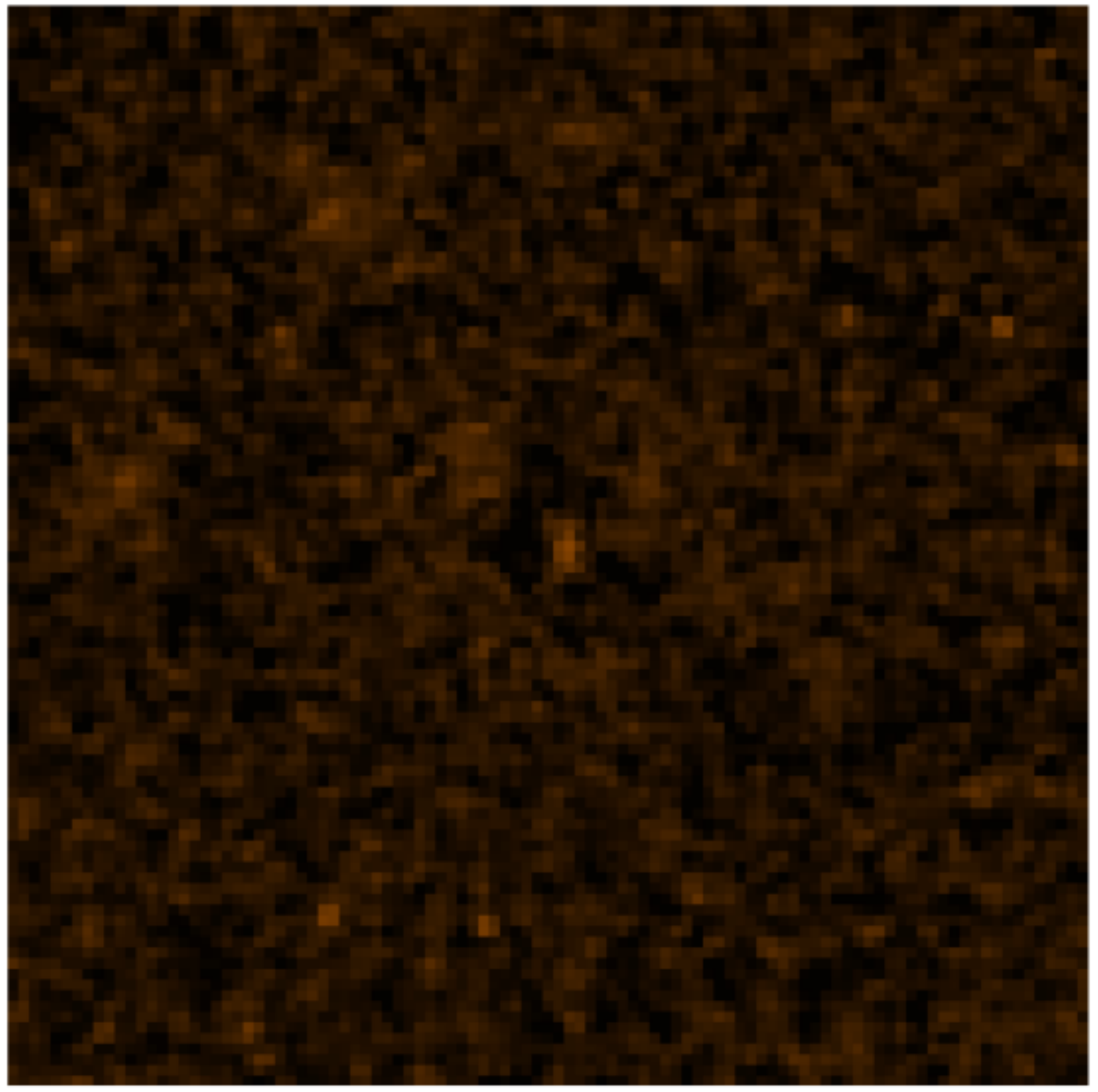}  &
\includegraphics[scale=0.15]{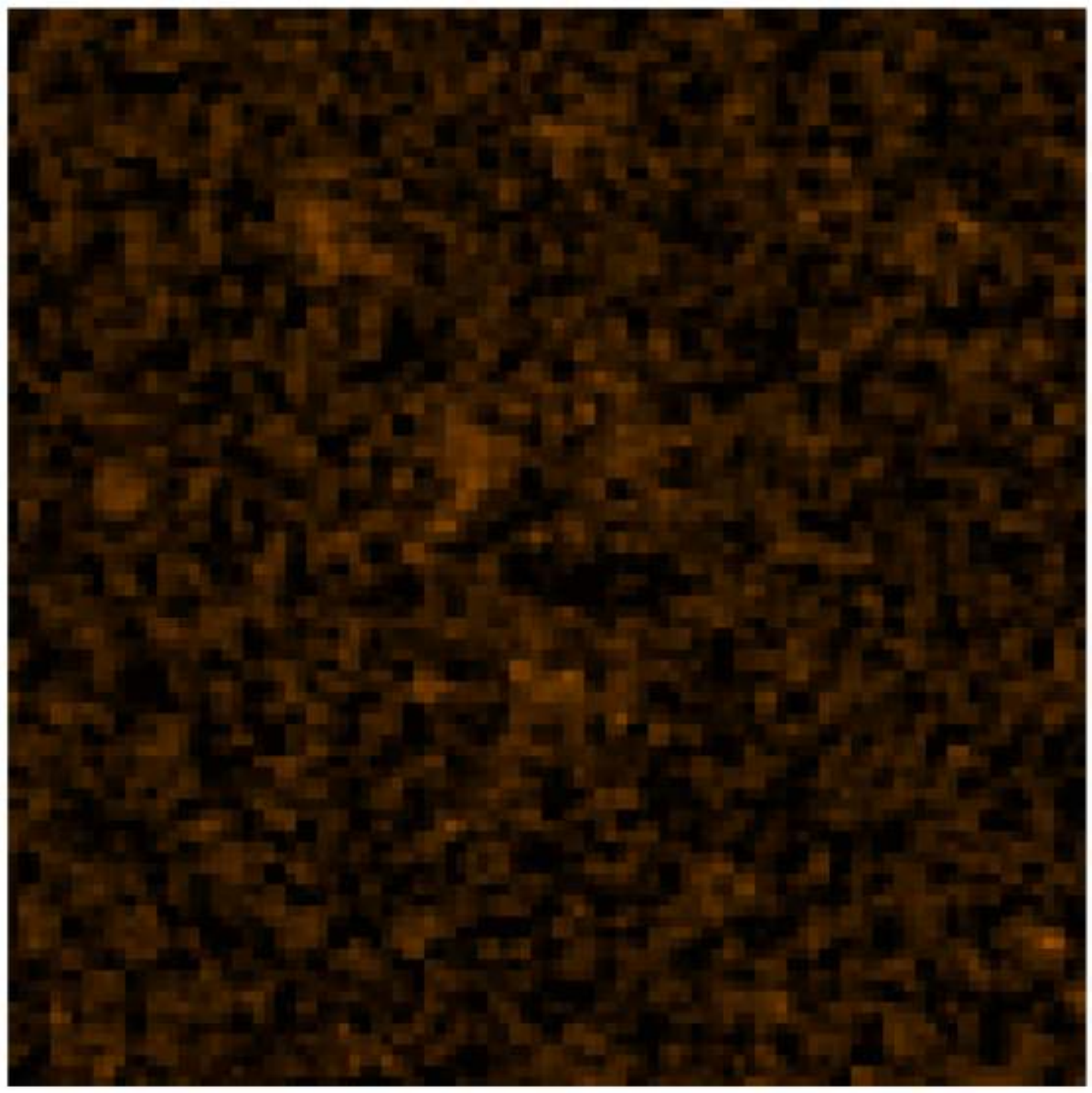}  &
\includegraphics[scale=0.15]{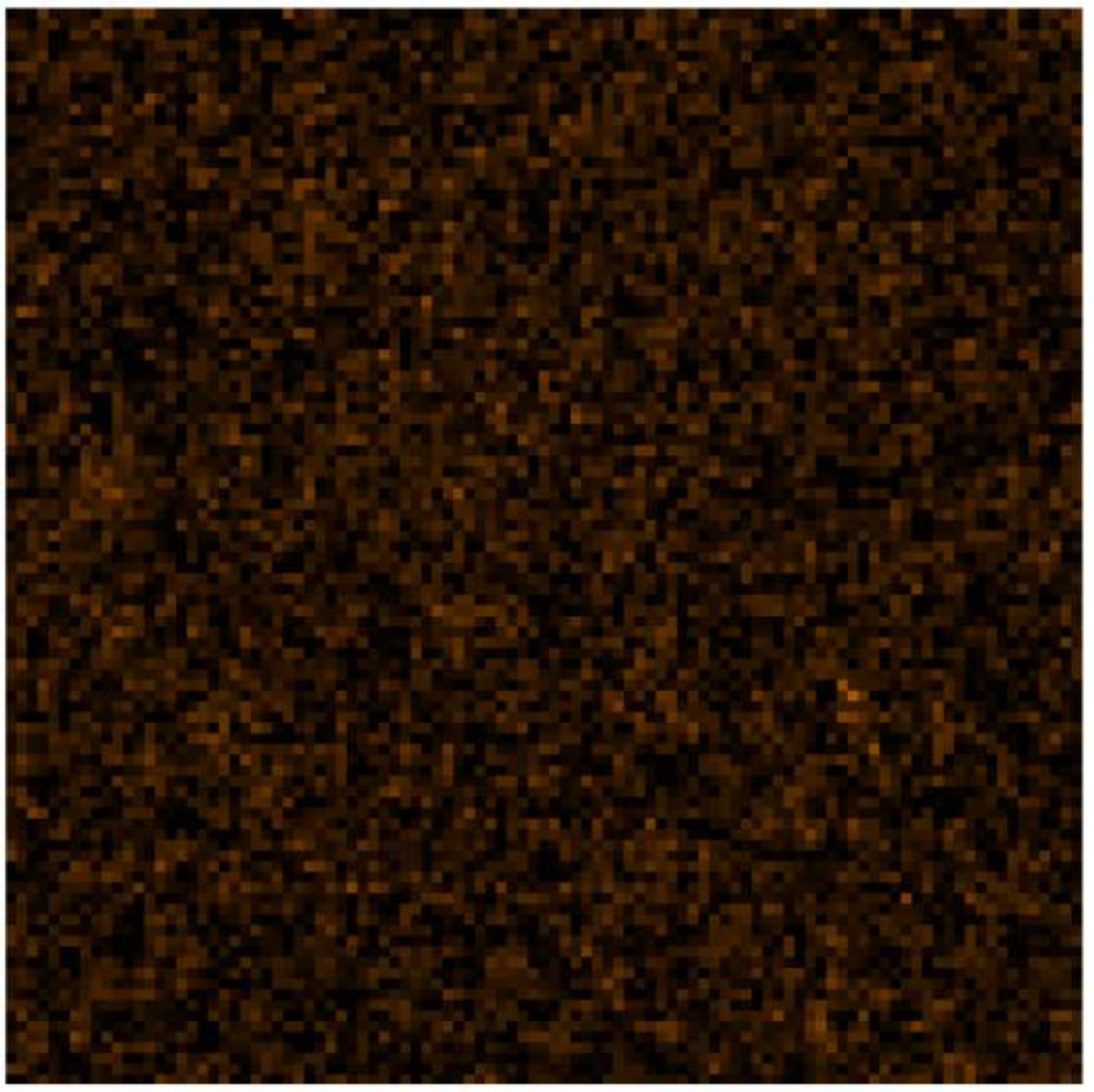}  &
\includegraphics[scale=0.15]{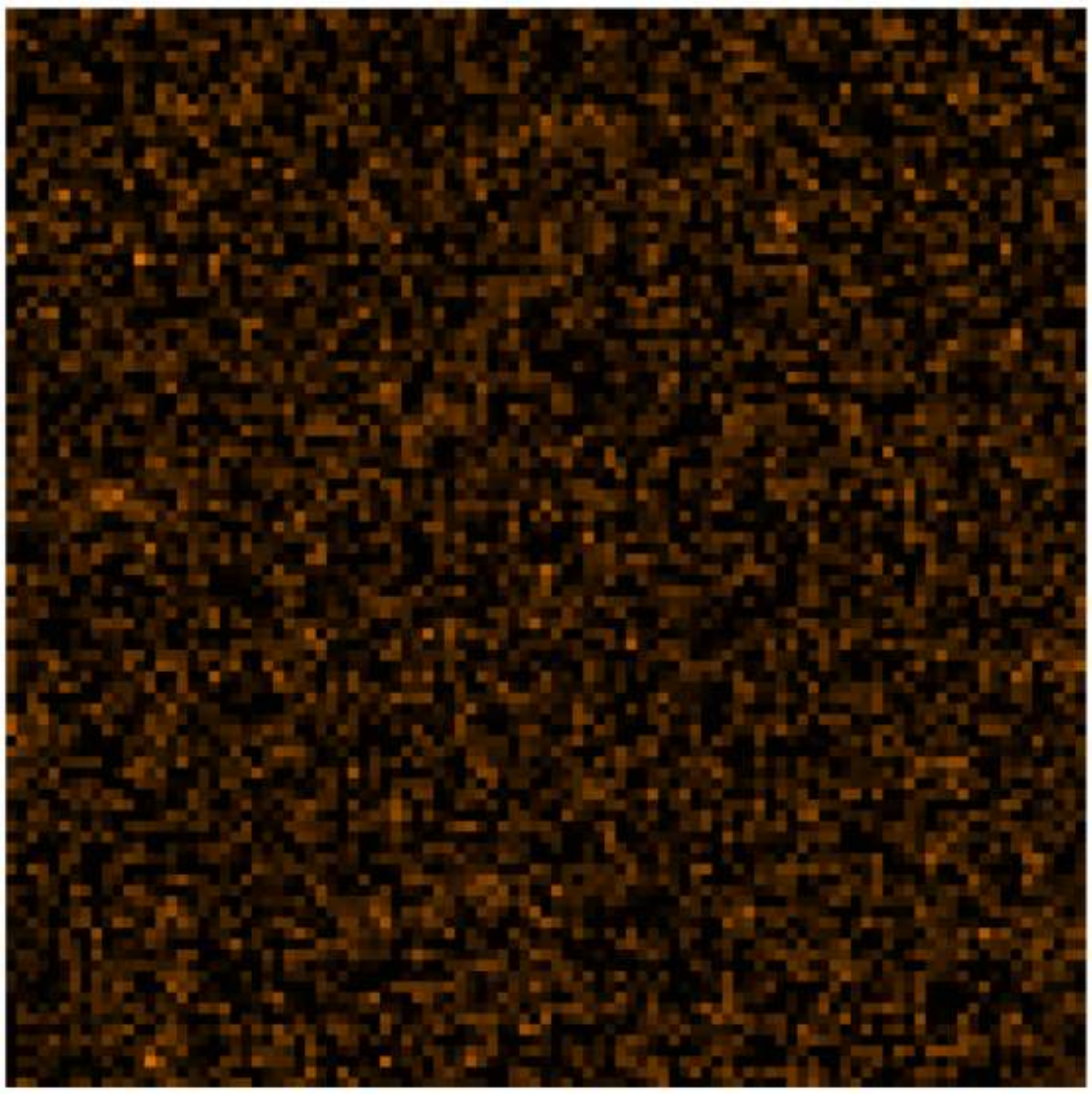}  \\
\end{tabular}
\caption[Image stamps of an example fit for a pure disk object.]{Image stamps of an example fit for a pure-disk object. As in Fig.\,\ref{fig:pure_bulge}, displayed are $6\times6$ arcsec image, model and residual stamps, ranked from left to right by decreasing wavelength. In this case, the image and model stamps from this fitting clearly illustrate that, despite being best fit by a pure-disk morphology, this component becomes fainter in the bluer bands. The residuals show that no additional structure becomes prominent at the bluer wavelengths.}
\label{fig:pure_disk}
\end{center}
\end{figure*}

\begin{figure*}
\begin{center}
\begin{tabular}{m{3cm}m{3cm}m{3cm}m{3cm}}
\includegraphics[scale=0.15 ]{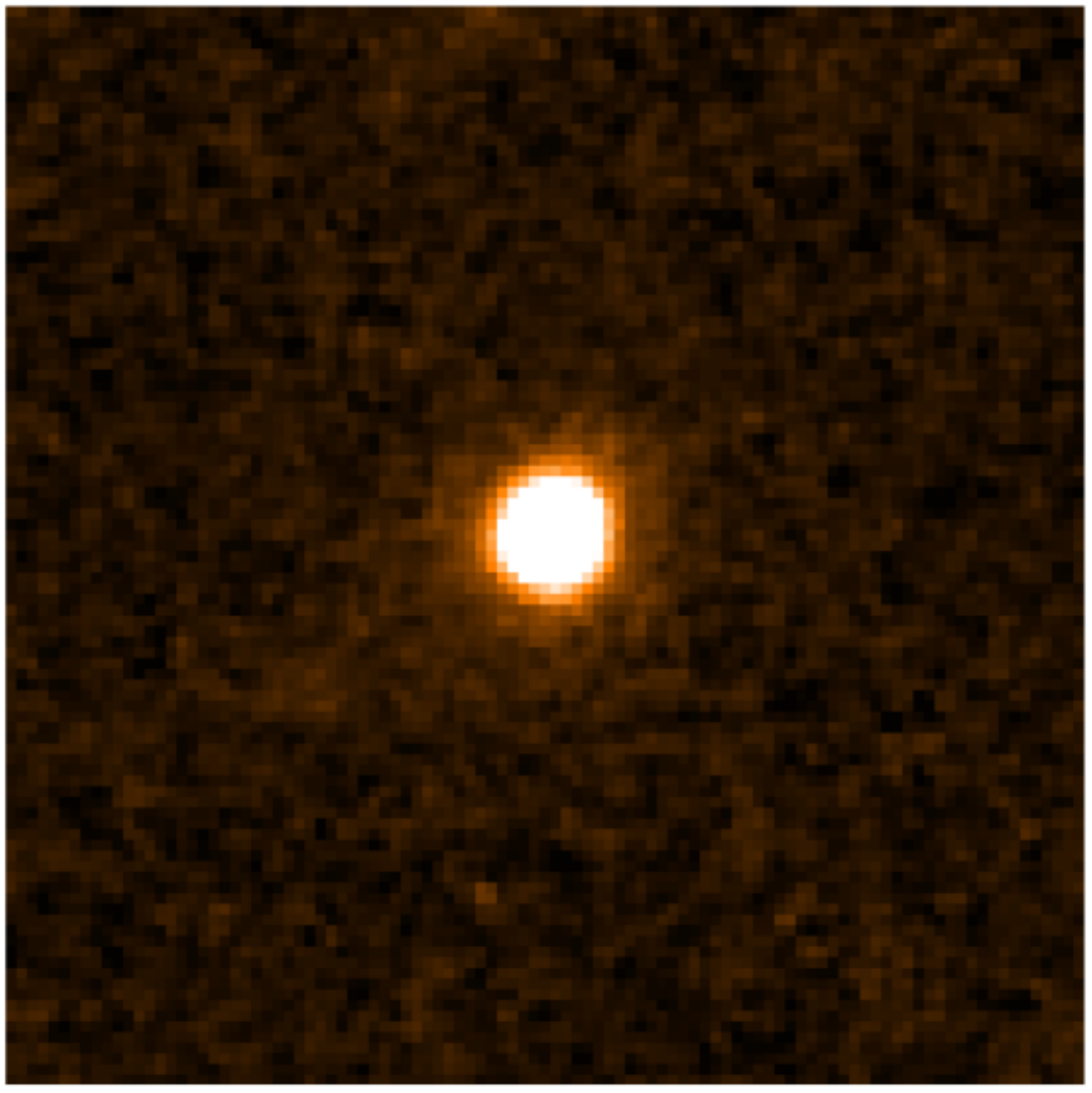}  &
\includegraphics[scale=0.15]{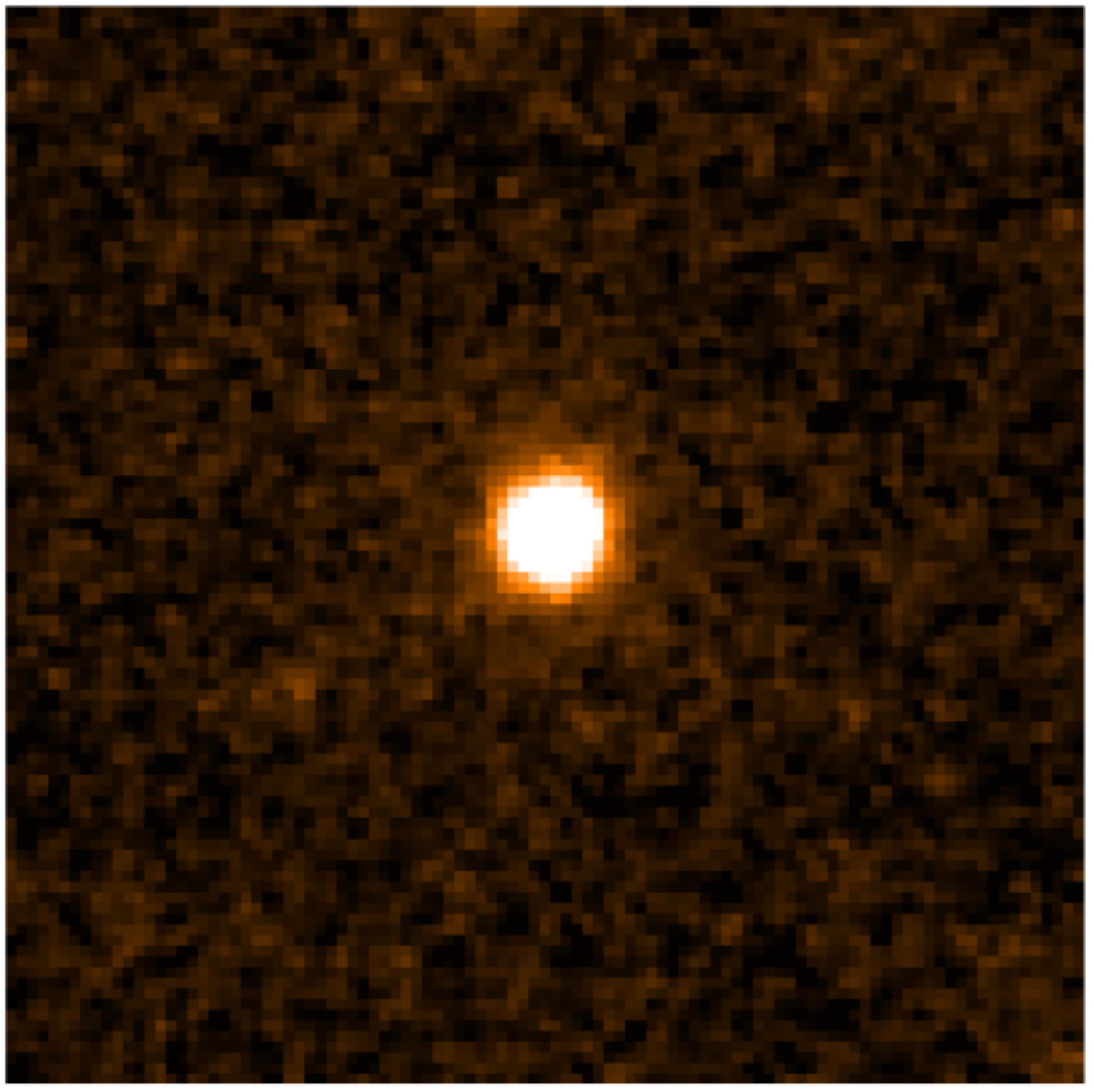}  &
\includegraphics[scale=0.15]{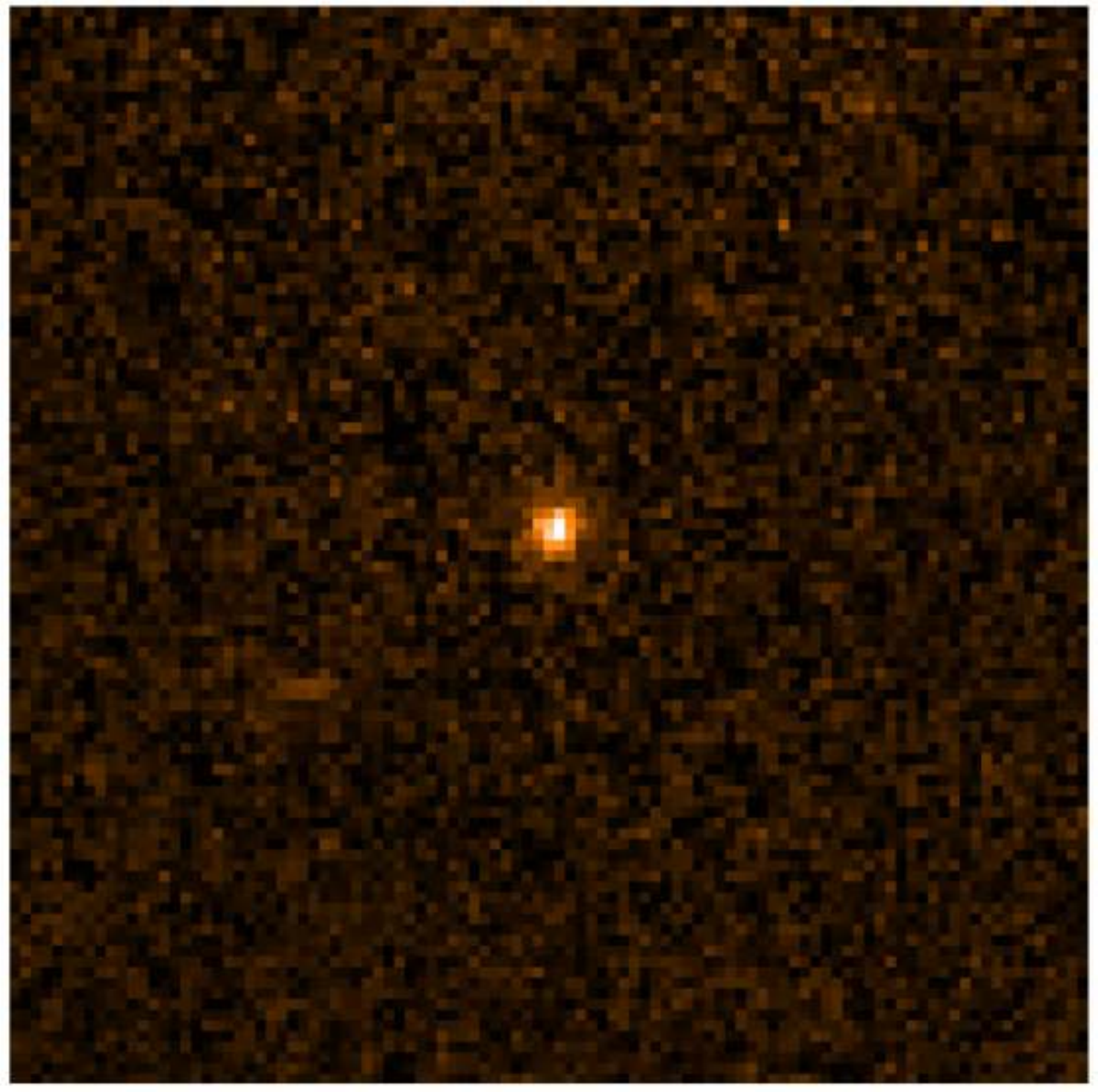}  &
\includegraphics[scale=0.15]{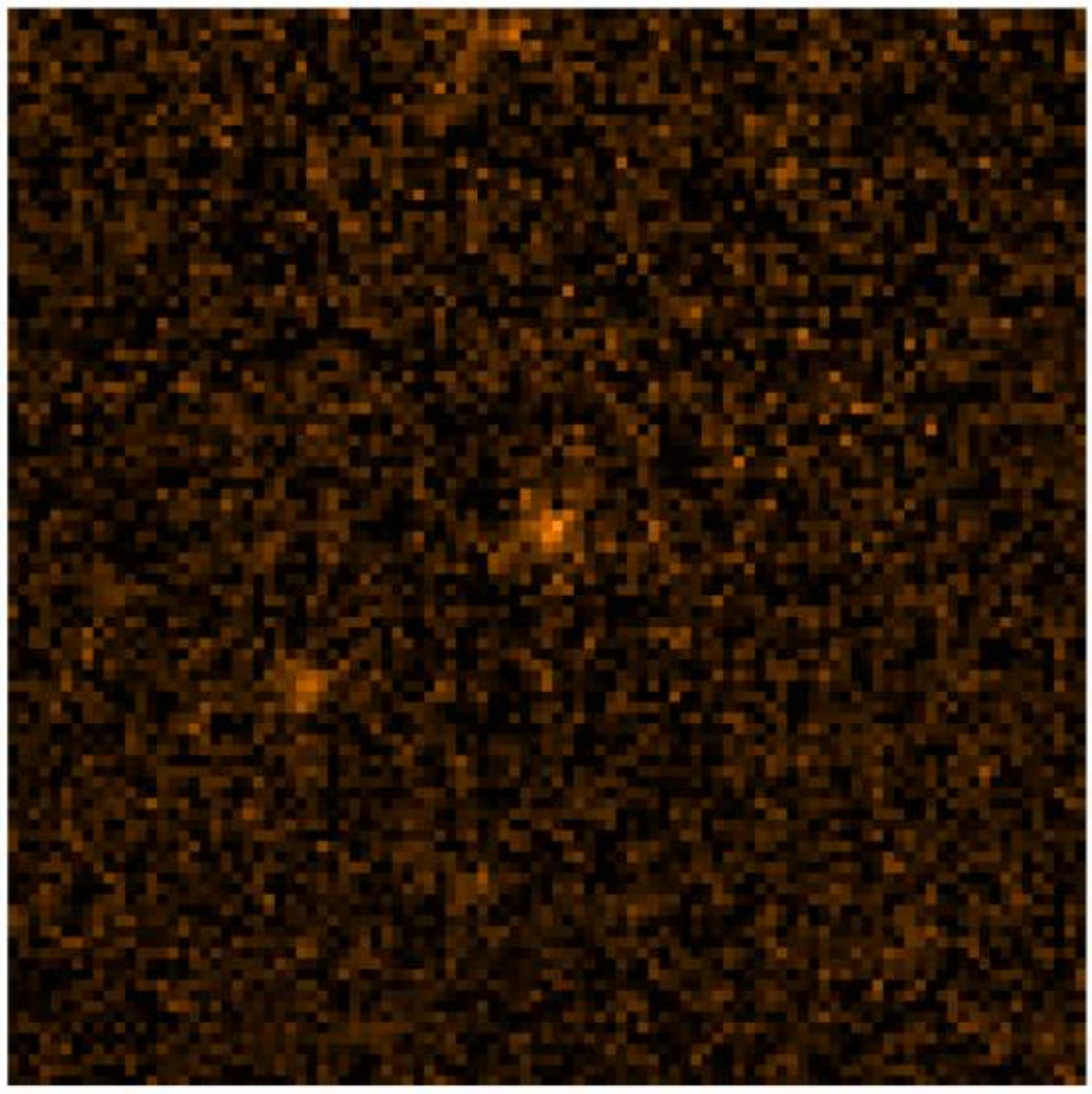}  \\
\includegraphics[scale=0.15]{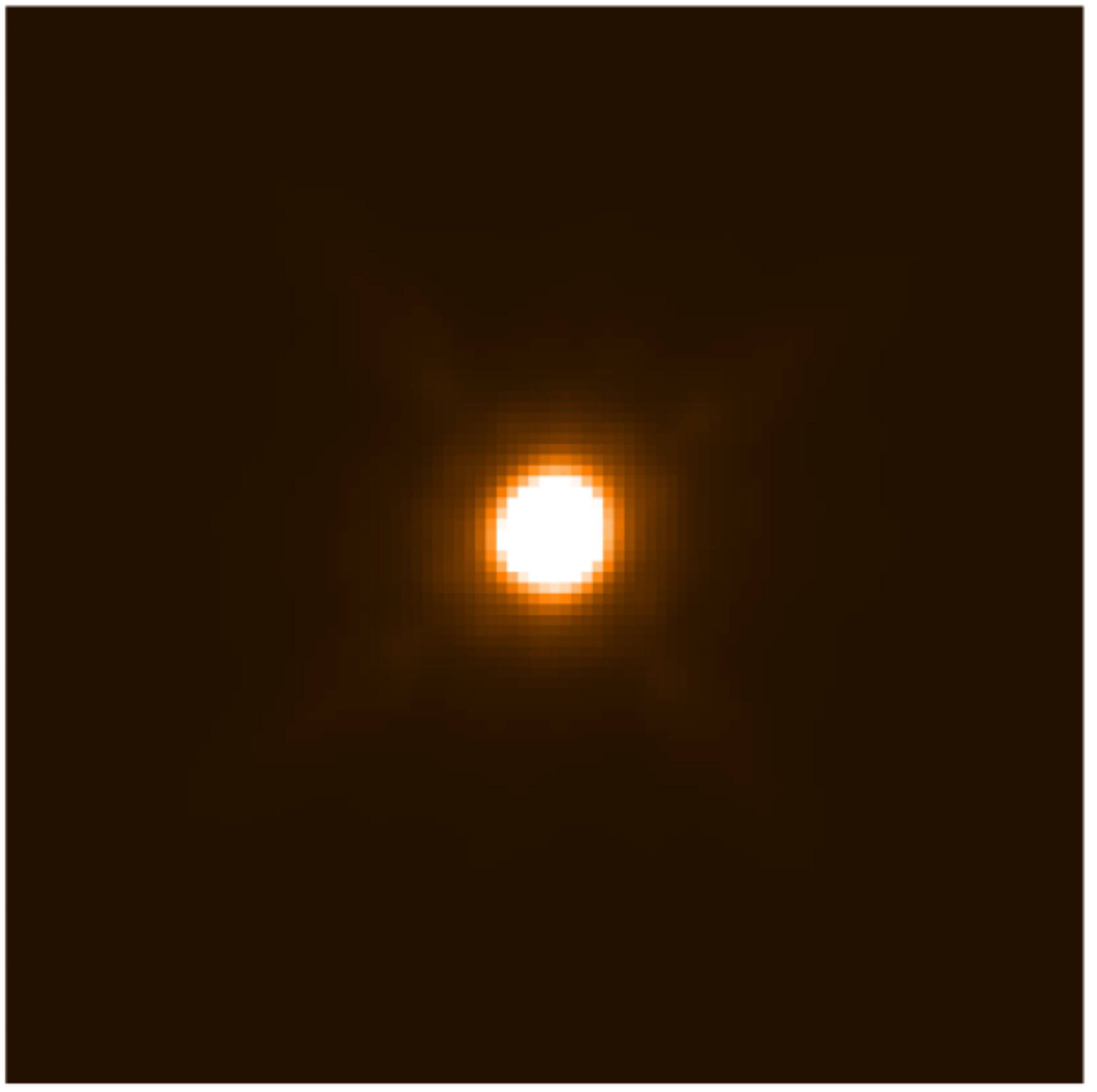}  &
\includegraphics[scale=0.15]{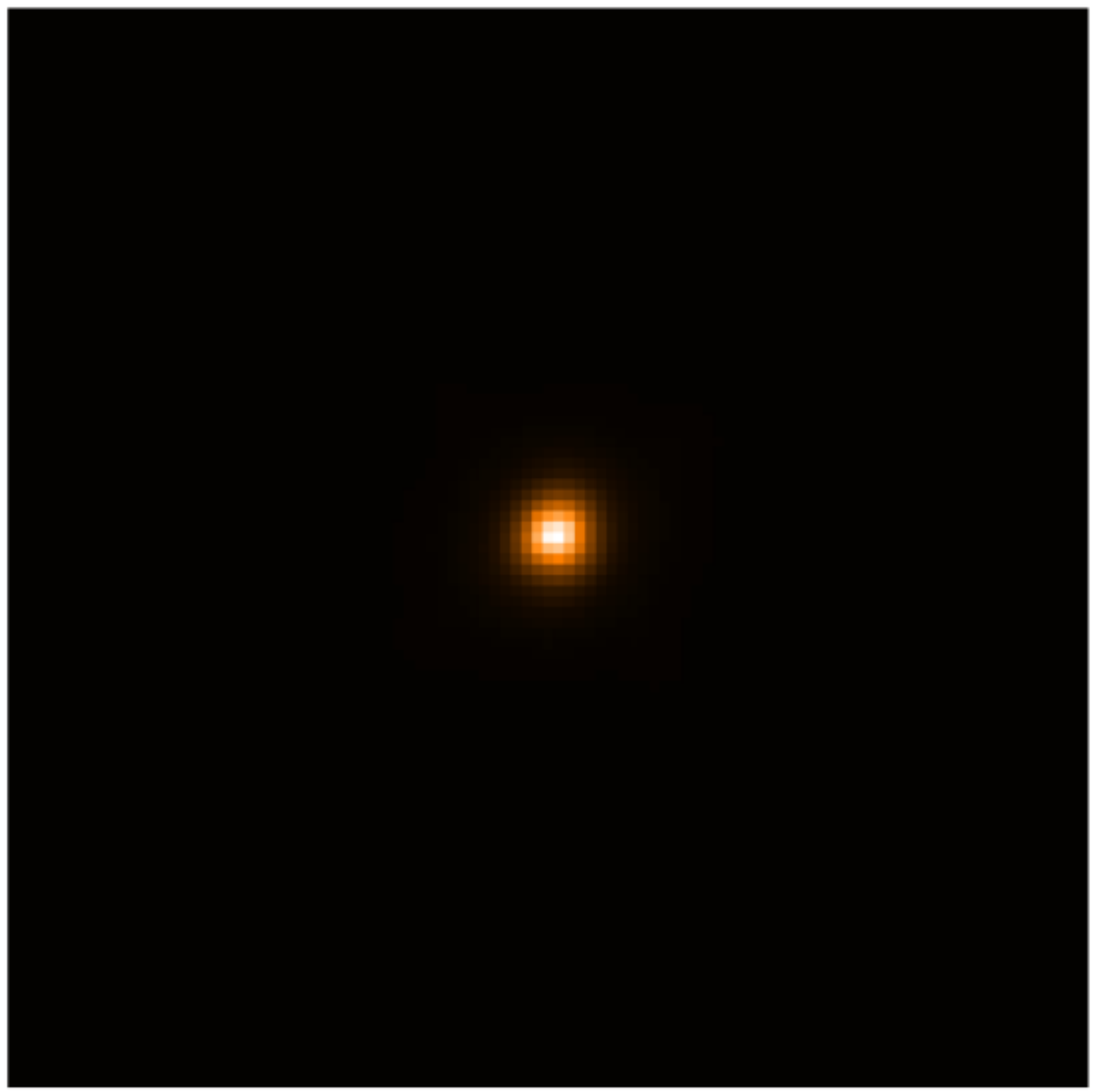}  &
\includegraphics[scale=0.15]{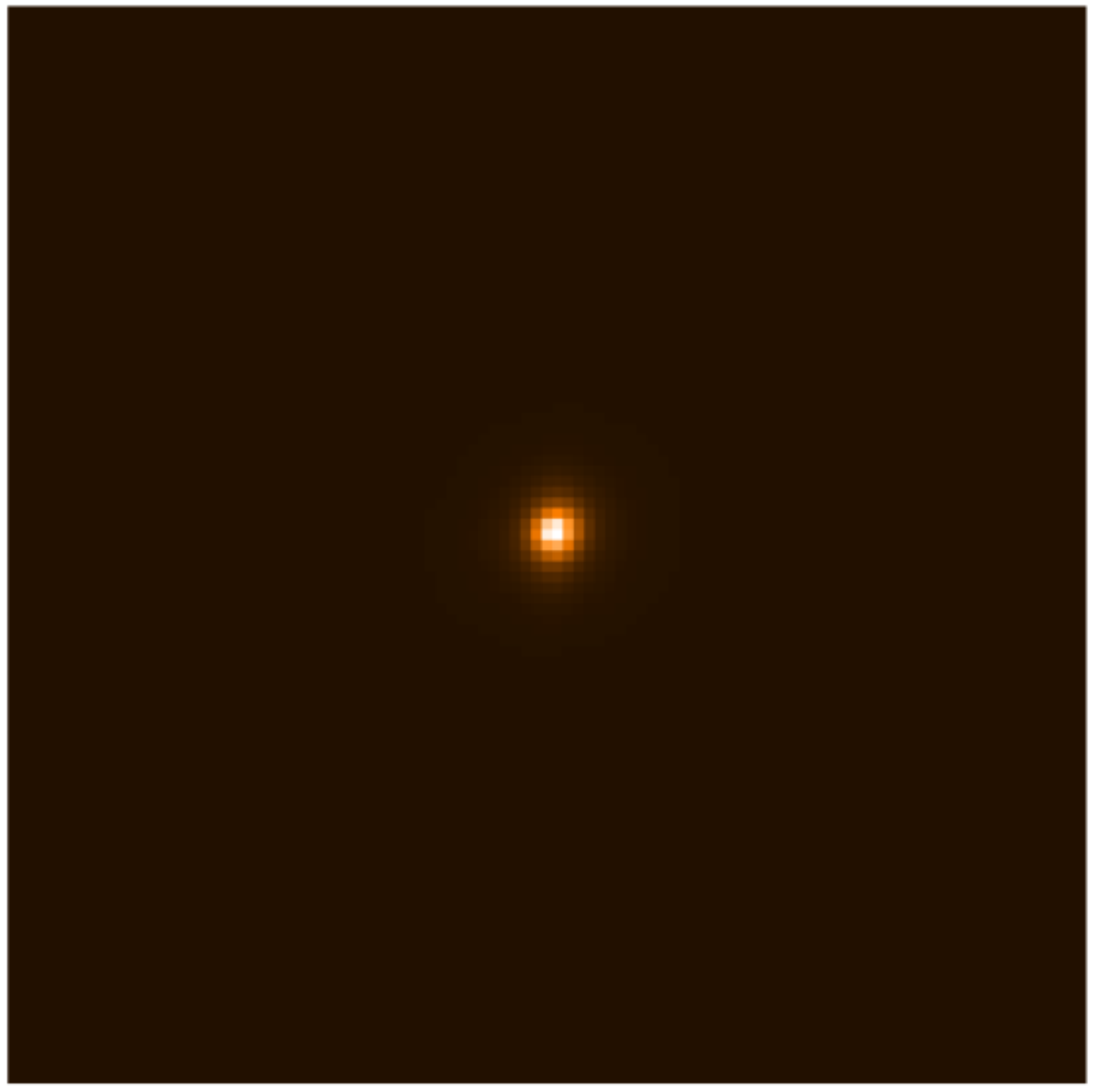}  &
\includegraphics[scale=0.15]{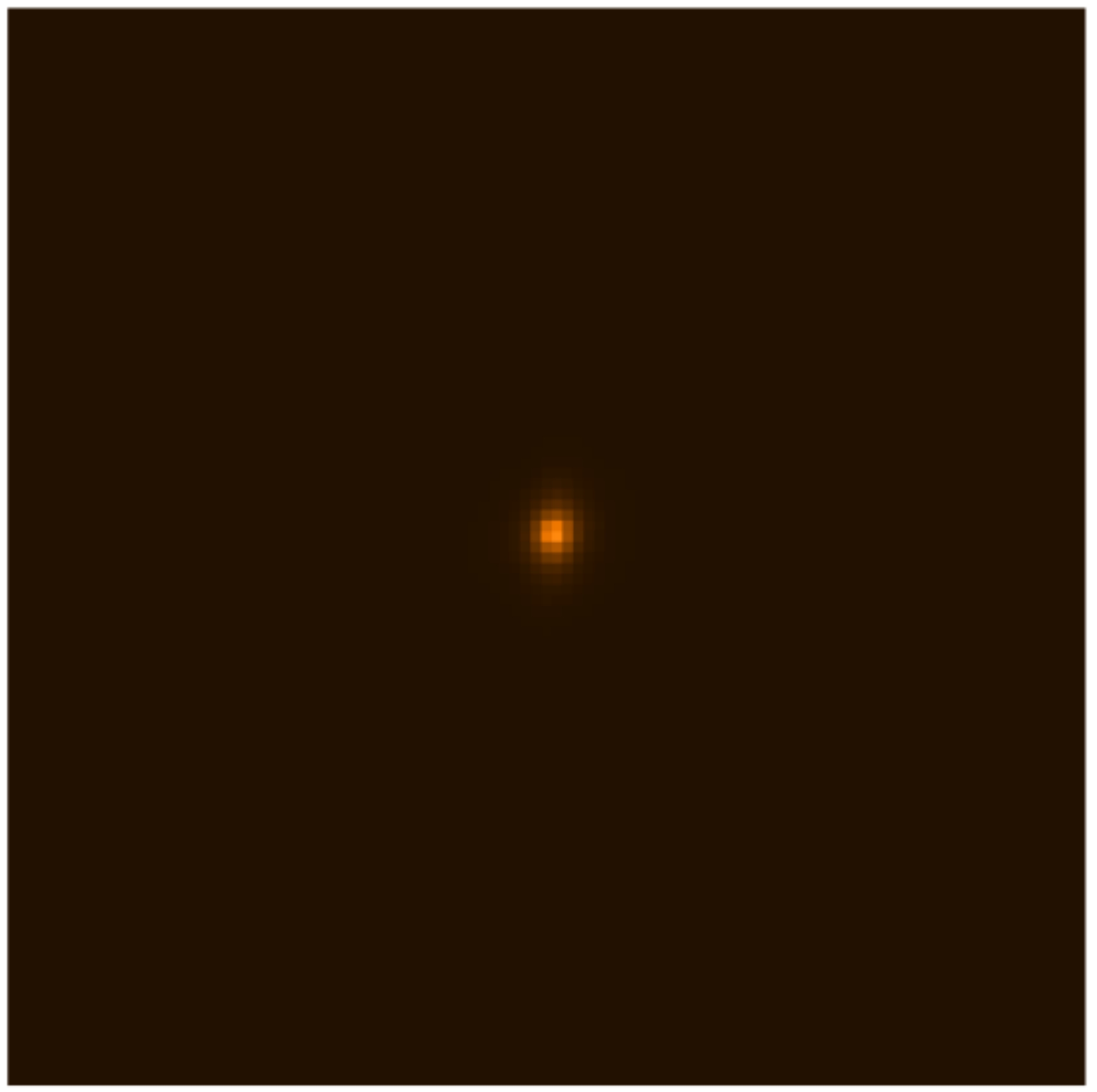}  \\
\includegraphics[scale=0.15]{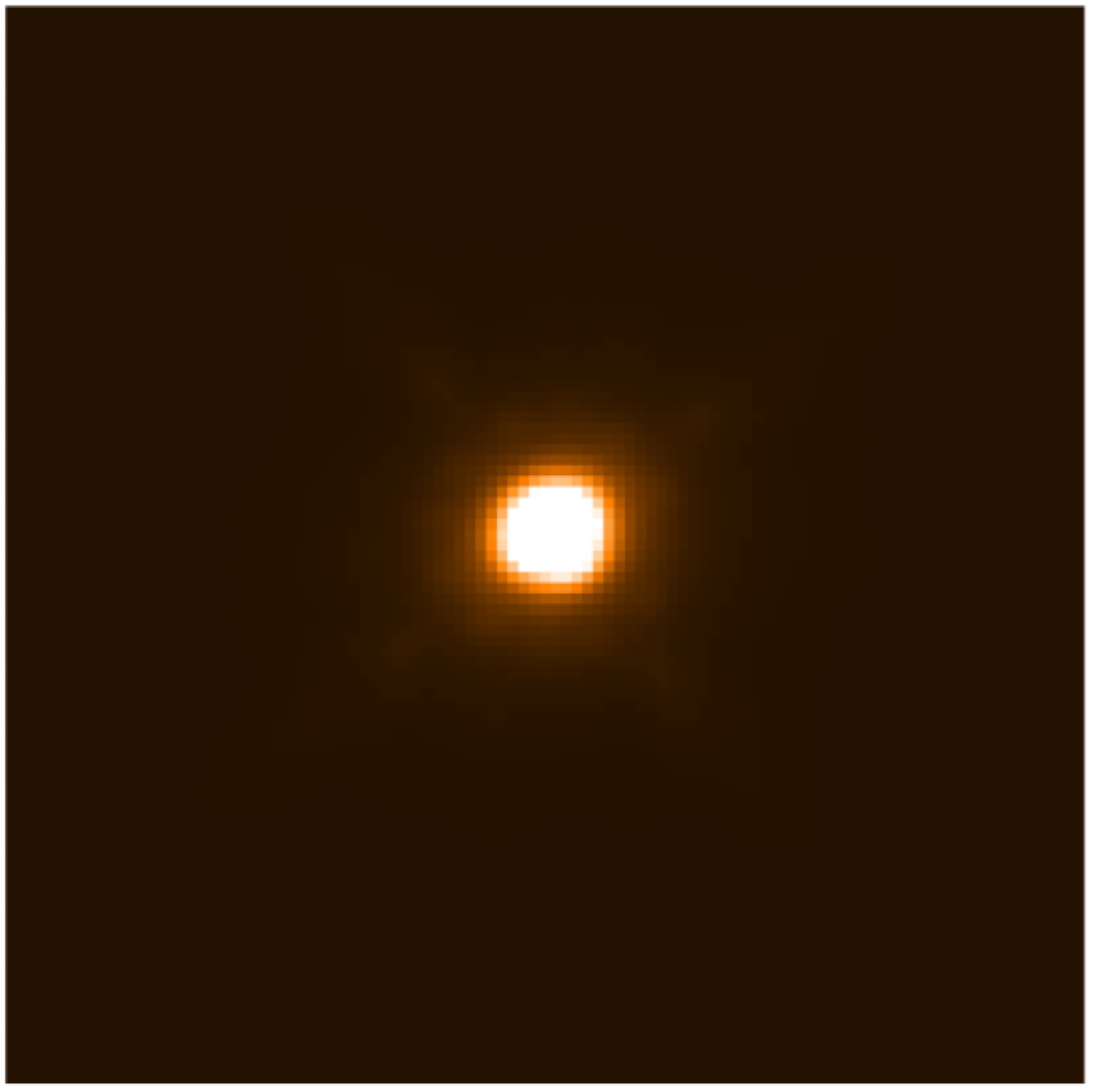}  &
\includegraphics[scale=0.15]{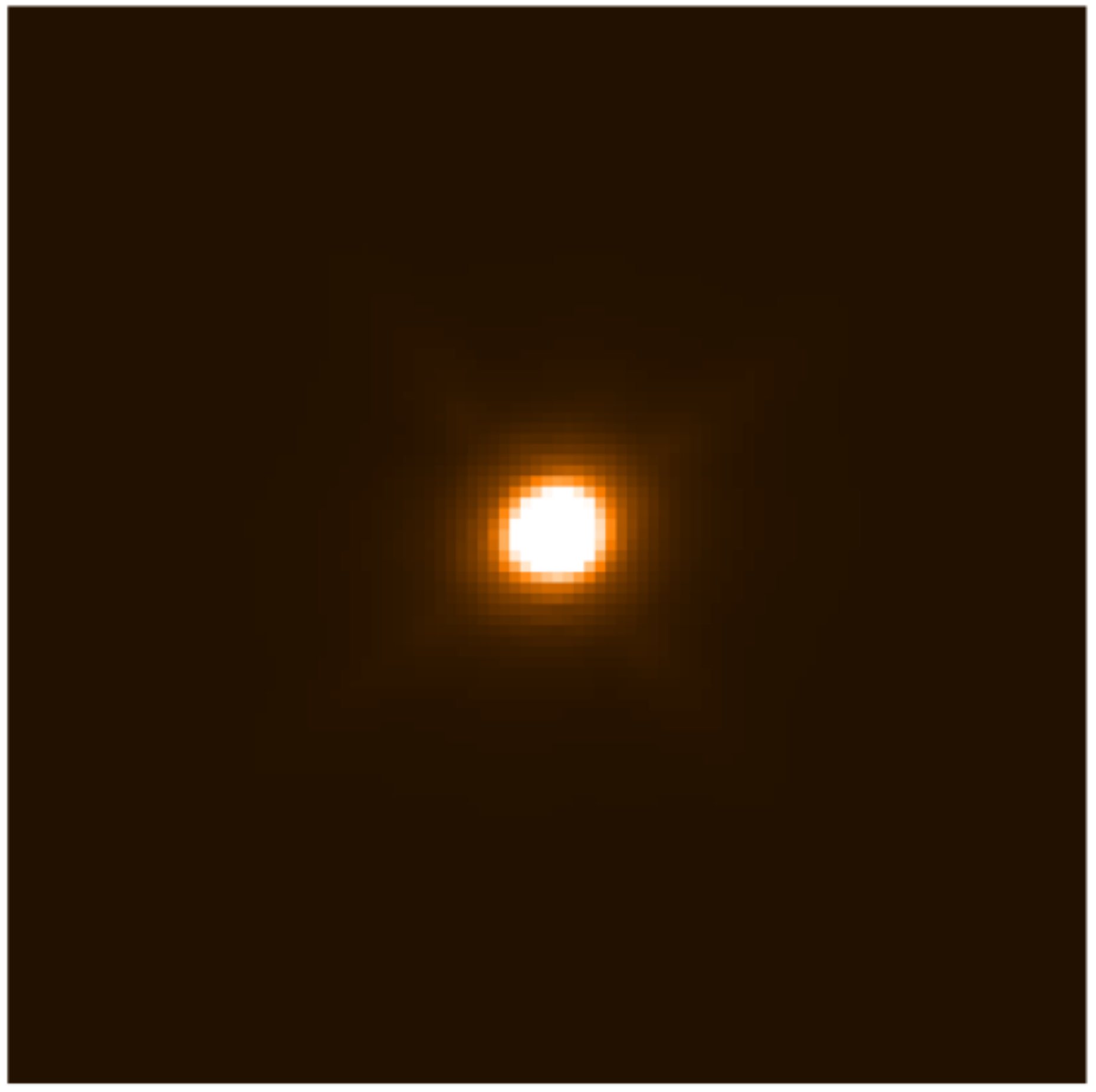}  &
\includegraphics[scale=0.15]{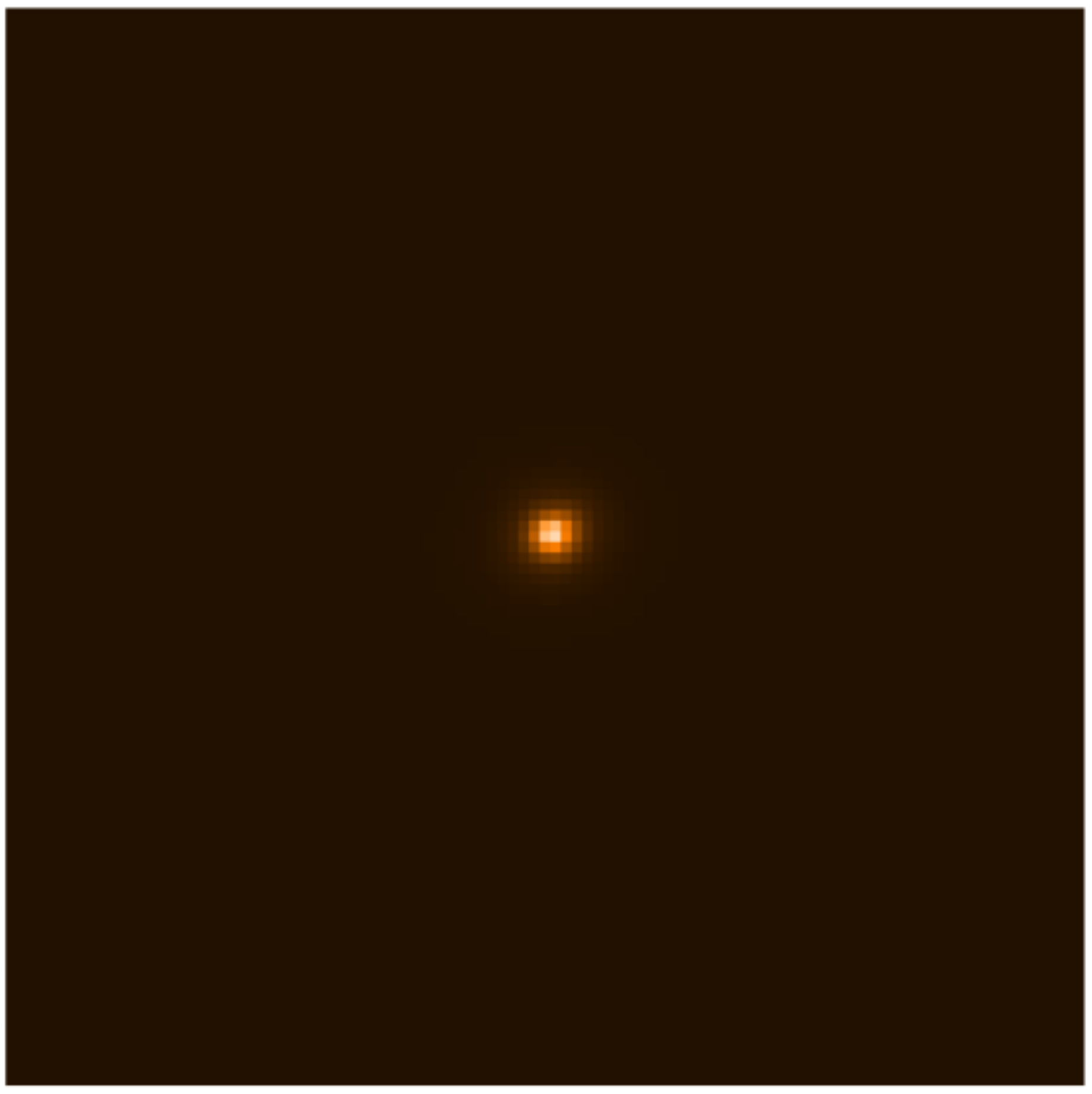}  &
\includegraphics[scale=0.15]{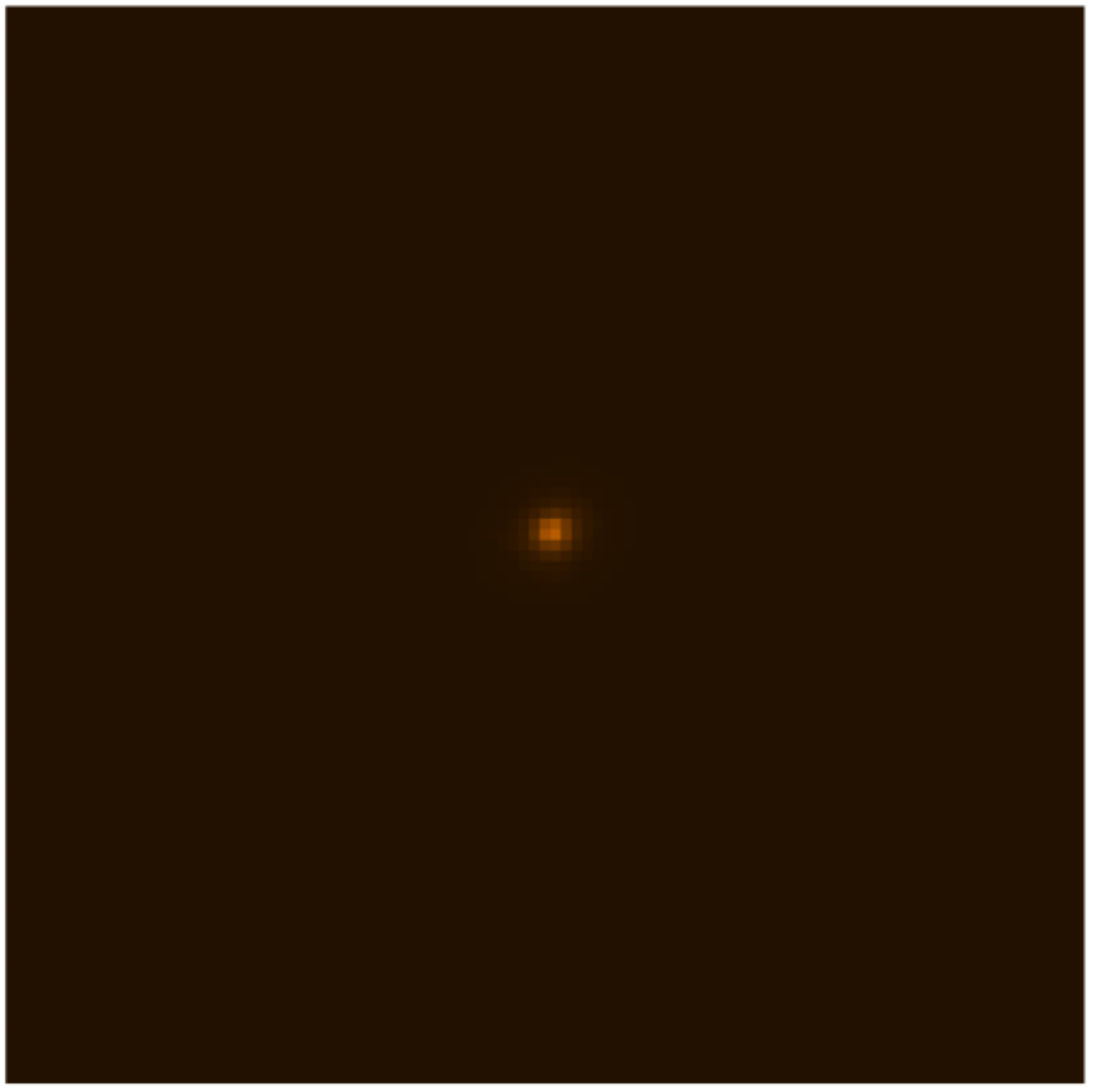}  \\
\includegraphics[scale=0.15]{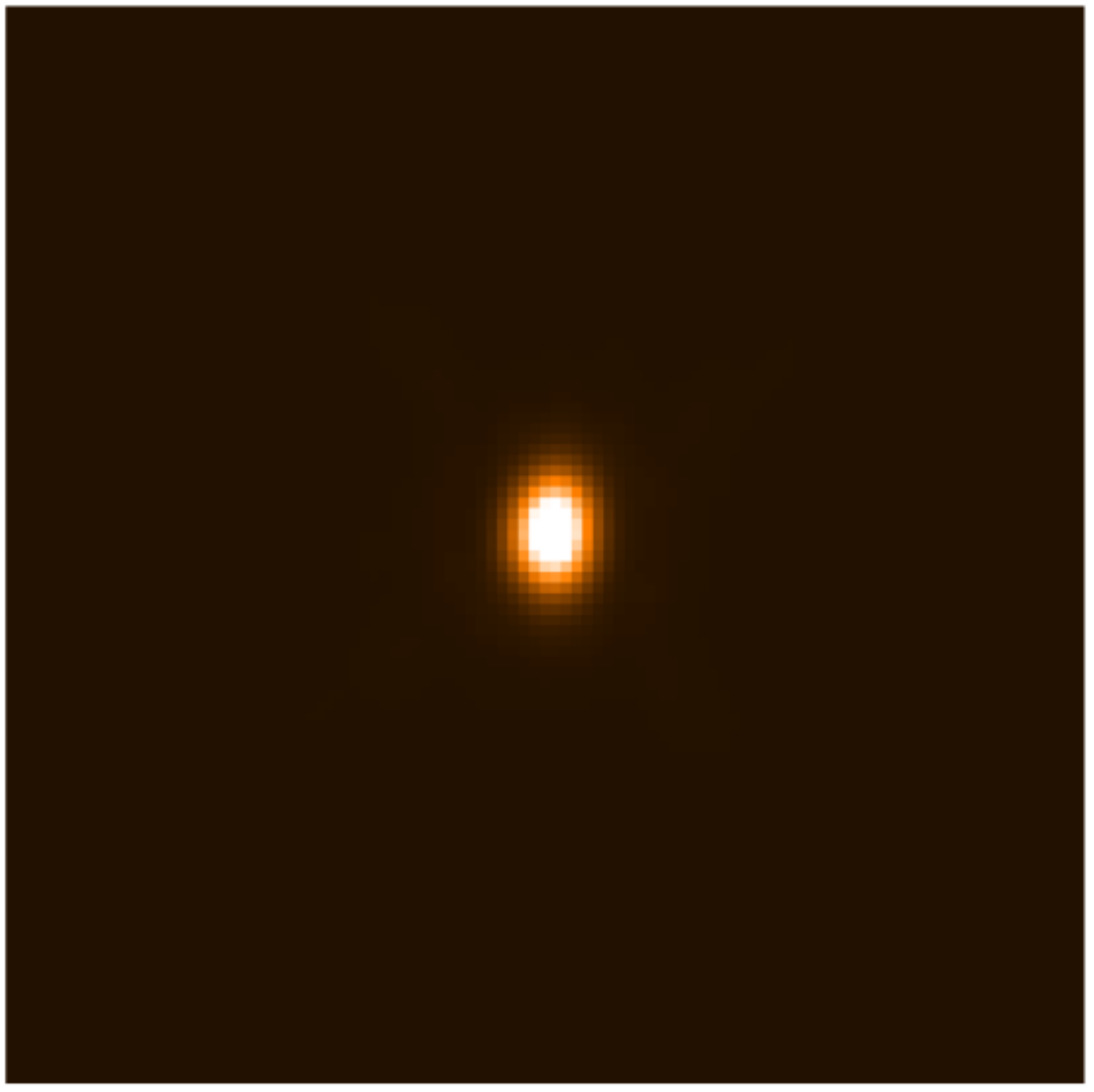}  &
\includegraphics[scale=0.15]{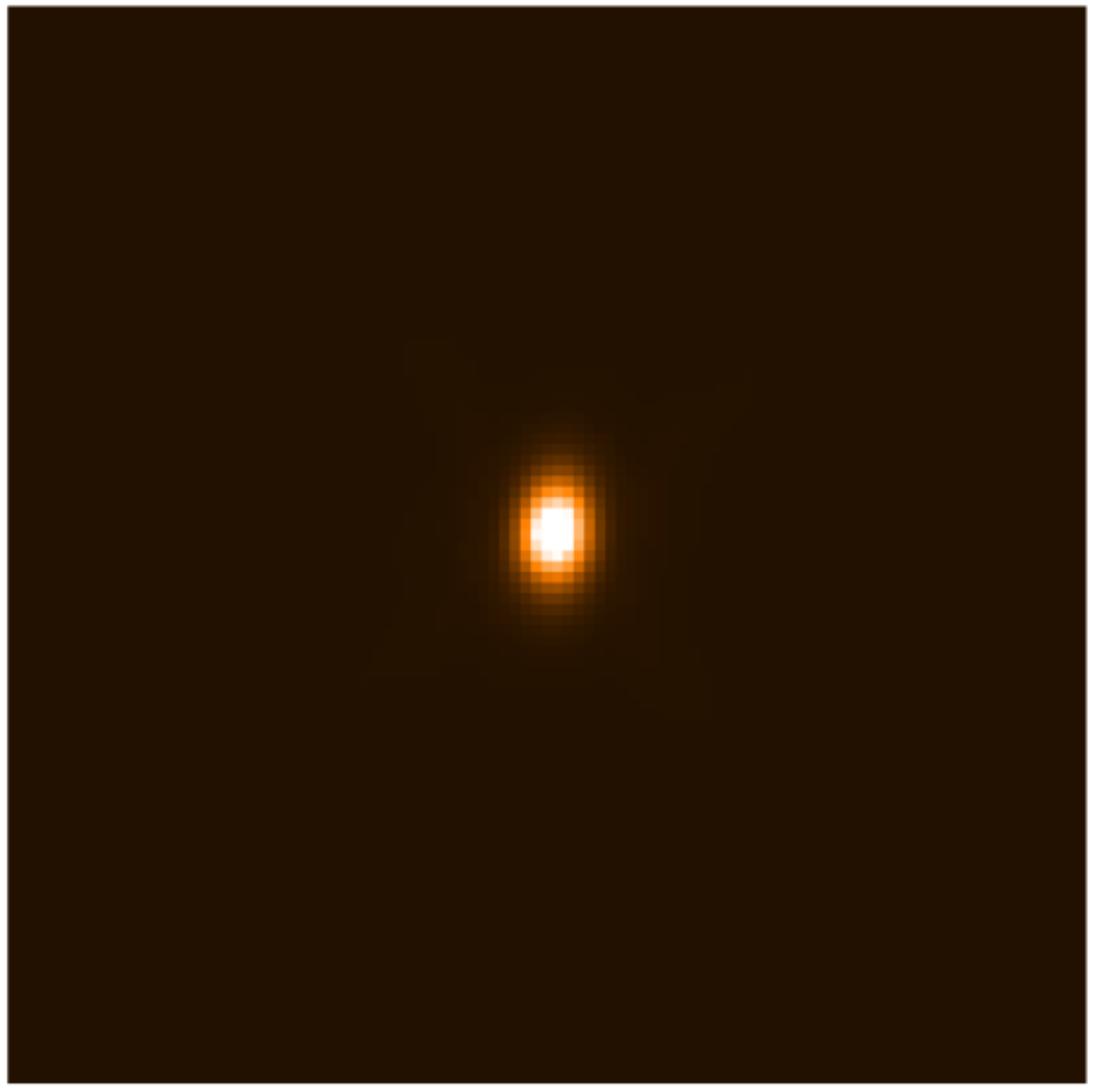}  &
\includegraphics[scale=0.15]{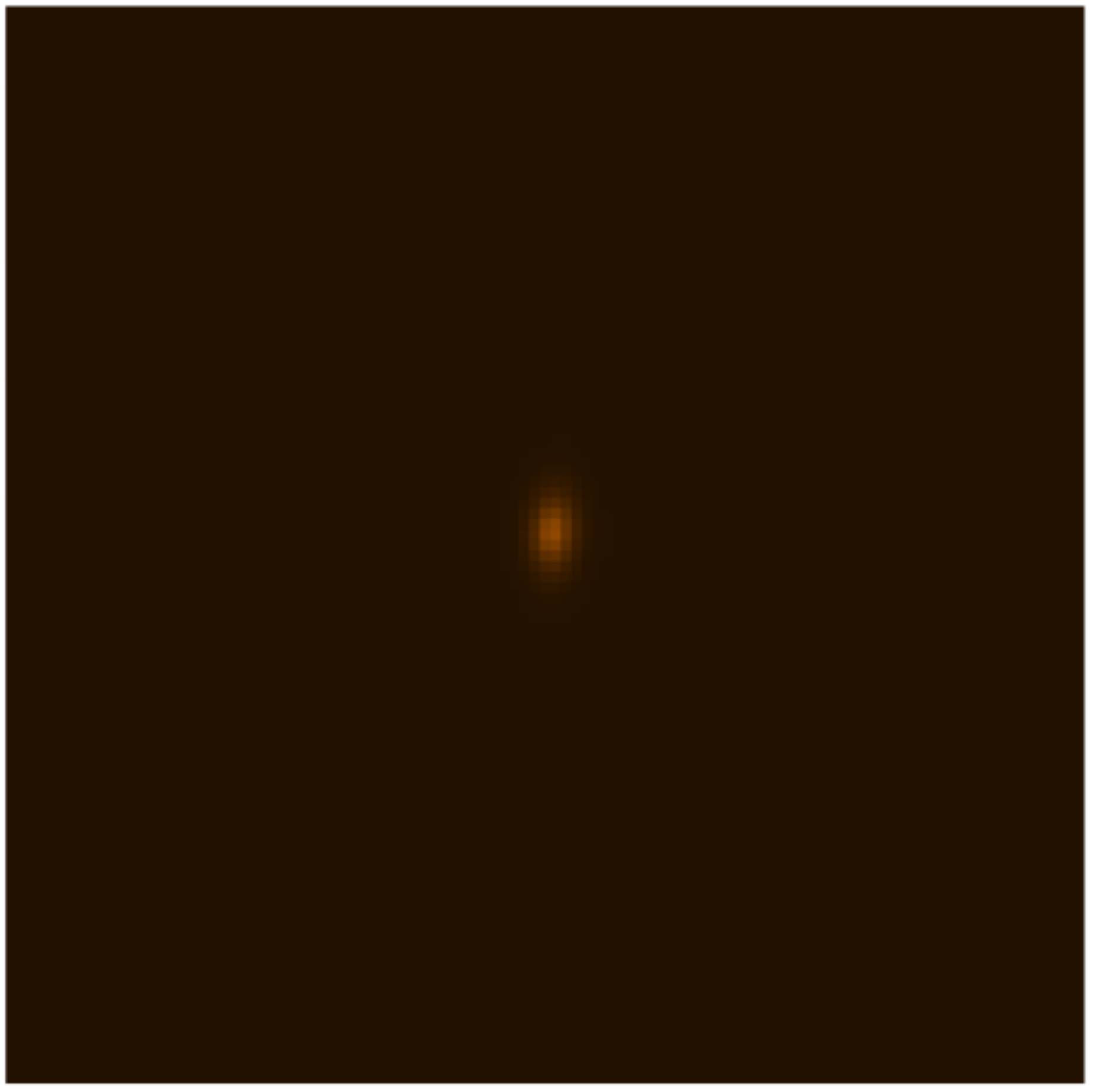}  &
\includegraphics[scale=0.15]{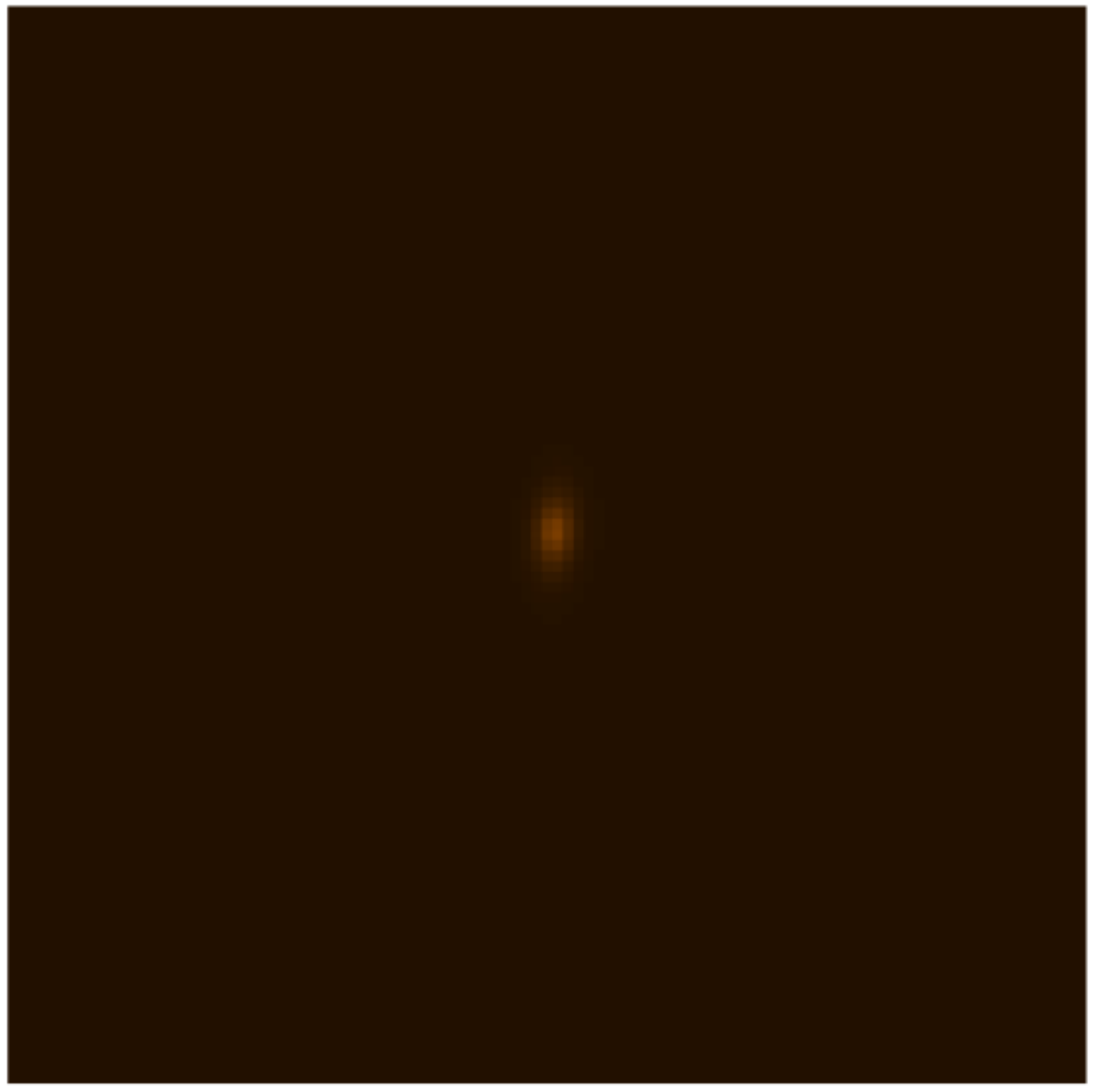}  \\
\includegraphics[scale=0.15]{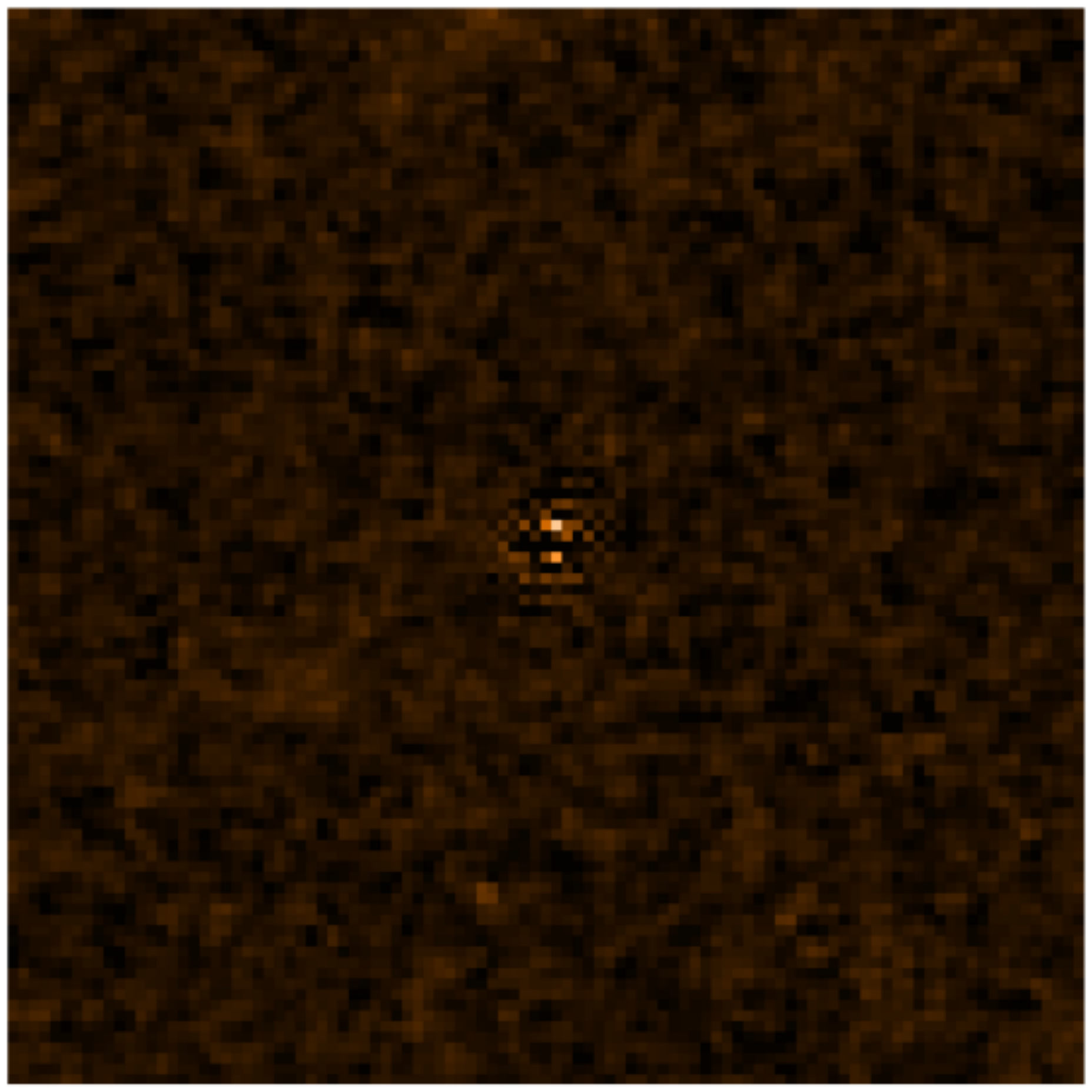}  &
\includegraphics[scale=0.15]{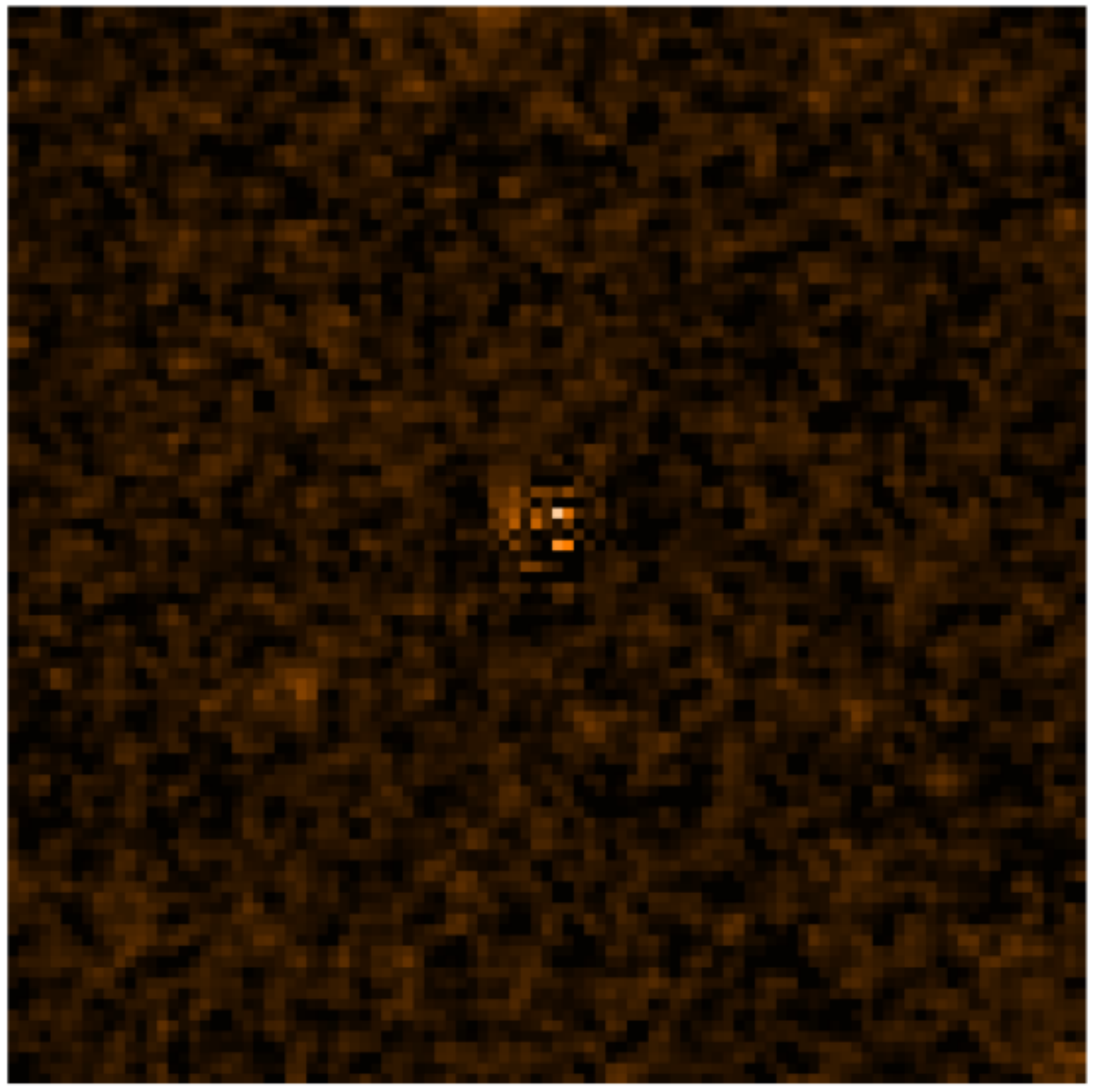}  &
\includegraphics[scale=0.15]{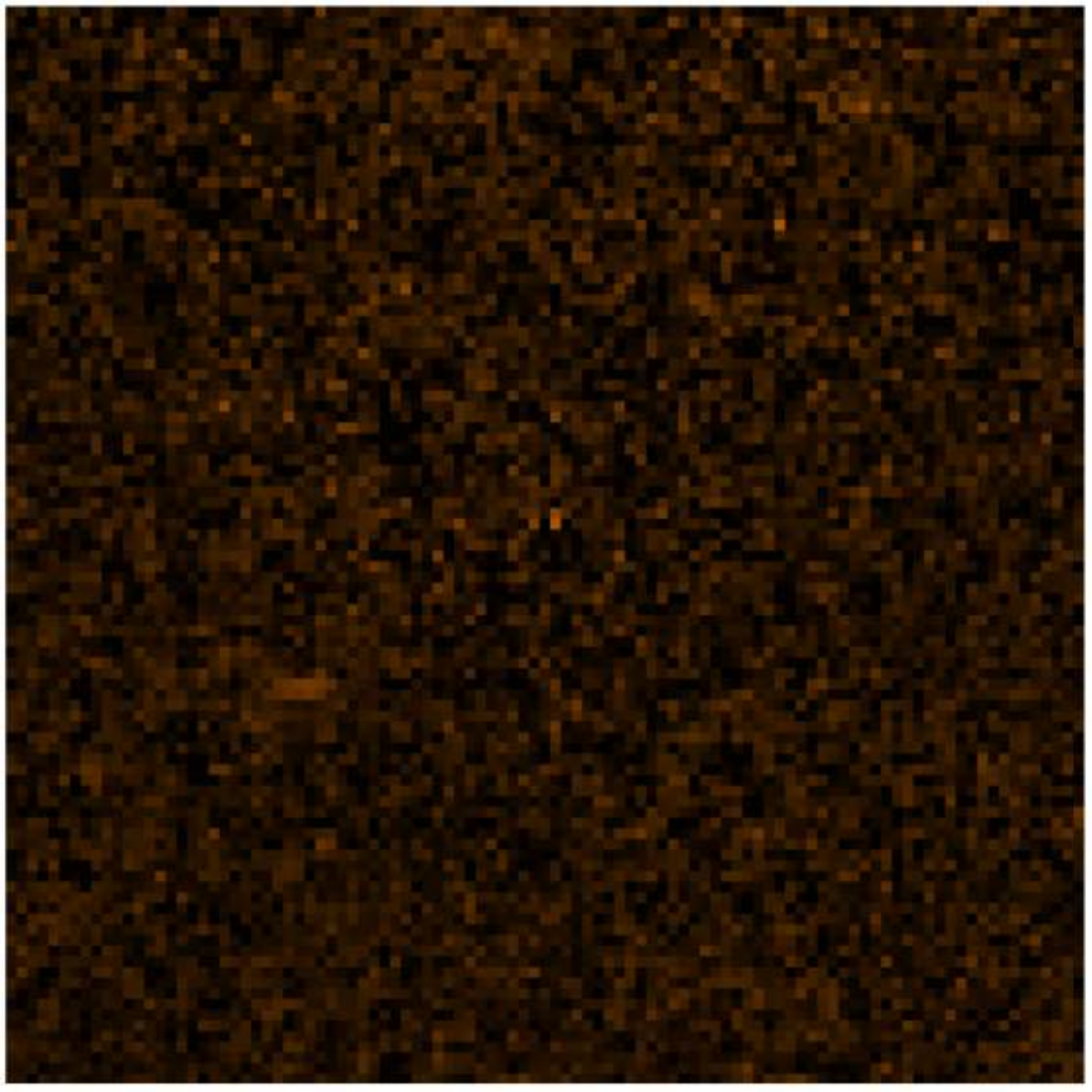}  &
\includegraphics[scale=0.15]{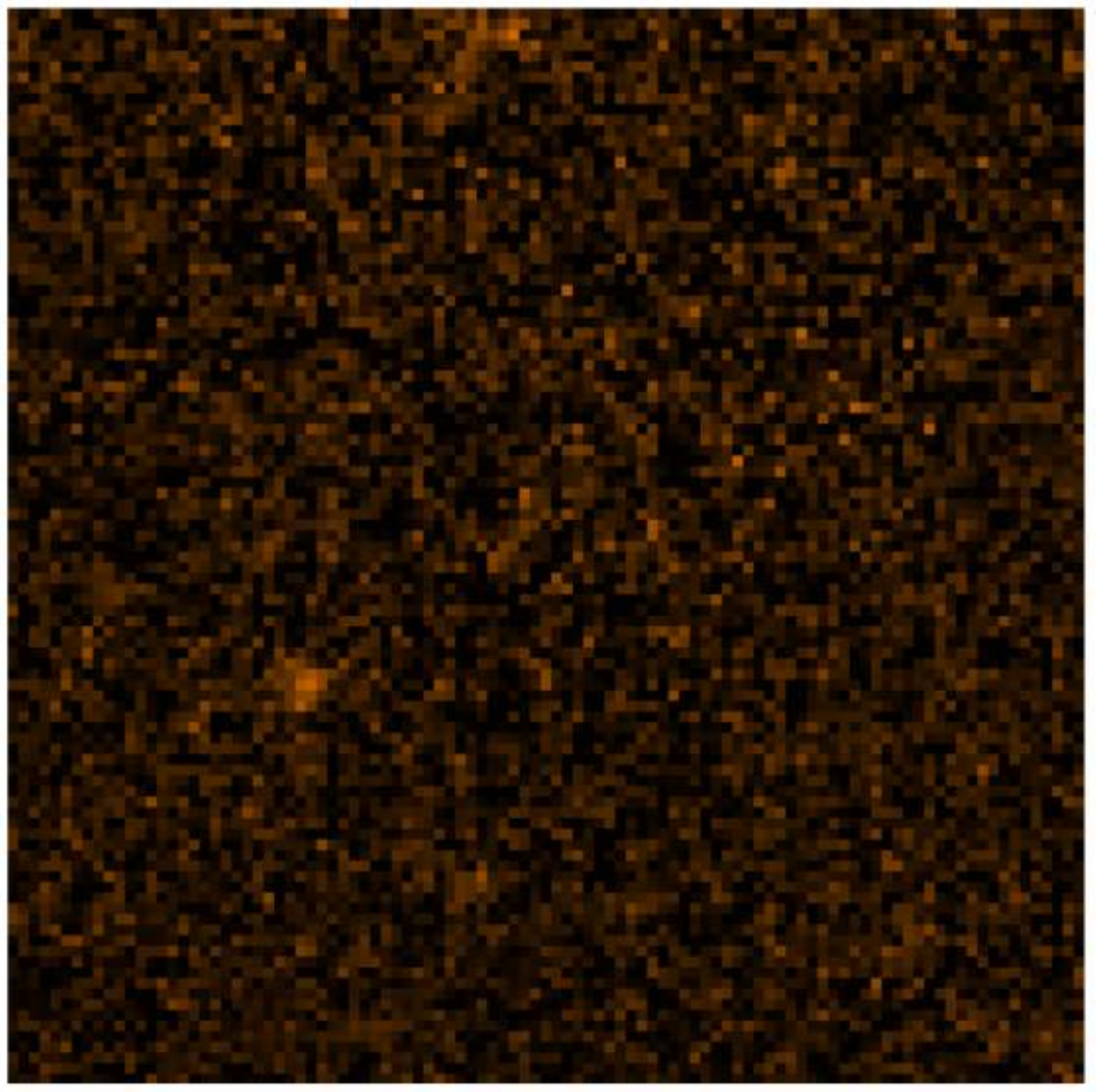}  \\
\end{tabular}
\caption[Image stamps of an example fit for a bulge+disk object with $B/T>0.5$.]{Image stamps of an example fit for a bulge+disk object with $D/T<0.5$. The configuration of these stamps is as follows. Left to right: $H_{160}$, $J_{125}$, $i_{814}$ and $v_{606}$. Top to bottom: images, best-fit combined bulge+disk models, individual best-fit bulge components, individual best-fit disk components and combined model residuals. In comparison to Fig.\,\ref{fig:bulge_less_50},  for this object the disk is the dominant component, although it can also be seen that the bulge component remains prominent even in the $v_{606}$-band.}
\label{fig:disk_less_50}
\end{center}
\end{figure*}

\begin{figure*}
\begin{center}
\begin{tabular}{m{3cm}m{3cm}m{3cm}m{3cm}}
\includegraphics[scale=0.15 ]{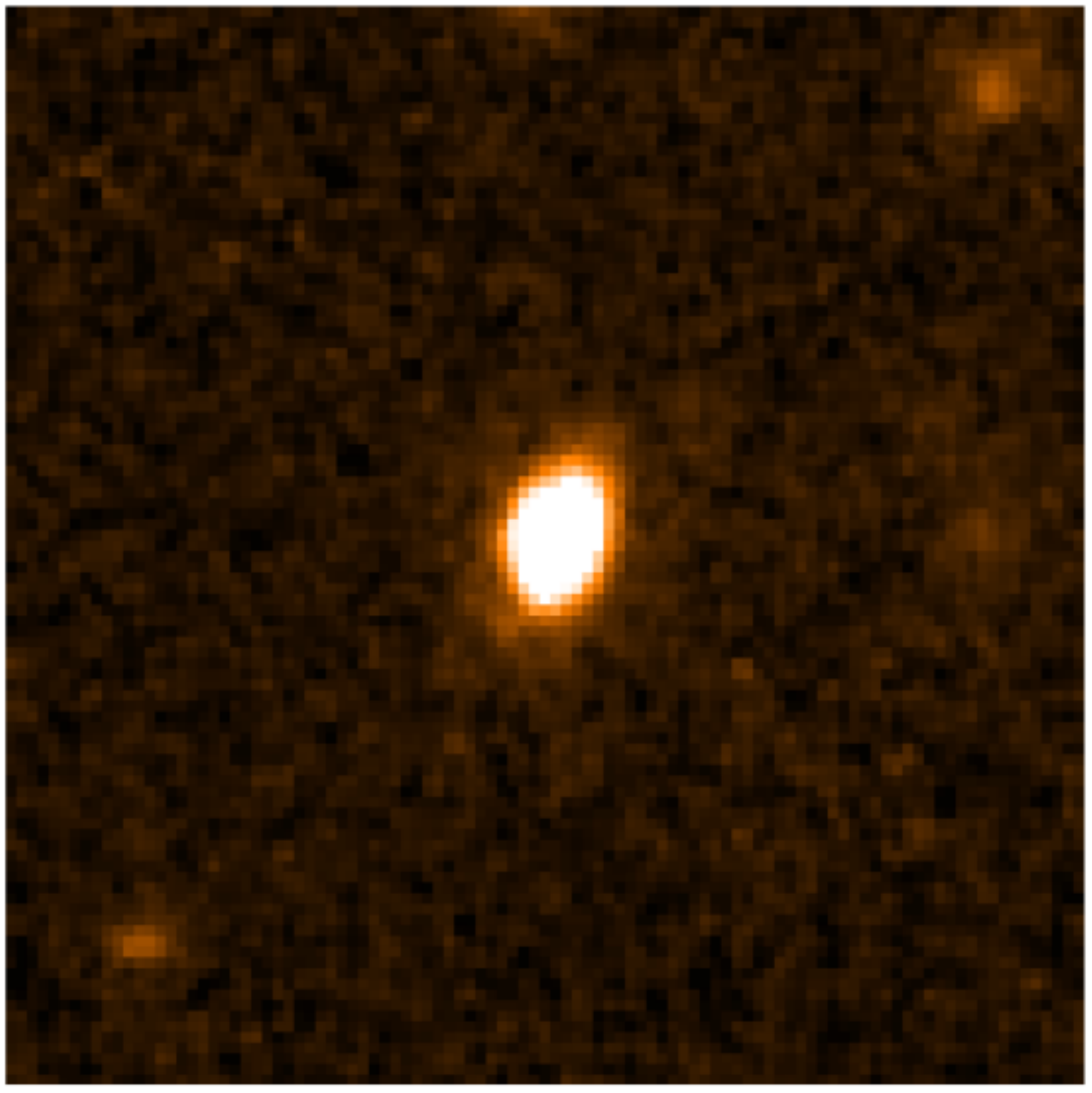}  &
\includegraphics[scale=0.15]{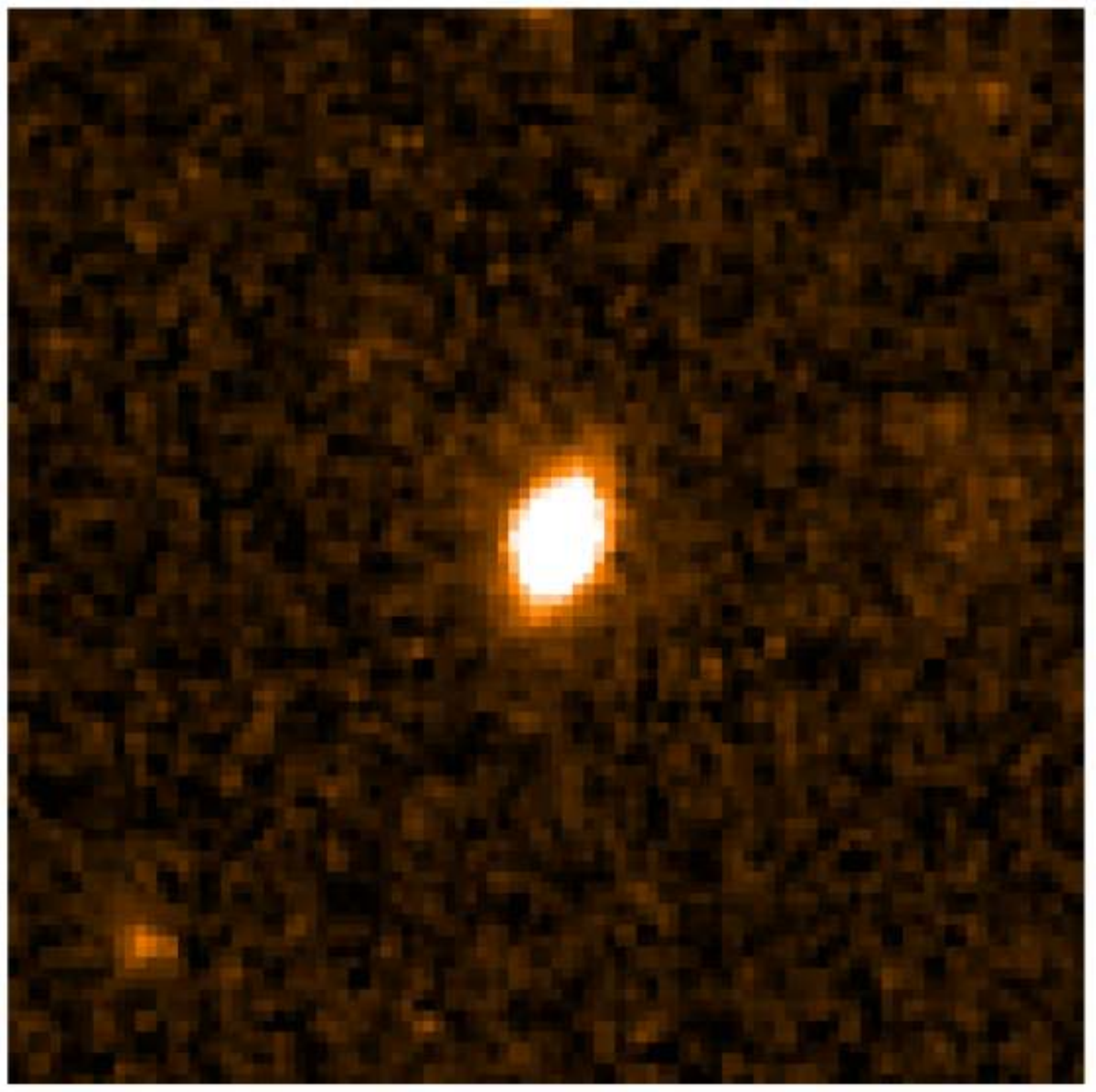}  &
\includegraphics[scale=0.15]{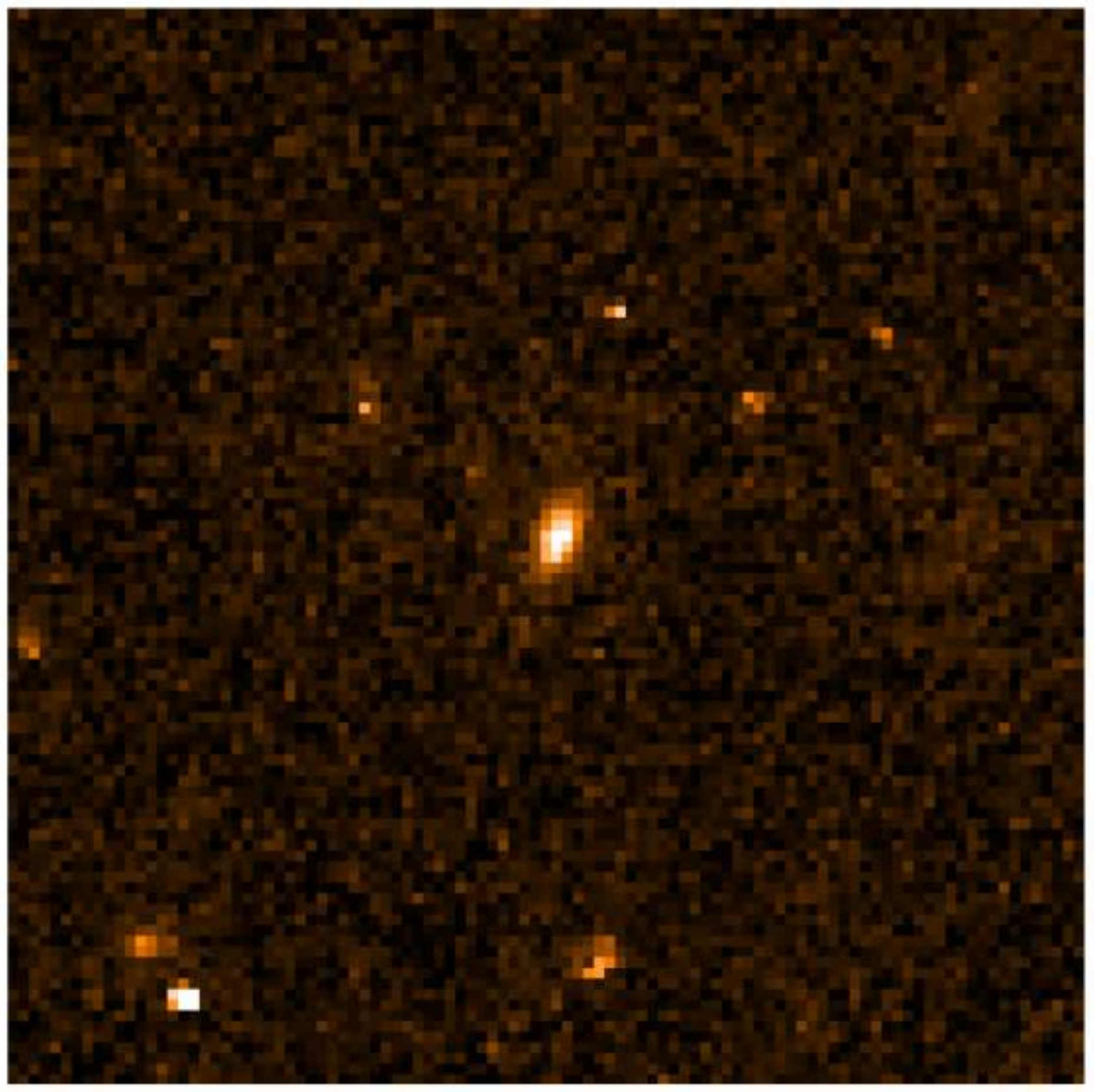}  &
\includegraphics[scale=0.15]{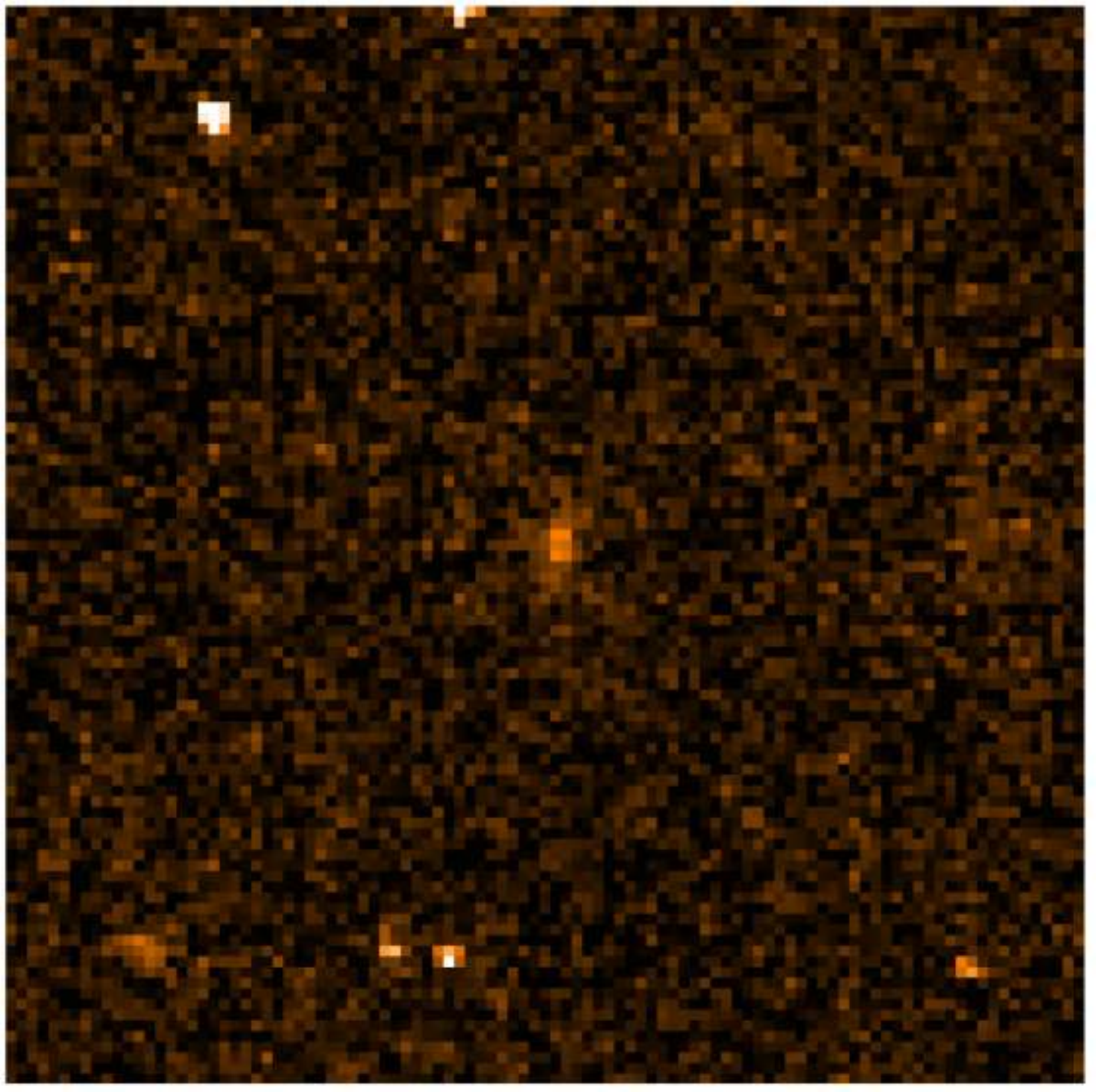}  \\
\includegraphics[scale=0.15]{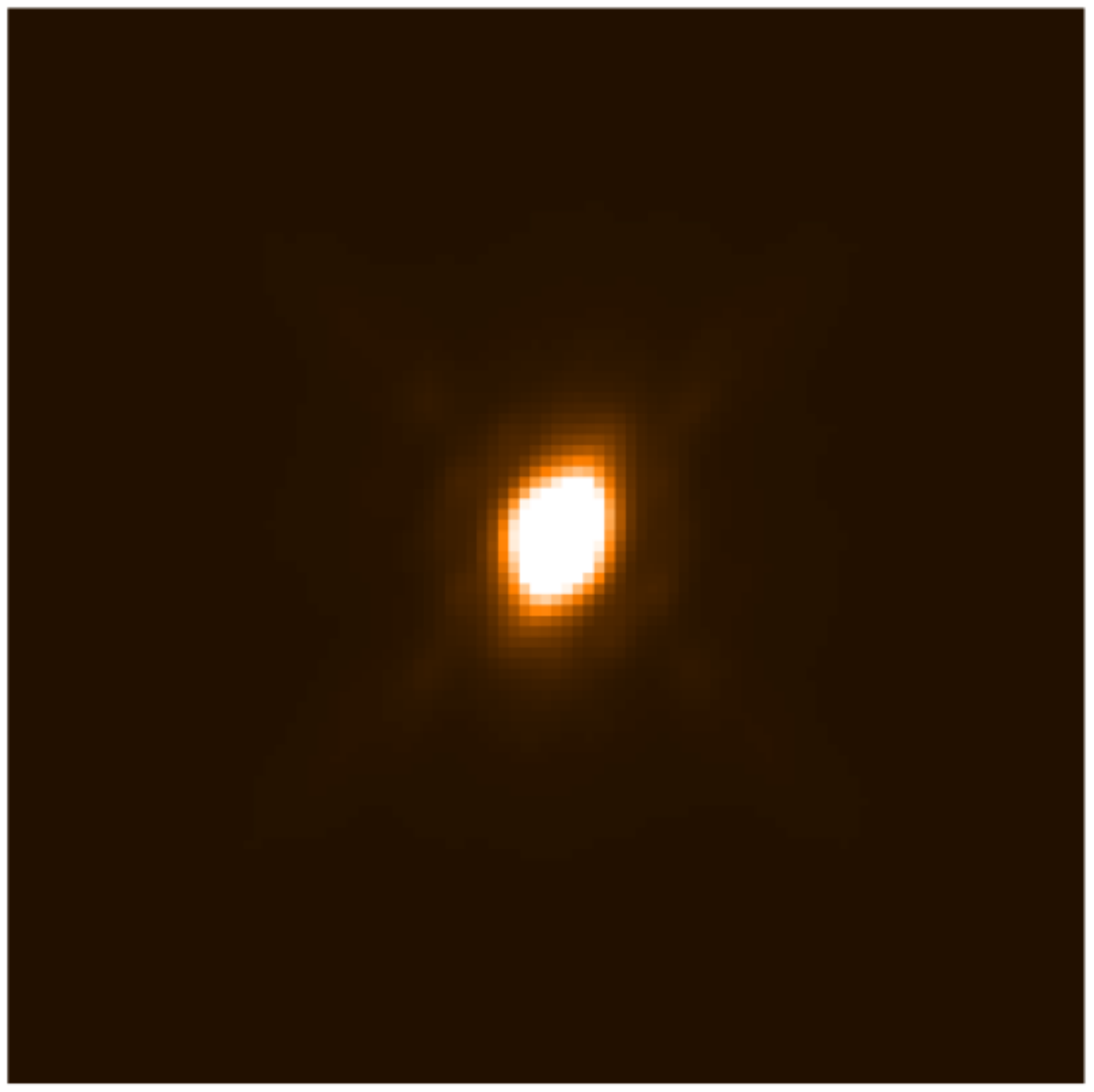}  &
\includegraphics[scale=0.15]{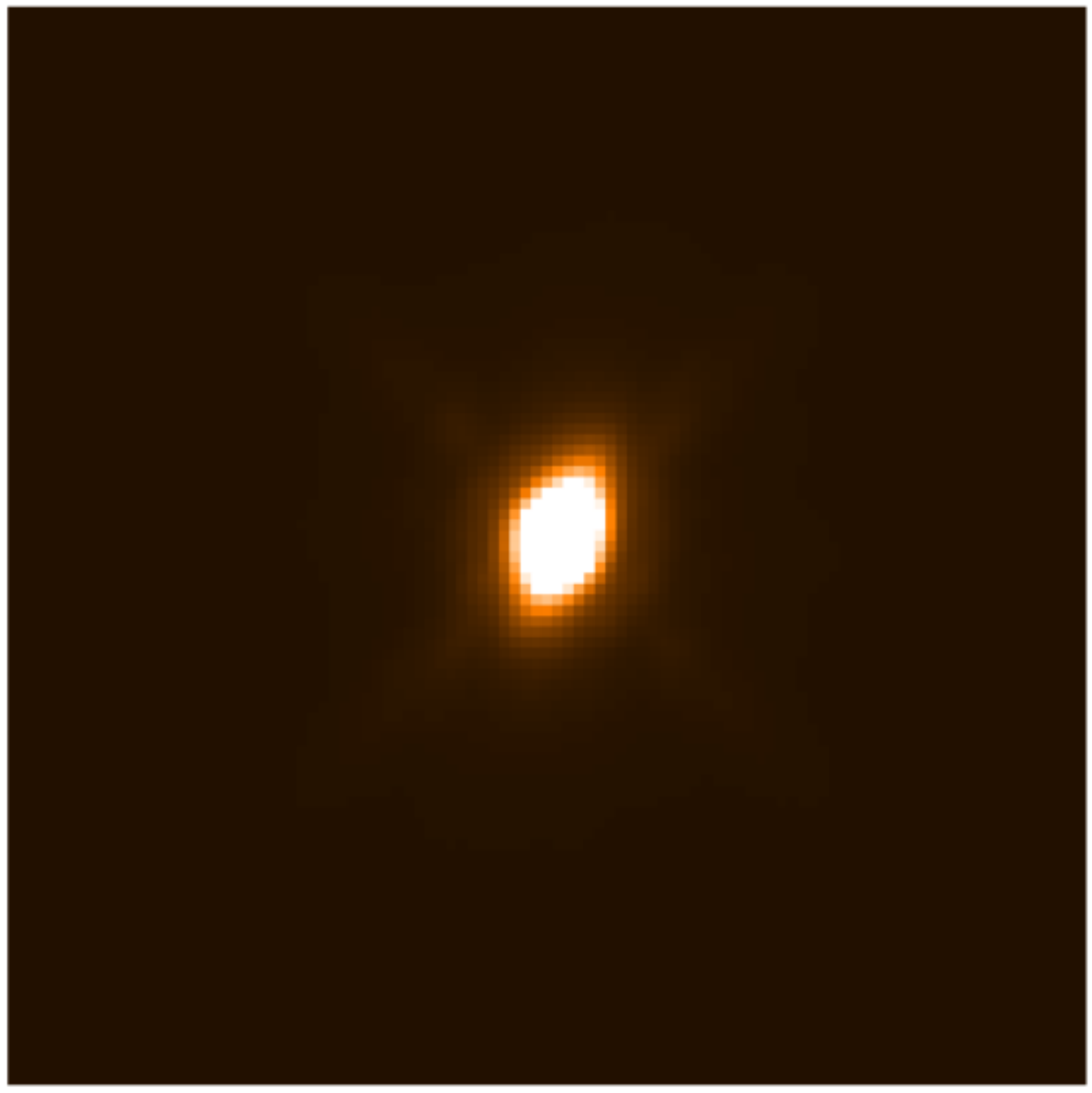}  &
\includegraphics[scale=0.15]{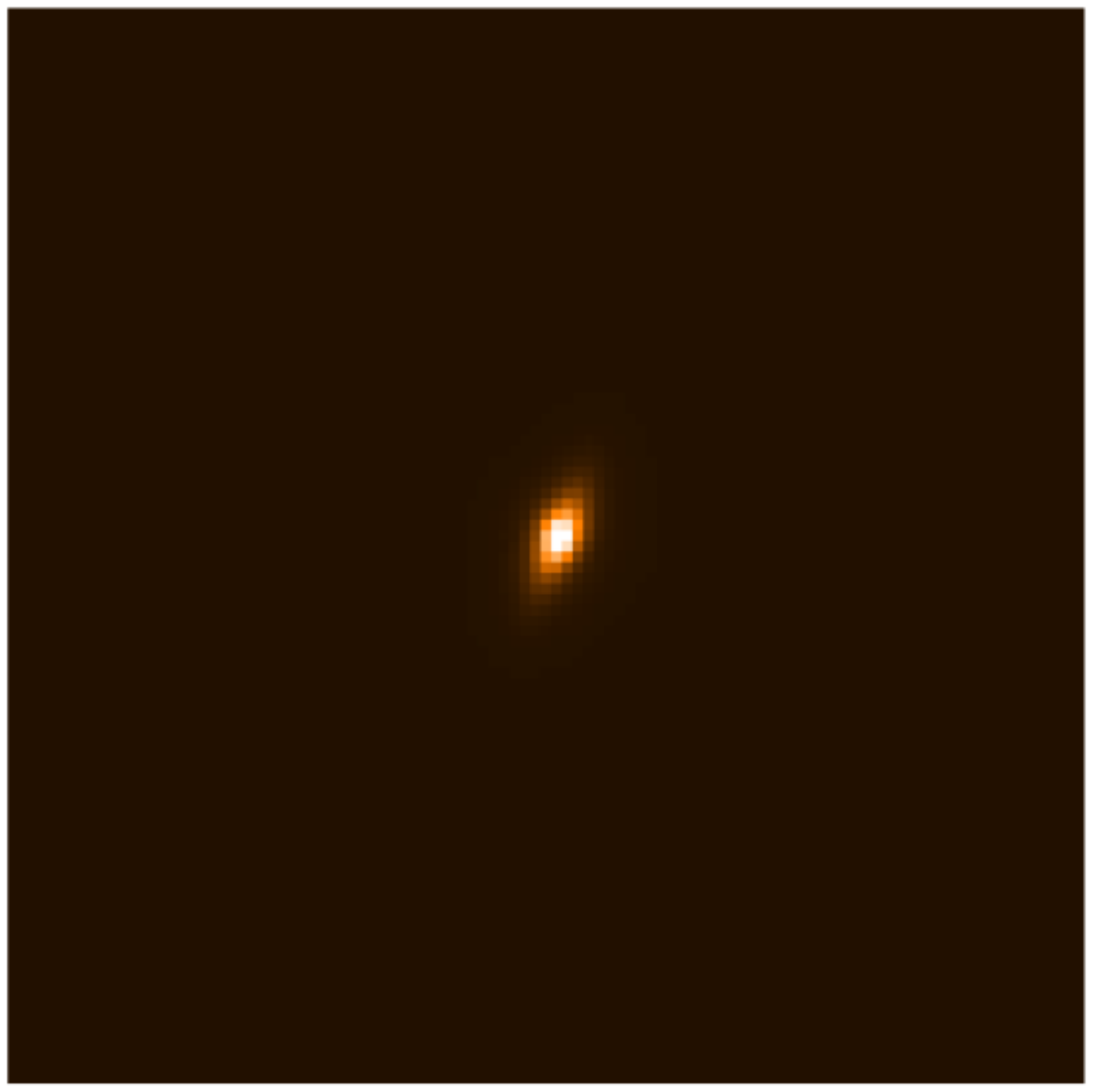}  &
\includegraphics[scale=0.15]{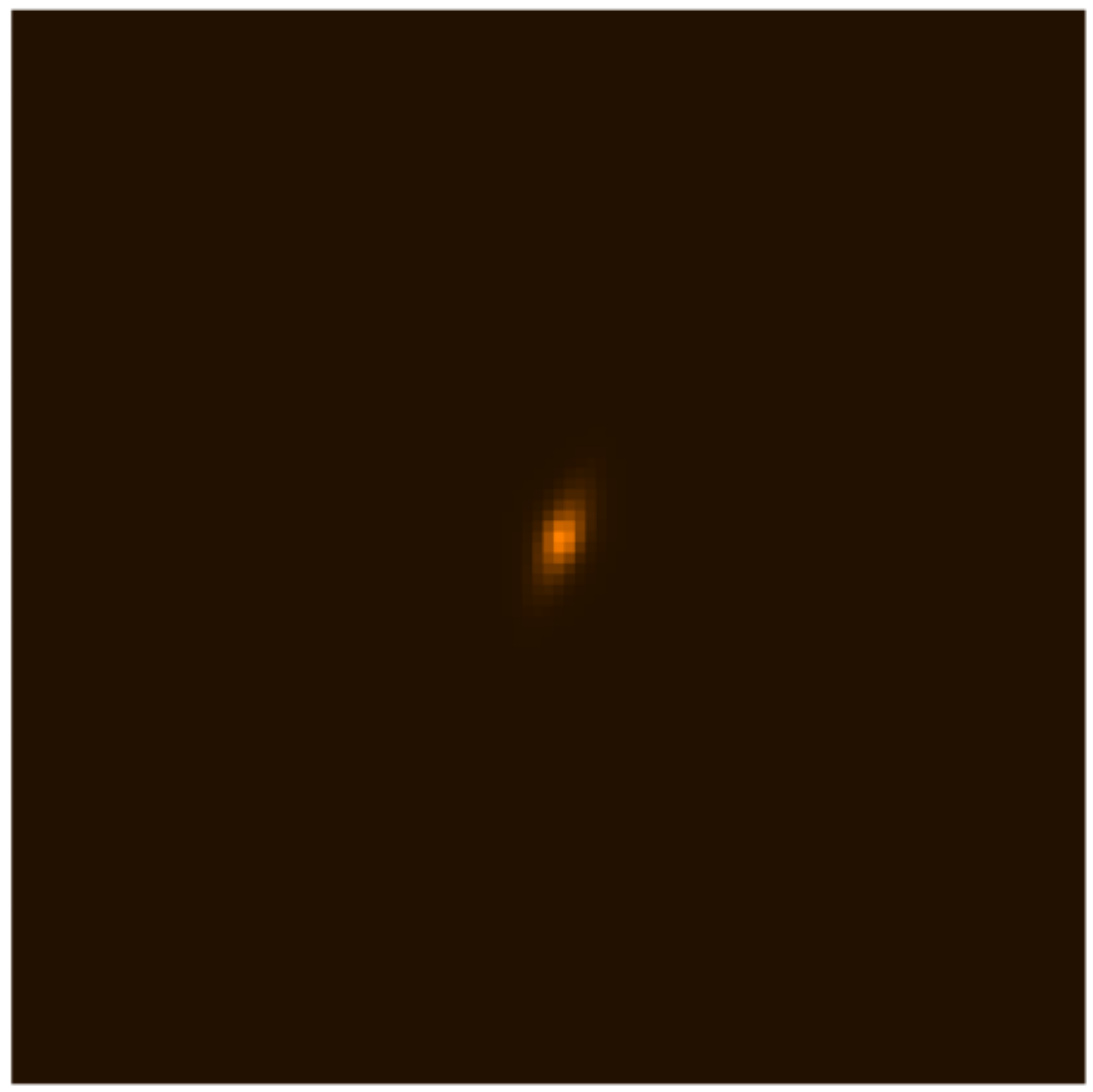}  \\
\includegraphics[scale=0.15]{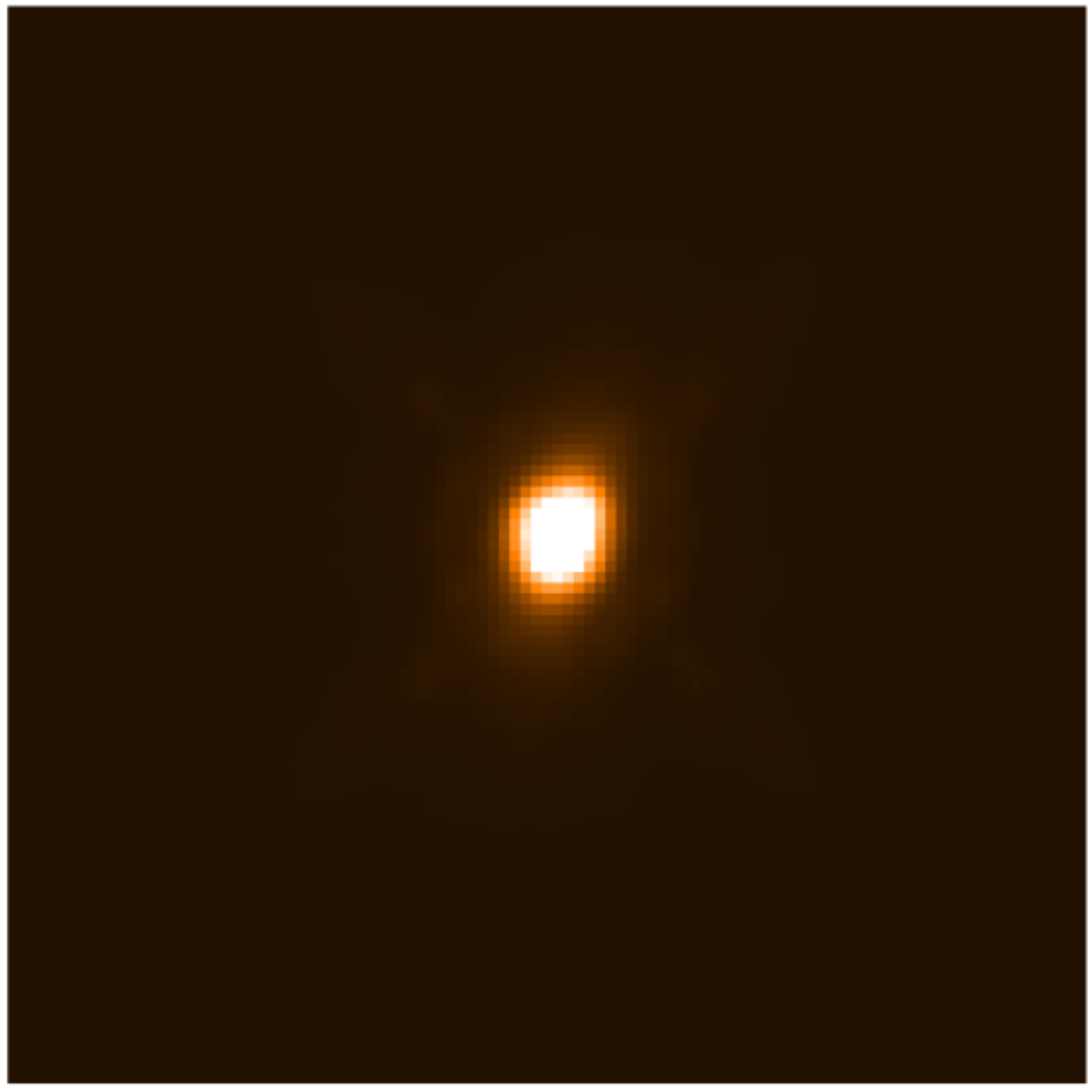}  &
\includegraphics[scale=0.15]{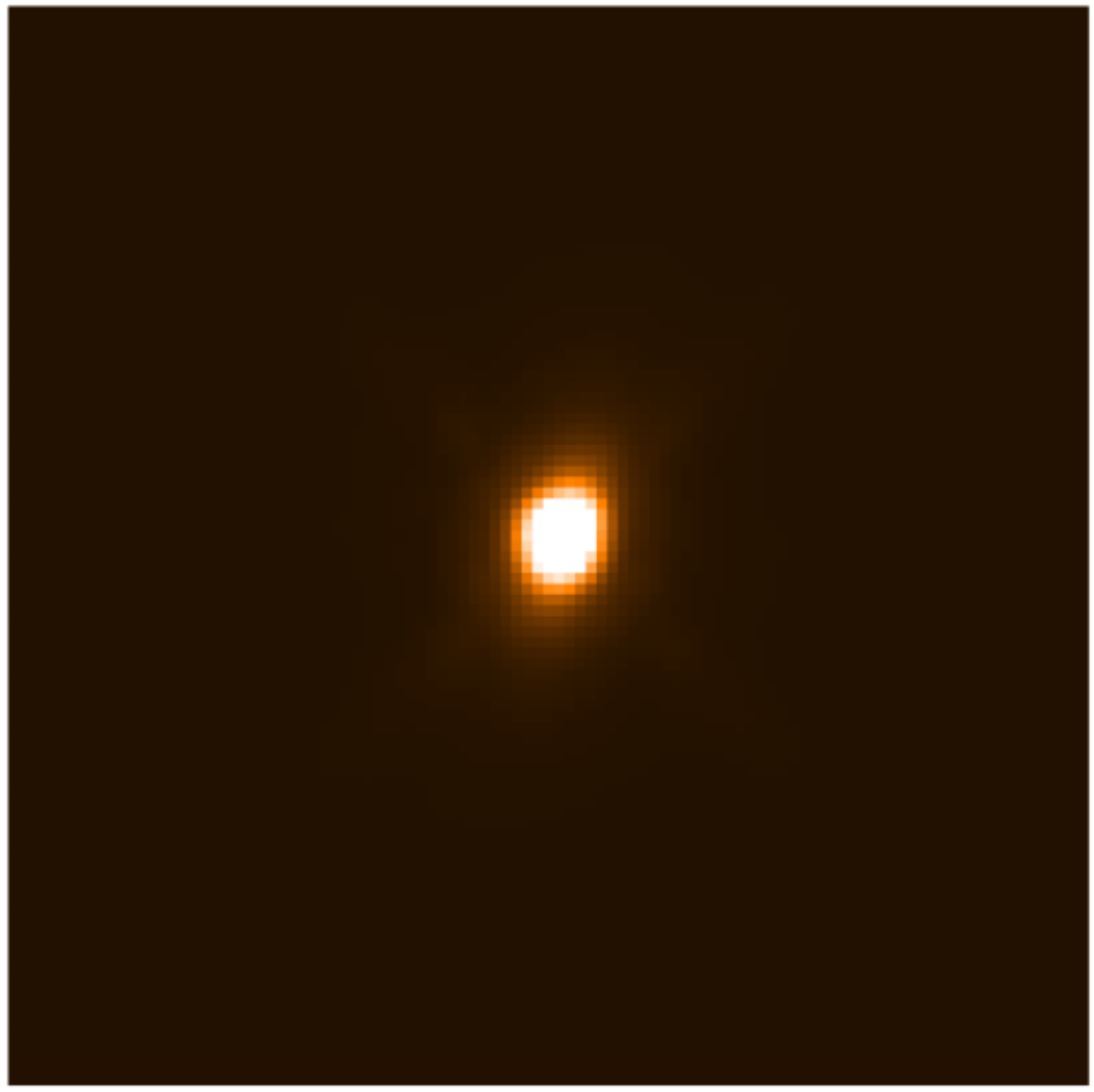}  &
\includegraphics[scale=0.15]{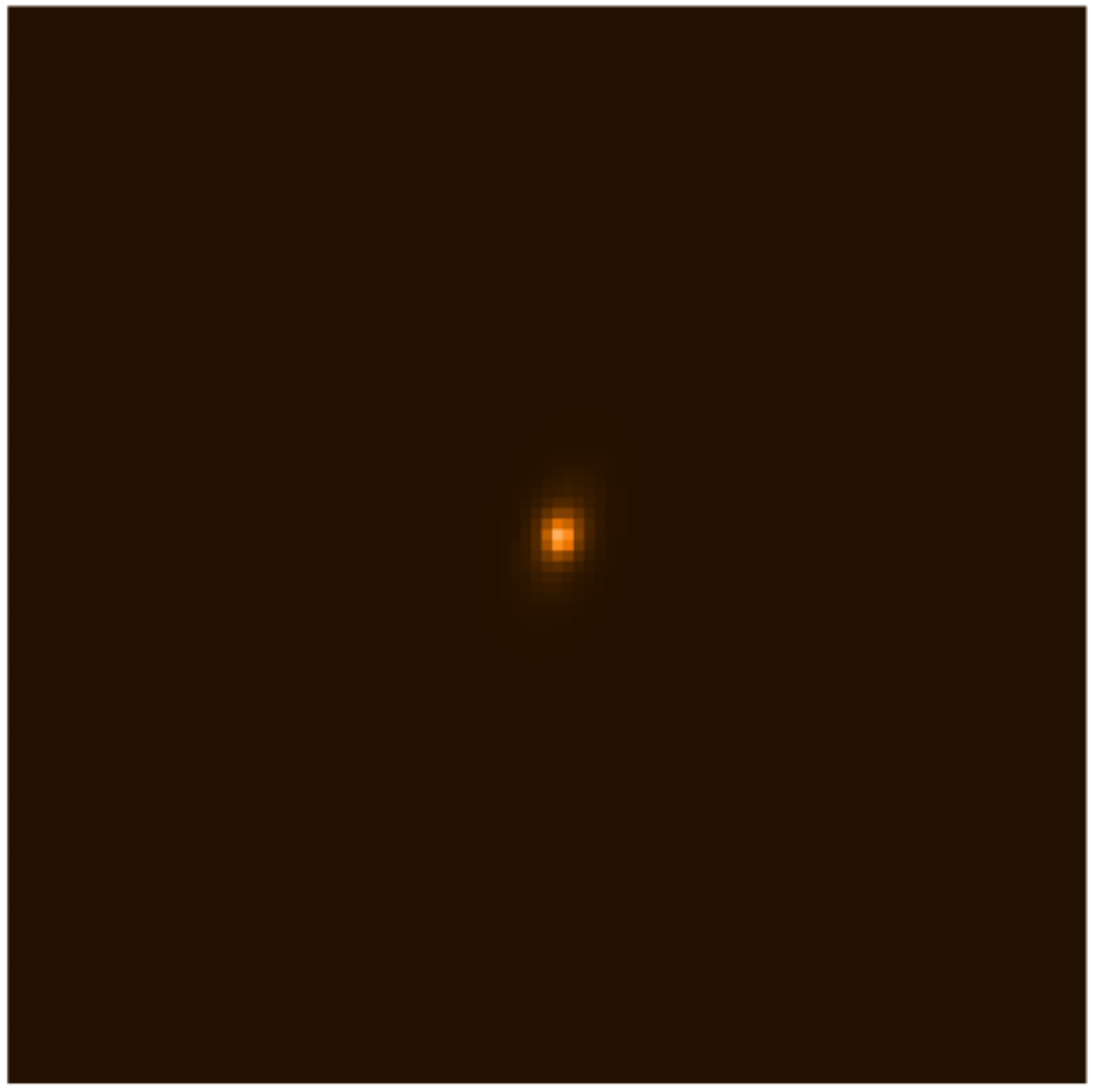}  &
\includegraphics[scale=0.15]{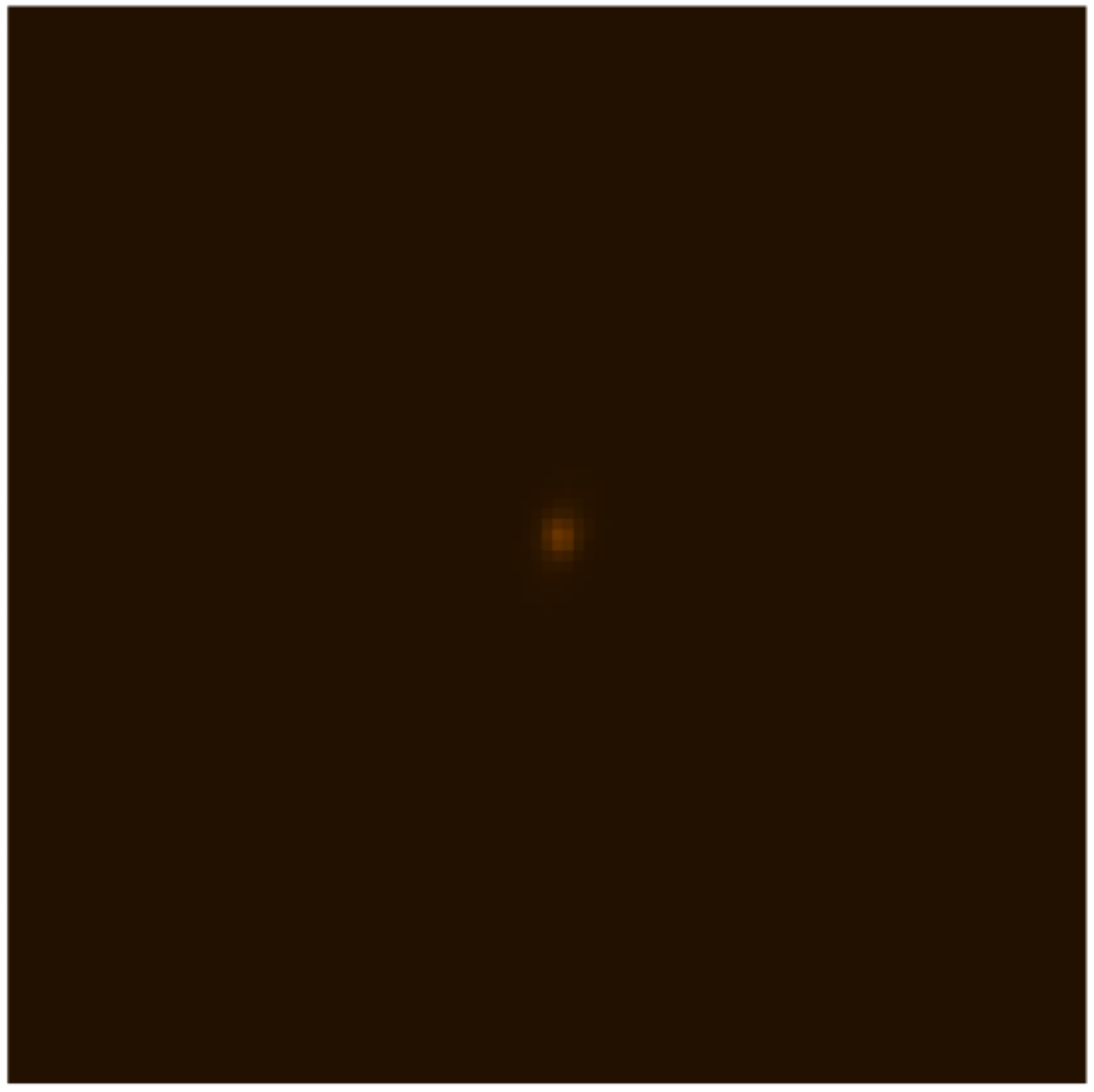}  \\
\includegraphics[scale=0.15]{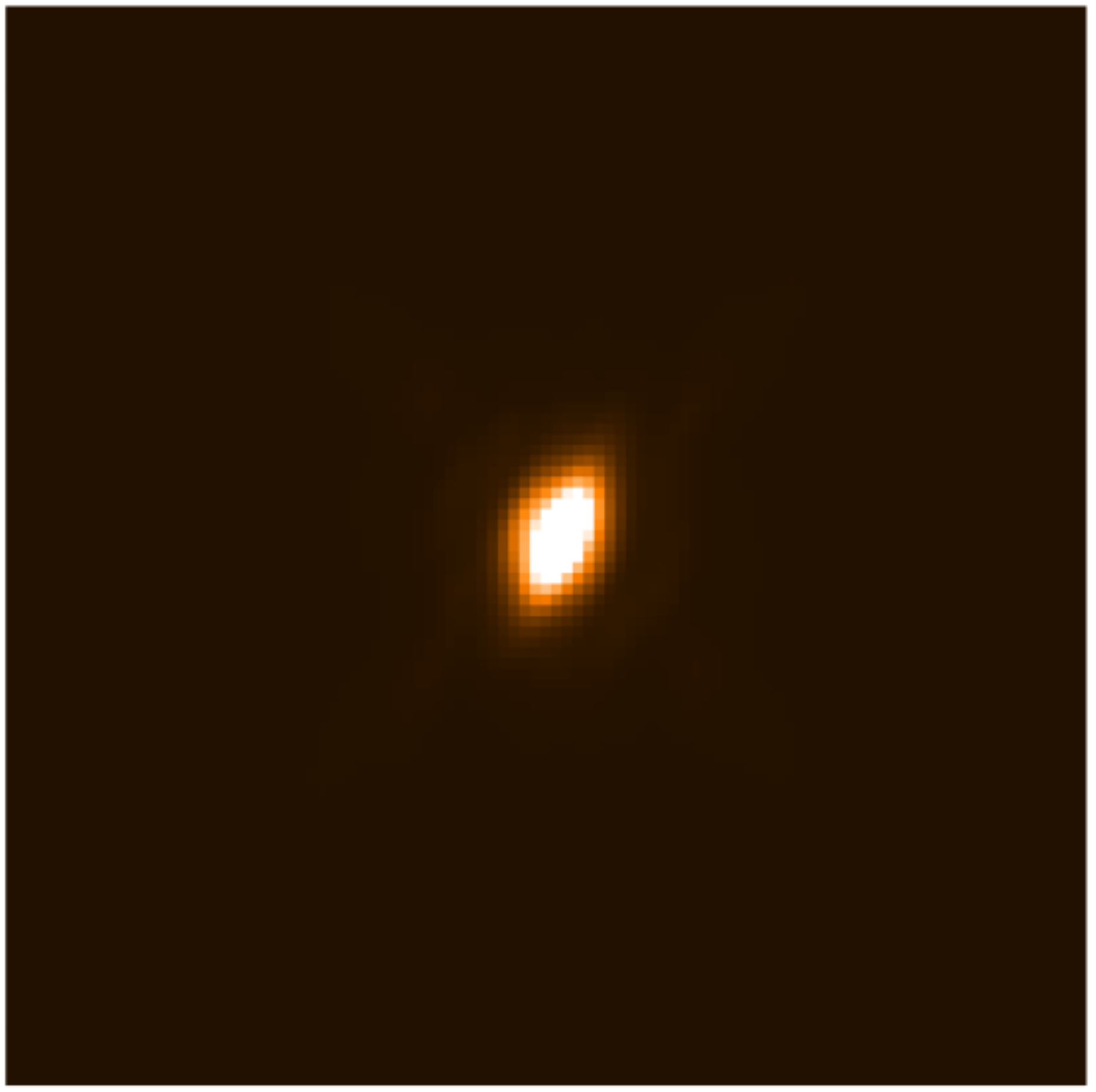}  &
\includegraphics[scale=0.15]{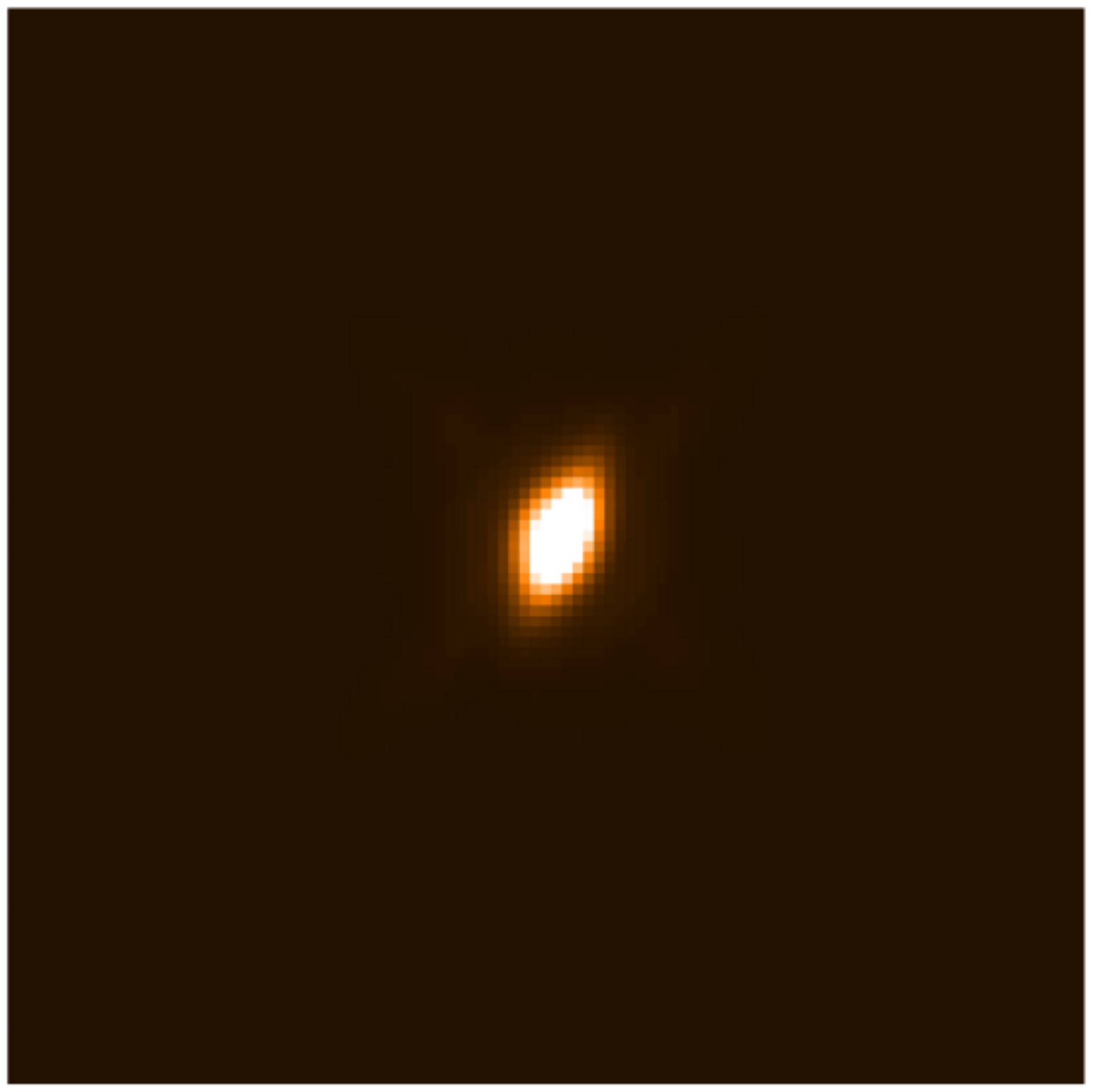}  &
\includegraphics[scale=0.15]{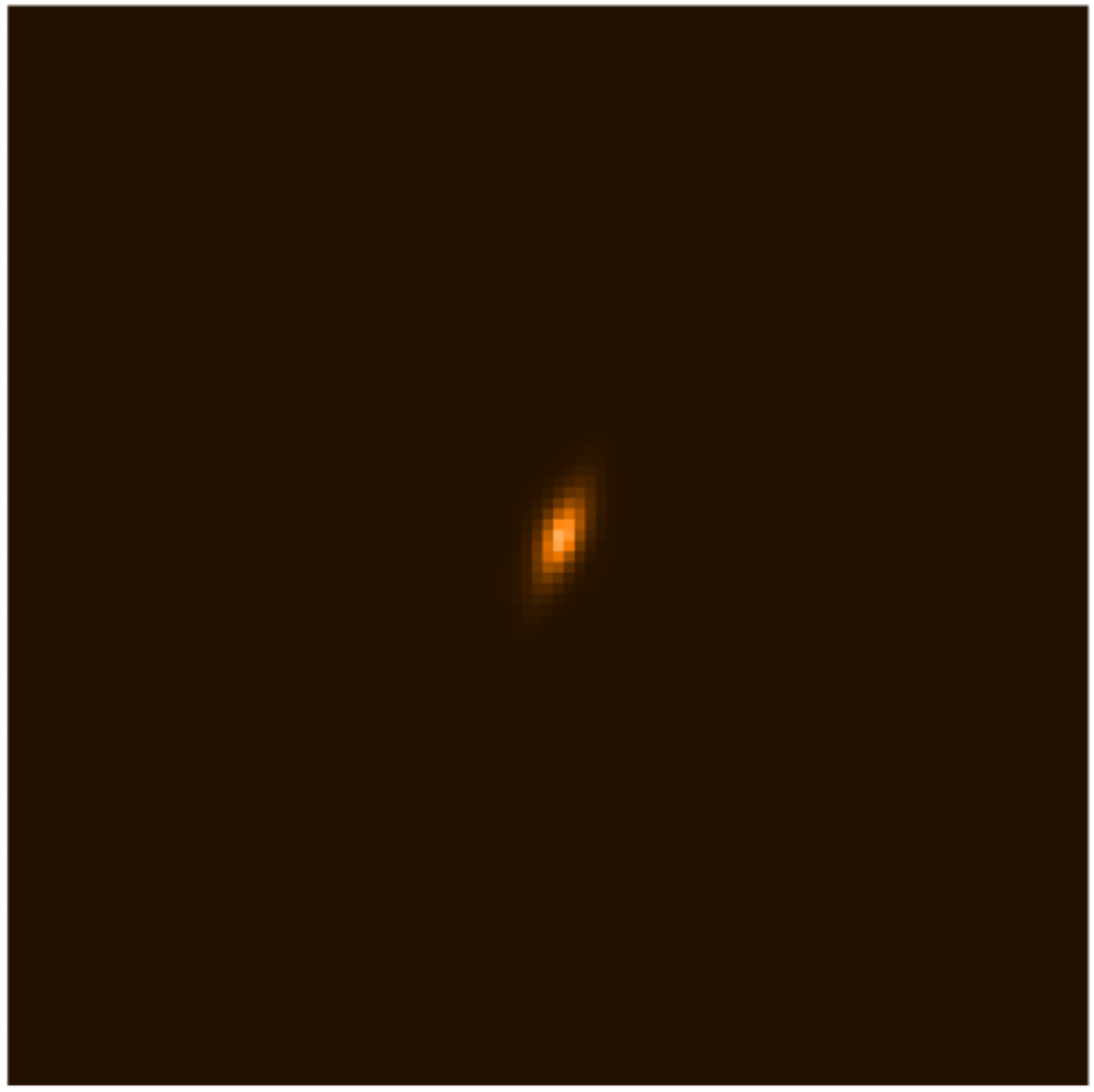}  &
\includegraphics[scale=0.15]{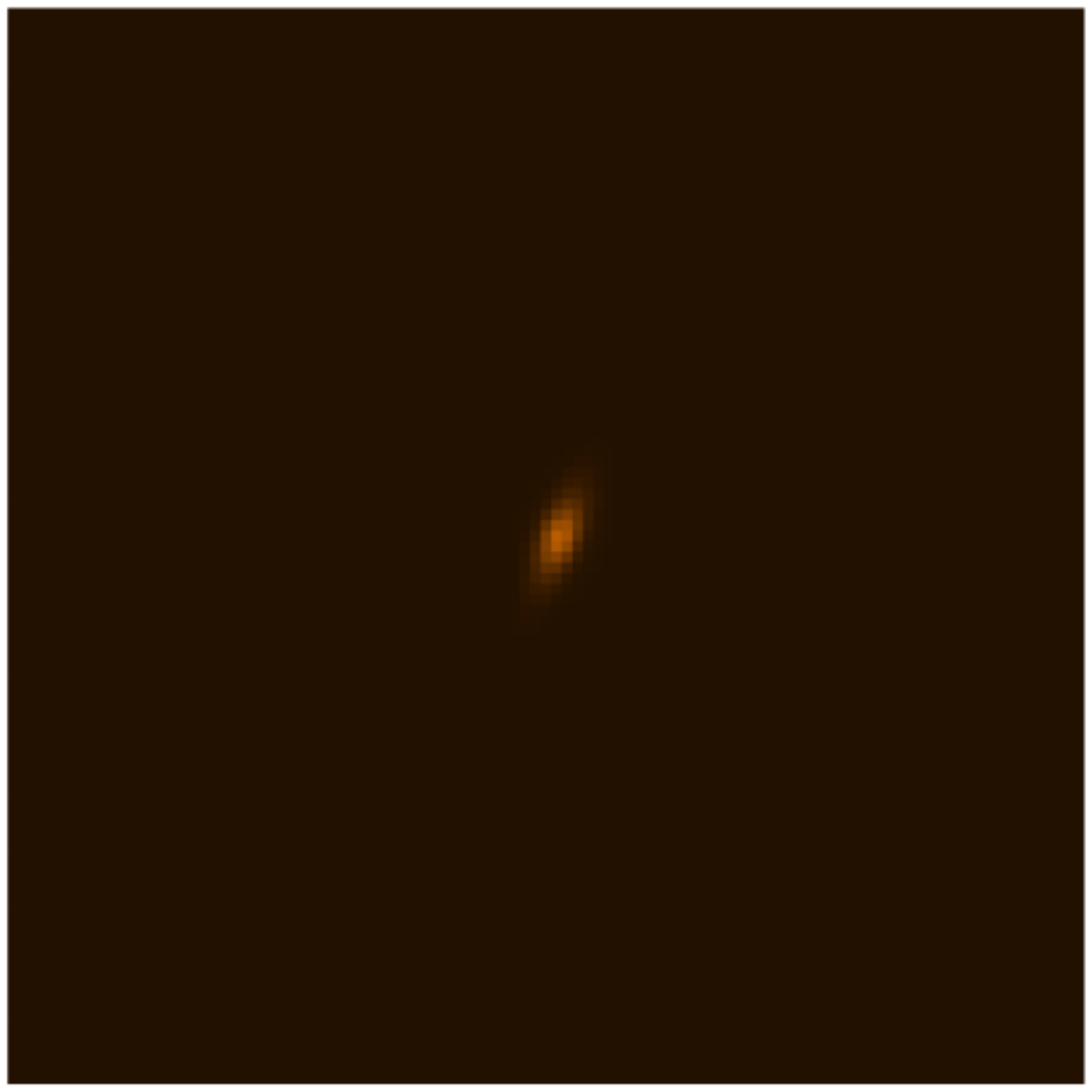}  \\
\includegraphics[scale=0.15]{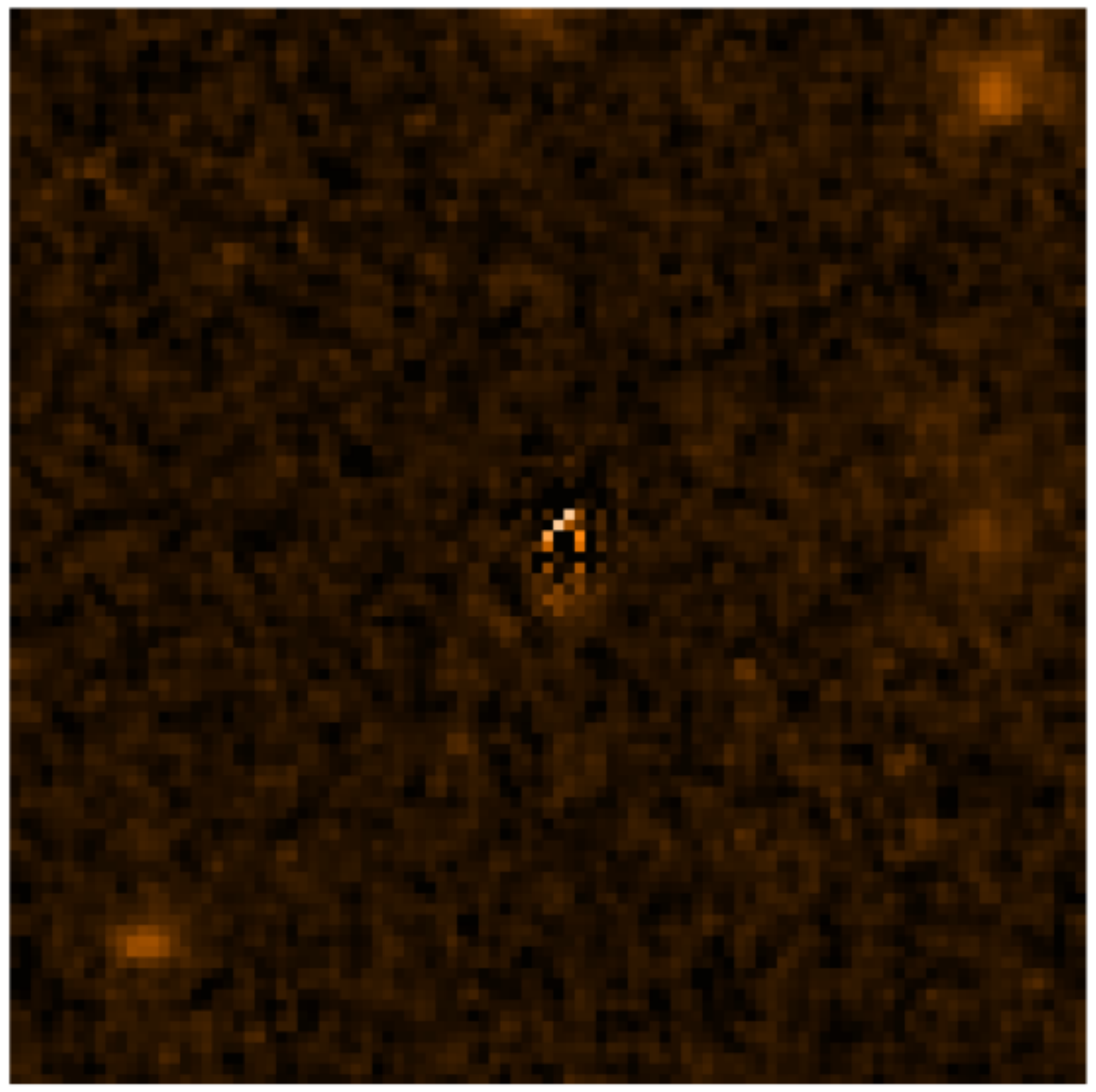}  &
\includegraphics[scale=0.15]{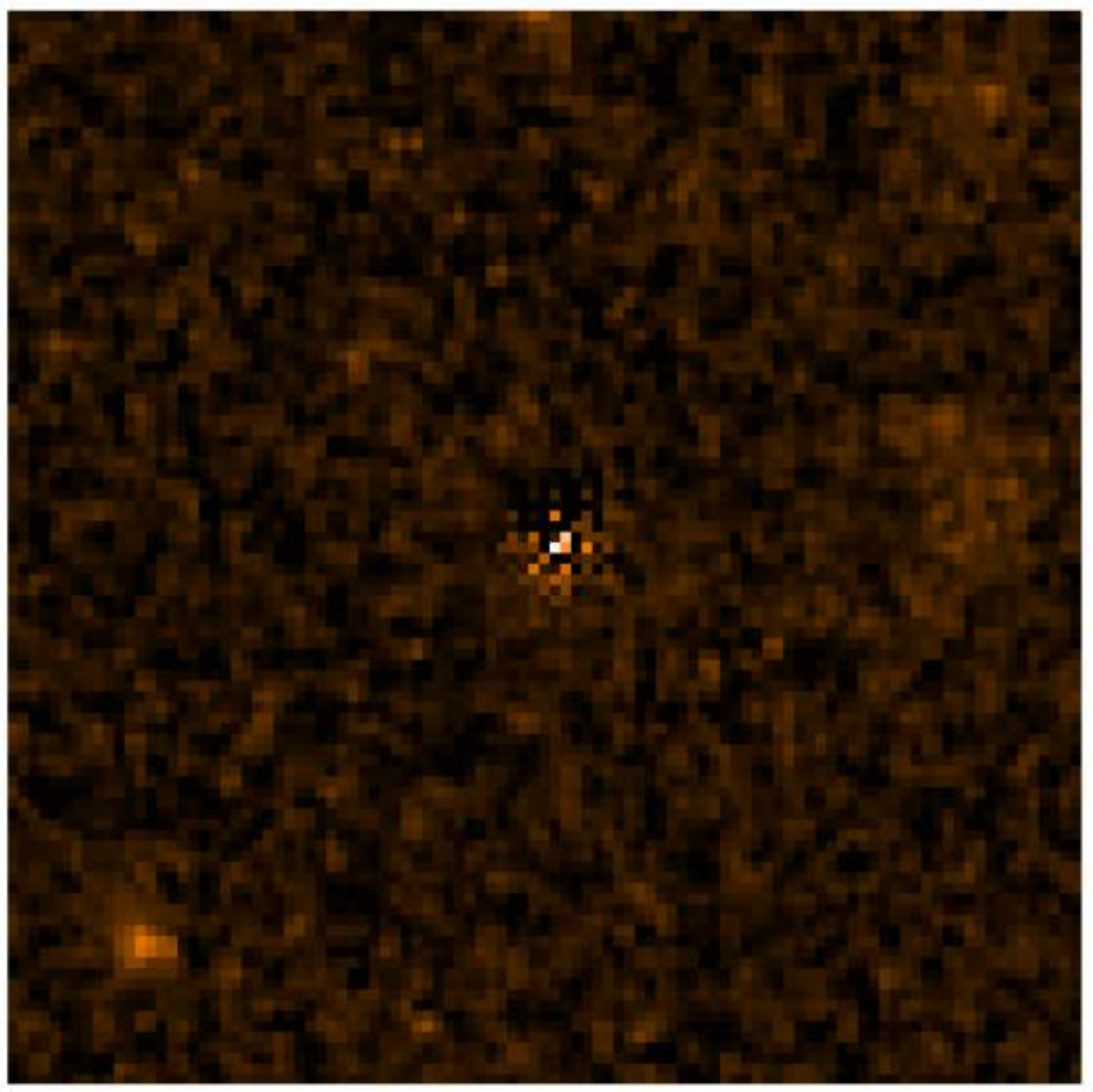}  &
\includegraphics[scale=0.15]{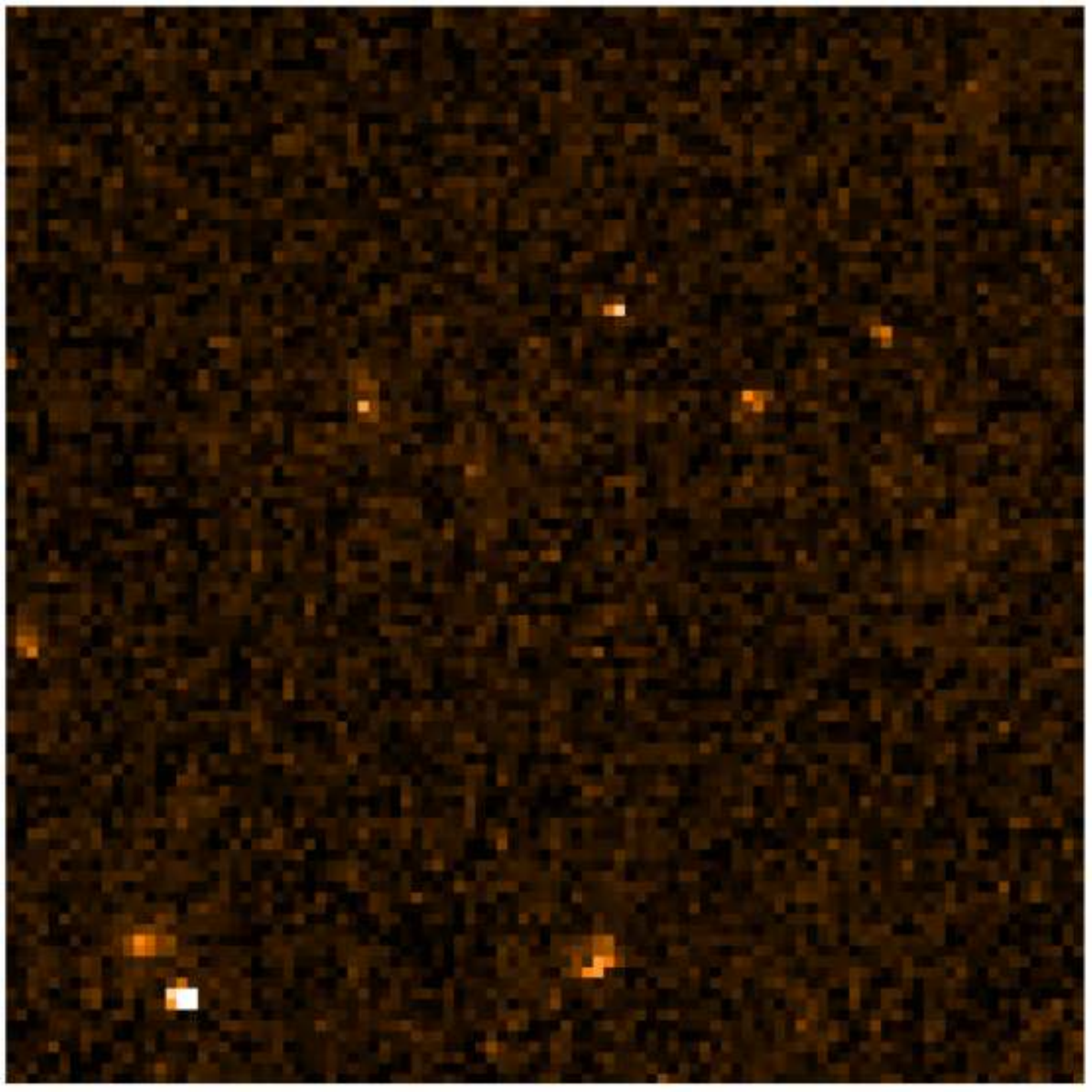}  &
\includegraphics[scale=0.15]{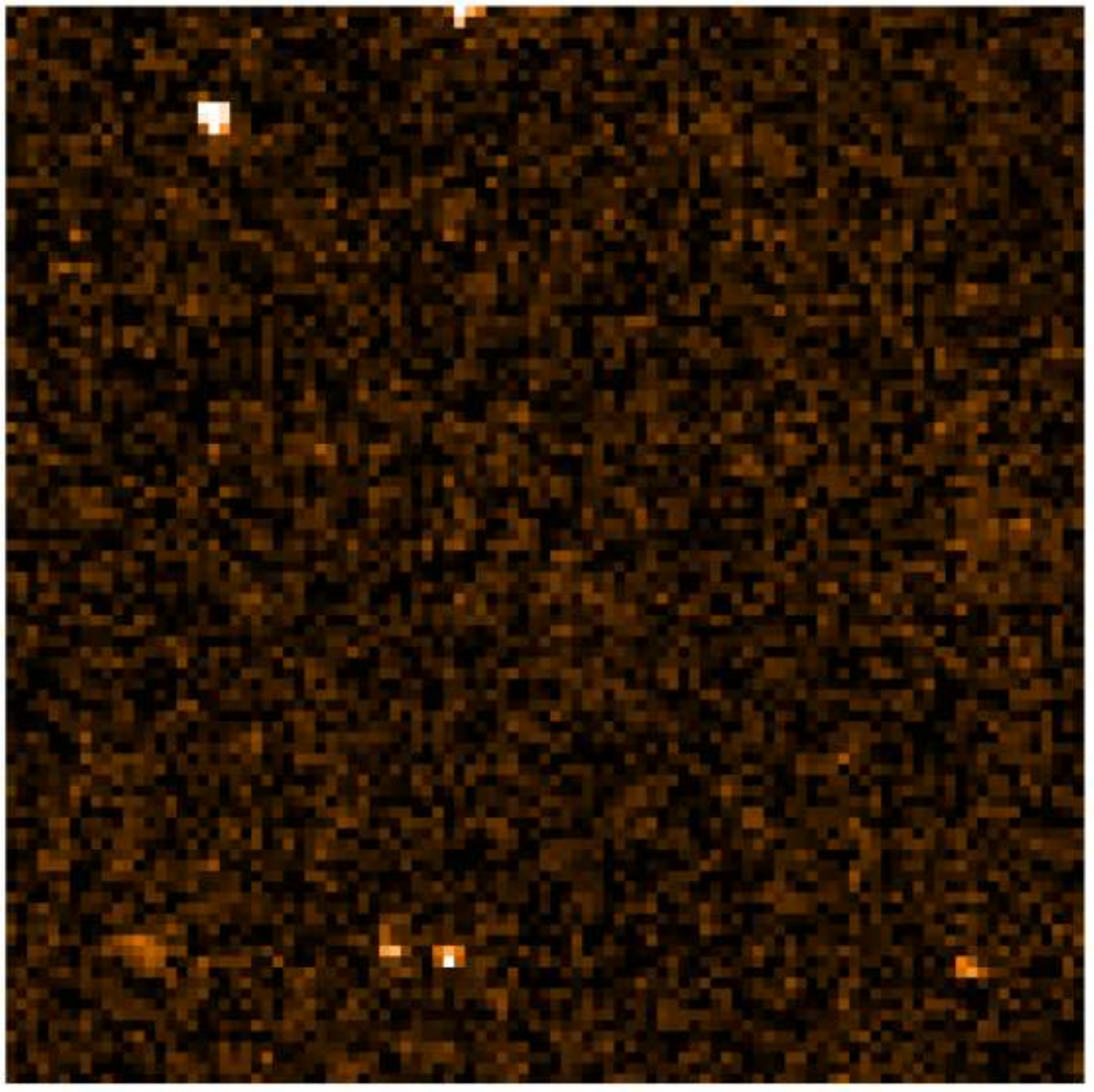}  \\
\end{tabular}
\caption[Image stamps of an example fit for a bulge+disk object with $B/T=0.5$.]{Image stamps of an example fit for a bulge+disk object with $B/T=0.5$, again following the placement of stamps described in the caption of Fig.\,\ref{fig:disk_less_50}. This object has a more equal contribution from the bulge and disk components. As with the previous examples, examination of the residual stamps for all 4 bands illustrates the good quality of the fits achieved by the adopted modelling technique, with no evidence of any additional structure in the bluer bands which has failed to be reproduced by the best-fitting models.}
\label{fig:bulge_disk_50}
\end{center}
\end{figure*}

Further example fits of representative objects across all four of the CANDELS bands are given in Figs.\,\ref{fig:pure_bulge},  \ref{fig:pure_disk}, \ref{fig:disk_less_50} and \ref{fig:bulge_disk_50}.

\clearpage
\clearpage

 \section{Double-Component SED Fitting}

 Further to the example given in Fig.\,\ref{fig:sedfit}, additional double-component SED fits for objects with $H_{160}$-band best fit morphologies with $B/T\sim0.5$ and $B/T>0.5$ are given here, to further illustrate how well the model photometry is fitted using the fixed $H_{160}$ parameter model approach with the extension to additional ground-based $\lambda<0.6\rm\mu$m and $\lambda>1.6\rm\mu$m data with re-measured 2.5\,arcsec aperture photometry in the blue bands. 
\begin{minipage}{18cm}
\begin{center}
\includegraphics[scale=0.7]{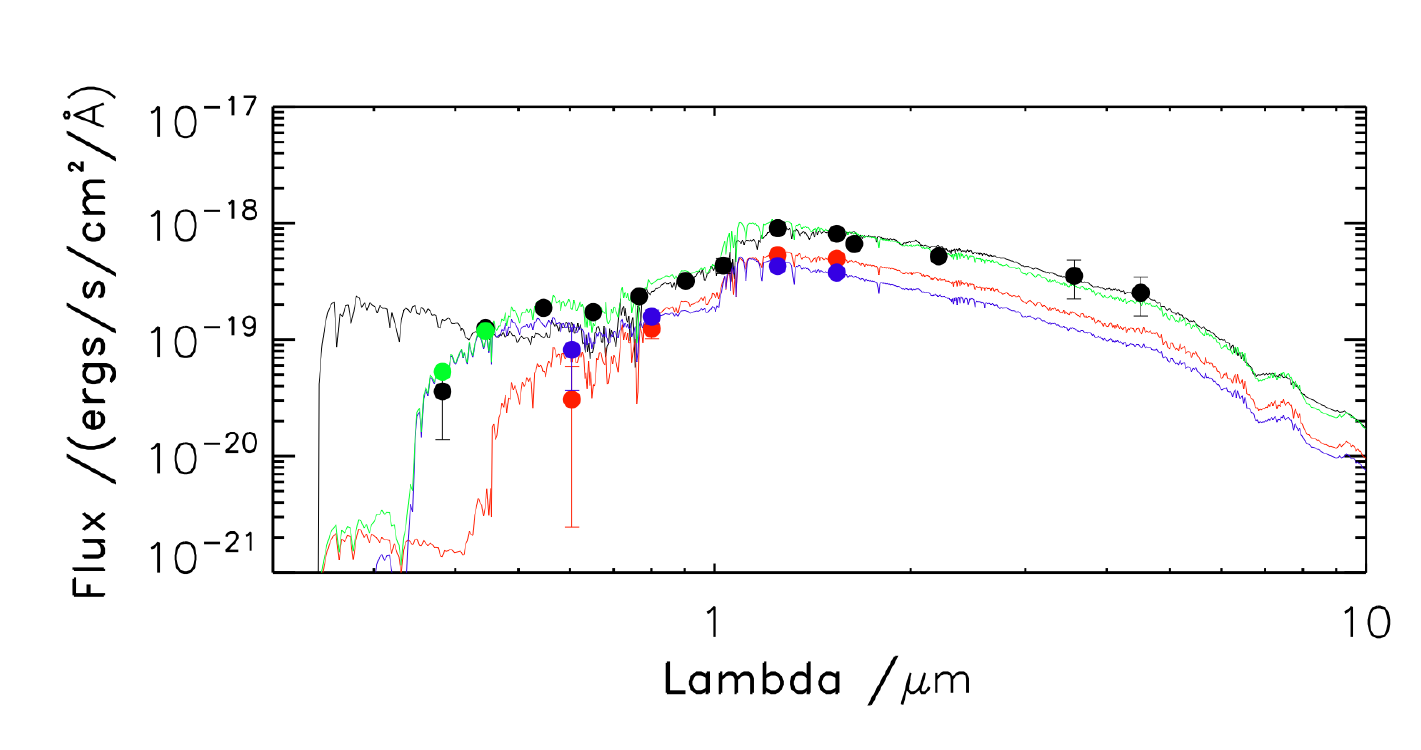}\\
\includegraphics[scale=0.7]{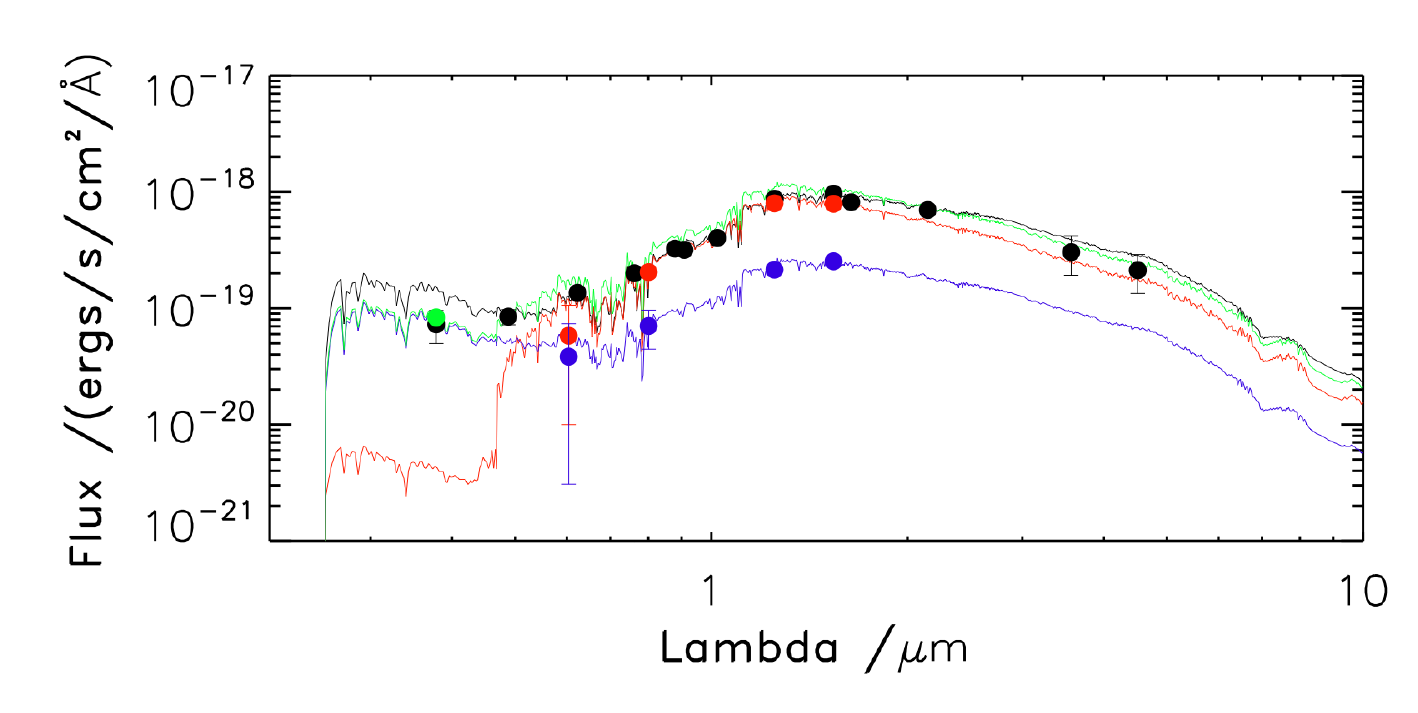}
\end{center}
\small{{\bf Figure B1.} SED fits for the example objects shown in Figs.\,\ref{fig:bulge_disk_50} and \ref{fig:disk_less_50}, respectively. Top panel for the $B/T\sim0.5$ fit, and bottom panel for the $B/T>0.5$ object. Plotted as black data-points and the solid black line is the total, overall galaxy photometry (with its associated error-bars) and the corresponding best-fit single-component SED. In blue is the modelled disk component photometry and the corresponding best-fit decomposed disk SED model, and in red is the modelled bulge photometry and the best-fit decomposed bulge SED model. Over-plotted in green is the sum of the best-fit bulge and disk SED models, which can be directly compared to the single-component fit in black and can be seen to be in better agreement with the overall galaxy photometry. Finally, the green points, and their error-bars, are the re-measured 2.5\,arcsec radius photometry for the blue bands. The top panel shows the fits to an object from the COSMOS field, where only u'-band photometry is available, whereas the bottom panel shows an object from the UDS field with re-measured photometry for both the $u'$ and $B$ bands.}
\end{minipage}

\clearpage
\clearpage
\section{Best-fit SED Templates}

The extra degrees of freedom introduced by the bulge+disk models also help to mitigate the need for limitations on the star-formation history templates and allowed ages of the models considered during the SED fitting. Having conducted several tests for different sub-sets of star-formation histories we find that there is no longer any need to restrict either component to a minimum age of $>50$Myrs or to exponentially star-formation histories in the range $0.3\leq \tau {\rm (Gyr)}\leq 5$, as a second younger burst, or exponentially decaying star-forming population can reasonably account for any continued or recent star-formation superimposed on an older, redder population.

\begin{figure}
\begin{center}
\includegraphics[scale=0.45]{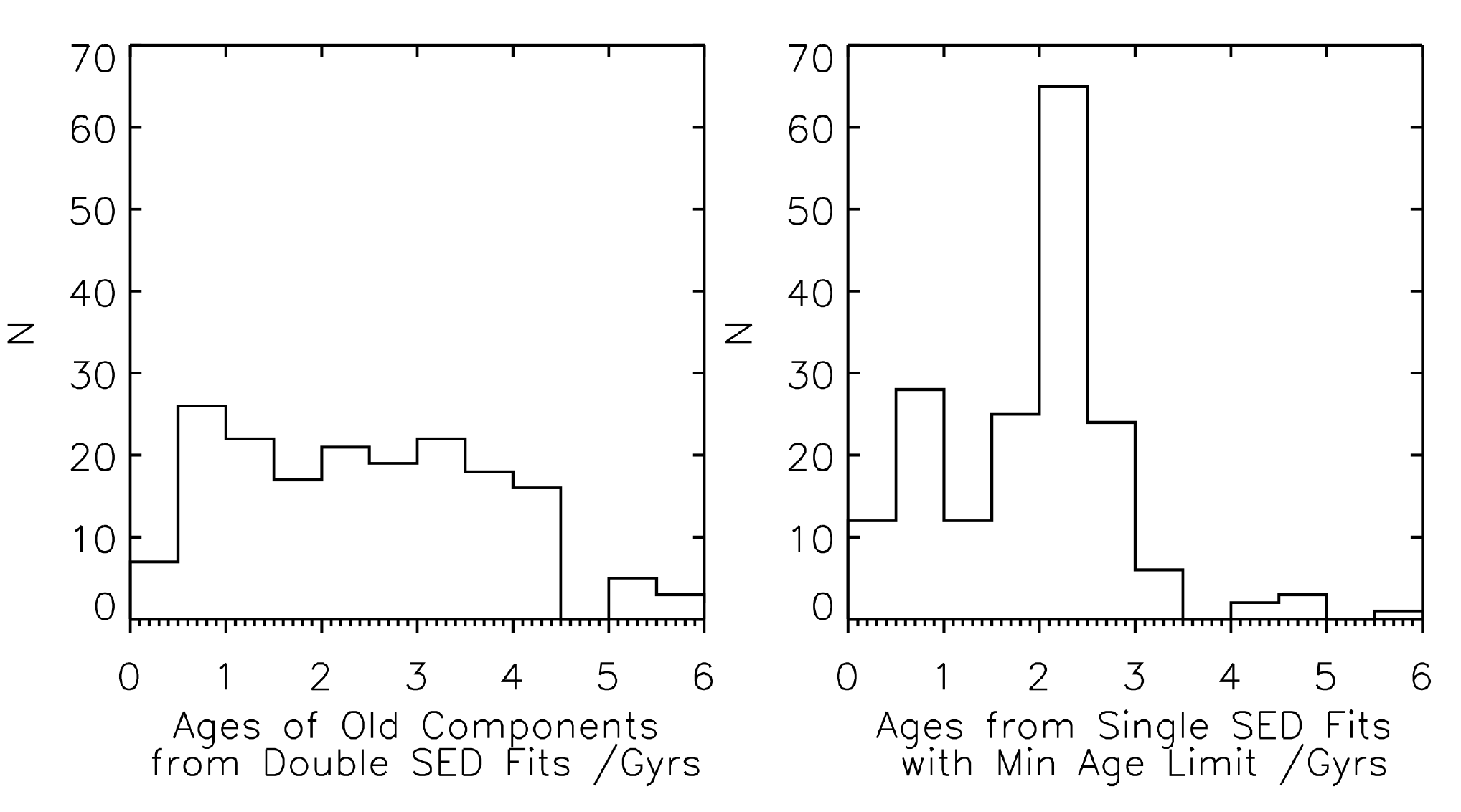} 
\end{center}
 \caption[Comparison between the age distributions of the single-component models and old population of the double-component models.]{Comparison between the age distributions of the single-component models and the older components of the double-component models. Left: the age distribution of the old component of double $0\leq \tau {\rm (Gyr)}\leq 5$ models. Right: distribution of the ages from single, limited $0.3\leq \tau {\rm (Gyr)}\leq 5$ and age-capped SED fits. The broad agreement between these distributions verifies that the additional degrees of freedom introduced by the double $0\leq \tau {\rm (Gyr)}\leq 5$ models naturally resolve the problem encountered with single SED fits sometimes becoming restricted to un-physically young ages.}
\label{fig:ageagree}
\end{figure}

In fact, in the 56 cases where the best-fit two-component models are double-burst models, 53 galaxies have at least one component with an age of $>50$ Myr and 49 galaxies have at least one component with an age of $>500$ Myr, which we use to provide a more robust measure of passivity. This trend also extends to the double-$\tau$ and burst$+\tau$ models, where none of the fits have both components with ages $<500$ Myr. Moreover, for the double-burst models there are only 21/56 objects where both components are older than $500$ Myr and the SED fits to the photometry find no evidence for ongoing star-formation. The agreement between the ages of the old component from the full $0\leq \tau {\rm (Gyr)}\leq 5$ double-component models and the ages of the single populations from the original single-component, limited $0.3\leq \tau {\rm (Gyr)}\leq 5$ and age capped, SED fitting is shown in Fig.\,\ref{fig:ageagree}.

\begin{figure}
\begin{center}
\includegraphics[scale=0.4]{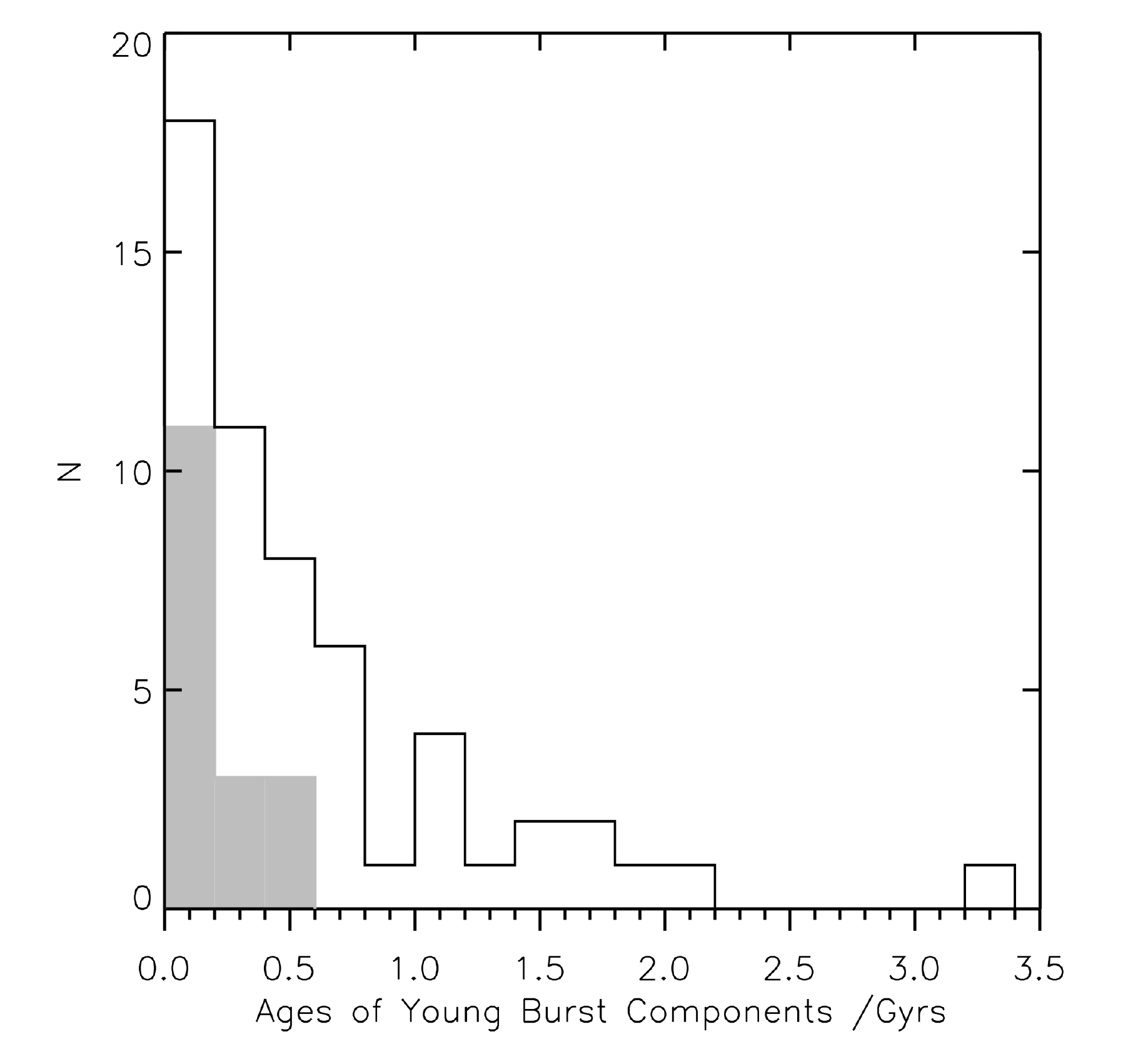} 
 \caption[Age distribution of the young components from the double-burst fits over-plotted with $24\mu$m.]{Age distribution of the young components of the double-burst fits, over-plotted by the shaded region with the objects which were found to have a $24\mu$m counterpart. The lack of objects with $24\mu$m counterparts and ages $>500$ Myr confirms that the $500$ Myr criterion for passivity is physically motivated, and that the double-burst models produce realistic ages and star-formation rates for the massive galaxies in our sample.}
\label{fig:age24}
 \end{center}
\end{figure}

As a secondary check, the ages of these double-burst fits for objects which have $24\mu$m counterparts from SpUDS and S-COSMOS have also been examined and are consistent with the fitting in that, for these double-burst models, there is always a young component which can account for star-formation and none of the galaxies with $24\mu$m counterparts have both components with ages  $>500$ Myr, as can be seen in Fig.\,\ref{fig:age24}.

As an additional check of this approach, it can also be seen from Fig.\,\ref{fig:age24} that our adoption of the $500$ Myr age boundary (above which we attribute none of the entire galaxy's star-formation to the component) is justified as there is no additional evidence from 24\,$\mu$m of on-going star formation in these components with older ages.  

\newpage

\label{lastpage}

\end{document}